\documentclass[12 pt]{article}
\usepackage{amsfonts}
\usepackage{amsmath}
\usepackage{amsthm}
\usepackage{braket}
\usepackage{graphicx}
\usepackage{hyperref}
\usepackage{color}
\usepackage{amssymb}
\usepackage{dsfont}
\usepackage{verbatim}
\usepackage{mathrsfs}
\usepackage{sidecap}
\usepackage[T1]{fontenc}
\usepackage[utf8]{inputenc}
\usepackage{authblk}
\usepackage[a4paper, total={6in, 9in}]{geometry}
\usepackage[nottoc,notlot,notlof]{tocbibind}
\usepackage{float}

\title{\textbf{Anyons and matrix product operator algebras}}
\author[a]{N.~Bultinck}
\author[a]{M.~Mari\"en}
\author[b]{D.~J.~Williamson}
\author[b]{M.~B.~Şahinoğlu}
\author[a]{J.~Haegeman}
\author[a,b]{F.~Verstraete}
\affil[a]{\small{\emph{Department of Physics and Astronomy, Ghent University}}}
\affil[b]{\small{\emph{Vienna Center for Quantum Technology, University of Vienna}}}

\begin{document}
  \maketitle

\begin{abstract}
Quantum tensor network states and more particularly projected entangled-pair states provide a natural framework for representing ground states of gapped, topologically ordered systems. The defining feature of these representations is that topological order is a consequence of the symmetry of the underlying tensors in terms of matrix product operators. In this paper, we present a systematic study of those matrix product operators, and show how this relates entanglement properties of projected entangled-pair states to the formalism of fusion tensor categories. From the matrix product operators we construct a $C^*$-algebra and find that topological sectors can be identified with the central idempotents of this algebra. This allows us to construct projected entangled-pair states containing an arbitrary number of anyons. Properties such as topological spin, the $S$ matrix, fusion and braiding relations can readily be extracted from the idempotents. As the matrix product operator symmetries are acting purely on the virtual level of the tensor network, the ensuing Wilson loops are not fattened when perturbing the system, and this opens up the possibility of simulating topological theories away from renormalization group fixed points. We illustrate the general formalism for the special cases of discrete gauge theories and string-net models.

\end{abstract}

\pagebreak
\tableofcontents

\section{Introduction} \label{sec:intro}
Since the conception of quantum mechanics, the quantum many-body problem has been of central importance. Due to a lack of exact methods one has to study these systems using approximate techniques such as mean field theory or perturbative expansions in a small parameter, or using an effective description obtained from symmetry or renormalization arguments. Another approach that has proven to be very fruitful is the use of toy models and trial wave functions. Tensor network states constitute a class of such trial wave functions that has emerged in past decades from the interplay of quantum information theory and condensed matter theory \cite{VerstraeteMurgCirac08}. The power of these states is two-sided. On the one hand, they can be used to study universal properties of quantum many-body systems, which makes them interesting objects from the theoretical perspective. On the other hand,  they allow for novel methods to simulate the many-body problem, which makes them interesting from the point of view of computational physics. For example, in one dimension Matrix Product States not only underpin the highly successful Density Matrix Renormalization Group algorithm \cite{DMRG,OstlundRommer}, but have also been used to completely classify all gapped phases of matter in quantum spin chains \cite{1Done,1Dtwo,SchuchGarciaCirac11}.

In this work we focus on two-dimensional tensor network states, so-called Projected Entangled-Pair States (PEPS) \cite{peps}. Because of their local structure these trial states serve as a window through which we can observe the entanglement properties of ground states of complex quantum many-body systems. We use this to study ground states of local two-dimensional Hamiltonians that have topological order, a kind of quantum order characterized by locally indistinguishable ground states and exotic excitations which can behave differently from bosons and fermions \cite{einarsson,Wen90}.

In recent years it became clear that topological order can be interpreted as a property of (a set of) states \cite{BravyiHastingsVerstraete}, the local Hamiltonian seems to be merely a tool to stabilize the relevant ground state subspace. It was realized that topological order manifests itself in entanglement properties such as the entanglement entropy \cite{KitaevPreskill,levinwenentanglement}. This has resulted in a definition of topological order via long-range entanglement \cite{LRE}. More recent works have shown that the ground state subspace on a torus contains information about the topological excitations \cite{MES} and that for chiral phases the so-called entanglement spectrum reveals the nature of the edge physics \cite{LiHaldane}. In Ref.~\cite{haah}, it was even shown that for a restricted class of Hamiltonians a single ground state contains sufficient information to obtain the $S$~matrix, an invariant for topological phases.

Utilizing the transparent entanglement structure of PEPS, we further examine this line of reasoning. We consider a class of PEPS with nonchiral topological order, which were introduced in \cite{Criticality,Ginjectivity,Buerschaper14,MPOpaper}. The intrinsic topological order in these states is characterized by a Matrix Product Operator (MPO) at the virtual level, which acts as a projector onto the virtual subspace on which the PEPS map is injective. This class of trial wave functions was shown to provide an exact description of certain renormalization group fixed-point models such as discrete gauge theories \cite{SPTpaper} and string-net models~\cite{stringnet1,stringnet2,MPOpaper}, but can also be perturbed away from the fixed point in order to study e.g. topological phase transitions \cite{transfermatrix,shadows}. We show that the entanglement structure of these `MPO-injective' PEPS enables a full characterization of the topological sectors of the corresponding quantum phase. In other words, the injectivity space of the tensors in a finite region of a single MPO-injective PEPS contains all information to fully determine the topological phase. More concretely, we show that the MPO that determines the entanglement structure of the PEPS allows one to construct a $C^*$-algebra whose central idempotents correspond to the topological sectors. A similar identification of topological sectors with central idempotents was made in \cite{Qalgebra,haah}. The advantage of the PEPS approach is that the idempotents can be used to explicitly write down generic wave functions containing anyonic excitations, which allows for a deeper understanding of how topological theories are realized in the ground states of local Hamiltonians. In addition, we obtain an intuitive picture of what happens in the wave function when these anyons are manipulated, and we can extract all topological information such as topological spins, the $S$ matrix, fusion properties and  braiding matrices. We would like to note that a very similar framework was recently discussed in the context of statistical mechanics \cite{Aasen}, where universal information about the CFT describing the critical point was obtained.

Section~\ref{sec:overview} starts with an overview of the paper. In Section~\ref{sec:MPO} we discuss general properties of projector MPOs and their connection to fusion categories. The construction of MPO-injective PEPS, as originally presented in \cite{MPOpaper}, is worked out in detail in Section~\ref{sec:MPOinjPEPS}. Section~\ref{sec:anyons} explains how to obtain the topological sectors and construct PEPS containing anyonic excitations. The corresponding anyon ansatz is illustrated for discrete gauge theories and string-nets in the examples of Section~\ref{sec:examples}. Section~\ref{sec:conclusions} contains a discussion of the results and possible directions for future work. The appendices contain several technical calculations and detailed results for some specific examples.

\section{Overview of the paper} \label{sec:overview}

In this section we convey the main ideas presented in this work, before obscuring them with technical details. We start by considering a large class of projector matrix product operators $P$ that can be written as $P = \sum_a w_a O_a$, where the $w_a$ are complex numbers and $O_a$ are injective (single block) matrix product operators with periodic boundary conditions:
\begin{align} 
O_a = \vcenter{\hbox{ 
\includegraphics[width=0.15\linewidth]{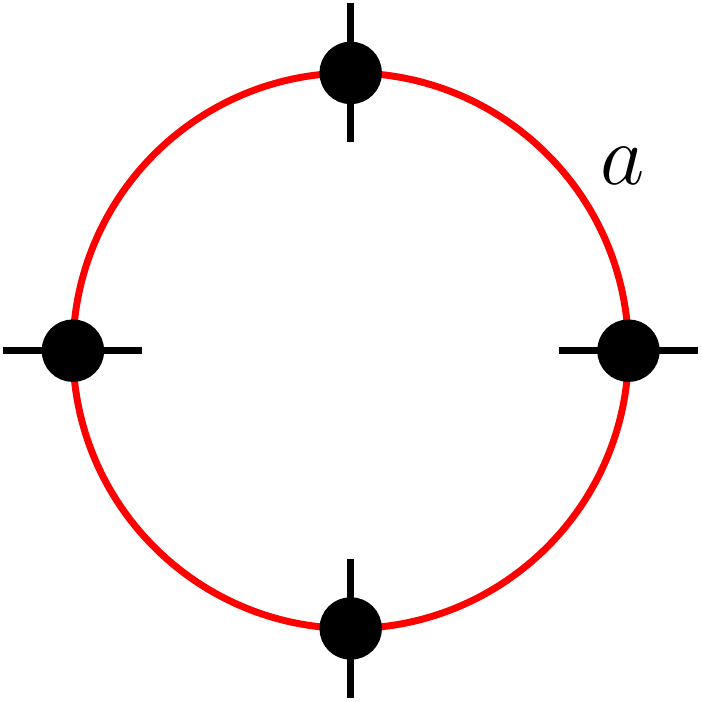}}}
\end{align}

Because we want $P$ to be a projector for every length it follows that $\{O_a\}$ forms the basis of a matrix algebra with positive integer structure coefficients: $O_aO_b = \sum_c N_{ab}^c O_c$, with $N_{ab}^c \in \mathbb{N}$. In section \ref{sec:MPO} we work out the details of this algebra and show that we can associate many concepts to it that are familiar from fusion categories.

In section \ref{sec:MPOinjPEPS} we turn to tensor network states on two-dimensional lattices, called Projected Entangled-Pair States (PEPS). We discuss how the MPO tensors can be used to construct a (family of) PEPS that satisfies the axioms of MPO-injectivity, i.e. the algebra $\{O_a\}$ constitutes the virtual `symmetry' algebra of the local PEPS tensors and the virtual support of the PEPS on any topologically trivial region corresponds to the subspace determined by the MPO projector $P$ along the boundary of that region. As shown in \cite{MPOpaper}, the axioms of MPO-injectivity allow us to prove that such PEPS are unique ground states of their corresponding parent Hamiltonians on topologically trivial manifolds. We can also explicitly characterize the degenerate set of ground states on topologically non-trivial manifolds. The most important axiom of MPO-injectivity is the `pulling through' property. To prove that it is satisfied by our construction, we need to impose that the MPO tensors satisfy the so-called \emph{zipper} condition, i.e. there must exist a three-leg `fusion' tensor X, which we depict as a grey box, such that the following identity holds:
\begin{align}\label{overviewzipper}
\vcenter{\hbox{
\includegraphics[width=0.35\linewidth]{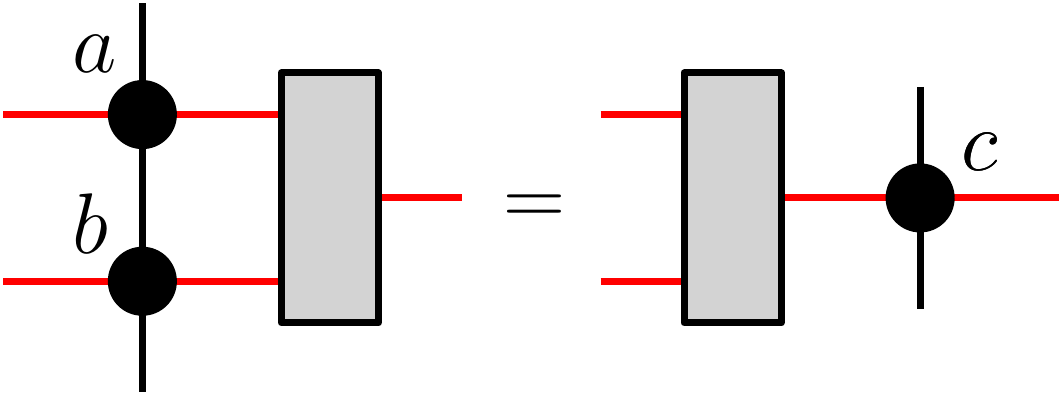}}}\,\,\, .
\end{align}

Remarkably, the same properties of the MPOs $O_a$ that guarantee the pulling through property to hold also allow us to construct a \emph{second type} of MPO algebra. The basis of this second algebra is bigger than that of the first one and its elements can be presented schematically as:
\begin{align}\label{algebra2}
\vcenter{\hbox{
\includegraphics[width=0.15\linewidth]{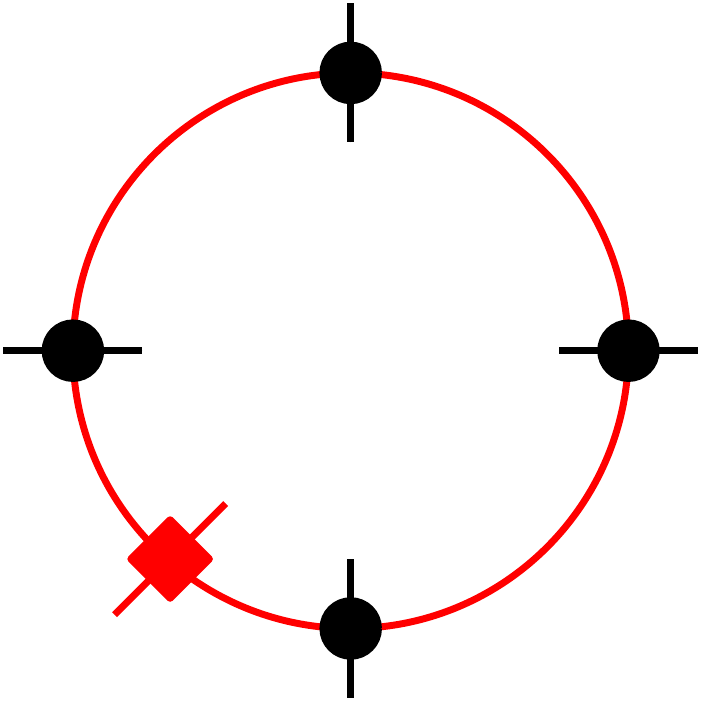}}}\,\,\, ,
\end{align}
where the red square is a new type of tensor, defined in the main text, that is completely determined by the MPOs $O_a$. We show that this second algebra is actually a $C^*$-algebra, hence it is isomorphic to a direct sum of full matrix algebras. We use this decomposition to identify the topological sectors with the different blocks, or equivalently, with the central idempotents that project onto these blocks. A large part of the paper is then devoted to show that, once one has identified these central idempotents (for which we give a constructive algorithm), one can construct MPO-injective PEPS containing anyonic excitations and study their topological properties. For example, the topological spin $h_i$ of an anyon $i$ can be obtained via the identity:
\begin{align}\label{overviewspin}
\vcenter{\hbox{
\includegraphics[width=0.18\linewidth]{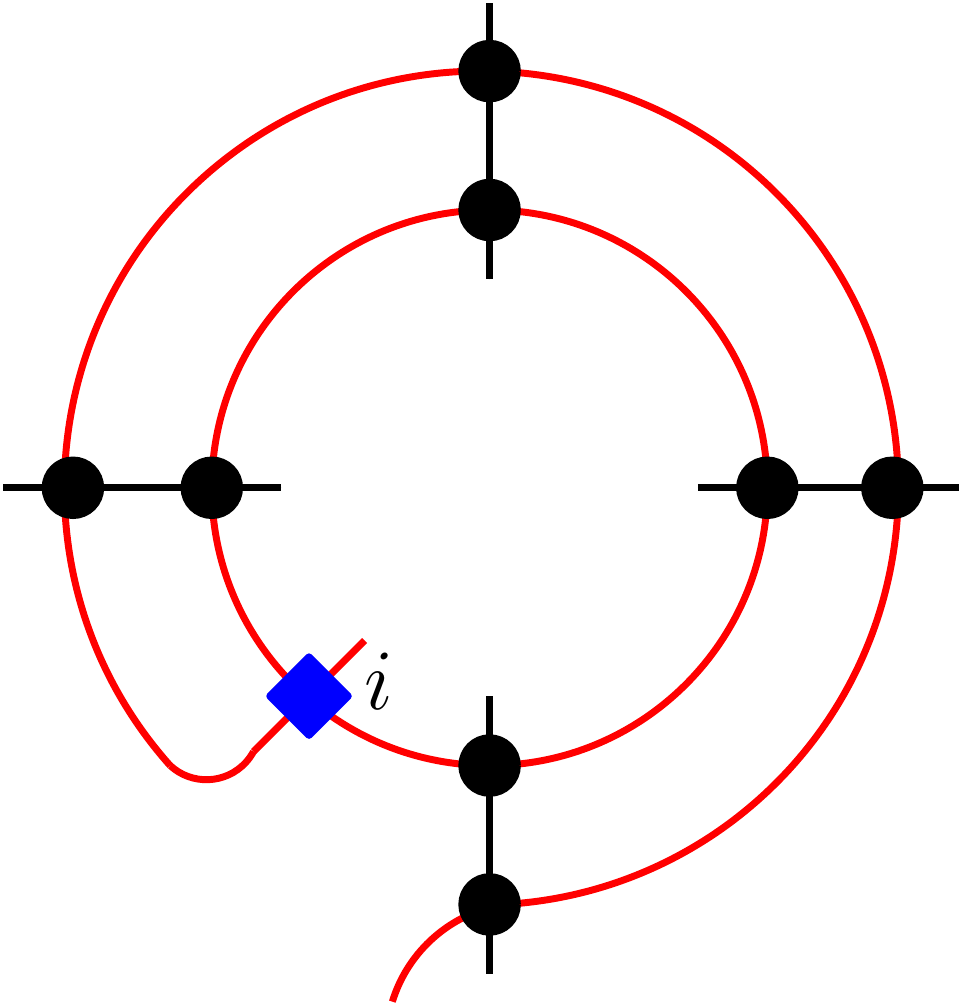}}} = e^{i2\pi h_i}
\vcenter{\hbox{
\includegraphics[width=0.13\linewidth]{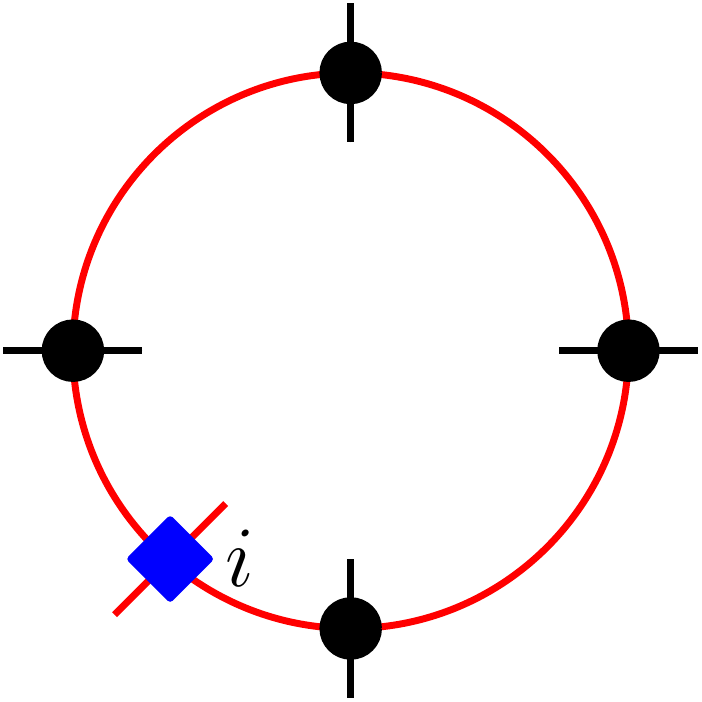}}}\,\,\, ,
\end{align}
where we used a blue square to denote a central idempotent (here the one corresponding to anyon $i$) as opposed to a red square, which denotes a basis element of the second algebra. In a similar manner one can extract the $S$-matrix, fusion relations and braiding matrices in a way that does not scale with the system size. 

Let us now illustrate this general scheme for the simplest example, namely Kitaev's Toric Code \cite{toriccode}. Note that the excitations in the Toric code are already completely understood in the framework of G-injective PEPS \cite{Ginjectivity}, which is a specific subset of the MPO-injective PEPS formalism with building blocks $O_a$ that are tensor products of local operators, i.e. the MPOs have virtual bond dimension 1. However, as a pedagogical example we would like to study the anyons in the general language introduced above. In the standard PEPS construction of the Toric code \cite{Criticality} the virtual indices of the tensors are of dimension two and have a $\mathbb{Z}_2$ symmetry, i.e. they are invariant under $\sigma_z^{\otimes 4}$. So in this case the symmetry algebra is really a group and is given by the two MPOs $O_1 = \mathds{1}^{\otimes 4}$ and $O_z = \sigma_z^{\otimes 4}$. Let us now introduce following tensor where all indices have dimension two:
\begin{align}
\vcenter{\hbox{
\includegraphics[width=0.35\linewidth]{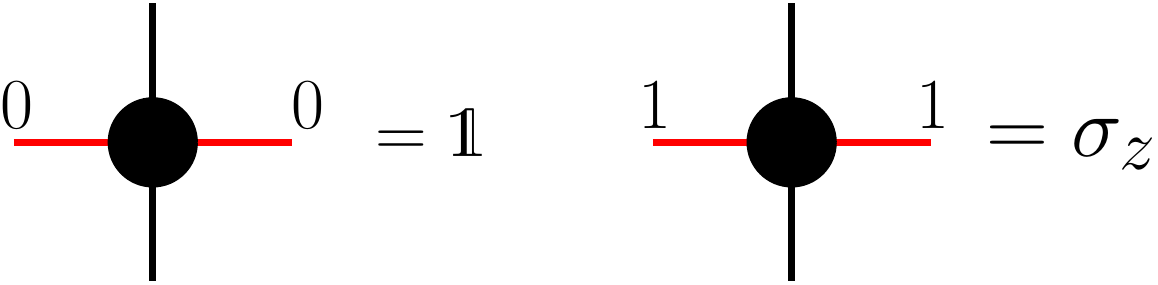}}}\, .
\end{align}
All components that are not diagonal in the red indices are zero. One can clearly construct the Toric code symmetry MPOs $O_1$ and $O_z$ using these tensors. By defining a fusion tensor where all indices have dimension two and with the following non-zero components:
\begin{align}
\vcenter{\hbox{
\includegraphics[width=0.5\linewidth]{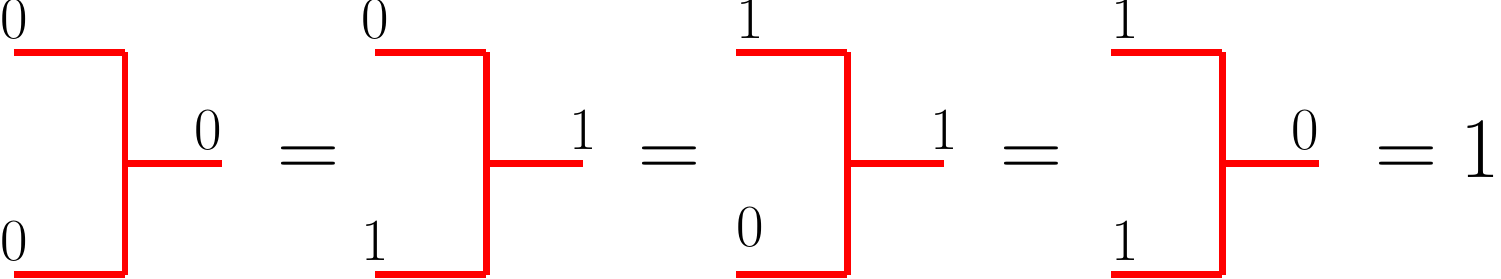}}}\, ,
\end{align}
one can verify that the zipper condition \eqref{overviewzipper} is trivially satisfied.

The second type of algebra is four-dimensional. The basis elements \eqref{algebra2} can be obtained by using one of following tensors:
\begin{align}
\vcenter{\hbox{
\includegraphics[width=0.55\linewidth]{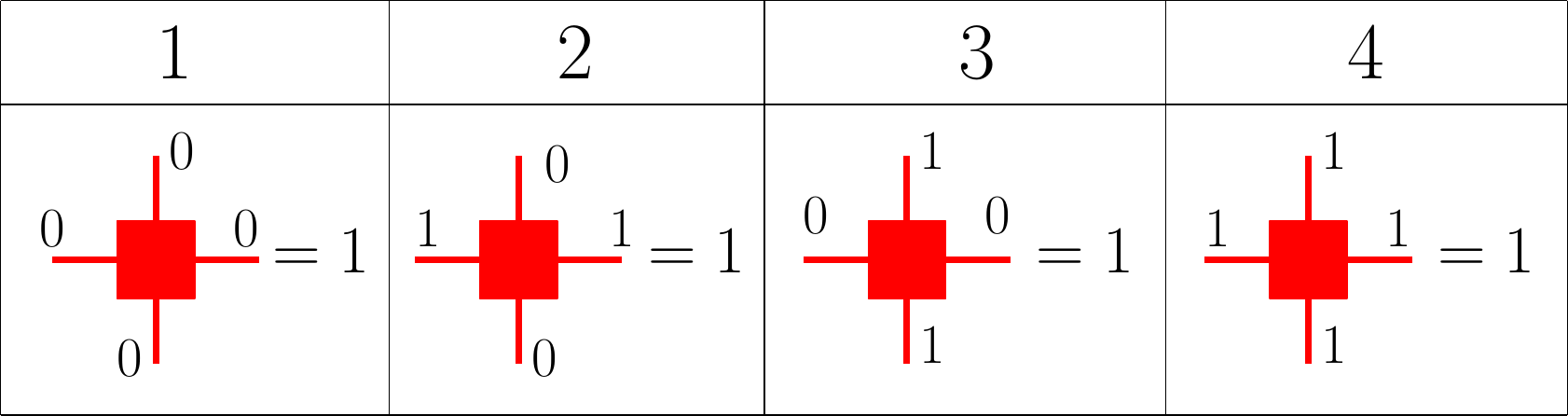}}}\, 
\end{align}
Each of these tensors has only one non-zero component, which is given in the table.

The central idempotents of this second algebra, labeled by the usual notation $\{1,e,m,em\}$, are now easily obtained by using following tensors:
\begin{align}
\vcenter{\hbox{
\includegraphics[width=0.35\linewidth]{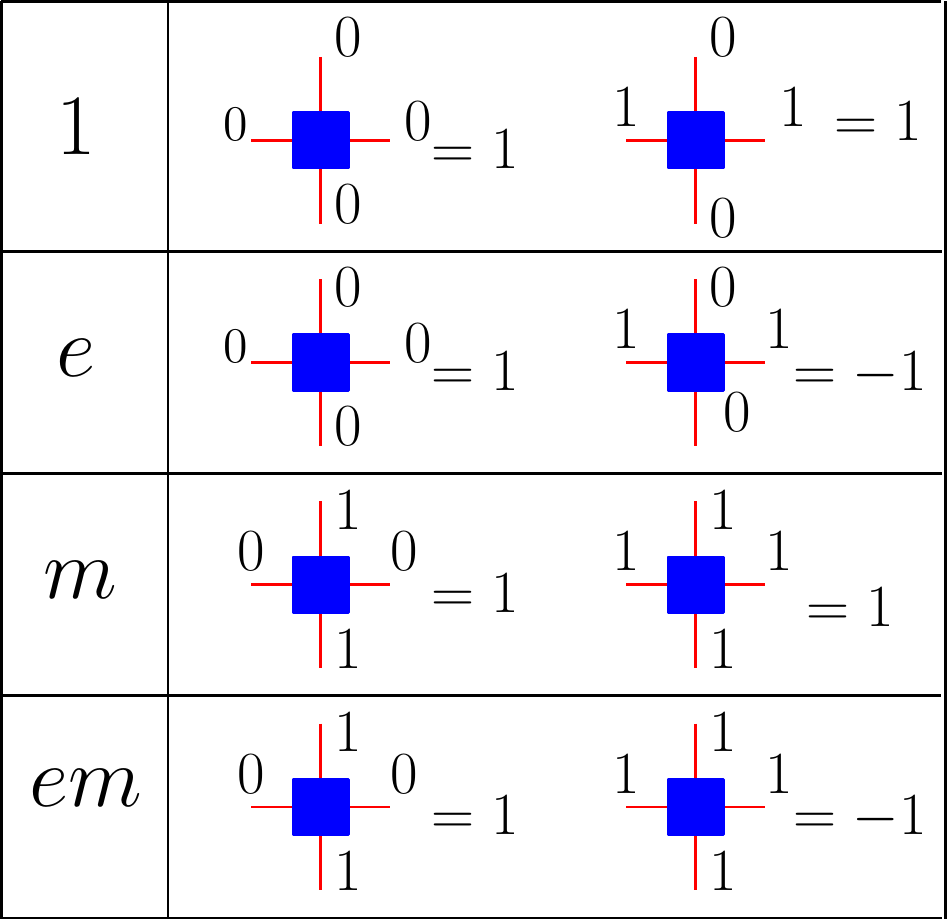}}}\, 
\end{align}
Again, we only denote non-zero elements. The vertical indices in the tensors above with value one indicate that a string of $\sigma_z$ is connected to these idempotents. This agrees with the $G$-injectivity construction of $m$ and $em$ anyons. From \eqref{overviewspin} one can now immediately see that the topological spins of these idempotents are $h_1 = h_e = h_m = 0$ and $h_{em} = 1/2$.

While our treatment of the Toric Code might seem overloaded, we will show in the remainder of the paper that it is in fact the correct language to describe anyons in general topological PEPS. We hope that this section can give some intuition and motivation to understand the more technical parts. 

\section{Projector Matrix Product Operators} \label{sec:MPO}
\vspace{3 mm}
\subsection{Definition}
\vspace{5 mm}
We start the general theory with a discussion of Projector Matrix Product Operators (PMPO), the fundamental objects of MPO-injectivity, and their connection to known concepts of category theory. We consider PMPOs $P_L$ that form translation invariant Hermitian projectors for every length $L$ and can be represented as
\begin{equation} \label{mpo}
P_L =\sum_{\{i\}, \{j\} = 1}^{D}\text{tr}(\Delta B^{i_1j_1}B^{i_2j_2}\dots B^{i_Lj_L})\ket{i_1 i_2\dots i_L}\bra{j_1 j_2 \dots j_L}\, ,
\end{equation}
where $B^{ij}$ are $\chi \times \chi$ matrices for fixed values of indices $i,j=1,\ldots,D$. We use this MPO to construct a PEPS in the next section, and $D$ will then become the bond dimension of the resulting PEPS. Furthermore, $\Delta$ is a $\chi \times \chi$ matrix such that the specific position where it is inserted is irrelevant; every position of $\Delta$ will result in the same PMPO $P_L$. We also assume that the insertion of $\Delta$ still allows for a canonical form of the MPO such that the tensors have the following block diagonal structure \cite{MPSrepresentations}
\begin{align}
B^{ij} = \bigoplus_{a = 1}^{\mathcal{N}} B_a^{ij}\\
\Delta = \bigoplus_{a = 1}^{\mathcal{N}} \Delta_a\, ,
\end{align}
with $B_a^{ij}$ and $\Delta_a$ $\chi_a \times \chi_a$ matrices such that $\sum_{a = 1}^{\mathcal{N}} \chi_a = \chi$. $P_L$ thus decomposes into a sum of MPOs
\begin{equation}
P_L =\sum_{a = 1}^{\mathcal{N}}\sum_{\{i\}, \{j\} = 1}^{D}\text{tr}(\Delta_a B_a^{i_1j_1}B_a^{i_2j_2}\dots B_a^{i_Lj_L})  \ket{i_1 i_2\dots i_L}\bra{j_1 j_2 \dots j_L}
\end{equation}
The resulting MPOs labelled by $a$ in this sum are injective, hence for each $a$ the matrices $\{B_a^{ij}; i,j=1,\ldots,D\}$ and their products span the entire space of $\chi_a \times \chi_a$ matrices.
Equivalently, the corresponding transfer matrices $\mathbb{E}_{a}=\sum_{i,j} B_{a}^{i,j}\otimes\bar{B}_{a}^{i,j}$ have a unique eigenvalue $\lambda_a$ of largest magnitude that is positive and a corresponding (right) eigenvector equivalent to a full rank positive definite matrix $\rho_a$. The PMPO $P_L$ can now only be translation invariant if the $\Delta_a$ commute with all the matrices $B_a^{ij}$. Injectivity of the tensors $B_a$ then implies that $\Delta_a=w_a \mathds{1}_{\chi_a}$, with $w_a$ some complex numbers.

\subsection{Fusion tensors} \label{subsec:fusiontensors}
\begin{figure}
  \centering
    (a) \includegraphics[width=0.1\textwidth]{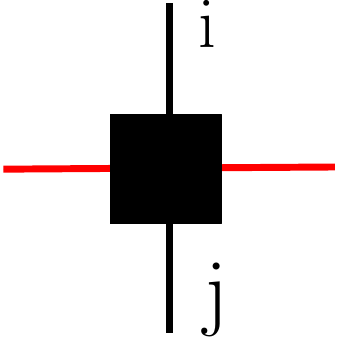}\;\; (b)
    \includegraphics[width=0.1\textwidth]{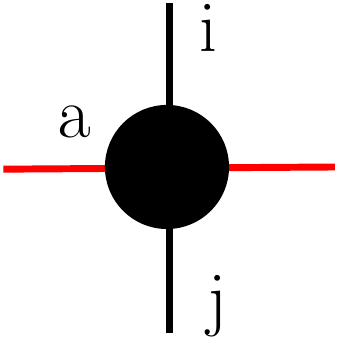}
    (c) \includegraphics[width=0.32\textwidth]{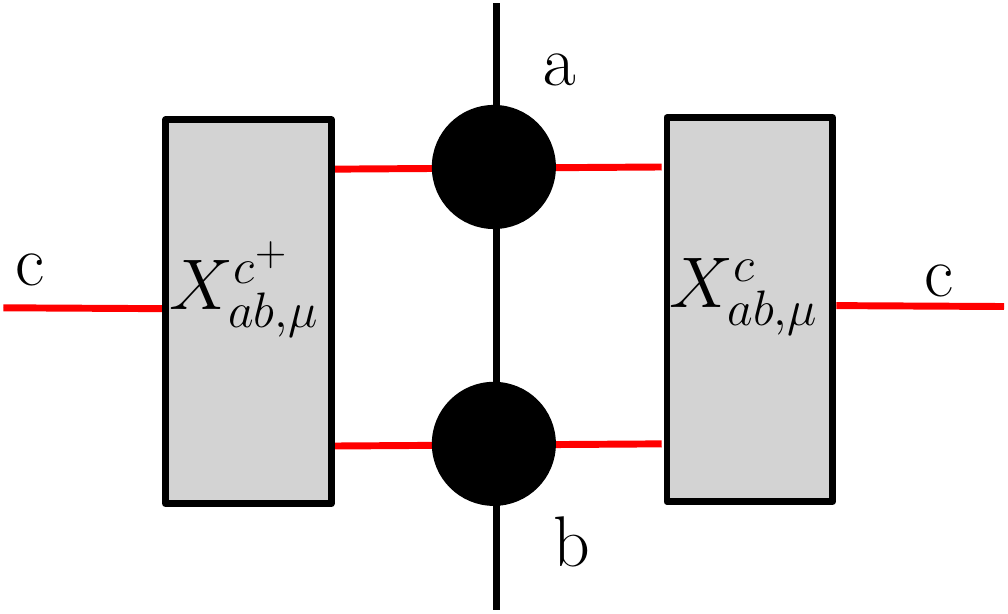}
\caption{(a) MPO tensor $B^{ij}$. (b) Injective MPO tensor $B^{ij}_a$. (c) Left hand side of equation \eqref{blocks}.}
\label{graphical}
\end{figure}

We thus arrive at the following form for $P_L$
\begin{equation}
\begin{split}
P_L & =  \sum_{a = 1}^{\mathcal{N}}w_a O^L_a \\
 & =   \sum_{a = 1}^{\mathcal{N}}w_a\sum_{\{i\}, \{j\} = 1}^{D} \text{tr}(B_a^{i_1j_1}B_a^{i_2j_2}\dots B_a^{i_Lj_L}) 
\ket{i_1 i_2\dots i_L}\bra{j_1 j_2  \dots j_L}
\end{split}
\end{equation}
Since $P_L$ is required to be a projector, we have that
\begin{equation}
P_L^2 = \sum_{a, b = 1}^{\mathcal{N}} w_a w_b O^L_a O^L_b =  \sum_{a = 1}^{\mathcal{N}}w_a O^L_a = P_L\, ,
\end{equation}
which has to hold for all $L$. One can show that this implies that $P_L$ and $P_L^2$ have the same blocks in their respective canonical forms\footnote{This essentially follows from the fact that $\lim_{L \rightarrow \infty}\text{tr}(O^L_aO_b^{L\dagger})/\left(\sqrt{\text{tr}(O^L_aO_a^{L\dagger})}\sqrt{\text{tr}(O^L_bO_b^{L\dagger})}\right) = \delta_{a,b}$ for two injective MPOs $O_a^L$ and $O^L_b$.} \cite{PerezGarciaRG}, leading to the following relations
\begin{align} \label{fusioncategory1}
& O^L_a O^L_b = \sum_{c = 1}^{\mathcal{N}} N_{ab}^c O^L_c\ ,\\
 & \sum_{a,b = 1}^{\mathcal{N}} N_{ab}^c w_a w_b = w_c\ ,
\end{align}
where $N_{ab}^c$ is a rank three tensor containing integer entrees. The theory of MPS representations \cite{MPSrepresentations} implies the existence of matrices $X_{ab,\mu}^c: \mathbb{C}^{\chi_a}\otimes  \mathbb{C}^{\chi_b} \rightarrow  \mathbb{C}^{\chi_c}$ for $\mu=1,\ldots,N_{ab}^{c}$ and left inverses $X_{ab,\mu}^{c^+}$ satisfying $X_{ab,\nu}^{d^+} X_{ab,\mu}^c = \delta_{de}\delta_{\mu\nu}\mathds{1}_{\chi_c}$, such that we have following identities on the level of the individual matrices that build up the injective MPOs $O^L_a$,
\begin{equation} \label{blocks}
X_{ab,\mu}^{c^+}\left(\sum_{j = 1}^D B_a^{ij}\otimes B_b^{jk}\right)X^{c}_{ab,\mu} =  B_c^{ik}.
\end{equation}
We call the set of rank three tensors $X^c_{ab,\mu}$ the \emph{fusion tensors}. These fusion tensors play an important role in constructing the anyon ansatz further on. From
\begin{equation}\label{gauge}
\bigoplus_{\mu = 1}^{N_{ab}^c}X_{ab,\mu}^{c^+}\left(\sum_{j = 1}^D B_a^{ij}\otimes B_b^{jk}\right)X^{c}_{ab,\mu} = \mathds{1}_{N_{ab}^c} \otimes B_c^{ik}\, ,
\end{equation}
we see that the $\mu$-label is arbitrary and the fusion tensors $X_{ab,\mu}^c$ are only defined up to a gauge transformation given by a set of invertible $N_{ab}^c\times N_{ab}^c$ matrices $Y_{ab}^c$; every transformed set of fusion tensors $X'^c_{ab,\mu} = \sum_{\nu = 1}^{N_{ab}^c} (Y^c_{ab})_{\mu\nu}X^c_{ab,\nu}$ also satisfies \eqref{gauge}. The MPO tensors and equation~\eqref{blocks} are represented in figure~\ref{graphical} in a graphical language that is used extensively throughout this paper. Note in particular the difference between the square for the full MPO tensor $B^{ij}$ with virtual indices (red lines) of dimension $\chi=\sum_{a} \chi_{a}$ and the disc for the injective MPO tensors $B^{ij}_a$ with virtual indices (red line with symbol $a$) of dimension $\chi_a$. 

Two complications are worth mentioning. First, the canonical form of $O^L_a O^L_b$ can contain diagonal block matrices which are identically zero. Therefore, the fusion matrices $X_{ab,\mu}^{c}$ do not span the full space and $\sum_{c,\mu} \chi_{c}$ can be smaller than $\chi_{a}\times\chi_{b}$. Correspondingly, $\sum_{c,\mu} X^c_{ab,\mu}X^{c^+}_{ab,\mu}$ is not necessarily the identity but only a projector on the support subspace of the internal MPO indices of $O^L_aO^L_b$.

Secondly, there can be nonzero blocks above the diagonal, i.e.\ $X_{ab,\mu}^{c^+}\left(\sum_{j = 1}^D B_a^{ij}\otimes B_b^{jk}\right)\\ X^{d}_{ab,\nu}  \neq 0$ for some $(c,\mu) < (d,\nu)$ (according to some ordering). These blocks do not contribute when the MPO is closed by the trace operation, but prevent us from writing
\begin{equation}\label{inversegaugeone}
\sum_{c = 1}^{\mathcal{N}}\sum_{\mu=1}^{N_{ab}^c} X_{ab,\mu}^c B_c^{ik} X_{ab,\mu}^{c+} = \sum_{j=1}^D B^{ij}_a\otimes B^{jk}_b\, .
\end{equation}
We noticed above that a set of fusion tensors is only defined up to a gauge transformation $Y$. For PMPOs without nonzero blocks above the diagonal we now argue that the converse is also true, i.e. two  collections of fusion tensors that satisfy equations \eqref{gauge} and \eqref{inversegaugeone} must be related by a gauge transformation $Y$. To see this, note that the absence of nonzero blocks above the diagonal is equivalent to the existence of an invertible matrix $X_{ab}$ such that
\begin{equation}
X_{ab}^{-1}\left(\sum_{j = 1}^D B_a^{ij}\otimes B_b^{jk}\right)X_{ab}=  \bigoplus_{c}\left(\mathds{1}_{N_{ab}^c} \otimes B_c^{ik}\right)\, .
\end{equation}
The fusion tensors that have the required properties are then simply the product of $X_{ab}$ and the projector on the appropriate block: $X^c_{ab,\mu} = X_{ab} P^c_{\mu}$. It is clear that these fusion tensors are unique up to a matrix in the commutant of $\bigoplus_{c}\left(\mathds{1}_{N_{ab}^c} \otimes B_c^{ik}\right)$. Since the $B^{ik}_c$ are injective their commutant consists of multiples of the identity matrix. From this we can indeed conclude that the only ambiguity in the definition of $X_{ab,\mu}^c$ is given by the gauge transformation $Y$.

Note that equation \eqref{inversegaugeone} is equivalent to
\begin{equation}\label{zippercondition2}
\left(\sum_{j=1}^D B^{ij}_a\otimes B^{jk}_b\right) X_{ab,\mu}^{c} = X_{ab,\mu}^{c} B^{ik}_{c}\, ,
\end{equation}
\begin{equation*}
X_{ab,\mu}^{c^+} \left(\sum_{j=1}^D B^{ij}_a\otimes B^{jk}_b\right)  =B^{ik}_{c} X_{ab,\mu}^{c^+} \, .
\end{equation*}
We refer to these last two equations as the \emph{zipper condition}. While we continue somewhat longer in the general setting, we will need to assume that the zipper condition holds for most of the results in the remainder of the paper.

\subsection{Hermiticity, duality and unital structure} \label{subsec:hermiticity}
If we also require $P_L$ to be Hermitian for all $L$, then we find that for every block $a$ there exists a unique block $a^*$ such that
\begin{eqnarray}
\bar{w}_a & = & w_{a^*} \label{realnumber} \\
O^{L\dagger}_a&  = & O^L_{a^*}\, ,
\end{eqnarray}
where the bar denotes complex conjugation. The tensor $N$ then obviously satisfies
\begin{equation} \label{pivotal3}
N_{ab}^c = N_{b^*a^*}^{c^*}\, .
\end{equation}
Note that in general the tensors $\bar{B}^{ji}_a$ and $B^{ij}_{a^*}$, which build up $O^{L\dagger}_a$ and $O^L_{a^*}$, are related by a gauge transformation: $\bar{B}^{ji}_a = Z^{-1}_aB^{ij}_{a^*}Z_a$ where $Z_{a}$ is defined up to a multiplicative factor. By applying Hermitian conjugation twice we find
\begin{eqnarray}
B^{ij}_{a} & = & \bar{Z}_a^{-1} \bar{B}^{ji}_{a^*} \bar{Z}_a \\
 & = & \bar{Z}_a^{-1} Z^{-1}_{a^*} B^{ij}_{a} Z_{a^*} \bar{Z}_a \; .
\end{eqnarray}
Combining the above expression with the injectivity of $B_a^{ij}$ we find $Z_{a}\bar{Z}_{a^\ast} = \gamma_a\mathds{1} = \bar{Z}_{a^*} Z_{a}$, with $\gamma_a=\bar{\gamma}_{a^*}$ a complex number. If $a\neq a^\ast$, we can redefine one of the two $Z$ matrices with an additional factor such that $\gamma_a=1$. If, on the other hand, $a=a^\ast$ we find that $\gamma_a$ must be real but we can at most absorb its absolute value in $Z_a \bar{Z}_a$ by redefining $Z_a$ with an extra factor $\lvert\gamma_a\rvert^{-1/2}$. The sign $\varkappa_a=\text{sign} (\gamma_a)$ cannot be changed by redefining $Z_a$. It is a discrete invariant of the PMPO which is analogous to the Frobenius-Schur indicator in category theory.

To recapitulate, Hermitian conjugation associates to every block $a$ a unique `dual' block $a^*$ in such a way that $(a^*)^ * = a$. In fusion category theory there is also a notion of duality, but it is defined in a different way. There, for every simple object $a$ the unique dual simple object $a^*$ is such that the tensor product of $a$ and $a^ *$ contains the identity object 1. The identity object is defined as the unique simple object that leaves all other objects invariant under taking the tensor product. Moreover, 1 appears only once in the decomposition of the tensor product of $a$ and $a^*$. We now show that if a PMPO contains a trivial identity block then our definition of duality inferred from Hermitian conjugation coincides with the categorical definition. To do so, let us first revisit the transfer matrices
\begin{align*}
\mathbb{E}_{a} &= \sum_{i,j} B^{ij}_a \otimes \bar{B}^{ij}_a \\
&= (\mathds{1}\otimes Z_{a}^{-1}) \sum_{i,j} B^{ij}_a \otimes B^{ji}_{a^*} (\mathds{1}\otimes Z_{a}).
\end{align*}
We can thus use the tensors $(\mathds{1}\otimes Z_{a}^{-1}) X_{aa^*;\mu}^{c}$ (and their left inverses) to bring $\mathbb{E}_{a}$ into a block form with nonzero blocks on and above the diagonal (upper block triangular). In particular, there are $N_{aa^*}^{c}$ diagonal blocks of size $\chi_c\times \chi_c$ that are given by $M_{c}=\sum_{i} B^{ii}_{c}$. They can be brought into upper triangular form by a Schur decomposition within the $\chi_c$-dimensional space, such that we can identify the eigenvalue spectrum of $\mathbb{E}_{a}$ with that of the different matrices $M_c$ for $c$ appearing in the fusion product of $a$ and $a^*$. Since $\mathbb{E}_{a}$ has a unique eigenvalue of largest magnitude $\lambda_a$, it must correspond to the unique largest eigenvalue of $M_{c_{a}}$ for one particular block $c_{a}$, for which also $N_{aa^*}^{c_a}=1$.

We now assume that there is a unique distinguished label $c$, which we choose to be $c=1$, such that the spectral radius of $M_{1}$ is larger than the spectral radius of all other $M_{c}$ for $c=2,\ldots,N$ (whose labeling is still arbitrary). We furthermore assume that $N_{aa^*}^{1}\neq 0$ for all $a$, i.e.\ $O_1^L$ appears in the product $O_a^L O_{a^*}^L$ for any $a$. This condition, as we now show, corresponds to imposing a unital structure and excludes cases where e.g.\ $P_L$ is actually a sum of independent orthogonal projectors, corresponding to a partition $A,B,...$ of the injective blocks that is completely decoupled (such that $N_{ab}^{c}=0$ for any $c$ if $a\in A$ and $b\in B$).

With this condition, we find that independent of $a$, $c_a = 1$ and all transfer matrices $\mathbb{E}_{a}$ have $\lambda_a=\lambda$ as unique largest eigenvalue, with $\lambda$ the largest magnitude eigenvalue of $M_1$. This immediately gives rise to the following consequences. Firstly, $N_{aa^*}^{1}=1$ and not larger. Secondly, the largest eigenvalue of $M_1$ is positive and non-degenerate. Thirdly, any $M_{a}$ for $a\neq 1$ has a spectral radius strictly smaller than $\lambda$. Fourthly, since the spectral radii of $M_{a}$ and $M_{a^*}$ are identical, it follows that $1^* = 1$. Furthermore, denoting the corresponding (right) eigenvector as $\mathbf{v}_{R}$ and using $\bar{M}_1 = Z_1^{-1} M_1 Z_{1}$, we find $Z_1 \overline{\mathbf{v}}_{R} \sim \mathbf{v}_{R}$, where we can absorb the proportionality constant into $Z_1$. Applying this relation twice reveals that $Z_1\bar{Z}_1 \mathbf{v}_{R} = \mathbf{v}_{R}$, such that label $1$ must have a trivial Frobenius-Schur indicator $\varkappa_1=1$.

In addition, it is well known from the theory of MPS (but here applied to the MPOs by using the Hilbert-Schmidt inner product for the operators $O_a^L$) that for two injective MPO tensors $B_a^{ij}$ and $B_{b}^{ij}$ that are both normalized such that the spectral radius $\rho(\mathbb{E}_{a}) = \rho(\mathbb{E}_{b}) = \lambda$, the spectral radius of $\sum_{ij} B_{a}^{ij}\otimes \bar{B}_{b}^{ij} = (\mathds{1}\otimes Z_{b}^{-1}) \sum_{i,j} B^{ij}_a \otimes B^{ji}_{b^*} (\mathds{1}\otimes Z_{b})$ is either $\lambda$ (in which case $O_a^L$ and $O_b^L$ are identical and the tensors are related by a gauge transform) or the spectral radius is strictly smaller than $\lambda$. Since we can now use the fusion tensors $X_{ab;\mu}^{c}$ to bring $\sum_{i,j} B^{ij}_a \otimes B^{ji}_{b}$ into upper block triangular form with diagonal blocks $M_c$ and thus to relate the spectra, this immediately shows that $1$ cannot appear in the fusion product of $a$ and $b^*$ unless $b=a$, i.e.\ $N_{ab^*}^{1} = \delta_{ab}$. We can continue along these lines to show some extra symmetry properties of the tensor $N$. If $N_{ab}^{c} \neq 0$, then $\sum_{ijk} B_{a}^{ij}\otimes B_{b}^{jk} \otimes \bar{B}^{ik}_{c}$ should have a largest magnitude eigenvalue $\lambda$ with degeneracy $N_{ab}^{c}$. But using the $Z$ matrices, and swapping the matrices in the tensor product, this also means that
\begin{equation}\label{pivotal}
N_{ab}^{c} = N^{a^*}_{b c^*} = N_{c^*a}^{b^*}\, ,
\end{equation}
which can further be combined with equation~\eqref{pivotal3}. In particular, this also shows that $N_{a1}^{b} = N_{1a}^{b} = \delta_{ab}$, such that the single block MPO $O_{1}^L$ indeed corresponds to the neutral object of our algebra.

\subsection{Associativity and the pentagon equation}\label{subsec:associativity}
\begin{figure}
  \centering
    \includegraphics[width=0.38\textwidth]{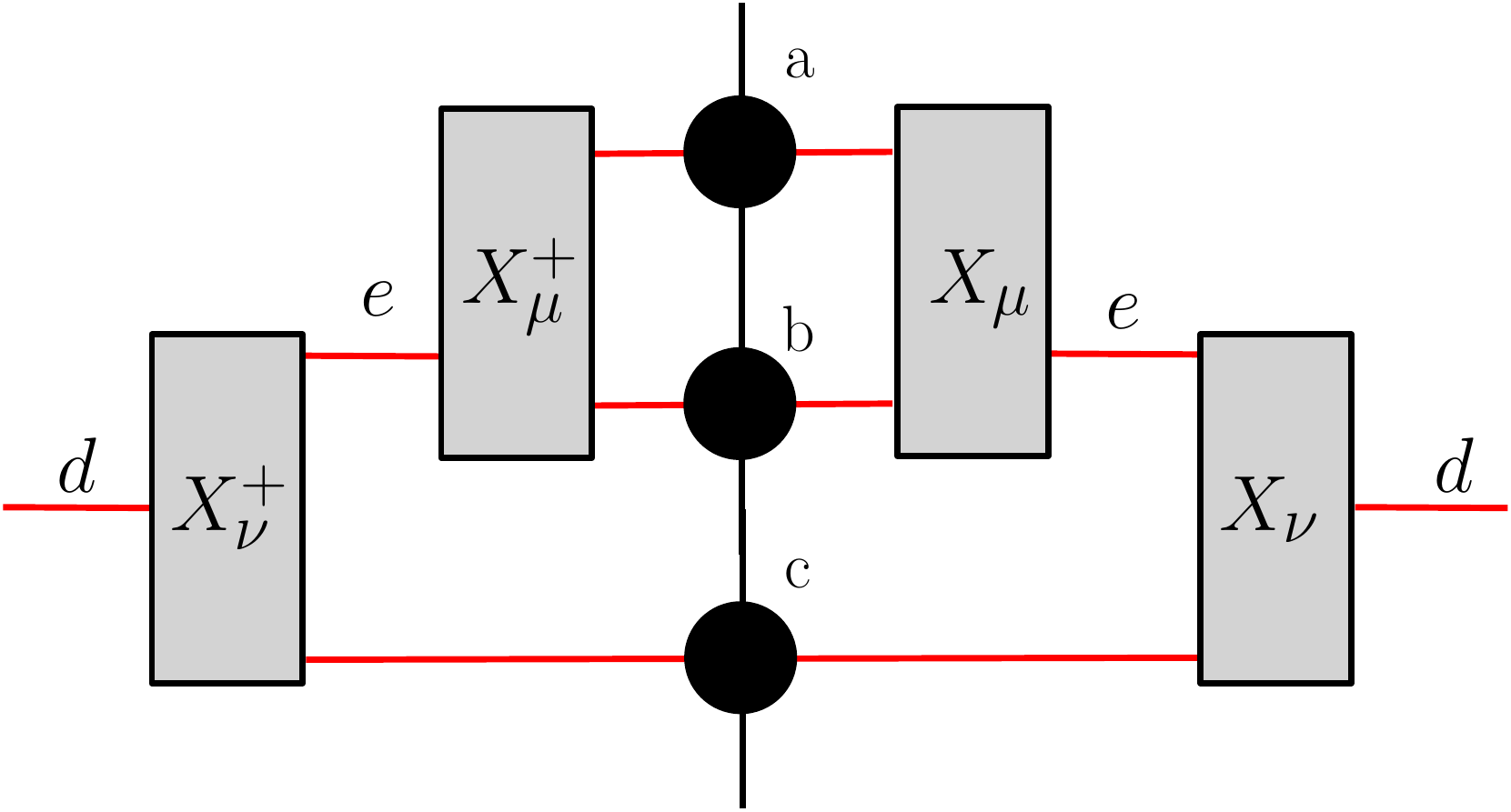}
    \includegraphics[width=0.055\textwidth]{n10}
    \includegraphics[width=0.38\textwidth]{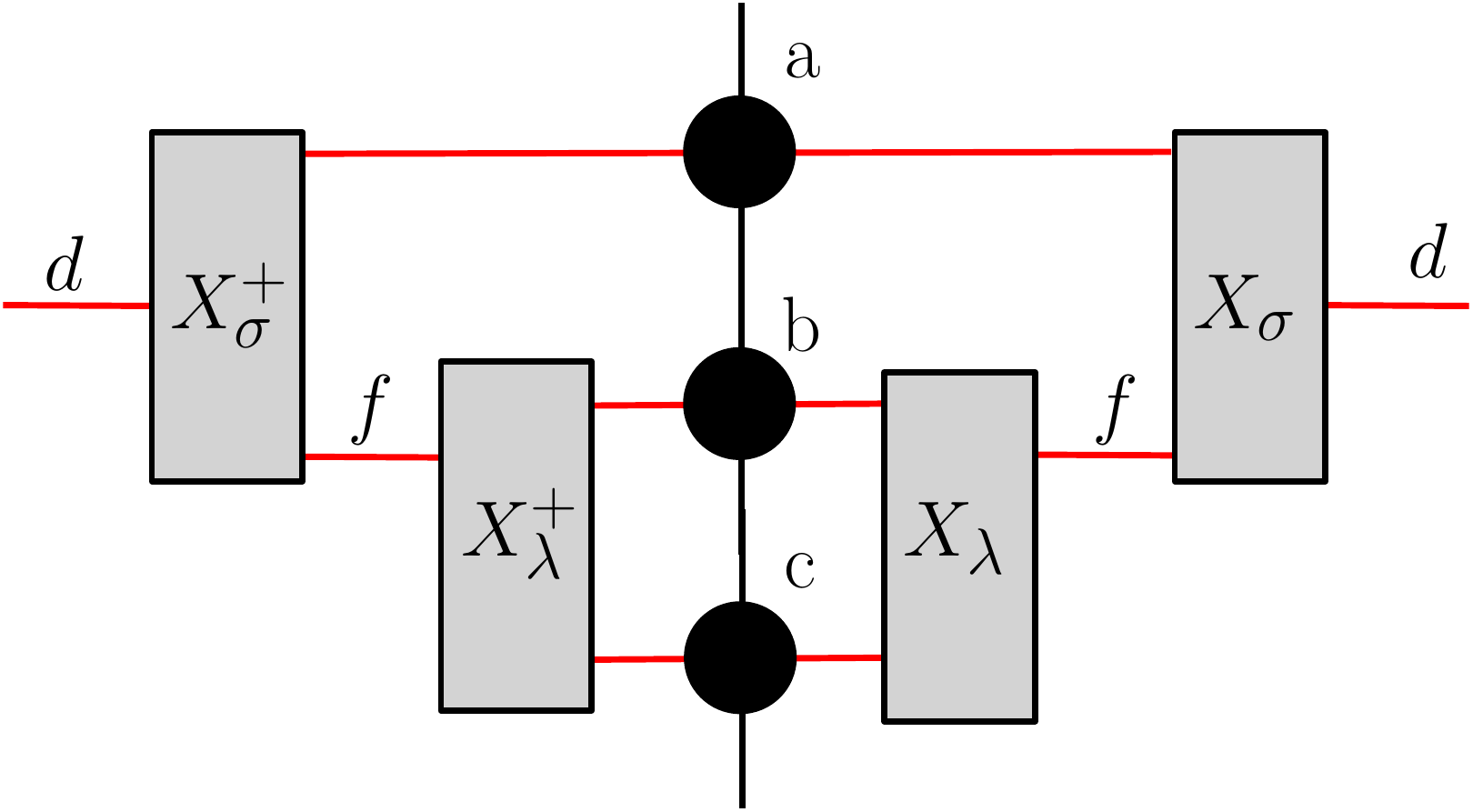}
\caption{Property of MPO and fusion tensors that follows from associativity of the multiplication of $O^L_a$, $O^L_b$ and $O^L_c$.}
\label{associativity}
\end{figure}

Associativity of the product $(O_a^L O_b^L) O_c^L = O_a^L (O_b^L O_c^L)$ implies that
\begin{equation}
\sum_e N_{ab}^{e} N_{ec}^{d} = \sum_{f} N_{af}^{d} N_{bc}^{f}.
\end{equation}
In addition, there are two compatible ways to obtain the block decomposition of $B_{abc}^{i,l}=\sum_{j,k} B_a^{i,j}\otimes B_{b}^{j,k} \otimes B_{c}^{k,l}$ into diagonal blocks of type $B_{d}^{i,l}$. Indeed, we have
\begin{align*}
X^{d^+}_{ec,\nu}\left(X^{e^+}_{ab,\mu}\otimes\mathds{1}_{\chi_c}\right)B_{abc}^{i,l}\left(X^{e}_{ab,\mu}\otimes\mathds{1}_{\chi_c}\right)X^{d}_{ec,\nu}&= B_{d}^{i,l} \\
{X^{d^+}_{af,\sigma}}\left(\mathds{1}_{\chi_a} \otimes X^{f^+}_{bc,\lambda}\right) B_{abc}^{i,l} \left(\mathds{1}_{\chi_a}\otimes X^{f}_{bc,\lambda}\right)X^{d}_{af,\sigma} &= B_{d}^{i,l}\, ,
\end{align*}
as illustrated in figure~\ref{associativity}. For PMPOs satisfying the zipper condition \eqref{zippercondition2} similar reasoning as in section \ref{subsec:fusiontensors} shows that for every $a,b,c,d$ there must exist a transformation
\begin{equation}\label{Fmove}
\left(X^e_{ab,\mu}\otimes\mathds{1}_{\chi_c}\right)X^d_{ec,\nu} =
\sum_{f=1}^{\mathcal{N}} \sum_{\lambda = 1}^{N_{bc}^f} \sum_{\sigma = 1}^{N_{af}^d}(F^{abc}_{d})^{f\lambda\sigma}_{e\mu\nu} \left(\mathds{1}_{\chi_a}\otimes X^f_{bc,\lambda}\right)X^d_{af,\sigma}\, ,
\end{equation}
where $F^{abc}_d$ are a set of invertible matrices. To see this, consider following identity, which follows from the zipper condition,

\begin{align}
\sum_{de\mu\nu}
\vcenter{\hbox{ 
\includegraphics[width=0.32\linewidth]{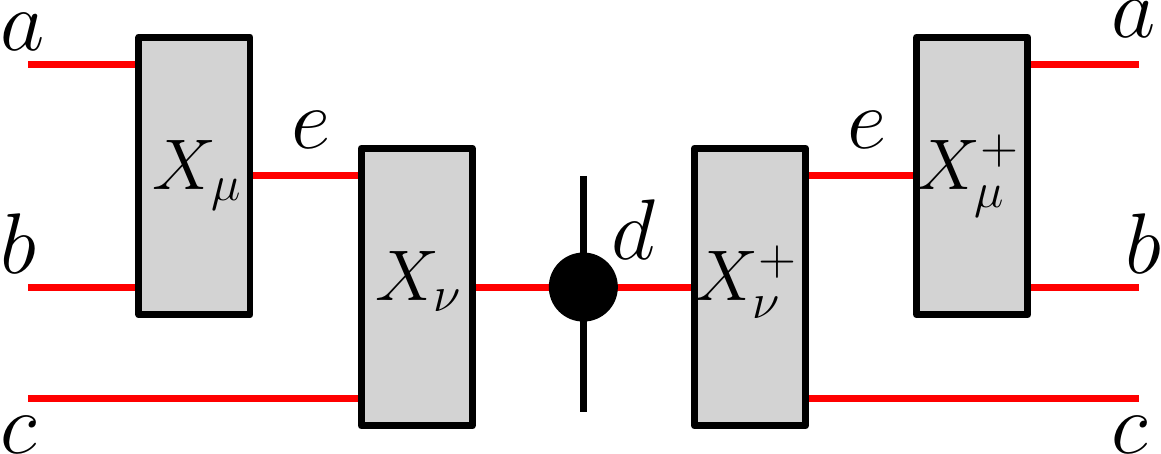}}} = \sum_{df\sigma\lambda}
\vcenter{\hbox{
\includegraphics[width=0.35\linewidth]{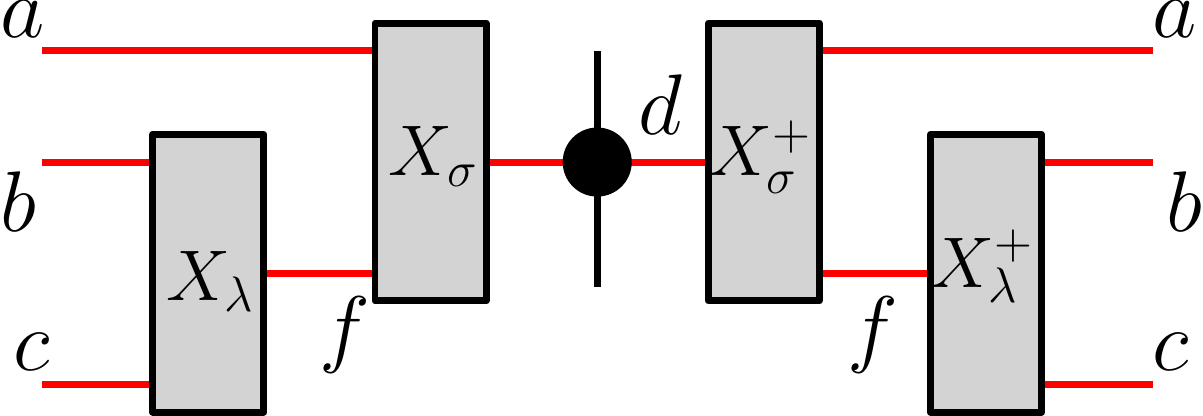}}}
\end{align}
Acting with fusion tensors on both sides of the equation gives

\begin{align}\label{pentagon3}
\vcenter{\hbox{ 
\includegraphics[width=0.2\linewidth]{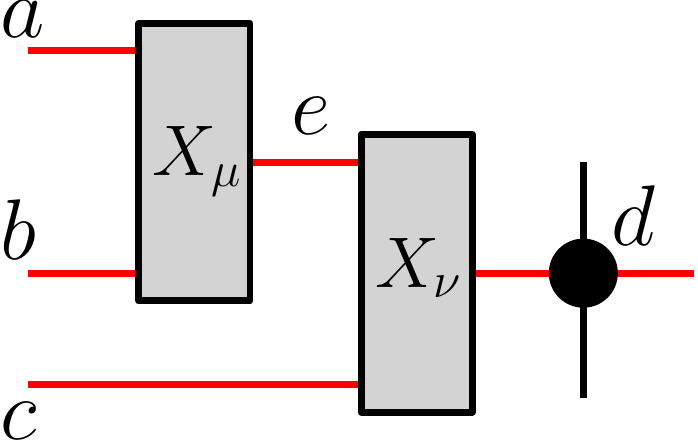}}} = \sum_{d'f\sigma\lambda}
\vcenter{\hbox{
\includegraphics[width=0.42\linewidth]{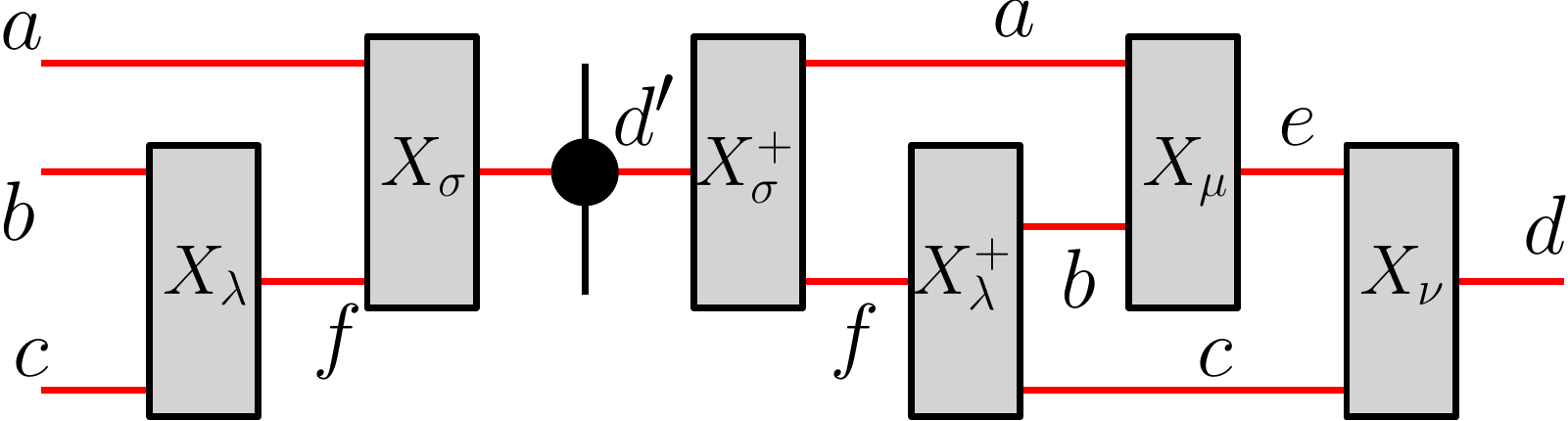}}}
\end{align}
As a final step we use injectivity of the single block MPO tensors. This property implies that $\left(B_d^{ij}\right)_{\alpha\beta}$, when interpreted as a matrix with rows labeled by $ij$ and columns by $\alpha\beta$, has a left inverse $B_d^{+}$ such that $B_d^+B_{d'} = \delta_{dd'}\mathds{1}_{\chi_d}\otimes\mathds{1}_{\chi_d}$. Applying this inverse on both sides of \eqref{pentagon3} leads to the desired expression

\begin{align}
\vcenter{\hbox{ 
\includegraphics[width=0.23\linewidth]{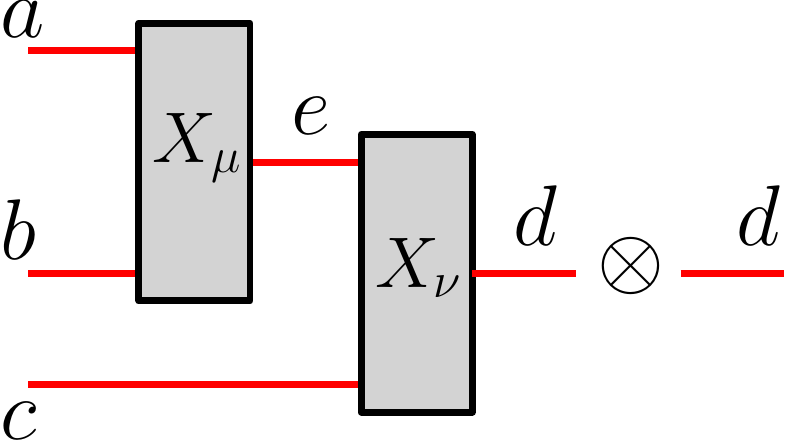}}} = \sum_{f\sigma\lambda}
\vcenter{\hbox{
\includegraphics[width=0.47\linewidth]{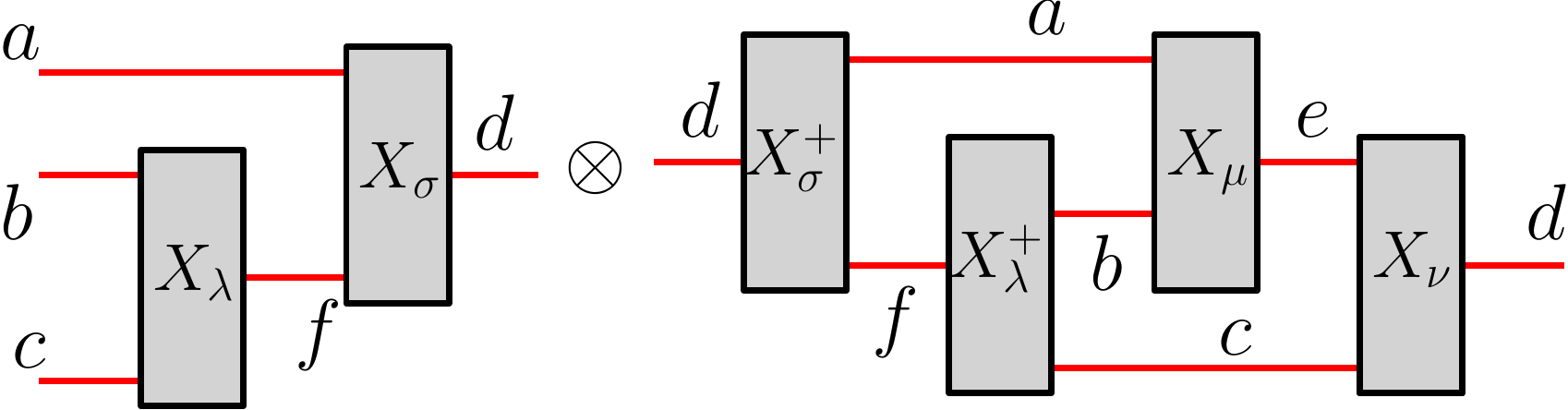}}}
\end{align}
The second tensor product factor on the right hand side is exactly $(F^{abc}_d)^{f\lambda\sigma}_{e\mu\nu} \mathds{1}_{\chi_d}$.

The $F$ matrices have to satisfy a consistency condition called the pentagon equation, which is well-known in category theory. It results from deriving the matrix that relates $(X_{ab,\mu}^f\otimes \mathds{1}_{\chi_c}\otimes \mathds{1}_{\chi_d})(X_{fc,\nu}^g\otimes\mathds{1}_{\chi_d})X_{gd,\rho}^e$ to $(\mathds{1}_{\chi_a}\otimes \mathds{1}_{\chi_d}\otimes X_{cd,\lambda}^h)(\mathds{1}_{\chi_a}\otimes X_{bh,\kappa}^i)X_{ai,\sigma}^e$ in two different ways and equating the two resulting expressions. Written down explicitly, the pentagon equation reads
\begin{equation}\label{pentagoneq}
\sum_{h,\sigma\lambda\omega}(F^{abc}_g)^{f\mu\nu}_{h\sigma\lambda}(F^{ahd}_e)^{g\lambda\rho}_{i\omega\kappa}(F^{bcd}_i)^{h\sigma\omega}_{j\gamma\delta} = 
\sum_\sigma (F^{fcd}_e)^{g\nu\rho}_{j\gamma\sigma}(F^{abj}_e)^{f\mu\sigma}_{i\delta\kappa}.
\end{equation}

\begin{figure}
  \centering
    \includegraphics[width=0.35\textwidth]{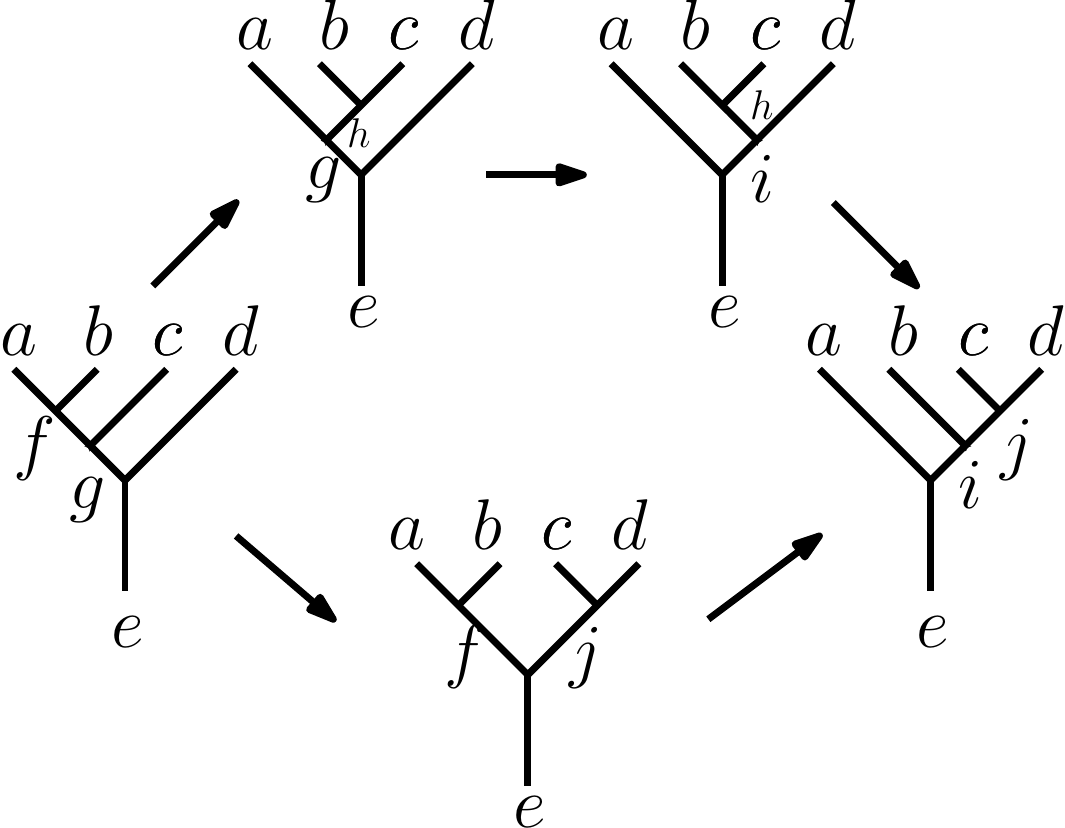}
\caption{Two paths giving rise to the pentagon equation \eqref{pentagoneq}.}
\label{pentagon}
\end{figure}
The two ways to obtain the same matrix leading to the pentagon equation are shown in figure \ref{pentagon}. A standard result in category theory,  called Mac Lane's coherence theorem, states that the pentagon equation is the only consistency relation that needs to be checked; once it is satisfied all other possible consistency conditions are also automatically satisfied \cite{Kitaev06,maclane}.

The complete set of algebraic data we have associated to a Hermitian PMPO $P_L$ that satisfies the zipper condition \eqref{zippercondition2} is $(N_{ab}^c, F^{abc}_d, \varkappa_a)$. Note that $(N_{ab}^c, F^{abc}_d)$ is (in many cases) known to be robust in the sense that every small deformation of the matrices $F^{abc}_d$ that satisfies the pentagon equation can be absorbed in the fusion tensors via a suitable gauge transformation $Y$. This remarkable property is called Ocneanu rigidity \cite{Kitaev06,ocneanu} and it shows that PMPOs satisfying the zipper condition naturally fall into discrete families.

$(N_{ab}^c, F^{abc}_d, \varkappa_a)$ is very similar to the algebraic data defining a fusion category. We argued in section \ref{subsec:hermiticity} that when a PMPO has a unital structure then the definition of duality as derived from Hermitian conjugation is equivalent to the categorical definition.
Similar kind of reasoning also shows that our definition of $\varkappa_a$ coincides with that of the Frobenius-Schur indicator in fusion categories for a large class of PMPOs with unital structure that satisfy the zipper condition. We elaborate on this and other connections to fusion categories in Appendix \ref{app:equivalence}. If the PMPO does not have a unital structure then the data $(N_{ab}^c, F^{abc}_d, \varkappa_a)$ defines a multi-fusion category, i.e. a kind of tensor category whose definition does not require the unit element to be simple. 

\section{MPO-injective PEPS} \label{sec:MPOinjPEPS}

Using the PMPOs introduced in the previous section we can now define a class of states on two-dimensional lattices called MPO-injective PEPS, as introduced in \cite{Ginjectivity,Buerschaper14,MPOpaper}. The importance of this class of PEPS is that it can describe topologically ordered systems. For example, it was shown in \cite{MPOpaper} that all string-net ground states have an exact description in terms of MPO-injective PEPS. In section \ref{subsec:zipper} we first impose some additional properties on the PMPOs, which are required in order to construct PEPS satisfying all MPO-injectivity axioms in section \ref{subsec:entangled}. In section \ref{subsec:virtualsupport} we review some properties of the resulting class of MPO-injective PEPS.

\subsection{Unitarity, zipper condition and pivotal structure}\label{subsec:zipper}
To be able to construct MPO-injective PEPS in section \ref{subsec:entangled} we have to impose three properties on the PMPOs we consider.

Firstly, we require is that there exists a gauge on the internal MPO indices such that the fusion tensors $X_{ab,\mu}^{c}$ are isometries --such that $X_{ab,\mu}^{c^{+}} = (X_{ab,\mu}^{c})^\dagger$-- and the gauge matrices $Z_a$, introduced in section \ref{subsec:hermiticity}, are unitary. This brings PMPOs into the realm of unitary fusion categories, which will be required for various consistency conditions throughout. We now devise a new graphical language where the matrices $Z_a$ are represented as
\begin{align}
\vcenter{\hbox{
\includegraphics[width=0.18\linewidth]{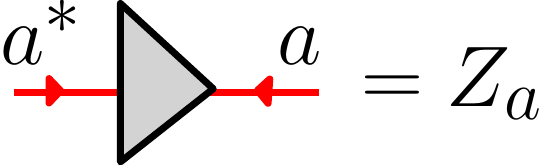}}} \hspace{10 mm}
\vcenter{\hbox{
\includegraphics[width=0.2\linewidth]{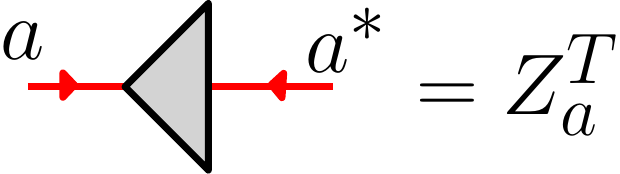}}}
\end{align}
\begin{align*}\hspace{4 mm}
\vcenter{\hbox{ 
\includegraphics[width=0.22\linewidth]{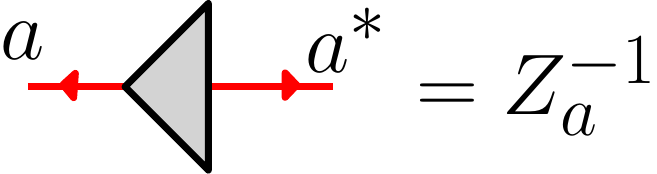}}} \hspace{6 mm}
\vcenter{\hbox{
\includegraphics[width=0.22\linewidth]{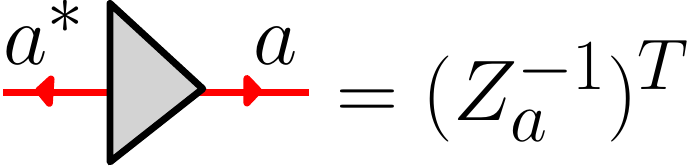}}}
\end{align*}
Note that absolute orientation of the symbols used to represent the matrices has no meaning, as we will be using those in a two-dimensional setting where the tensors will be rotated. Rotating the first figure by $180^\circ$ exchanges the row and column indices of the matrix and is thus equivalent to transposition, which is compatible with the graphic representation of $Z_a^T$. Because of unitarity, $(Z_a^{-1})^{T}= \bar{Z}_a$ and complex conjugation of the tensor simply amounts to reversing the arrows. The definition of the Frobenius-Schur indicator $Z_a\bar{Z}_{a^*} = \varkappa_a\mathds{1}$ can now also be written as
\begin{align}\label{FSindicator}
\vcenter{\hbox{
\includegraphics[width=0.3\linewidth]{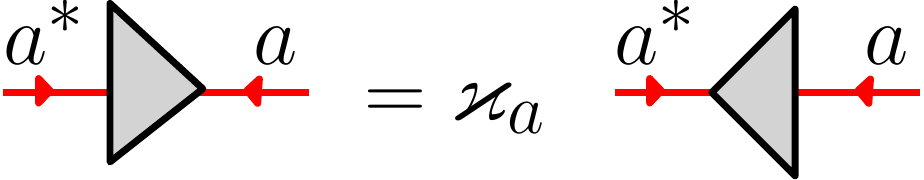}}}\,.
\end{align}

The second requirement is that the zipper condition \eqref{zippercondition2} holds:
\begin{align} \label{zippercondition}
\vcenter{\hbox{
\includegraphics[width=0.43\linewidth]{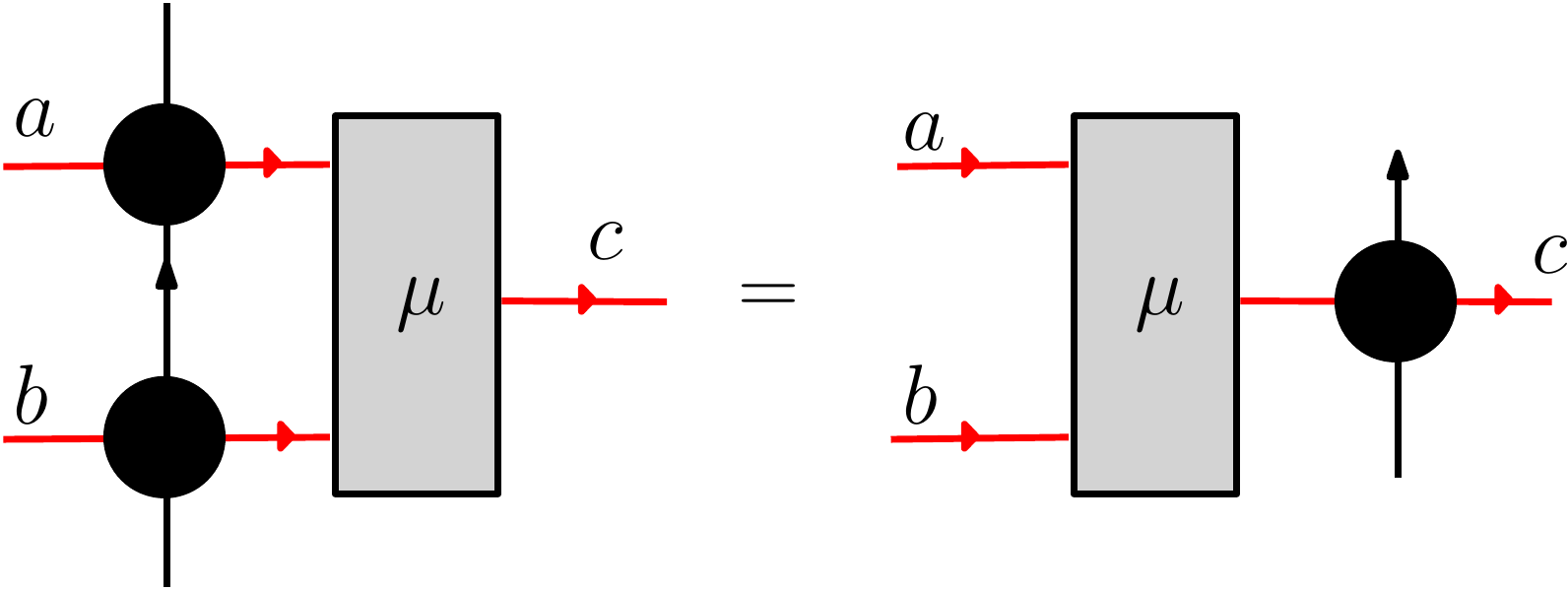}}}
\end{align}
\begin{align*}
\vcenter{\hbox{
\includegraphics[width=0.43\linewidth]{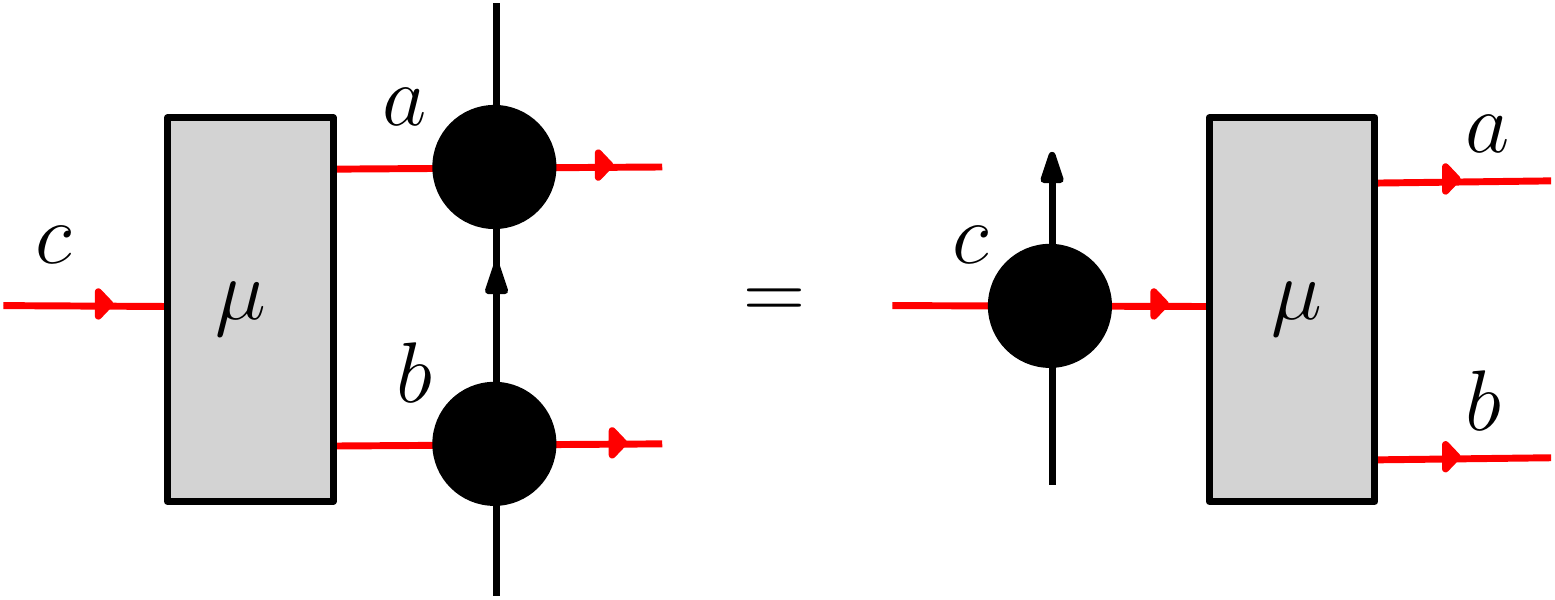}}}
\end{align*}
As already mentioned in section \ref{subsec:fusiontensors} this corresponds to the absence of blocks above the diagonal. In this new graphical notation, we no longer explicitly write $X$ on the fusion tensors, but only the degeneracy label $\mu$. A normal fusion tensor $X_{ab,\mu}^{c}$ has two incoming arrows and one outgoing, while its left inverse $X_{ab,\mu}^{c^{+}} = (X_{ab,\mu}^{c})^\dagger$ has two outgoing arrows and one incoming. In order to determine the difference between e.g.\ $X_{ab}^{c}$ and $X_{ba}^{c}$, any fusion tensor in a graphical diagram always has to be read by rotating it back to the above standard form; note that one should not flip (mirror) any symbol. Consistent use of the arrows is also indispensable in the graphical notation for MPO-injective PEPS in the next section.

The third and final requirement for the PMPO is that the fusion tensors satisfy a property which is closely related to the \emph{pivotal structure} in fusion category theory:
\begin{align}\label{pivotalnew}
\vcenter{\hbox{
\includegraphics[width=0.47\linewidth]{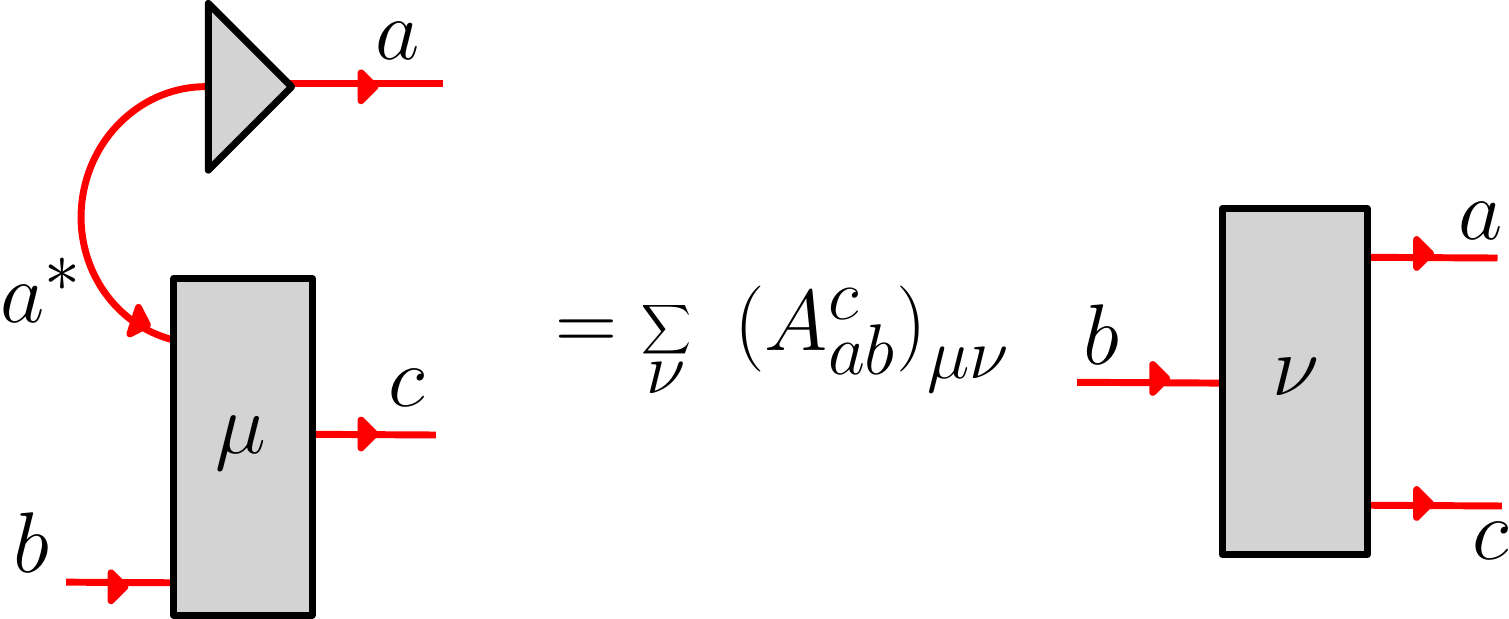}}} \, ,
\end{align}
where the square matrices $A_{ab}^c$ satisfy $\left(A_{ab}^c\right)^\dagger A_{ab}^c = \frac{w_c}{w_b}\mathds{1}$. A similar property holds if we bend the lower $b$ index on the left hand side of (\ref{pivotalnew}), with a set of invertible matrices $A'^c_{ab}$ satisfying $\left(A'^c_{ab}\right)^\dagger A'^c_{ab} = \frac{w_c}{w_a}\mathds{1}$. Note that this is only possible when all the numbers $w_a$ have the same phase. Using equation (\ref{realnumber}) this implies that all $w_a$ are either positive or negative real numbers. From $\sum_{a,b = 1}^{\mathcal{N}} N_{ab}^c w_a w_b = w_c$ and the fact that $N$ consists of nonnegative entries it then follows that all $w_a$ must be positive. Furthermore, the pivotal property requires that the tensor $N$ satisfies
\begin{equation}
N_{a^*b}^c = N_{ac}^b\,
\end{equation}
which is indeed satisfied by combining the equalities \eqref{pivotal3} and \eqref{pivotal} from Section~\ref{subsec:hermiticity}. While we do believe that the pivotal property \eqref{pivotalnew} follows from the zipper condition and the unitary/isometric property of the gauge matrices and the fusion tensors, the proof falls beyond the scope of this paper and we here impose it as an extra requirement.

By repeated application of the pivotal property, we obtain the following relation between the fusion tensors $X_{ab,\mu}^c$ and the gauge matrices $Z_a$:
\begin{align} \label{pivotalthree}
\vcenter{\hbox{
\includegraphics[width=0.53\linewidth]{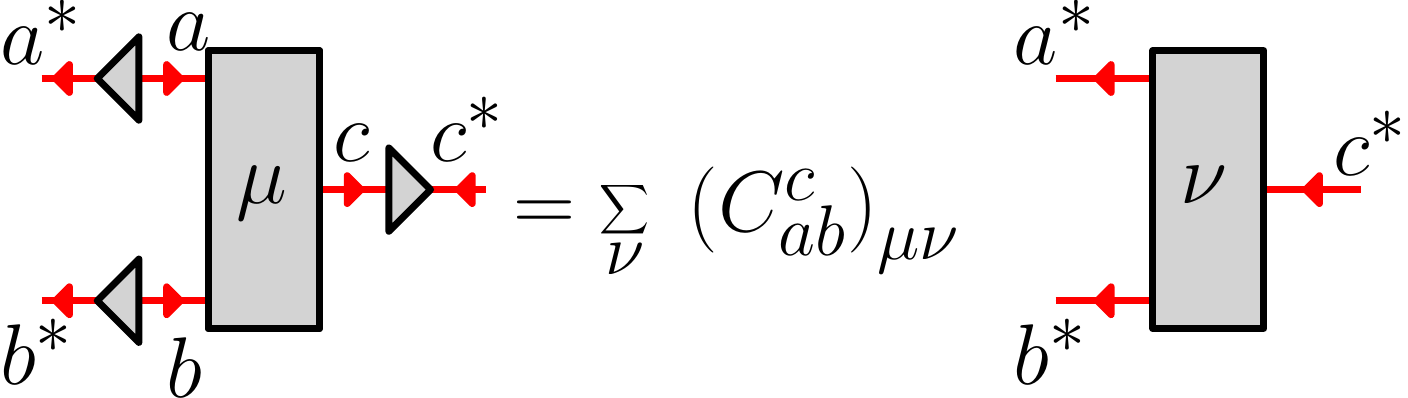}}} \, ,
\end{align}
where $C_{ab}^c= A_{a^* b}^c \bar{A}'^b_{a^*c^*} A^{a^*}_{b^*c^*}$ can be verified to be a unitary matrix. This relation also holds on more general grounds for any Hermitian PMPO satisfying the zipper condition, although with non-unitary $C_{ab}^c$ in general.

Now that we have collected all the necessary properties for the relevant PMPOs we can turn to tensor network states on two-dimensional lattices in the next section. Note that the PMPOs that satisfy the properties discussed in this section could be thought of as classifying anomalous one-dimensional topological orders, i.e. the gapped topological orders that can be realized on the boundary of a two-dimensional bulk \cite{KongWen}.

\subsection{Entangled subspaces}\label{subsec:entangled}
The first step in our construction of a MPO-injective PEPS is to introduce two different types of MPO tensors. For the right handed type,
\begin{align}
\vcenter{\hbox{
\includegraphics[width=0.25\linewidth]{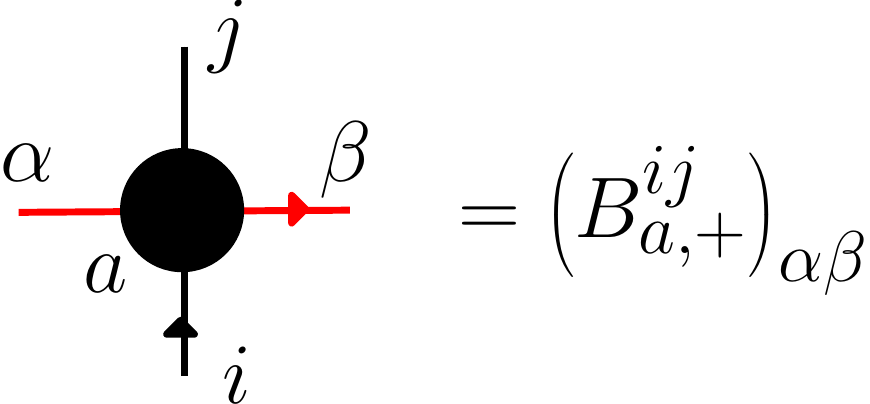}}},
\end{align}
we use the original tensors of the Hermitian PMPO we started from. The left handed type,
\begin{align}
\vcenter{\hbox{
\includegraphics[width=0.25\linewidth]{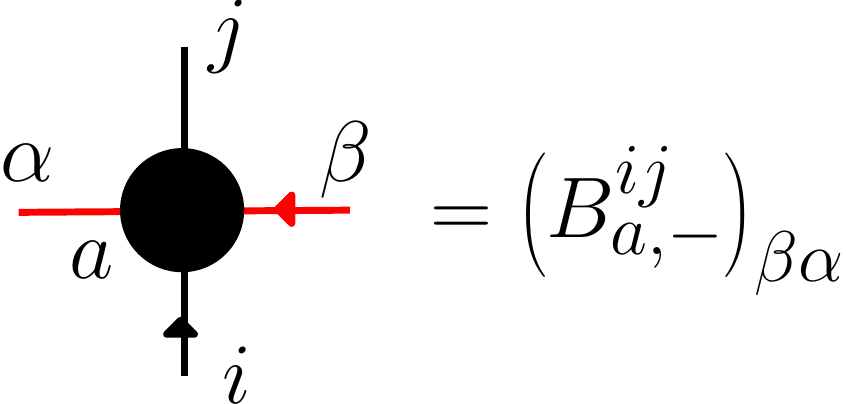}}},
\end{align}
is defined by complex conjugating $B_{a,+}$, which reverses the arrows, and then transposing $i$ and $j$, i.e.
\begin{equation}\label{eq:lefthandedtensor}
\left( B^{ij}_{a,-}\right)_{\beta\alpha} = \left(\bar{B}_{a,+}^{ji} \right)_{\alpha\beta}\, .
\end{equation}
This is exactly the tensor that is obtained by applying Hermitian conjugation to the resulting MPO, as discussed in section \ref{subsec:hermiticity}. We can thus relate both tensors using the gauge matrices $Z_a$, which we now depict using the graphical notation as
\begin{align}\label{leftright}
\vcenter{\hbox{
\includegraphics[width=0.35\linewidth]{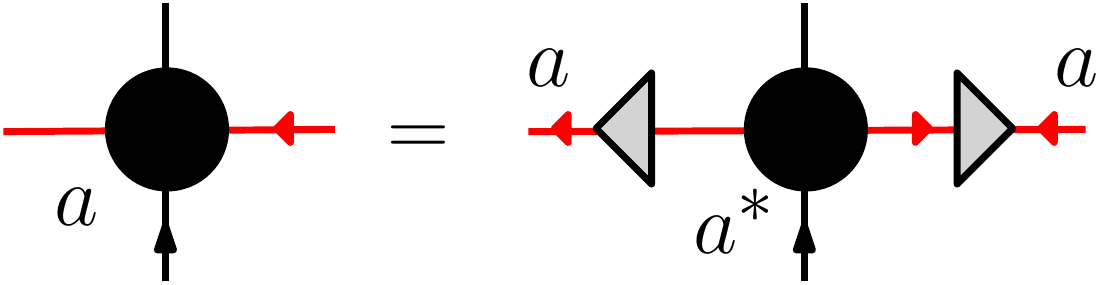}}}\,\,\, \\
\vcenter{\hbox{
\includegraphics[width=0.35\linewidth]{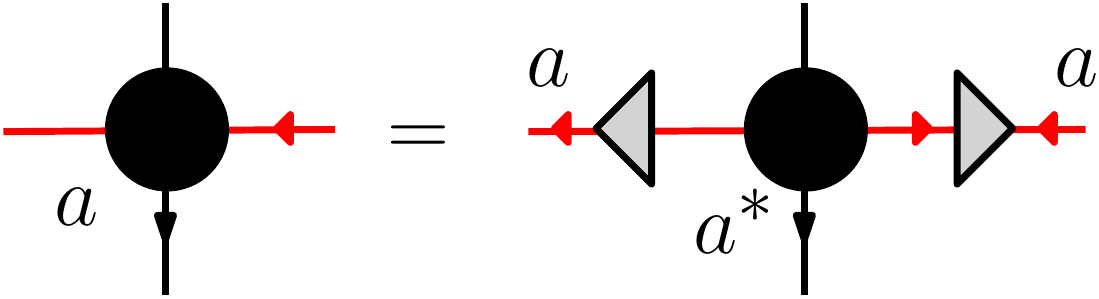}}} \,.
\end{align}
Here we also used that the matrices $Z_a$ are unitary and the identity in Eq.~\eqref{FSindicator}.

With these tensors, MPO-injective PEPS can be constructed on arbitrary lattices. Firstly, assign an orientation to every edge of the lattice. Now define the MPO $\tilde{P}_{C_v}$ at every vertex $v$, with $C_v$ the coordination number of $v$, as follows: assign a counterclockwise orientation to $v$. At every edge connected to $v$, place a right handed or left handed MPO tensor depending on the global orientation and the edge orientation. The MPO $\tilde{P}_{C_v}$ is then obtained by contracting the $C_v$ tensors along the internal MPO indices. With the unitary constraints on the gauge and fusion tensors, the original PMPO $P_L$ that we started from allows for a unitary gauge freedom on the virtual indices of every MPO tensor $B^{ij}_a$. Note, however, that the choice of $B^{ij}_{a,\pm}$ and the transformation behavior of the gauge matrices $Z_a$ is such that this is also a gauge freedom of the newly constructed $\tilde{P}_{C_v}$.

One can furthermore check that since the fusion tensors are isometries the resulting MPO $\tilde{P}_L$ is a Hermitian projector just like $P_L$ when the same weights $w_a$ for the blocks are used. Note that reversing the internal orientation of a single block MPO in $\tilde{P}_L$ amounts to taking the Hermitian conjugate and the weights satisfy $w_a = w_{a^*}$, so reversing the orientation of the internal index of $\tilde{P}_{C_v}$ is equivalent to Hermitian conjugation, which leaves $\tilde{P}_{C_v}$ invariant. So the counterclockwise global internal orientation on $\tilde{P}_{C_v}$ is completely arbitrary.

\begin{figure}
  \centering
    (a) \includegraphics[width=0.21\textwidth]{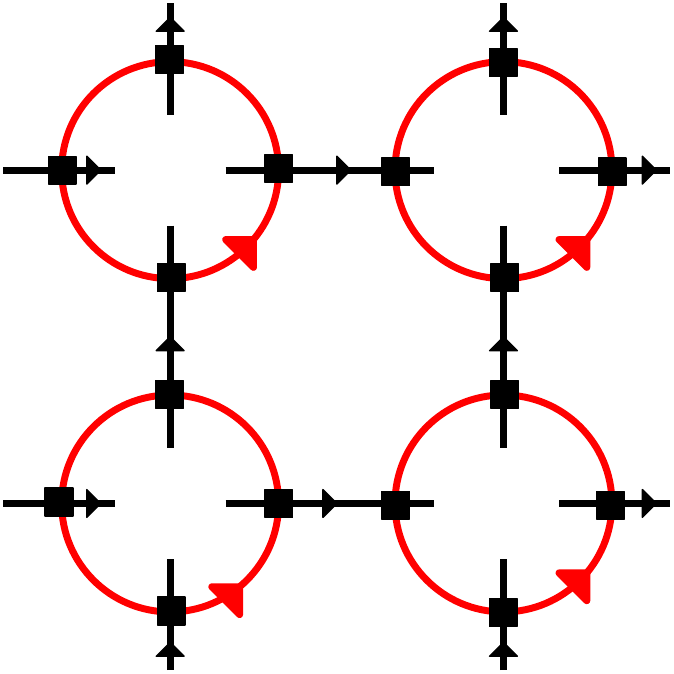}\;\; (b)
    \includegraphics[width=0.22\textwidth]{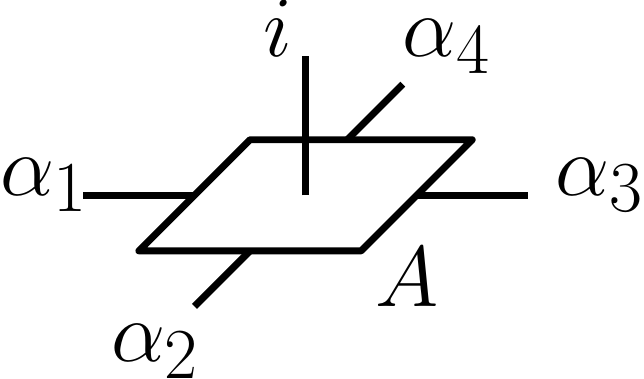}
\caption{(a) A MPO-injective PEPS on a 2 by 2 square lattice with open boundaries. We assigned an orientation to every edge as indicated by the black arrows and an orientation to the internal MPO index represented by the red arrow. (b) A tensor $A$ that can be used to complete the PEPS on the square lattice.}
\label{fig:squarepeps}
\end{figure}

Now that we have the Hermitian projectors $\tilde{P}_{C_v}$ at every vertex we place a maximally entangled qudit pair $\sum_{i=1}^D \ket{i}\otimes\ket{i}$ on all edges of the lattice. We subsequently act at every vertex $v$ with $\tilde{P}_{C_v}$ on the qudits closest to $v$ of the maximally entangled pairs on the neighboring edges. In this way we entangle the subspaces determined by $\tilde{P}_{C_v}$ at each vertex. The resulting PEPS is shown in figure \ref{fig:squarepeps} for a $2$ by $2$ patch out of a square lattice. More general PEPS are obtained by placing an additional tensor $A[v] = \sum_{i = 1}^d \sum_{\{\alpha\} = 1}^D A[v]^i_{\alpha_1\alpha_2\dots \alpha_{C_v}}\ket{i}\bra{\alpha_1\alpha_2\dots \alpha_{C_v}}$ at each vertex which maps the $C_v$ indices on the inside of every MPO loop to a physical degree of freedom in $\mathbb{C}^d$. As long as $A[v]$ is injective as a linear map from $\mathbb{C}^{D^{C_v}}$ to $\mathbb{C}^d$ (which requires $d \geq D^{C_v}$) the resulting PEPS satisfies the axioms of MPO injectivity as defined in \cite{MPOpaper}, which we show below. For the particular case where each $A[v]$ is an isometry, the resulting network is an \emph{(MPO)-isometric PEPS}. Throughout the remainder of this paper we ignore the tensors $A[v]$ as we will argue that the universal properties of the quantum phase of the PEPS are completely encoded in the entangled injectivity subspaces $\tilde{P}_{C_v}$.

We can now prove the following central identity, which we henceforth refer to as the \emph{pulling through equation}:
\begin{align}\label{pullingthrough2}
\vcenter{\hbox{
\includegraphics[width=0.39\linewidth]{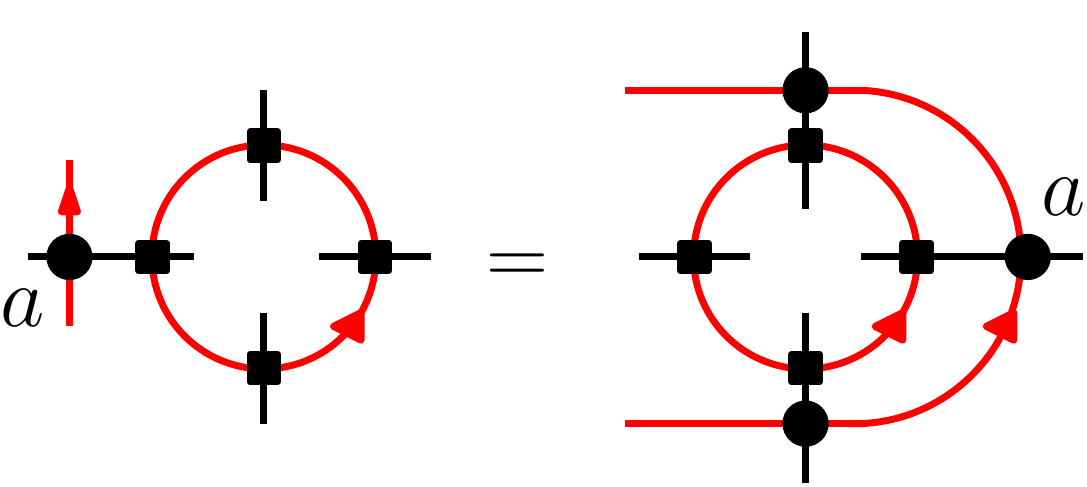}}}.
\end{align}
Note again the difference between squares that denote the superposition of the different injective MPOs $a=1,\ldots,N$ with suitable coefficients $w_a$ and the discs that represent a single block MPO tensor of type $a$.

Using the zipper condition (\ref{zippercondition}) we can write (\ref{pullingthrough2}) as
\begin{align} \label{pullingthrough3}
\vcenter{\hbox{
\includegraphics[width=0.68\linewidth]{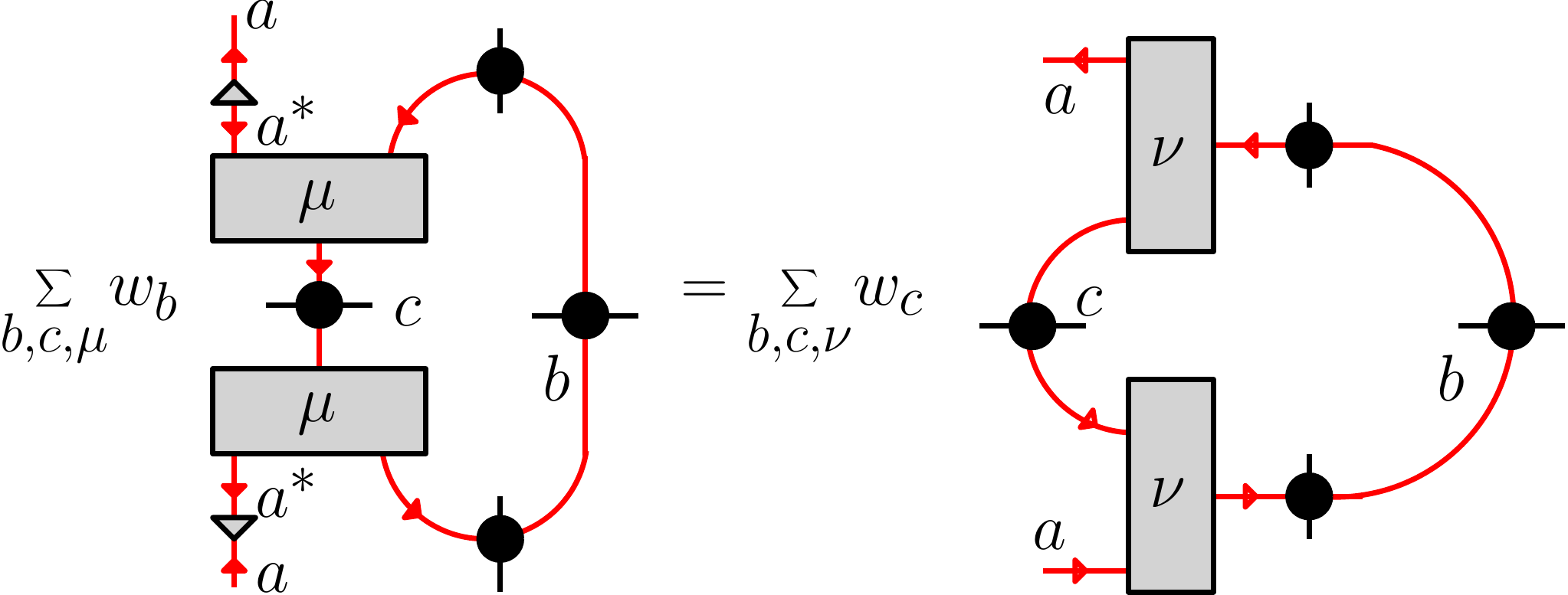}}} \, .
\end{align}
From the pivotal property (\ref{pivotalnew}) and the fact that $A_{ab}^c$ satisfy $\left(A_{ab}^c\right)^\dagger A_{ab}^c = \frac{w_c}{w_b}\mathds{1}$ one can then indeed check the validity of (\ref{pullingthrough3}). 

Two additional notes are in order. Firstly, one could easily imagine different simple generalizations of the MPO-injectivity formalism. But as they are not necessary to understand the fundamental concepts we wish to illustrate, we keep the presentation simple and do not consider them here. However, in the string-net example later on we come across such a simple generalization and see how it leads to a slightly modified form of condition (\ref{pivotalnew}).

Secondly, rather than starting from a PMPO and using it to construct a PEPS tensor, we could have followed the reverse strategy and started from the set of injective MPOs that satisfy the pulling through equation for a given PEPS tensor. This set also forms an algebra (the product of two such MPOs is an MPO satisfying the pulling through equation and can therefore be decomposed into a linear combination of injective MPOs from the original set) and the PEPS tensor would naturally be supported on the virtual subspace corresponding to the central idempotent of that algebra, which corresponds to our PMPO $\tilde{P}_{C_v}$. Both approaches are of course completely equivalent.

\subsection{Virtual support and parent Hamiltonians}
\label{subsec:virtualsupport}
\begin{figure}
  \centering
    \includegraphics[width=0.33\textwidth]{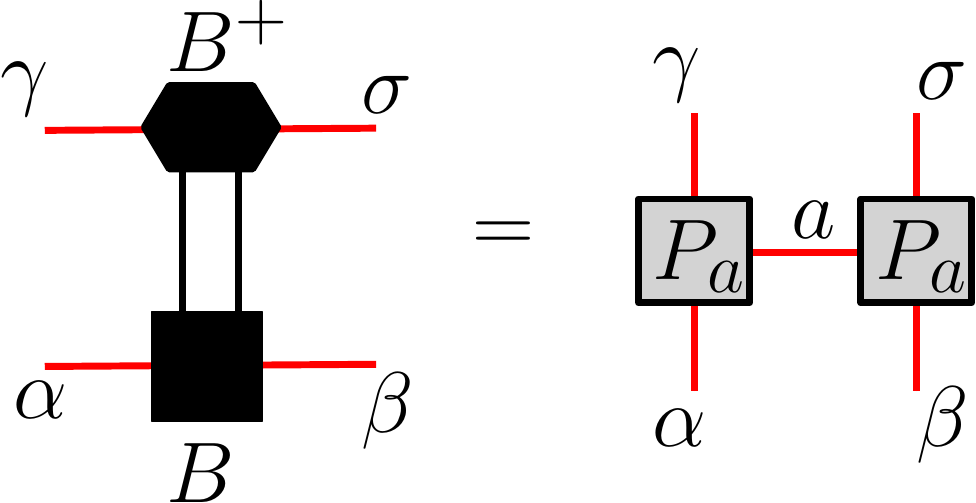}
\caption{Acting with pseudo-inverse $B^+$ on the MPO tensor $B$ gives a projector on block-diagonal matrices with $\mathcal{N}$ blocks of dimension $\chi_a$.}
\label{generalizedinverse}
\end{figure}

The pulling through equation (\ref{pullingthrough2}), which we have proven in the previous subsection, is the first property required for a PEPS to be MPO-injective according to the definition in Ref.~\cite{MPOpaper}. The second property, or an alternative to it, is obtained automatically if we assume that the MPO has already been blocked such that the set of tensors ${B^{ij},\forall i,j=1,\ldots,D}$ already span the full space $\oplus_{a=1,\ldots,N} \mathbb{C}^{\chi_a\times \chi_a}$ (corresponding to the injective blocks $a$ in the canonical form) without having to consider any products. We then have the relation
\begin{equation} \label{blockprojector}
\sum_{i,j = 1}^D B^{+ij}_{\gamma\sigma}B^{ij}_{\alpha\beta} = \sum_{a=1}^{\mathcal{N}} (P_a)_{\alpha\gamma} (P_a)_{\beta\sigma}\, ,
\end{equation}
where $B^{+ij}_{\gamma\sigma}$ is the pseudo-inverse of $B^{ij}_{\alpha\beta}$ interpreted as $D^2 \times \chi^2$ matrix. $P_a$ are a set of $\mathcal{N}$ projectors on the $\chi_a$-dimensional subspaces labeled by $a$. The right hand side of \eqref{blockprojector} thus represents the projector on block diagonal matrices with $\mathcal{N}$ blocks, labeled by $a$, of dimension $\chi_a$. Equation \eqref{blockprojector} is also shown graphically in figure \ref{generalizedinverse} and requires $D^2 > \sum_{a} \chi_a^2$. It essentially means that by acting with $B^+$ on a tensor $B$ we can `open up' the virtual indices of a closed projector MPO. 

With these two properties, we can show that the virtual support of the PEPS map on every contractible region with boundary of length $L$ is given by the PMPO $\tilde{P}_L$ surrounding that region. Indeed, using the pulling through property we can grow the PMPO of a single tensor (note that we no longer explicitly indicate the orientation of every edge):
\begin{align}\label{pullingthrough1}
\vcenter{\hbox{
\includegraphics[width=0.53\linewidth]{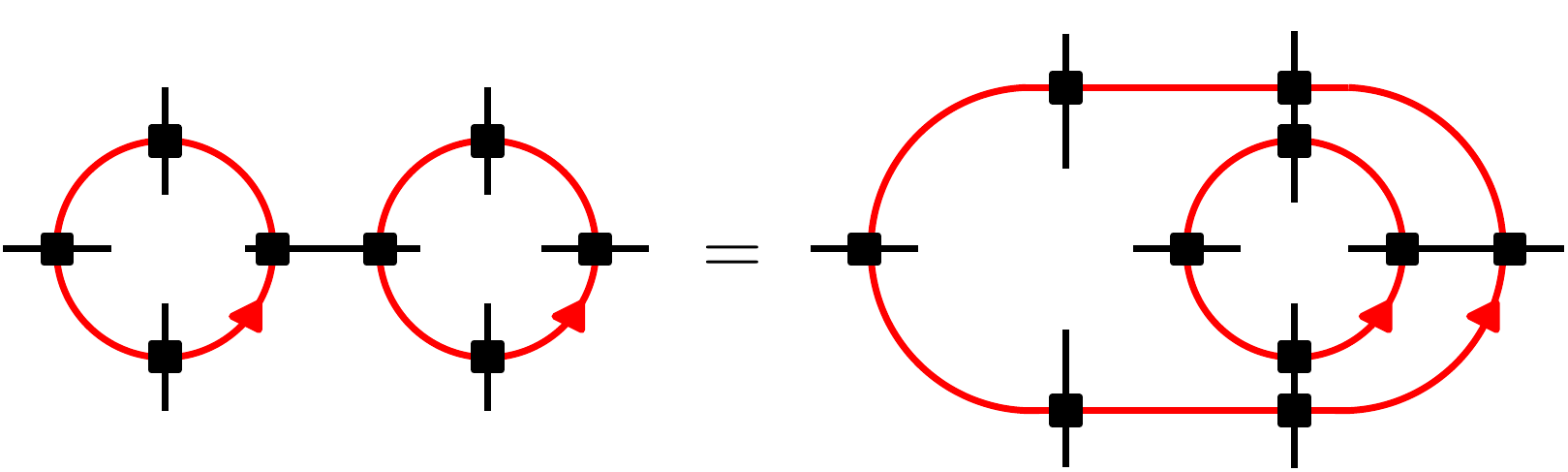}}}
\end{align}
Then we can act with $B^+$ on the inner MPO loop to open up the indices and make it act on the entire boundary:
\begin{align}
\vcenter{\hbox{
\includegraphics[width=0.53\linewidth]{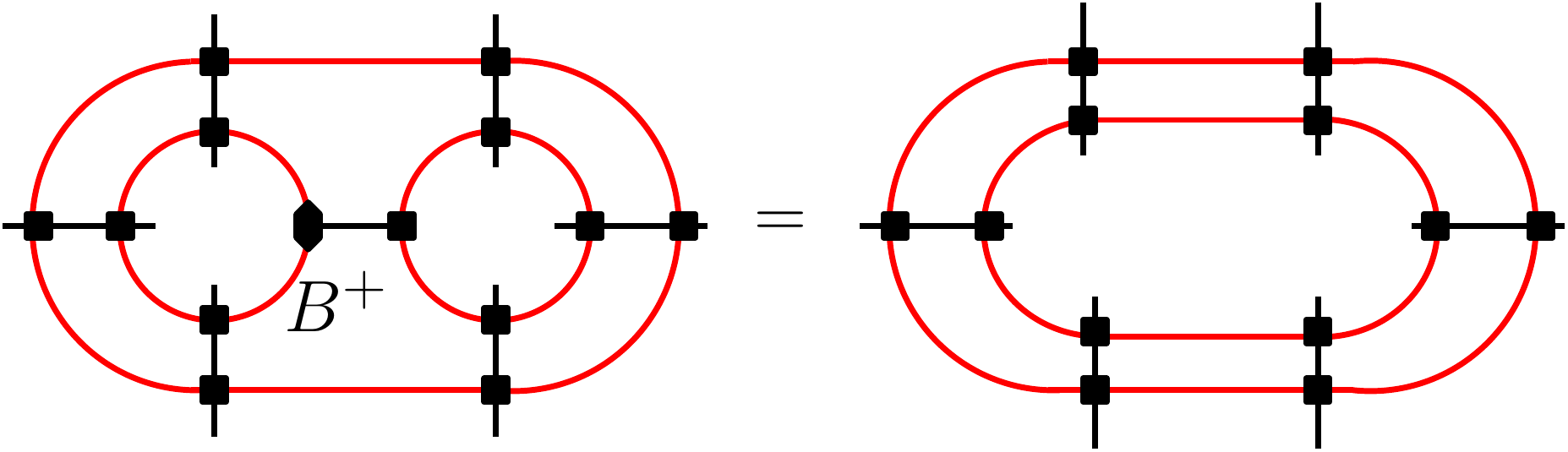}}}
\end{align}
By repeating this trick we can indeed grow the PMPO to the boundary of any contractible region.

These arguments also imply that the rank of the reduced density matrix of the physical degrees of freedom in some finite region with a boundary of length $L$ is equal to the rank of the PMPO $\tilde{P}_L$. Therefore, the zero Renyi entropy of a region with a boundary containing $L$ virtual indices is
\begin{equation}
\begin{split}
S_0 = \log(\text{Tr}(\tilde{P}_L)) = \log\text{tr}\left(\left(\sum_{i = 1}^D B^{ii}\right)^L\right) = \log\left(\sum_{a=1}^{\mathcal{N}}w_a\text{tr}\left( M_a^L\right)\right)\, ,
\end{split}
\end{equation}
where $\text{Tr}$ denotes the trace over external MPO indices, $\text{tr}$ denotes the trace over internal MPO indices and the matrices $M_a=\sum_{i} B_a^{ii}$ were defined in Section~\ref{subsec:hermiticity}. Using the eigenvalue structure of the matrices $M_a$, we find that if the PMPO has a unital structure the zero Renyi entropy for large regions scales as
\begin{equation}
S_0 \approx \lambda_1 L - \log\left(\frac{1}{w_1}\right)\, .
\end{equation}
For fixed point models this constant correction will also appear in the von Neumann entropy, thus giving rise to the well-known topological entanglement entropy \cite{KitaevPreskill}. It was shown in \cite{SPTpaper} that if the PMPO has a unique block, i.e. if $\mathcal{N} = 1$, there is no topological entanglement entropy. In the next section, we show explicitly that using a single blocked PMPO in our PEPS construction does indeed not allow for the existence of different topological sectors.

The constructed MPO-injective PEPS corresponds to the exact ground state of a local,  frustration free Hamiltonian. The so-called parent Hamiltonian construction is identical to that of standard injective PEPS \cite{GarciaVerstraeteWolfCirac08} and takes the form
\begin{equation}\label{parent}
H = \sum_{p} h_p\, ,
\end{equation}
where the sum is over all plaquettes of the lattice and $h_p$ is a positive semi-definite operator whose kernel corresponds to the image (physical support) of the PEPS map on that plaquette. For the square lattice, this is the image of the PEPS map shown in Fig.~\ref{fig:squarepeps}(a), interpreted as a matrix from the outer $8$ to the inner $16$ indices. Typically, $h_p$ is defined as the projector onto the orthogonal complement of the physical support of the PEPS map. In \cite{MPOpaper} the pulling through property was shown to be sufficient to prove that all the ground states of the parent Hamiltonian \eqref{parent} on a closed manifold are given by MPO-injective PEPS states whose virtual indices along the non-contractible cycles are closed using the same MPOs connected by a so-called ground state tensor $Q$. Because of the pulling through property these MPO loops can be moved freely on the virtual level of the PEPS, implying that all ground states are indistinguishable by local operators.

\pagebreak

\section{Anyon sectors in MPO-injective PEPS} \label{sec:anyons}

Having gathered all the concepts and technical tools of PMPOs and MPO-injective PEPS we can now turn to the question of how to construct topological sectors in these models. As argued in the previous section and shown in \cite{MPOpaper}, MPO-injective PEPS give rise to degenerate ground states on nontrivial manifolds that are locally indistinguishable. Systems with this property that are defined on a large but finite open region are believed to have a low-energy eigenstate basis that can be divided into a finite number of topological superselection sectors, such that it is possible to measure in which sector a state is by acting only locally in the bulk of the region, but to go from a state in one sector to a state in another sector one necessarily has to act on both bulk and boundary. The elementary excitations in each sector are called \emph{anyons} and can be seen as a generalization of bosons and fermions \cite{wilczek,LeinaasMyrheim}. In this section we show that the entanglement structure of the ground state PEPS as determined by the PMPO $\tilde{P}_{C}$ contains all necessary information to find the anyonic sectors and their topological properties.

\subsection{Topological charge}\label{subsec:ansatz}

To find the topological sectors we start by looking at a patch of the ground state PEPS on an annulus. It was shown in \cite{MPOpaper} that the support of the ground state tensors in the annulus is equal to the support of the following tensor when we interpret it as a collection of matrices from the indices outside the annulus to the indices inside:
\begin{align}\label{annulus}
 \vcenter{\hbox{
 \includegraphics[width=0.24\linewidth]{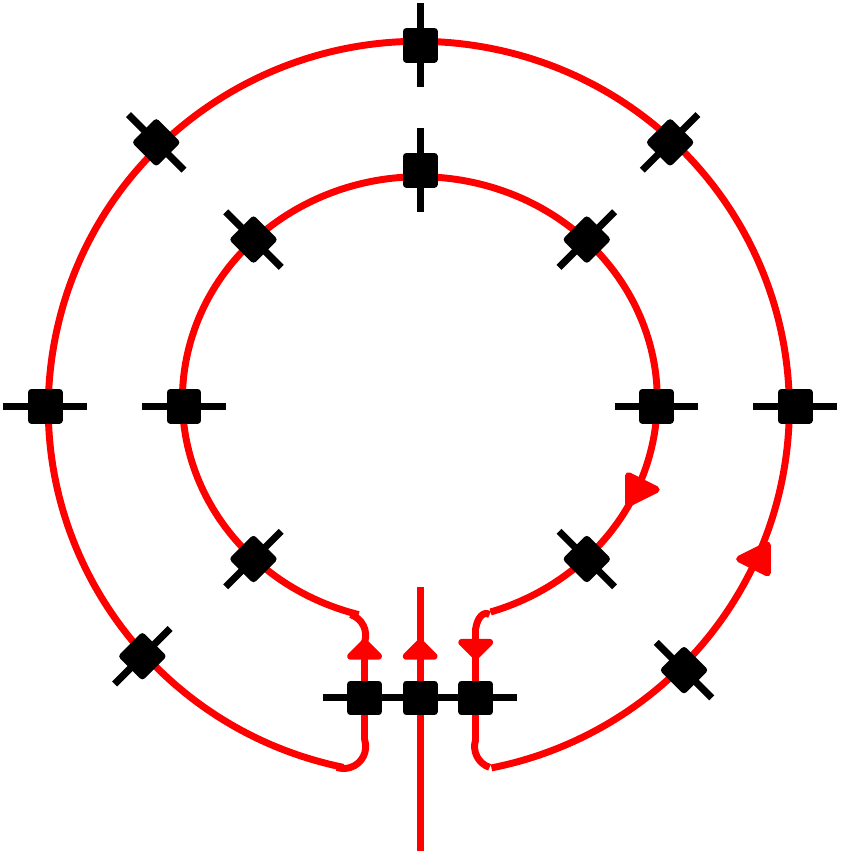}}}
\end{align}
We now use following equality, which follows from the zipper condition \eqref{zippercondition} and \eqref{leftright},
\begin{align}\label{reduction}
 \vcenter{\hbox{\hspace{0 mm}
 \includegraphics[width=0.7\linewidth]{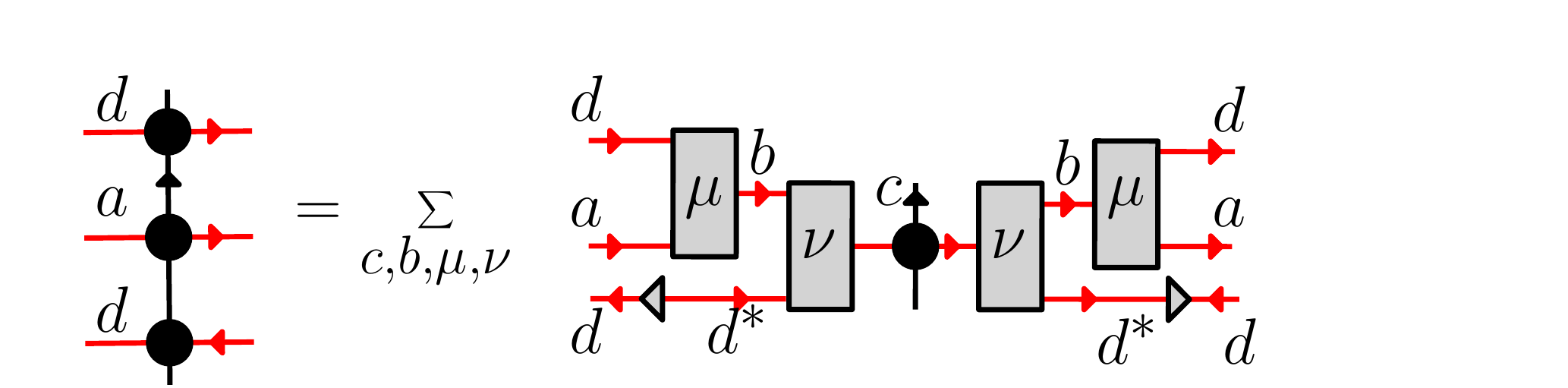}}}
\end{align}
to see that whatever tensor we put in the hole of the annulus, its relevant support space is given by the support of following tensors when interpreted as matrices from the outer indices to the inner ones
\begin{align}\label{algebraobject}
 A_{abcd,\mu\nu}\;\; =\;\; \vcenter{\hbox{
 \includegraphics[width=0.23\linewidth]{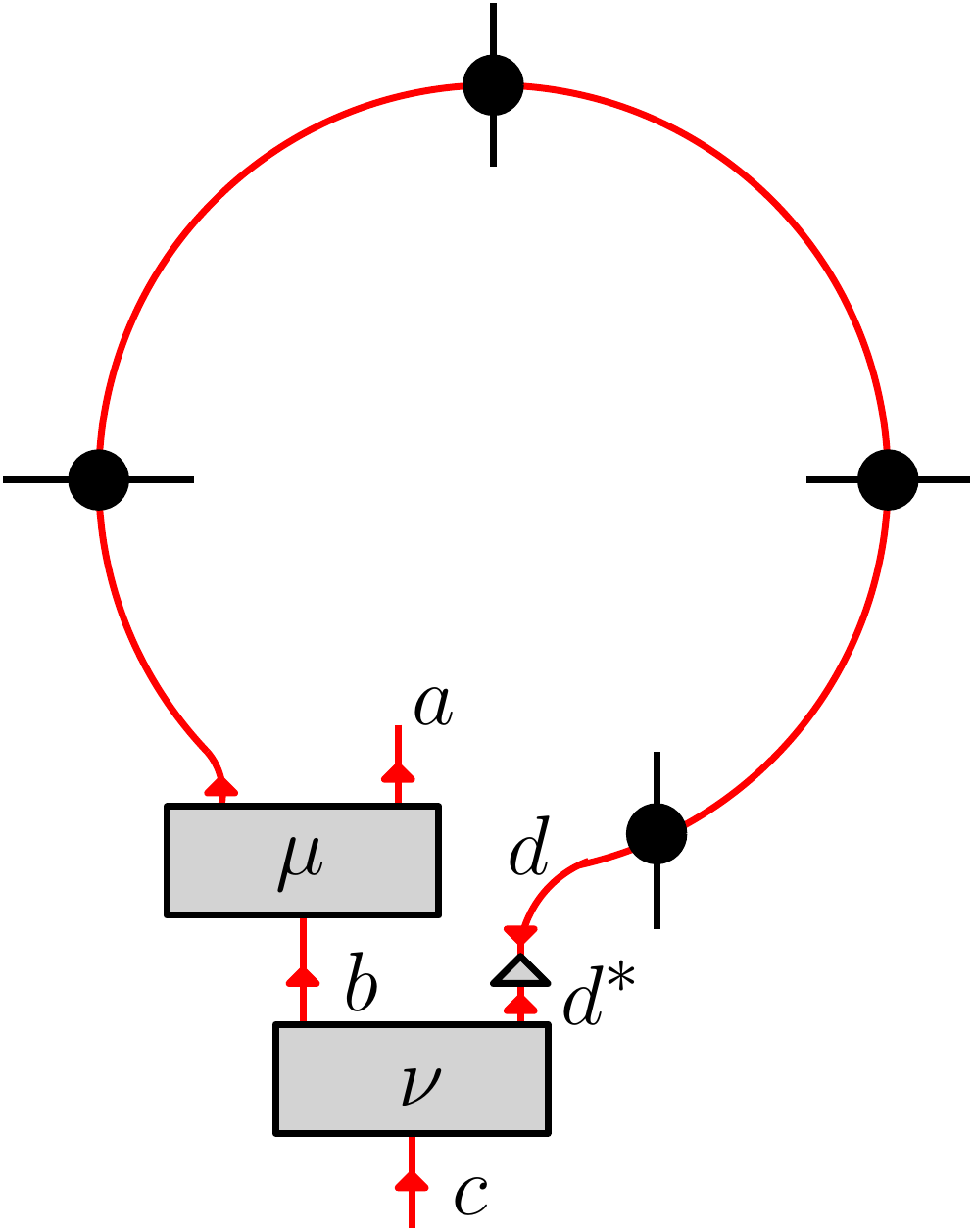}}}
\end{align}
A crucial observation is now that the matrices $A_{abcd,\mu\nu}$ form a $C^*$-algebra, i.e. we have that
\begin{equation}\label{eq:multiplicationalgebra}
A_{hegf,\lambda\sigma}A_{abcd,\mu\nu}=\delta_{ga}\sum_{ij,\rho\tau}\Omega_{(hegf,\lambda\sigma)(abcd,\mu\nu)}^{(hjci,\rho\tau)} A_{hjci,\rho\tau}
\end{equation}
and
\begin{equation}\label{eq:conjugationalgebra}
A_{abcd,\mu\nu}^\dagger = \sum_{e,\lambda\sigma}(\Theta_{abcd,\mu\nu})^{e\sigma\lambda}A_{cead^*,\sigma\lambda}
\end{equation}
We show this explicitly in appendix \ref{app:algebra}. It is a well-known fact that every finite dimensional $C^*$-algebra is isomorphic to a direct sum of simple matrix algebras. We now claim that the topological sectors correspond to the different blocks in this direct sum decomposition of the algebra. This means we relate an anyon sector $i$ to every minimal central idempotent $\mathcal{P}_i$ satisfying  $\mathcal{P}_i\mathcal{P}_j = \delta_{ij}\mathcal{P}_i$, $\mathcal{P}_i^\dagger = \mathcal{P}_i$ and $A_{abcd,\mu\nu}\mathcal{P}_i = \mathcal{P}_i A_{abcd,\mu\nu}$, where $\mathcal{P}_i$ takes the form
\begin{equation}
\mathcal{P}_i = \sum_{abd,\mu\nu} c^i_{abd,\mu\nu}A_{abad,\mu\nu}.
\end{equation}

Because we want to be able to measure the topological charge of an excitation it is a well-motivated definition to associate topological sectors to orthogonal subspaces. We note that in \cite{Qalgebra} a similar identification of anyons in string-net models with central idempotents was given. This idea dates back to the tube algebra construction of Ocneanu \cite{dyonicspectrum,tubealgebra}. In the remainder of this paper we represent the minimal central idempotents $\mathcal{P}_i$ graphically as
\begin{align}\label{idempotent}
 \mathcal{P}_i = \vcenter{\hbox{
 \includegraphics[width=0.25\linewidth]{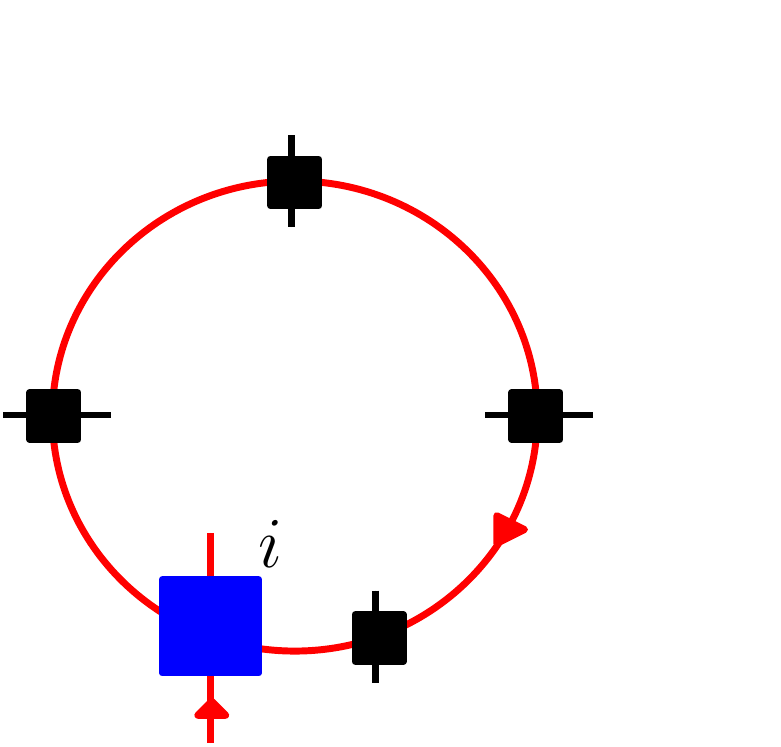}}}
\end{align}
One can easily see that the coefficients $c^i_{abd,\mu\nu}$ give the same central idempotent $\mathcal{P}_i$ independent of the number of MPO tensors used to define $A_{abad,\mu\nu}$, i.e. the $\mathcal{P}_i$ are projectors for every length. In appendix \ref{app:idempotents} we give a numerical algorithm to explicitly construct the central idempotents $\mathcal{P}_i$.

\subsection{Anyon ansatz}\label{sec:ansatz}

Having identified the topological sectors we would now like to construct MPO-injective PEPS containing anyonic excitations. To do so we first need to take a closer look at the $C^*$-algebra spanned by the elements $A_{abcd,\mu\nu}$.

A general central idempotent $\mathcal{P}_i$ consists of a sum of idempotents $P^{a_\alpha}_i$ that are not central:
\begin{equation}\label{eq:centralidempotentdecomposition}
\mathcal{P}_i = \sum_{a=1}^{D_i}\sum_{\alpha = 1}^{d_{i,a}} P^{a_\alpha}_i\, ,
\end{equation}
where the index $a$ refers to the $D_i$ MPO block labels that appear in $\mathcal{P}_i$, which we gave an arbitrary ordering. The $P^{a_\alpha}_i$ satisfy $P^{a_\alpha}_iP^{b_\beta}_j = \delta_{i,j}\delta_{a,b}\delta_{\alpha,\beta}P^{a_\alpha}_i$ and $P^{a_\alpha\dagger}_i = P_i^{a_\alpha}$. We also take the $P_i^{a_\alpha}$ to be simple, i.e. they cannot be decomposed further as an orthogonal sum of idempotents. A central idempotent with $r_i \equiv \sum_{a=1}^{D_i}d_{i,a} > 1$ is called a higher dimensional central idempotent. From the algebra structure \eqref{eq:multiplicationalgebra} and \eqref{eq:conjugationalgebra} we see that the simple idempotents have a diagonal block label, i.e. they can be expressed in terms of the basis elements as
\begin{equation}
P^{a_\alpha}_i = \sum_{bd,\mu\nu} t^{a_\alpha,i}_{bd,\mu\nu}A_{abad,\mu\nu}\, .
\end{equation}
The dimension of the algebra, i.e. the total number of basis elements $A_{abcd,\mu\nu}$, equals $\sum_i r_i^2$: for every $\mathcal{P}_i$ the algebra also contains $r_i(r_i - 1)$ nilpotents $P^{a_\alpha,b_\beta}_i$. The nilpotents satisfy $\left(P_i^{a_\alpha,b_\beta}\right)^\dagger = P_i^{b_\beta,a_\alpha}$ and $P_i^{a_\alpha,b_\beta}P_j^{c_\gamma,d_\delta} = \delta_{i,j} \delta_{b,c}\delta_{\beta,\gamma}P_i^{a_\alpha,d_\delta}$. Note that $P^{a_\alpha,a_\alpha}_i \equiv P^{a_\alpha}_i$ is not a nilpotent but one of the non-central simple idempotents. Combining the simple idempotents and nilpotents we can define the projectors
\begin{equation}
\Pi_i^{[x]} = \sum_{a_\alpha,b_\beta = 1}^{r_i}  x^i_{a_\alpha b_\beta} P_i^{a_\alpha,b_\beta}\, ,
\end{equation}
where
\begin{equation}
\bar{x}^i_{a_\alpha b_\beta} = x^i_{b_\beta a_\alpha}\,\;\;\;\;\text{and }\;\;\;\;\; \sum_{b_\beta} x^i_{a_\alpha b_\beta}x^i_{b_\beta d_\delta} = x^i_{a_\alpha d_\delta}\, .
\end{equation}
If $x^i_{a_\alpha b_\beta} = \bar{v}^i_{a_\alpha}v^i_{b_\beta}$ we call $\Pi^{[x]}_i$ a `rank one' projector (note that $\Pi_i^{[x]}$ does not have to be a rank one matrix, but this terminology refers to the $C^*$-algebra structure.)

Let us now return to the MPO-injective PEPS. As explained above, every ground state tensor has virtual indices which are supported in the subspace $\tilde{P}$. To introduce anyonic excitations in the tensor network we need a new type of PEPS tensor. If we want to place an anyon at vertex $v$ with coordination number $C_v$ this new type of tensor has $C_v$ virtual indices of dimension $D$, one virtual index of dimension $\chi$ and one physical index of dimension $d$. On a square lattice for example, we depict this tensor as

\begin{align}
I = \vcenter{\hbox{
 \includegraphics[width=0.1\linewidth]{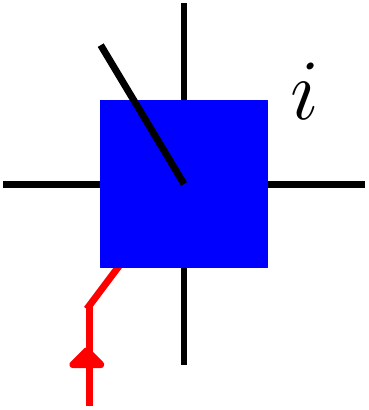}}}\, ,
\end{align}
where the physical index points to the top left corner and the $\chi$-dimensional index is on the bottom left corner. The label $i$ and the name of the tensor $I$ refer to the topological sector. A necessary condition for his tensor to describe an excitation with topological charge $i$ is that its virtual indices are supported in the subspace determined by $\mathcal{P}_i$:

\begin{align}\label{anyonsupport}
 \vcenter{\hbox{
 \includegraphics[width=0.17\linewidth]{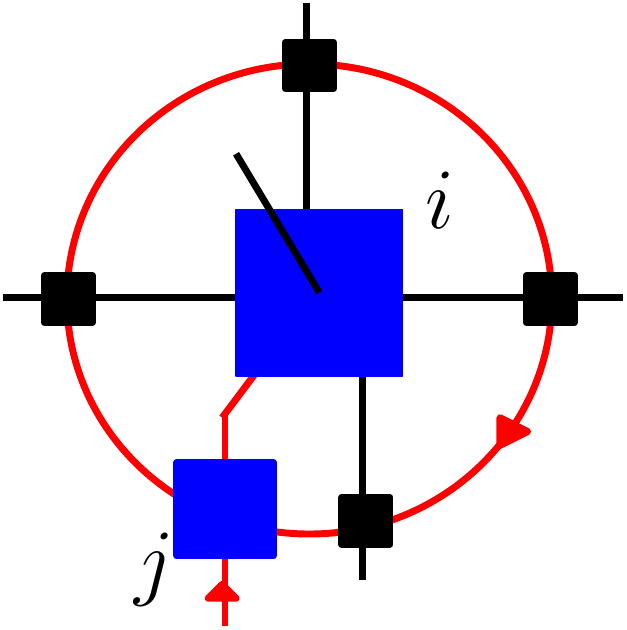}}} = \delta_{i,j}
\vcenter{\hbox{
 \includegraphics[width=0.1\linewidth]{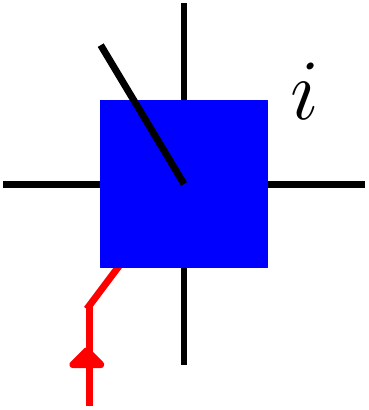}}} \equiv \delta_{i,j} I
\end{align}
This property provides a heuristic interpretation of topological sectors in terms of entanglement. For isometric PEPS the virtual indices along the boundary of a region label the Schmidt states of the physical indices and can therefore be interpreted as the `entanglement degrees of freedom'. Topological sectors are then characterized by entanglement degrees of freedom that live in orthogonal subspaces, so they are really the degrees of freedom that contain the topological information. When we go away from the fixed point the interpretation of virtual degrees of freedom as entanglement degrees of freedom starts to break down and one can understand how the construction fails beyond the phase transition.

\begin{figure}
  \centering
    \includegraphics[width=0.24\textwidth]{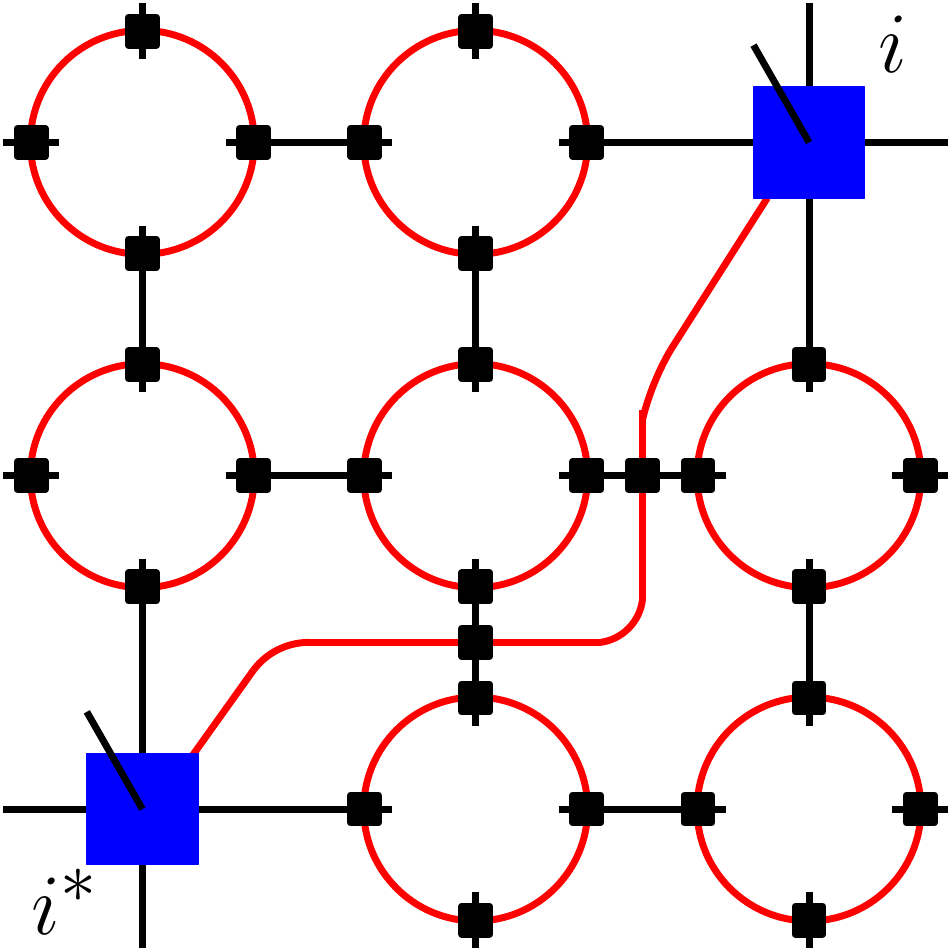}
\caption{A MPO-injective PEPS on a 3 by 3 square lattice containing an anyon pair $(i,i^*)$ located at the lower left and upper right corner.}
\label{fig:anyonpair}
\end{figure}

Property \eqref{anyonsupport} is not sufficient to obtain a good anyonic excitation tensor. To see what additional properties it should have we construct a MPO-injective PEPS containing an anyon pair $(i,i^*)$, where $i^*$ will be defined in a moment. We can do this by  starting from the ground state PEPS, replacing the tensors at two vertices by the excitation tensors corresponding to sectors $i$ and $i^*$ and then connecting the $\chi$-dimensional indices of the excitation tensors with the appropriate MPO on the virtual level. See figure \ref{fig:anyonpair} for an example on the 3 by 3 square lattice. Note that the position of the virtual MPO is irrelevant since it can be moved by using the pulling through property.

If we interpret $I$ as a matrix with the rows labeled by the physical index and the columns by the virtual indices then we can write \eqref{anyonsupport} as $I \mathcal{P}_j= \delta_{i,j}I$. To construct the PEPS containing the $(i,i^*)$ anyon pair we do not use the full tensor $I$ but project its virtual indices in the space corresponding to the simple idempotent $P_i^{a_\alpha}$: $I^{a_\alpha} \equiv IP_i^{a_\alpha}$, where $a$ is one of the MPO block labels appearing in \eqref{eq:centralidempotentdecomposition}. Let us assume that we have connected the tensors $I^{a_\alpha}$ and $I^{*a_\alpha}$ on the virtual level with a single block MPO $a$. We now want the total anyon pair to be in the vacuum sector. We can impose this by surrounding the tensors with the vacuum projector $\tilde{P}$. If we ignore the ground state PEPS environment we can represent the vacuum anyon pair graphically as

\begin{align}\label{vacuumpair1}
\vcenter{\hbox{
 \includegraphics[width=0.33\linewidth]{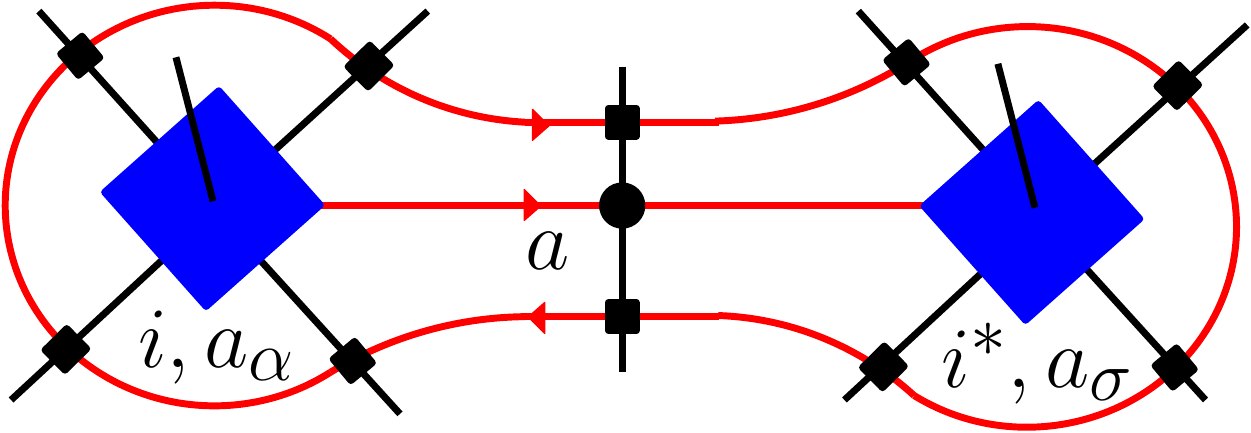}}}\, .
\end{align}
The anyon type $i^*$ is defined as the unique topological sector such that \eqref{vacuumpair1} is non-zero for a fixed $i$ (see section \ref{sec:fusion} for more details). Using \eqref{reduction} we can rewrite \eqref{vacuumpair1} as

\begin{align}\label{vacuumpair2}
\sum_{bcd,\mu\nu}w_d \vcenter{\hbox{
 \includegraphics[width=0.48\linewidth]{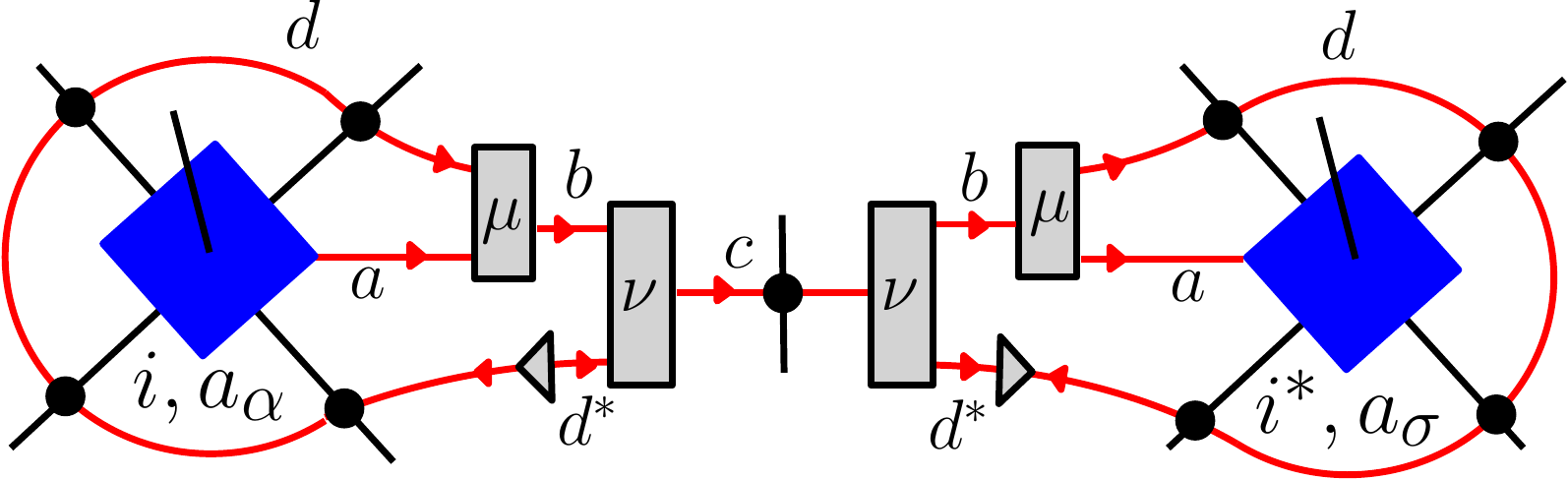}}}\, .
\end{align}
Here we again recognize the algebra basis elements $A_{abcd,\mu\nu}^\dagger$ and $A_{abcd,\mu\nu}$ (see Appendix \ref{subsec:hermitianconjugation} for more details on how to identify the Hermitian conjugate). We now make use of the fact that the basis elements can be written as
\begin{eqnarray}
A_{abcd,\mu\nu} & = & \sum_{i,\alpha\gamma}  u^{i,a_\alpha c_\gamma}_{abcd,\mu\nu} P_i^{a_\alpha,c_\gamma}\, , \\
A_{abcd,\mu\nu}^\dagger  & = & \sum_{i,\alpha\gamma}  \bar{u}^{i,c_\gamma a_\alpha}_{abcd,\mu\nu} P_i^{a_\alpha,c_\gamma}\, .
\end{eqnarray}
Note that once we have found the idempotents we can easily obtain the coefficients $u^{i,a_\alpha c_\gamma}_{abcd,\mu\nu}$ by $P_i^{a_\alpha} A_{abcd,\mu\nu}P_i^{c_\gamma} = u^{i,a_\alpha c_\gamma}_{abcd,\mu\nu} P^{a_\alpha,c_\gamma}_i$. If we represent the simple idempotents and nilpotents as

\begin{align}\label{nilpotent}
P_i^{a_\alpha,b_\beta} = \vcenter{\hbox{
 \includegraphics[width=0.17\linewidth]{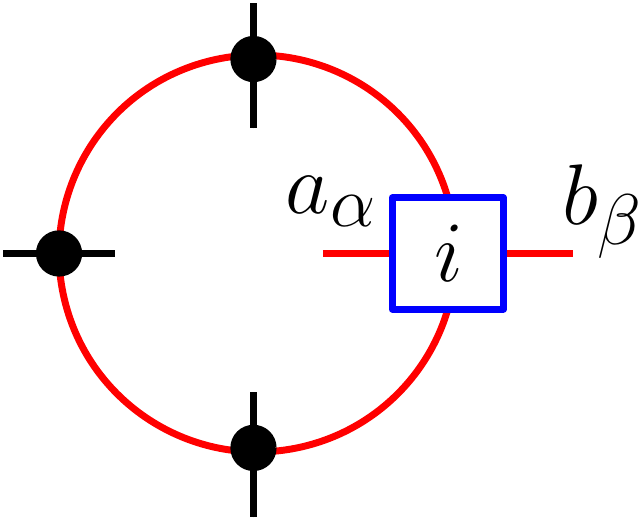}}}\, ,
\end{align}
then we can write \eqref{vacuumpair2} as

\begin{align}\label{vacuumpair3}
\sum_{c,\gamma\lambda} \left( \sum_{bcd,\mu\nu} u^{i,a_\alpha c_\gamma}_{abcd,\mu\nu} \bar{u}^{i^*,c_\lambda a_\sigma}_{abcd,\mu\nu}  w_d\right)  \vcenter{\hbox{
 \includegraphics[width=0.35\linewidth]{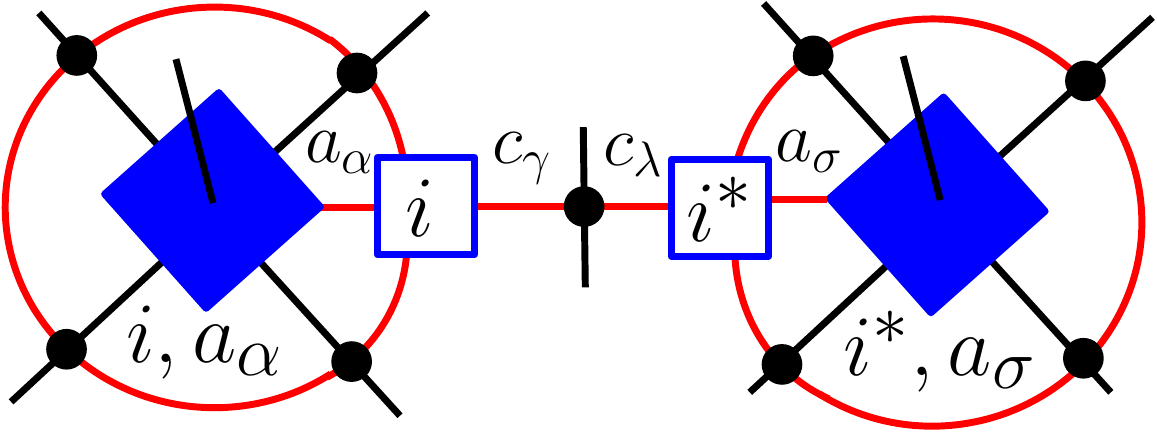}}}\, .
\end{align}
So we see that in order to be able to take the tensors $I$ and $I^*$ and connect them on the virtual level such that they are in the vacuum state it should hold that

\begin{equation} \label{injectivitystructure}
I^{a_\alpha} P_i^{a_\alpha,c_\gamma} \propto I^{c_\gamma}\, .
\end{equation}
Equation \eqref{injectivitystructure} tells us something about the injectivity property of $I$. Suppose that $I$ would be injective on the subspace corresponding to $\mathcal{P}_i$ when we interpret it as a matrix from the virtual to the physical indices, i.e. there is a left inverse $I^+$ such that $I^+I = \mathcal{P}_i$. (Note that we know from non-topological tensor networks that excitation tensors are generically not injective \cite{Haegeman,Vanderstraeten}. However, they could become injective by blocking them with multiple surrounding ground state tensors. Even if this would not be the case we want our formalism to hold irrespective of the specific Hamiltonian so we can focus on the extreme cases where the excitation tensors are injective.) Acting with $I^+$ on both sides of equation \eqref{injectivitystructure} would give $P_i^{a_\alpha,b_\beta} \propto P_i^{b_\beta}$, which is clearly an inconsisteny. For this reason we consider the more general case
\begin{equation}
I^+I = \Pi^{[x]}_i\, ,
\end{equation}
where we need to determine the coefficients $x^i_{a_\alpha b_\beta}$. Now we get from \eqref{injectivitystructure}
\begin{equation}
\sum_{c_\gamma} x^i_{c_\gamma a_\alpha} P_i^{c_\gamma ,b_\beta} \propto \sum_{c_\gamma} x^i_{c_\gamma b_\beta} P^{c_\gamma ,b_\beta}\,\;\;\forall b_\beta\, .
\end{equation}
This shows that all columns of the matrix $[x]$ are proportional, implying that $\Pi_i^{[x]}$ is rank one, i.e. $x^i_{a_\alpha b_\beta} = \bar{v}^i_{a_\alpha} v^i_{b_\beta}$ and $\sum_{a_\alpha} \bar{v}^i_{a_\alpha}v^i_{a_\alpha} = 1$. This is consistent with the fact that it should not be possible to differentiate between the $(i,i^*)$-pair before and after vacuum projection via the physical indices of only $I$ or $I^*$. That $(i,i^*)$ are together in the vacuum state is a global, topological property and this information should not be accessible by looking at only one anyon in the pair.  Anyons also detect each other's presence via nonlocal, topological interactions that can be thought of as a generalized Aharonov-Bohm effect \cite{AharonovBohm}. In section \ref{sec:braiding} we elaborate on how the virtual MPO strings implement this topological interaction. The fact that the virtual MPO strings implement topological interactions is consistent with the rank one property of $\Pi_i^{[x]}$ because the physical indices of a single excitation tensor should not allow one to deduce which virtual MPO string is connected to the tensor on the virtual level.

With the rank one $[x]$ we get from \eqref{injectivitystructure}

\begin{equation}
I^{a_\alpha}P^{a_\alpha,c_\gamma} = \frac{v^i_{a_\alpha}}{v^i_{c_\gamma}}I^{c_\gamma}\, .
\end{equation}
So we can always choose tensors $I^{a_\alpha}$ such that
\begin{equation}
I^{a_\alpha}P^{a_\alpha,c_\gamma} =I^{c_\gamma}\;\;\;\;\text{ and }\;\;\;\; I^+I=\left(\sum_{a_\alpha}P_i^{a_\alpha,x}\right)\left( \sum_{b_\beta} P^{x,b_\beta}_i \right)\, .
\end{equation}
Using this choice for $I$ we can write the equality of equations \eqref{vacuumpair1} and \eqref{vacuumpair3} as

\begin{eqnarray}
\left(I^{a_\alpha}\otimes_a I^{a_\sigma} \right)\tilde{P} & =& \sum_{c,\gamma\lambda} \left( \sum_{bcd,\mu\nu} u^{i,a_\alpha c_\gamma}_{abcd,\mu\nu} \bar{u}^{i^*,c_\lambda a_\sigma}_{abcd,\mu\nu}  w_d\right) I^{c_\gamma} \otimes_c I^{c_\lambda}  \\
 & \equiv & \sum_{c,\gamma\lambda} M_{c\gamma\lambda,a\alpha\sigma} I^{c_\gamma} \otimes_c I^{c_\lambda} \, , \label{Mmatrix}
\end{eqnarray}
where we used $\otimes_a$ to denote the tensor product of two anyonic excitation tensors connected with the single block MPO $a$ and introduced the matrix $M$. We get from \eqref{Mmatrix}
\begin{equation}
\left(\sum_{a\alpha\sigma} y_{a\alpha\sigma} I^{a_\alpha}\otimes_a I^{a_\sigma} \right)\tilde{P} = \sum_{c\gamma\lambda}\left(\sum_{a\alpha\sigma} M_{c\gamma\lambda, a\alpha\sigma}y_{a\alpha\sigma}\right) I^{c_\gamma} \otimes_c I^{c_\lambda}
\end{equation}
So in order for the $(i,i^*)$ pair to be in the vacuum we should choose $y_{a\alpha\sigma}$ to be an eigenvector of $M$:
\begin{equation}
\sum_{a\alpha\sigma} M_{c\gamma\lambda, a\alpha\sigma}y_{a\alpha\sigma}  = y_{c\gamma\lambda}\, .
\end{equation}
Because $\tilde{P}$ is a projector the matrix $M$ is also a projector and therefore has eigenvalues one or zero. Note that in general the vector $y_{c\gamma\lambda}$ will be entangled in the indices $\gamma$ and $\lambda$. However, this is purely `virtual' entanglement that cannot be destroyed by measurements on only one anyon in the pair because of the rank-one injectivity structure of $I$. 

As a final remark we would like to stress that we only looked at the universal properties of the anyonic excitation tensors. These tensors of course also contain a lot of degrees of freedom that one needs to optimize over using a specific Hamiltonian in order to construct eigenstates of the system. This one can do using similar methods as for non-topological PEPS \cite{Haegeman,Vanderstraeten}.

\subsection{Ground states on the torus and the $S$ matrix}\label{sec:Smatrix}

\begin{figure}
  \centering
    \includegraphics[width=0.35\textwidth]{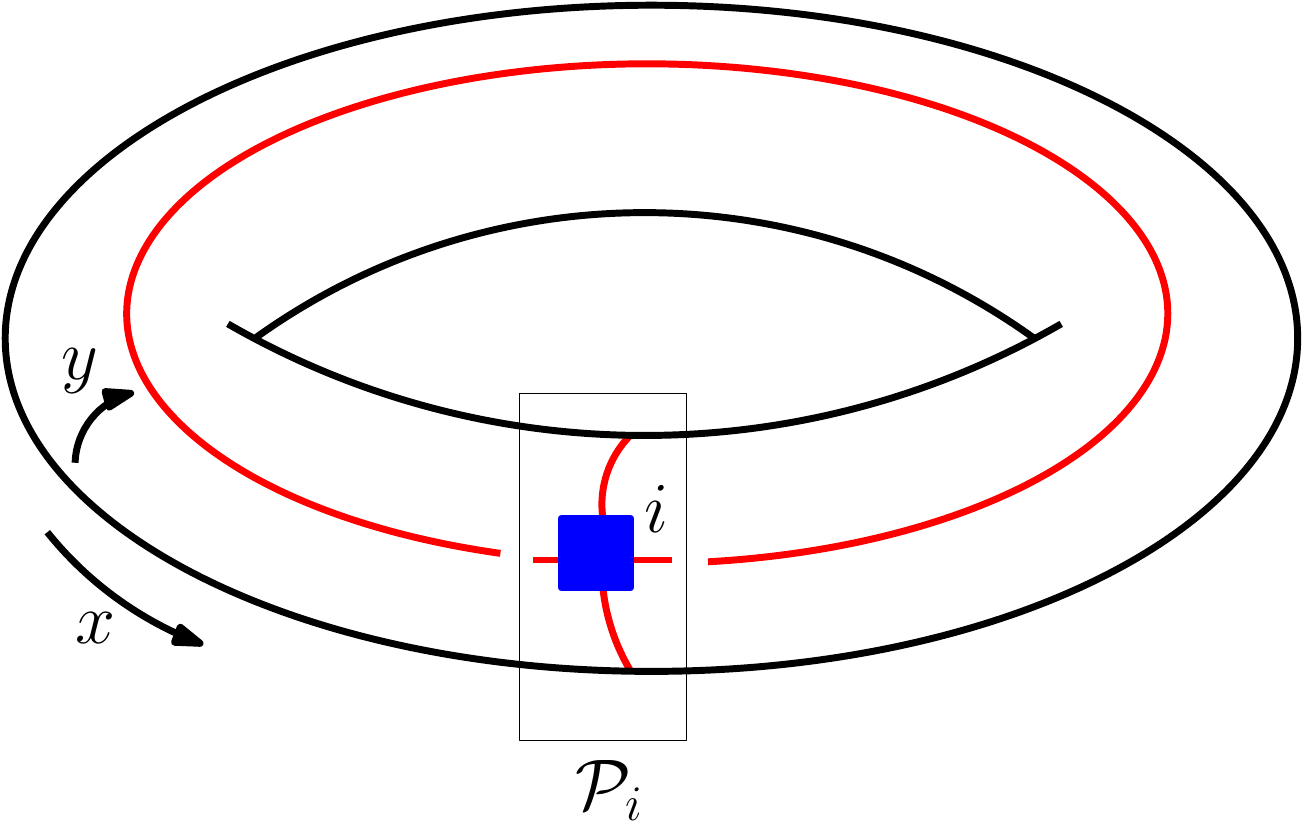}
\caption{A schematic representation of the minimally entangled state $\ket{\Xi}^x_i$ with anyon flux $i$ through the hole in the $x$-direction. It is obtained by placing the projector $\mathcal{P}_i$ on the virtual level of the tensor network on the torus along the non-contractible loop in the $y$-direction and connecting the open indices with a MPO along the $x$-direction.}
\label{fig:MES}
\end{figure}

The projectors $\mathcal{P}_i$ automatically allow one to construct the Minimally Entangled States (MES) on a torus \cite{MES};  one can simply put $\mathcal{P}_i$ along the non-contractible loop in the $y$-direction and close the `inner' and `outer' indices of $\mathcal{P}_i$ with an MPO along the non-contractible loop in the orthogonal $x$-direction. See figure \ref{fig:MES} for a schematic representation. The resulting structure on the virtual level of the tensor network can be moved around freely because of the pulling through property and is therefore undetectable via local operators, implying we have constructed a ground state $\ket{\Xi^x_i}$ with an anyon flux of type $i$ threaded through the hole in the $x$-direction. A similar construction also allows one to construct a MES $\ket{\Xi^y_i}$ with an anyon flux through the hole in the $y$-direction. Since for $\ket{\Xi^x_i}$ $\mathcal{P}_i$ lowers the rank of the reduced density matrix of a segment of the torus obtained by cutting along two non-contractible loops in the $y$-direction it indeed implies (for fixed-point models) that we have minimized the entanglement entropy. In \cite{MES} the topological entanglement entropy for such a bipartition in a MES $\ket{\Xi^x_i}$ was found to be $\gamma_i = 2(\log D - \log d_i)$, where $D$ is the so-called total quantum dimension and $d_i$ is the quantum dimension of anyon type $i$. The PEPS construction then shows that the topological entanglement entropy for any bipartition in a low-energy excited state with a contractible boundary surrounding an anyon in sector $i$ will be given by $\gamma'_i = \log D - \log d_i$.

This identification of the MES gives direct access to the $S$ matrix, which is defined as the unitary matrix that implements the basis transformation from one minimally entangled basis $\{\ket{\Xi^x_i} \}$ to the other $\{\ket{\Xi^y_i} \}$. The advantage of the MPO-injectivity formalism is that we can compute the $S$ matrix in a way that does not scale with the system size. For this we take a single PMPO $\tilde{P}_{C_4}$ with four tensors and use it to construct the smallest possible `torus' by contracting the virtual indices along the $x$ direction and the ones along the $y$ direction. Defining $T_i^x$ as the vector obtained by putting central idempotent $\mathcal{P}_i$ along the $y$-direction on the virtual level of the minimal torus and $T_i^y$ by putting $\mathcal{P}_i$ along the $x$-direction we then have
\begin{equation}
T_i^y = \sum_{j = 1} S_{ij} T_j^x\, .
\end{equation}
We have numerically verified the validity of this expression for all examples below and found that it indeed holds.

Recall that the central idempotents can be written as a sum of simple idempotents
\begin{equation}
\mathcal{P}_i = \sum_{a=1}^{D_i}\sum_{\alpha = 1}^{d_{i,a}} P^{a_\alpha}_i\, ,
\end{equation}
where each of the $P^{a_\alpha}_i$ satisfies $P^{a_\alpha}_iP^{b_\beta}_i = \delta_{a,b}\delta_{\alpha,\beta}P^{a_\alpha}_i$ and $P^{a_\alpha \dagger}_i = P_i^{a_\alpha}$. In principle one could use each of the $P^{a_\alpha}_i$ to construct a ground state on the torus, in a similar way as explained above for $\mathcal{P}_i$. For the examples below we have numerically verified that each $P^{a_\alpha}_i$ for fixed $i$ gives the same ground state, implying that the ground state degeneracy on the torus is indeed given by the number of central idempotents. 

\subsection{Topological spin}\label{sec:topspin}

Even in the absence of rotational symmetry an adiabatic $2\pi$ rotation of the system should not be observable. Normally, we would conclude from this that the $2\pi$ rotation acts as the identity times a phase (called the Berry phase in continuous systems \cite{Berry}) on the total Hilbert space: $R(2\pi) = e^{i\theta}\mathds{1}$. However, the existence of topological superselection sectors changes this conclusion \cite{WickWightmanWigner}. Because there are no local, i.e. physical, operators that couple states in different sectors the $2\pi$ rotation could produce a different phase $e^{i2\pi h_i}$ in each sector and still be unobservable. The number $h_i$ in a particular sector is generally called the \emph{topological spin} of the corresponding anyon.

To see this kind of behavior in MPO-injective PEPS it is important to realize that to define a $2\pi$ rotation one has to specify a specific (discrete) path of states, in the same way one has to define a continuous family of states in order to obtain a Berry phase. For example, in the case of a square lattice we can define the path using 4 successive rotations over $\pi/2$. When dealing with a non-regular lattice we have to use a family of different lattices along the path. We can now consider a region of MPO-injective PEPS in the sector defined by $\mathcal{P}_i$. This region has an open internal MPO-index along the boundary that cannot be moved freely. We show that one can obtain the topological spin associated to sector $\mathcal{P}_i$ by rotating the PEPS on a finite region while keeping the virtual boundary conditions fixed. After a $2\pi$ rotation $\mathcal{P}_i$ surrounding the PEPS region is transformed to

\begin{align}\label{rotation}
 \vcenter{\hbox{
 \includegraphics[width=0.18\linewidth]{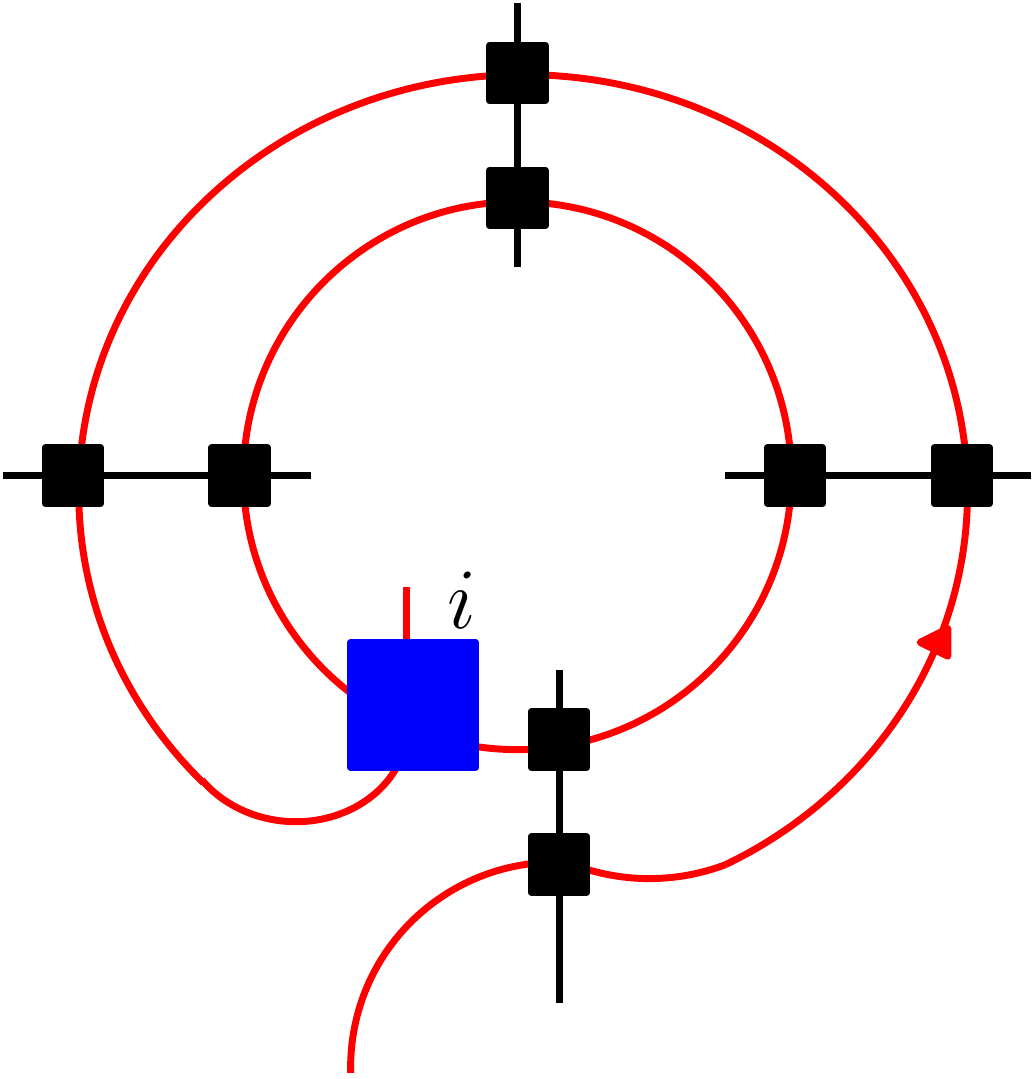}}}
\end{align}
Equation \eqref{rotation} can be interpreted as $\mathcal{P}_i$ acting on the matrix $\mathcal{R}_{2\pi}$ defined by

\begin{align}
 \mathcal{R}_{2\pi} = \vcenter{\hbox{
 \includegraphics[width=0.16\linewidth]{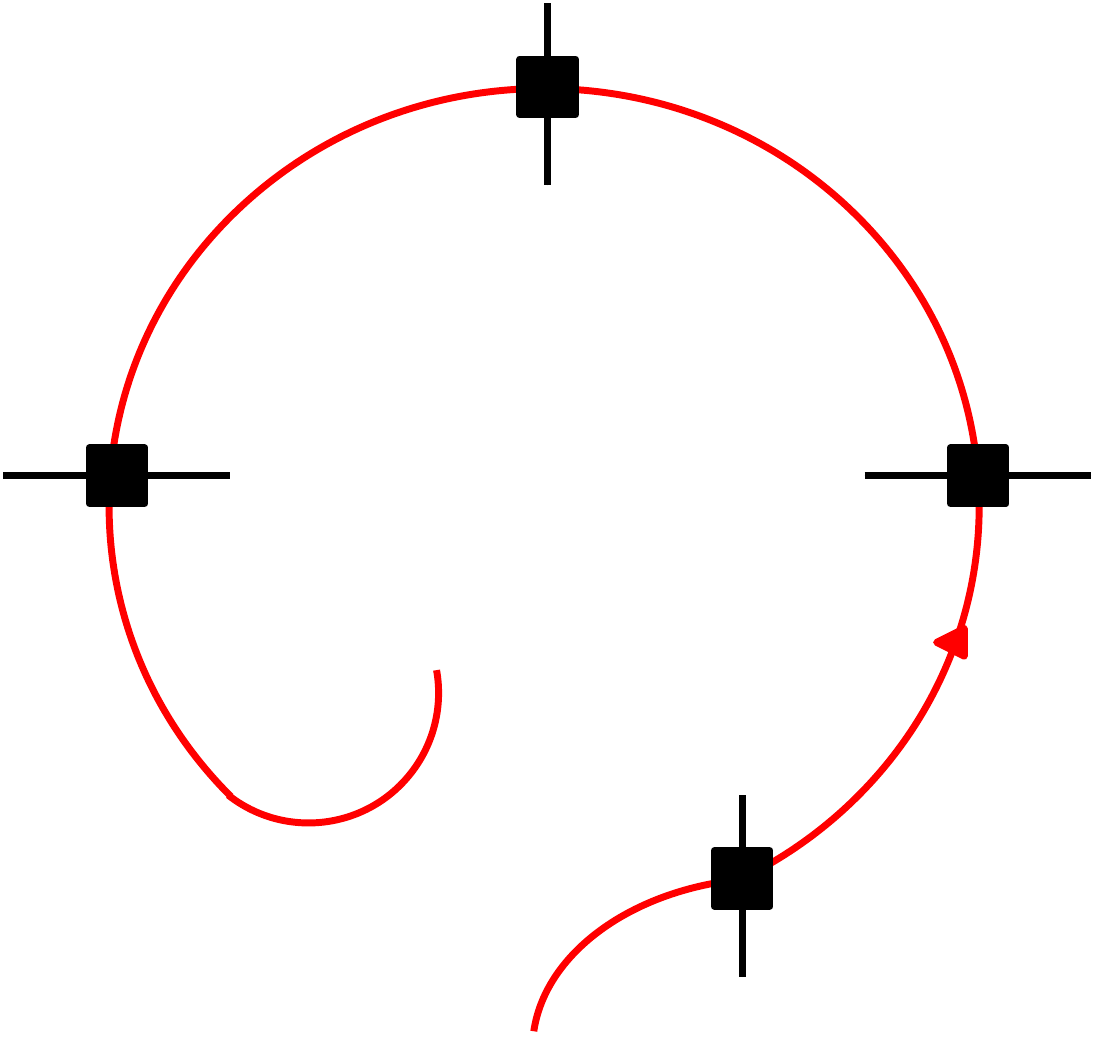}}}
\end{align}
By looking at the graphical expression for $\mathcal{R}_{2\pi}^\dagger\mathcal{R}_{2\pi}$

\begin{align}
 \mathcal{R}_{2\pi}^\dagger\mathcal{R}_{2\pi} = \vcenter{\hbox{
 \includegraphics[width=0.17\linewidth]{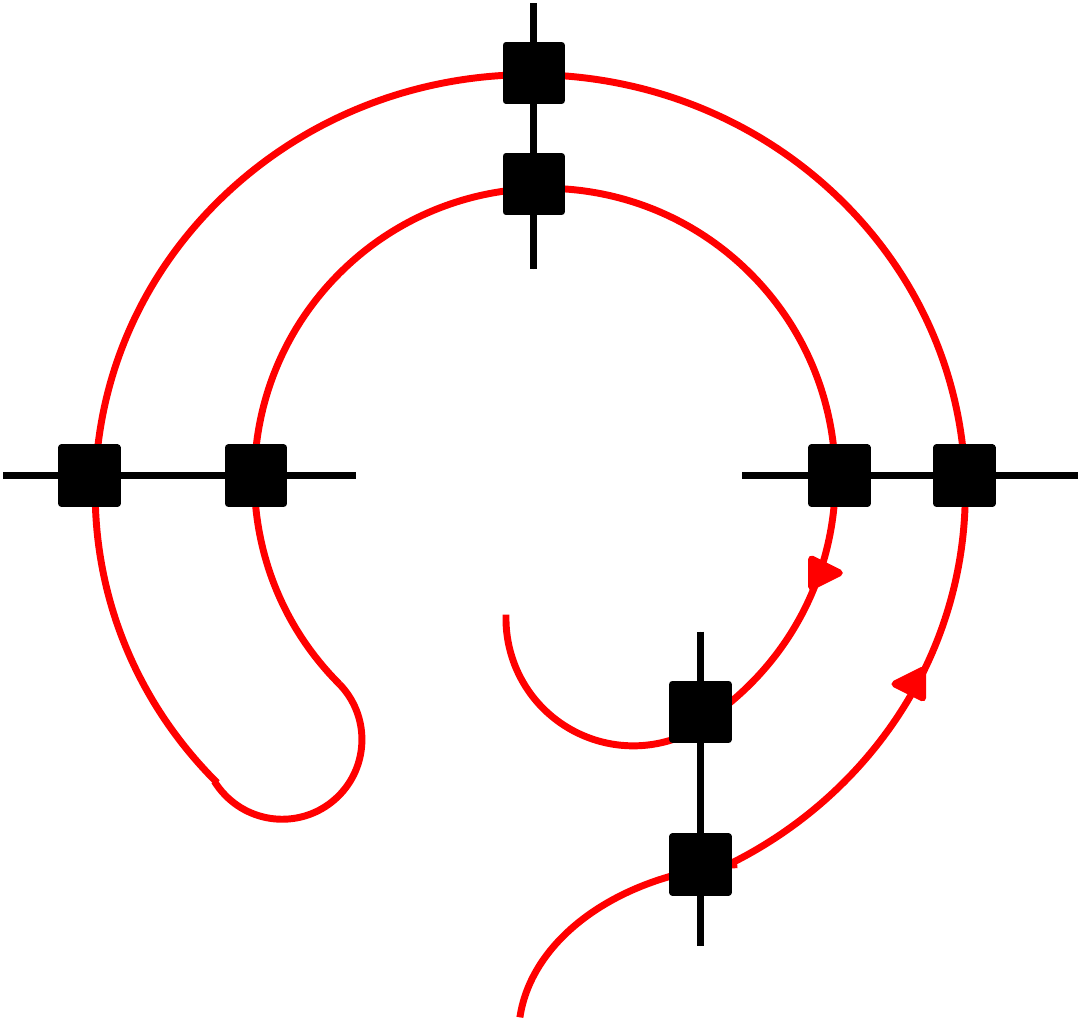}}} \, ,
\end{align}
one can easily see by embedding it in a MPO-injective PEPS and using the pulling through property (\ref{pullingthrough2}) that we can reduce it to a trivial action, implying $\mathcal{R}_{2\pi}$ is unitary on the relevant subspace. Using the zipper condition \eqref{zippercondition}, the pivotal property \eqref{pivotalnew} and again the pulling through property one can show via some graphical calculus that the following identity holds
\begin{equation}
\mathcal{R}^\dagger_{2\pi} A_{abcd,\mu\nu} \mathcal{R}_{2\pi} = A_{abcd,\mu\nu}
\end{equation}
on the relevant subspace for all elements $A_{abcd,\mu\nu}$ in the algebra. Schur's lemma thus allows us to conclude that $\mathcal{R}_{2\pi} = \sum_i \theta_i\mathcal{P}_i$, with $\theta_i$ some phases because of the unitarity of $\mathcal{R}_{2\pi}$. We thus arrive at the desired result, i.e.
\begin{equation}
\mathcal{P}_i \mathcal{R}_{2\pi} = \theta_i \mathcal{P}_i\, ,
\end{equation}
where $\theta_i = e^{i2\pi h_i}$ gives the topological spin of the anyon in sector $i$.

\subsection{Fusion} \label{sec:fusion}

We can associate an algebra, called the fusion algebra, to the topological sectors. Consider a state which has the ground state energy everywhere except for two spatially separated regions. Using operators that surround one of those individual regions, we can measure the topological charge within both regions. Say these measurements reveal topological charges $i$ and $j$. By considering the two different regions and a part of the ground state between them as one big region and using loop operators surrounding this bigger region, we can similarly measure the total topological charge. This measurement will typically have several outcomes, i.e. the total state is in a superposition of different topological sectors. The sectors appearing in this superposition for every $i$ and $j$ determine the integer rank three tensor $\mathscr{N}_{ij}^k$ and we formally write the fusion algebra as $i\times j = \sum_{k}\mathscr{N}_{ij}^k\,k$. It is also clear that this algebra is by construction commutative, i.e. $\mathscr{N}_{ij}^k = \mathscr{N}_{ji}^k$. Assuming that all states in the same topological sector are connected via local operators we should be able to move an anyon $i$ from one place to another using a string operator \cite{ReadChakraborty,Kitaev03,Kitaev06,LevinWen05}. Applying this string operator to a region that does not contain an excitation will create a pair $(i,i^*)$ of anyons, where $i^*$ is the unique dual/anti particle of anyon $i$. From this we see that $\mathscr{N}_{ij}^1 = \delta_{j,i^*}$.

\begin{figure}[H]
  \centering
     (a) \includegraphics[height=0.28\textwidth] {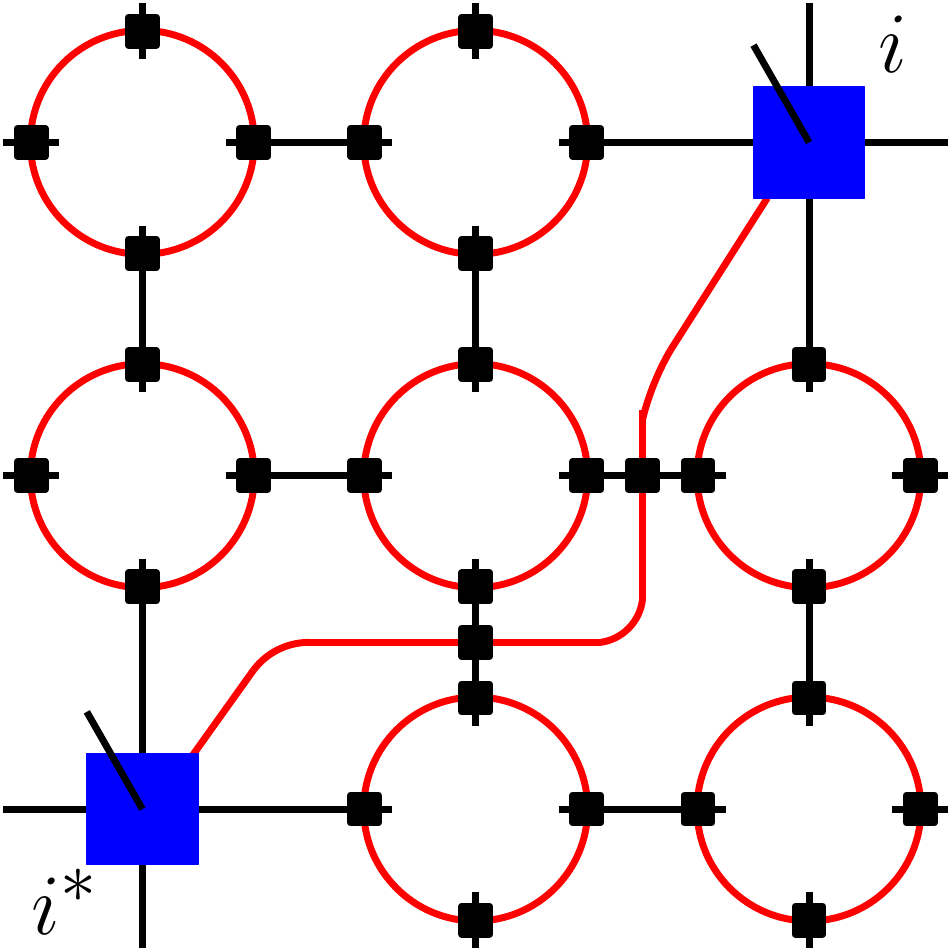}
\qquad     (b) \includegraphics[height=0.28\textwidth]{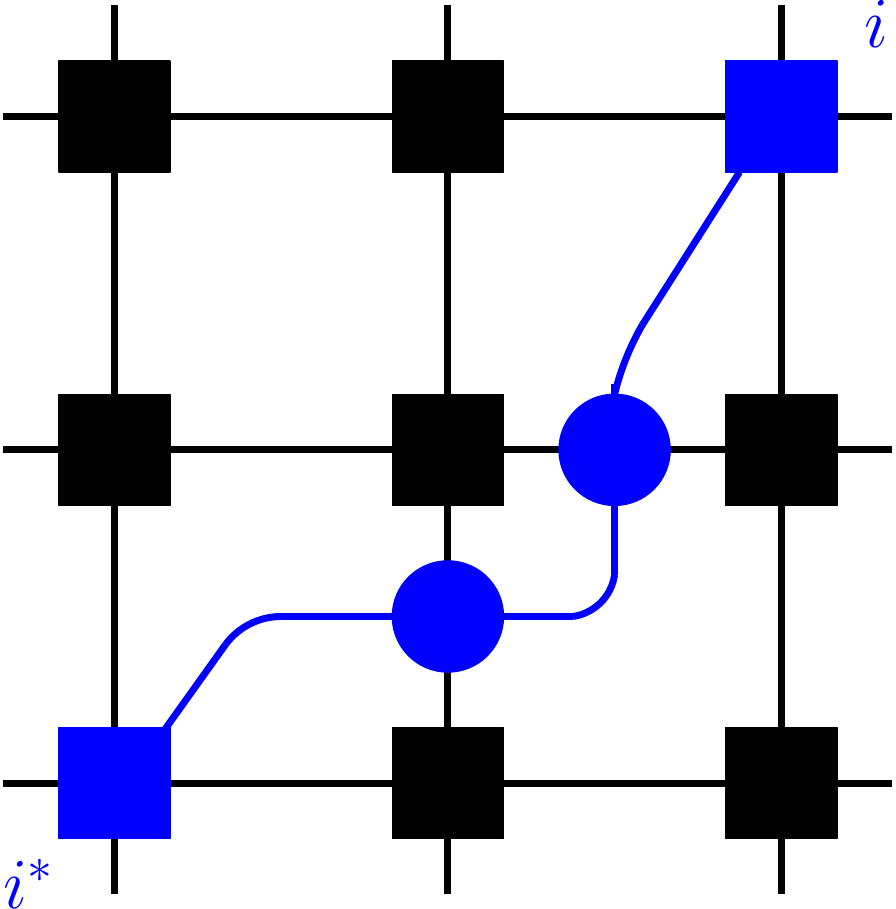}

\caption{(a) The original tensor diagram for the ground state with an anyon pair $(i,i^*)$ in the corners of the lattice. (b) Simplified tensor diagram for the state. In the remainder of the paper we will only use simplified diagrams. The ground state tensors are denoted by black squares and the physical indices are omitted. The blue squares describe an anyon of type $i,i^*$ living on the respective sites. The blue tensors are supposed to be invariant under the virtual action of the idempotent corresponding to the label $i$ or $i^*$. We use blue and red to denote sites containing an anyon, whereas other colors such as grey are reserved for fusion product of MPOs or anyons.}
\label{fig:groundstateandexcitation}
\end{figure}

\begin{figure}[H]
  \centering
 \includegraphics[width=0.42\linewidth]{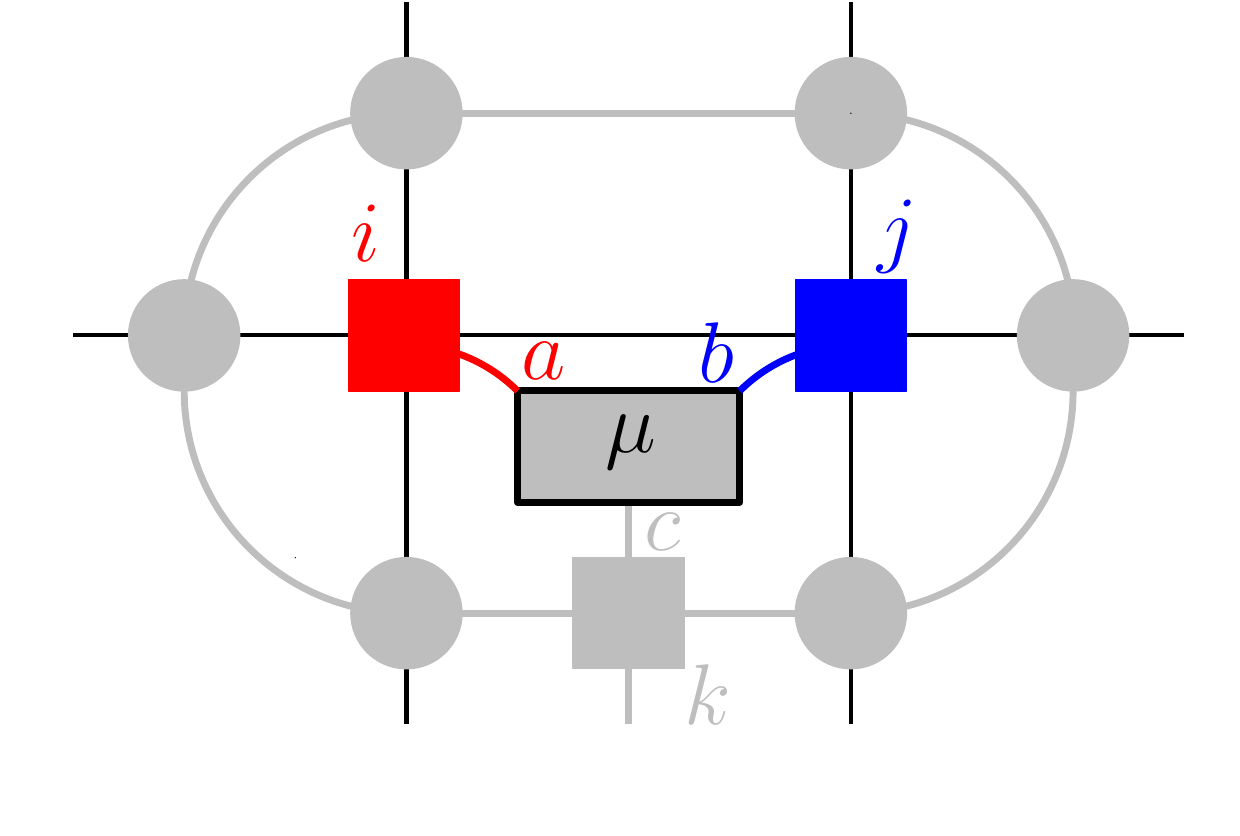}
    \caption{The procedure to determine the fusion product of two anyons in the new, simplified graphical notation (see also figure \ref{fig:groundstateandexcitation}). The anyons are given by the red and blue idempotents $\mathcal{P}_i, \mathcal{P}_j$. We first fuse their outgoing strings $a,b$ to all possible products $c$. We can now measure the fusion product of the anyons by projecting the result on the subspaces determined by the idempotents $\mathcal{P}_k$. The idempotents that give rise to a non zero projection correspond to the possible fusion products $(k)$ of the red $(i)$ and blue $(j)$ anyons. Importantly, the sum over all grey idempotents $\mathcal{P}_k$ acts as the identity on the virtual labels.}
 \label{fig:fusionidempotents}
\end{figure}

This fusion algebra is very easily and explicitly realized in MPO-injective PEPS. In the simplest case, we just place two single-site idempotents $\mathcal{P}_i$ and $\mathcal{P}_j$, next to each other on neighboring lattice sites. We can then fuse together the MPO strings emanating from $\mathcal{P}_i$ and $\mathcal{P}_j$ into one string. Looking at an annular ground state region surrounding the two anyons and using similar reasoning as in section \ref{subsec:ansatz} we find that the sum of all idempotents $\sum_k \mathcal{P}_k$ surrounding both anyons acts as a resolution of the identity on the relevant subspace. We can easily determine the subspaces $\mathcal{P}_k$ on which the combination of both anyons are supported. These subspaces correspond to the possible fusion products of $\mathcal{P}_i$ and $\mathcal{P}_j$. We illustrate this in figure \ref{fig:fusionidempotents}, which uses a new, simplified diagrammatic notation that is defined in figure \ref{fig:groundstateandexcitation}. From now on we shall denote the ground state by a tensor network consisting of black colored sites, omitting the physical indices. A site that contains an excitation is colored blue or red. Note that the procedure of figure \ref{fig:fusionidempotents} does not allow one to determine fusion multiplicities, i.e. it only tells whether $\mathscr{N}_{ij}^k$ is non-zero. The multiplicities --the specific values of $\mathscr{N}_{ij}^k$-- are in general harder to obtain directly since they arise from the number of linearly independent ways the MPO strings emanating from the idempotents can be connected on the virtual level. One could of course also just calculate the fusion multiplicities from the $S$ matrix using the Verlinde formula \cite{verlinde}.

Note that a projective measurement of the topological charge in some region via the physical PEPS indices greatly depends on the details of the tensors $A$ and $A'$ used to complete the tensor network. This is to be expected since the physical measurement is determined by the specific microscopic realization of the quantum phase.

\subsection{Braiding} \label{sec:braiding}

When introducing the PEPS anyon ansatz in section \ref{sec:ansatz} we mentioned that anyons detect each others presence in a non-local, topological way. We will now make this statement more precise. To every fixed configuration of anyons in the plane we can associate a collection of quantum many-body states. This set of states forms a representation of the colored braid group. This means that when we exchange anyons or braid them around each other this induces a non-trivial unitary transformation in the subspace corresponding to the configuration.  If there is only one state that we can associate to every anyon configuration then we only get one-dimensional representations. This situation is commonly referred to as Abelian statistics and the anyons are called Abelian anyons. With non-Abelian anyons we can associate multiple orthogonal states to one or more anyon configurations and these will form higher dimensional representations of the colored braid group. 

One can obtain a basis for the subspace associated to a certain anyon configuration by assigning an arbitrary ordering to the anyons and projecting the first two anyons in a particular fusion state. One subsequently does the same for the fusion outcome of the first two anyons and the third anyon. This can be continued until a final projection on the vacuum sector is made. So the degeneracy of an anyon configuration is given by the number of different ways an ordered array of anyons can fuse to the vacuum.

Just as in relativistic field theories there is a spin-statistics relation for anyons, connecting topological spin and braiding. It is expressed by the so-called `pair of pants' relation, which we show graphically using the same set-up as presented in figure \ref{fig:fusionidempotents}:

\begin{align}
\vcenter{\hbox{
 \includegraphics[width=0.17\linewidth]{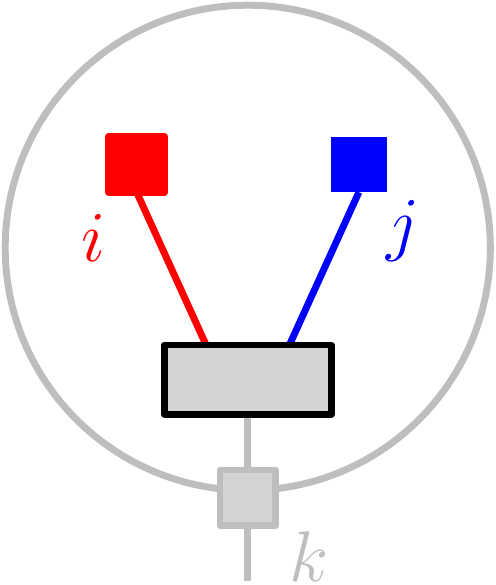}}} \xrightarrow{R}
\vcenter{\hbox{
 \includegraphics[width=0.17\linewidth]{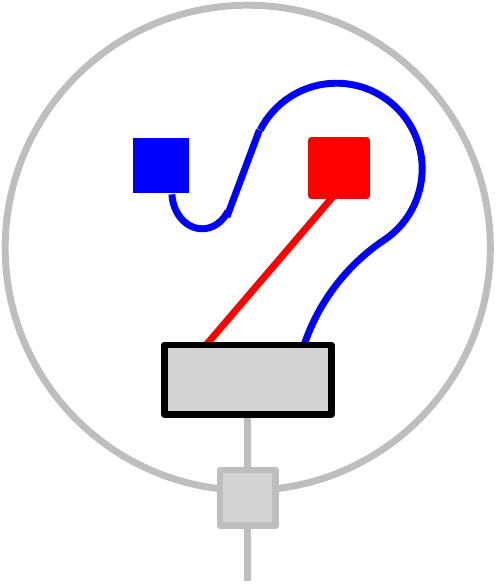}}} \xrightarrow{R}
\vcenter{\hbox{
 \includegraphics[width=0.17\linewidth]{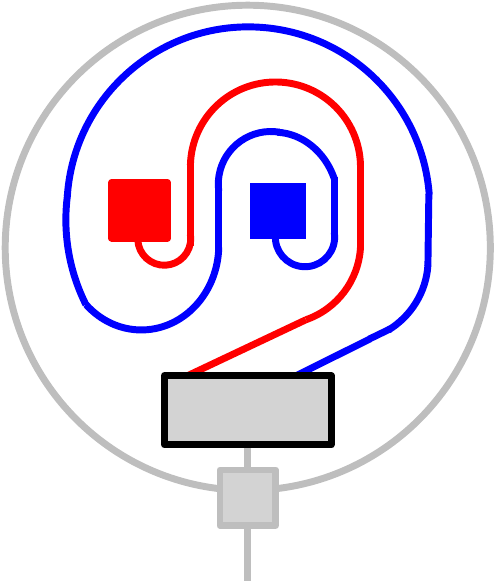}}} \nonumber \\ = 
\vcenter{\hbox{
 \includegraphics[width=0.17\linewidth]{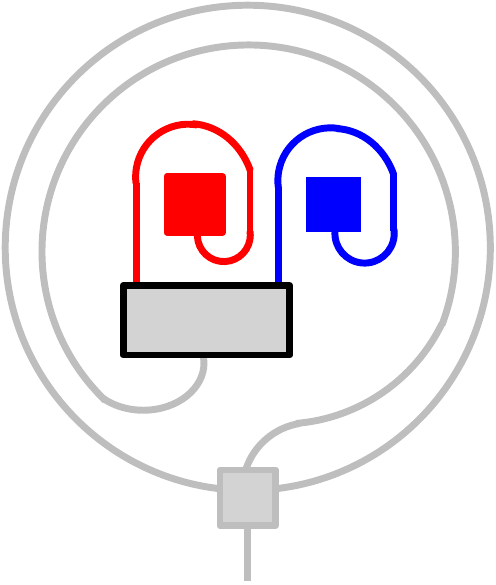}}} = e^{2\pi i(h_k - h_i - h_j)}
\vcenter{\hbox{
 \includegraphics[width=0.17\linewidth]{pants1}}}
\end{align}
The pair of pants relation shows that braiding acts diagonally on two anyons that are in a particular fusion state, which is realized in the figures by the grey idempotent $k$ surrounding $i$ and $j$. Because the topological spins can be shown to be rational numbers \cite{Vafa}, the spin-statistics connection reveals that every anyon configuration provides a representation of the truncated colored braid group, i.e. there exists a natural number $n$ such that $R^{2n} = \mathds{1}$. 

To describe the exchange and braiding of two anyons that are not in a particular fusion state we look for a generalization of the pulling through condition  \eqref{pullingthrough1}, \eqref{pullingthrough2}. The goal is to obtain tensors $\mathcal{R}_{\mathcal{P}_i,b}$ that describe the pulling of a MPO string of type $b$ through a site that contains an anyon corresponding to $\mathcal{P}_i$ according to the defining equation 
\begin{align}\label{Rmatrix}
\vcenter{\hbox{
 \includegraphics[width=0.28\linewidth]{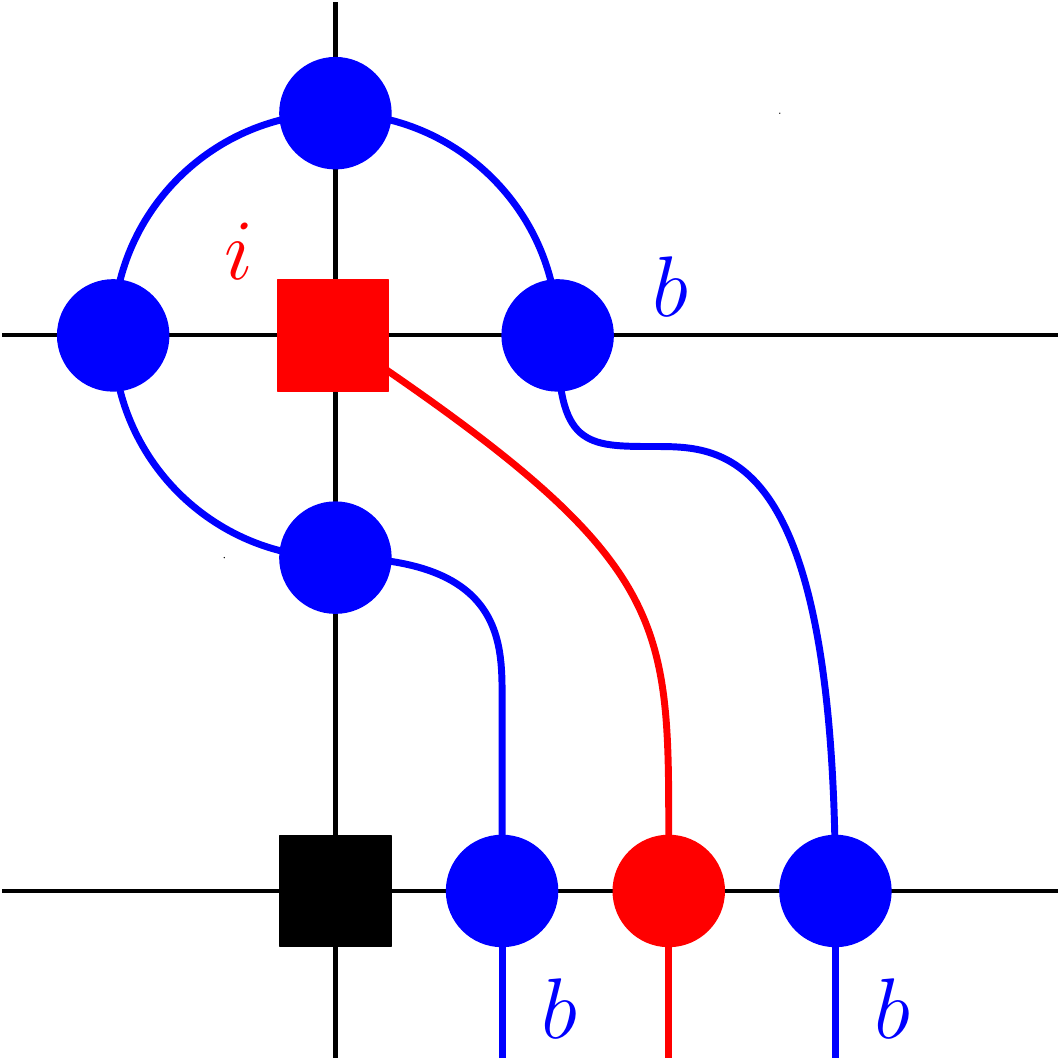}}} =
\vcenter{\hbox{
 \includegraphics[width=0.28\linewidth]{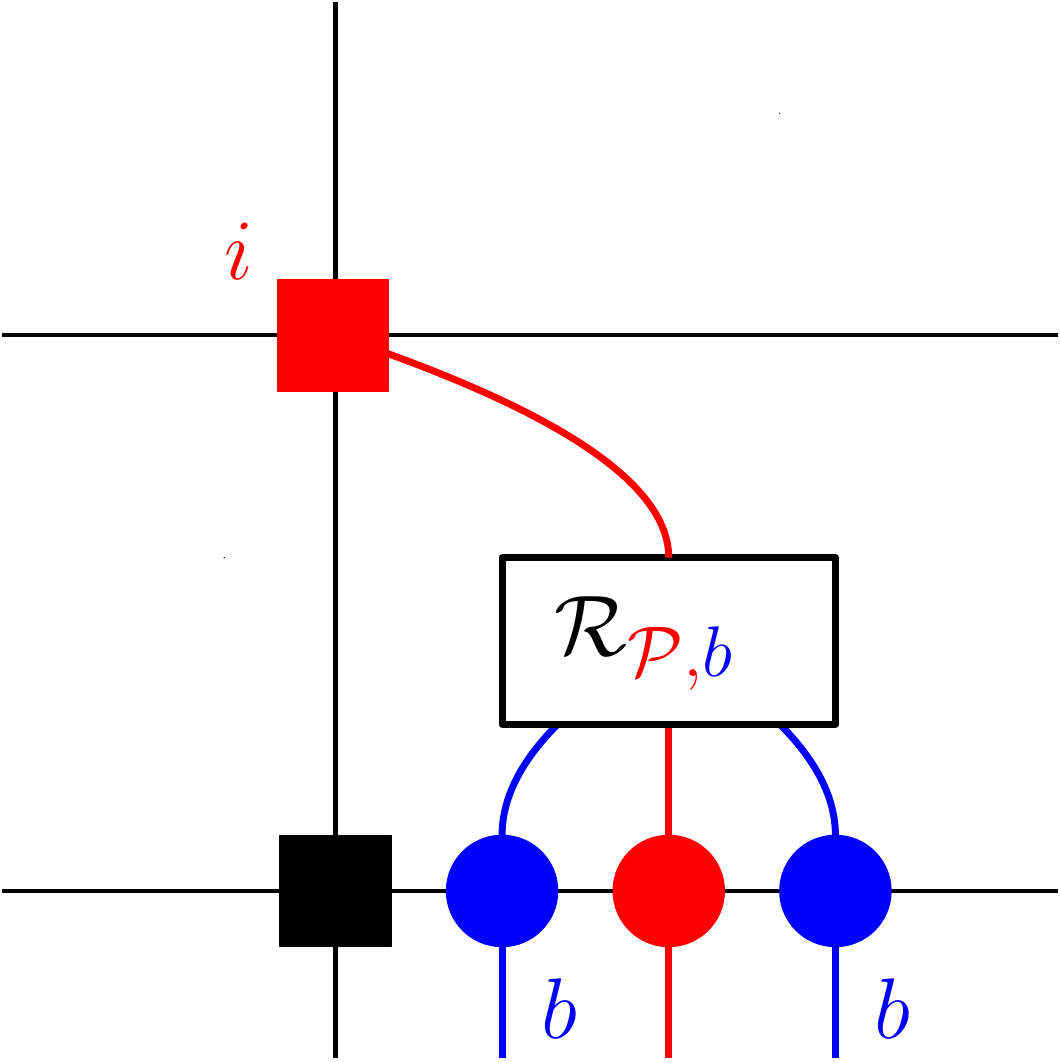}}}
\end{align}
If there is no anyon on the site we consider, i.e. the idempotent on this site is $\mathcal{P}_1$ corresponding to the trivial anyon, the  operator $\mathcal{R}_{\mathcal{P}_i,b}$ is equal to the identity on the MPO indices as follows from the pulling through property (\ref{pullingthrough2}).

While in practice one could solve the equation that determines $\mathcal{R}$ numerically, we can in fact obtain the tensors $\mathcal{R}_{\mathcal{P},b}$ analytically also for a nontrivial idempotent $\mathcal{P}_i$ with $i\neq 1$. We thereto rewrite the left hand side of \eqref{Rmatrix} by using relation \eqref{zippercondition} as follows (Note that we do not depict the required orientations on the indices and that we omit the corresponding gauge transformations $Z_a$ to keep the presentation simple. These issues will also not have to be taken into account for the string-net examples further on.),
\begin{align}\label{4Vs}
\vcenter{\hbox{
 \includegraphics[width=0.28\linewidth]{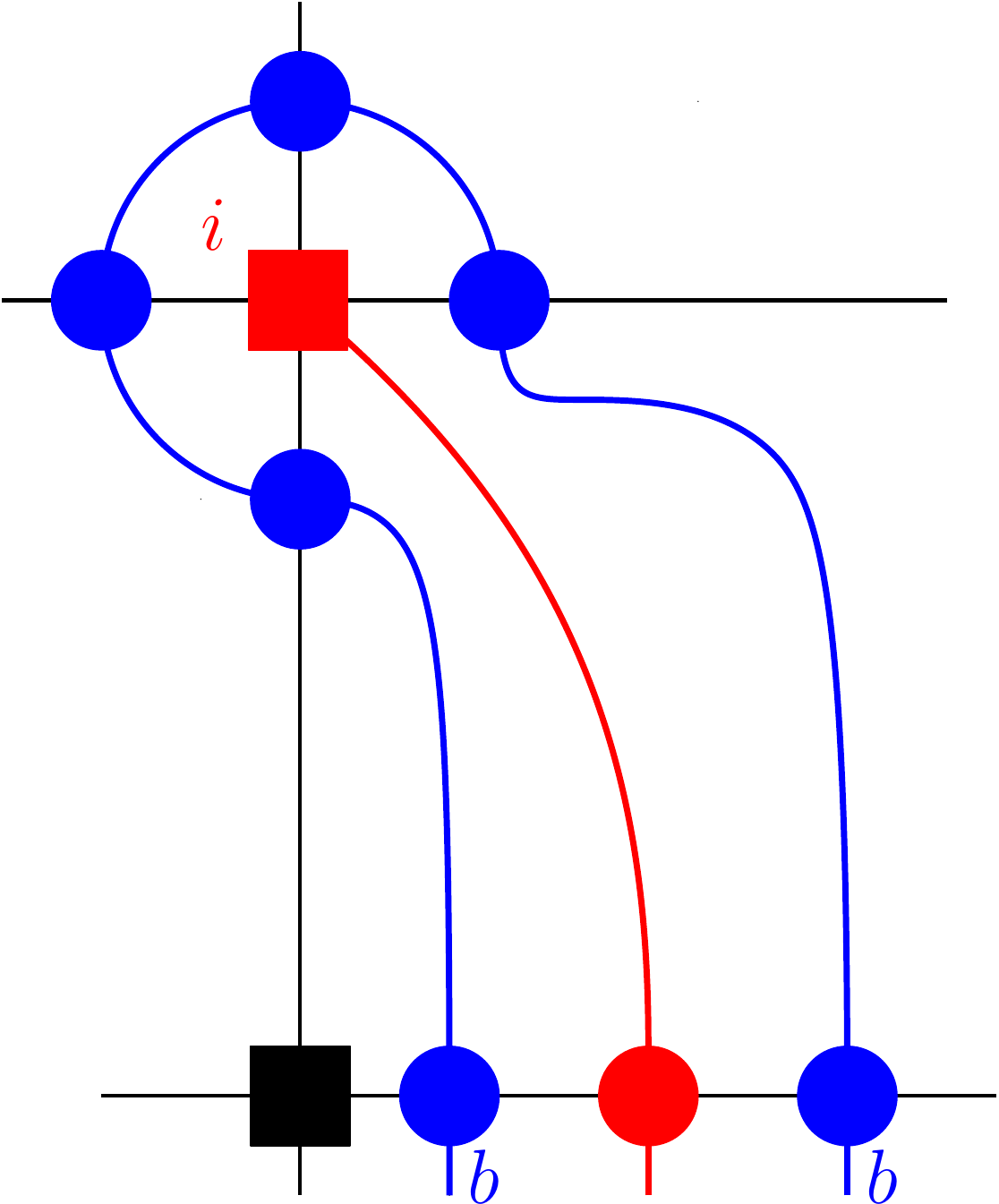}}} = \sum_{acd\mu\nu}
\vcenter{\hbox{
 \includegraphics[width=0.28\linewidth]{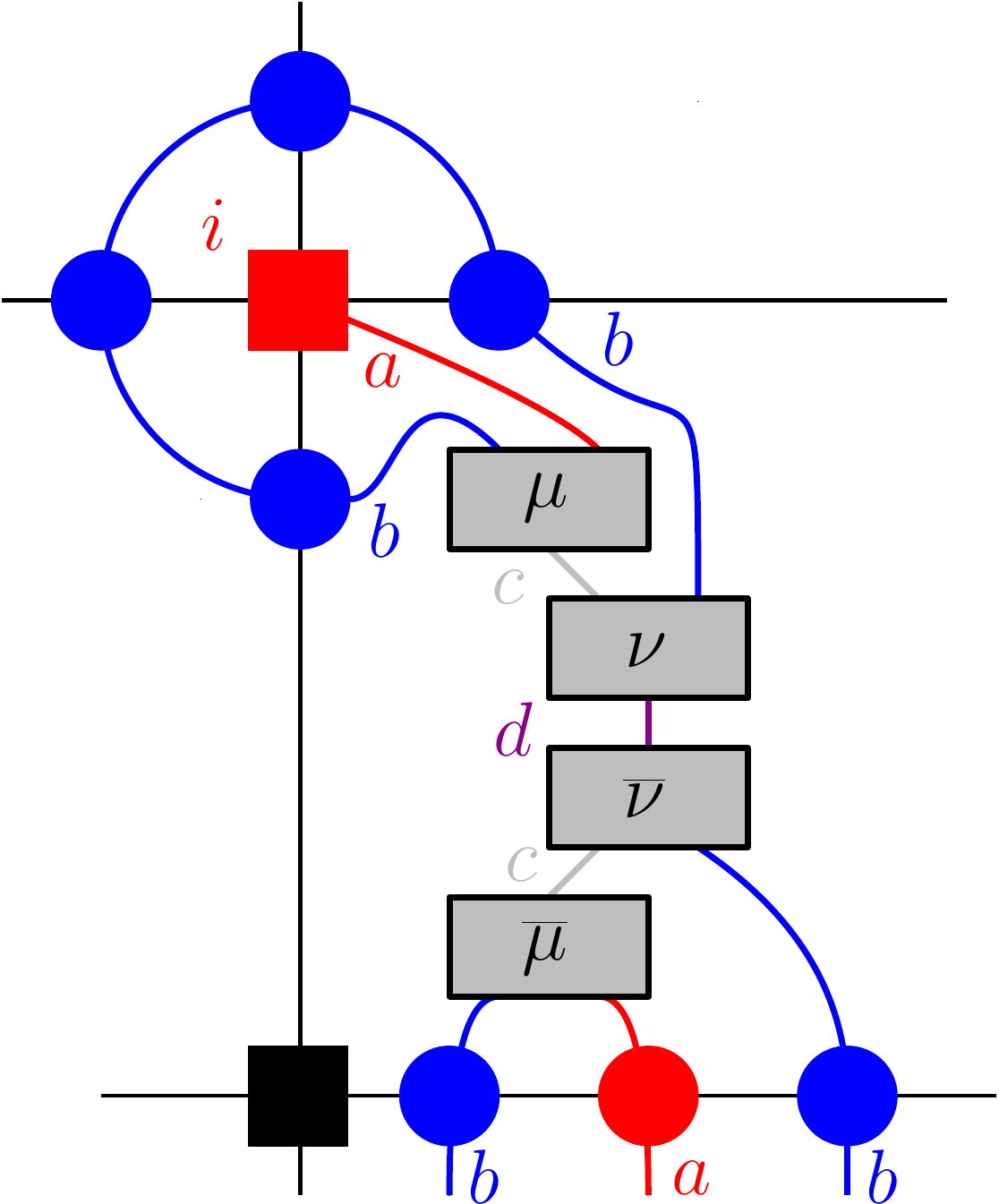}}}
\end{align}
If by $\mathcal{P}_iA_{abcd}$ we denote the multiplication of $\mathcal{P}_i$ and $A_{abcd}$ in the anyon algebra defined in \eqref{eq:multiplicationalgebra}, we find that
\begin{align} \label{braidend}
\vcenter{\hbox{
 \includegraphics[width=0.28\linewidth]{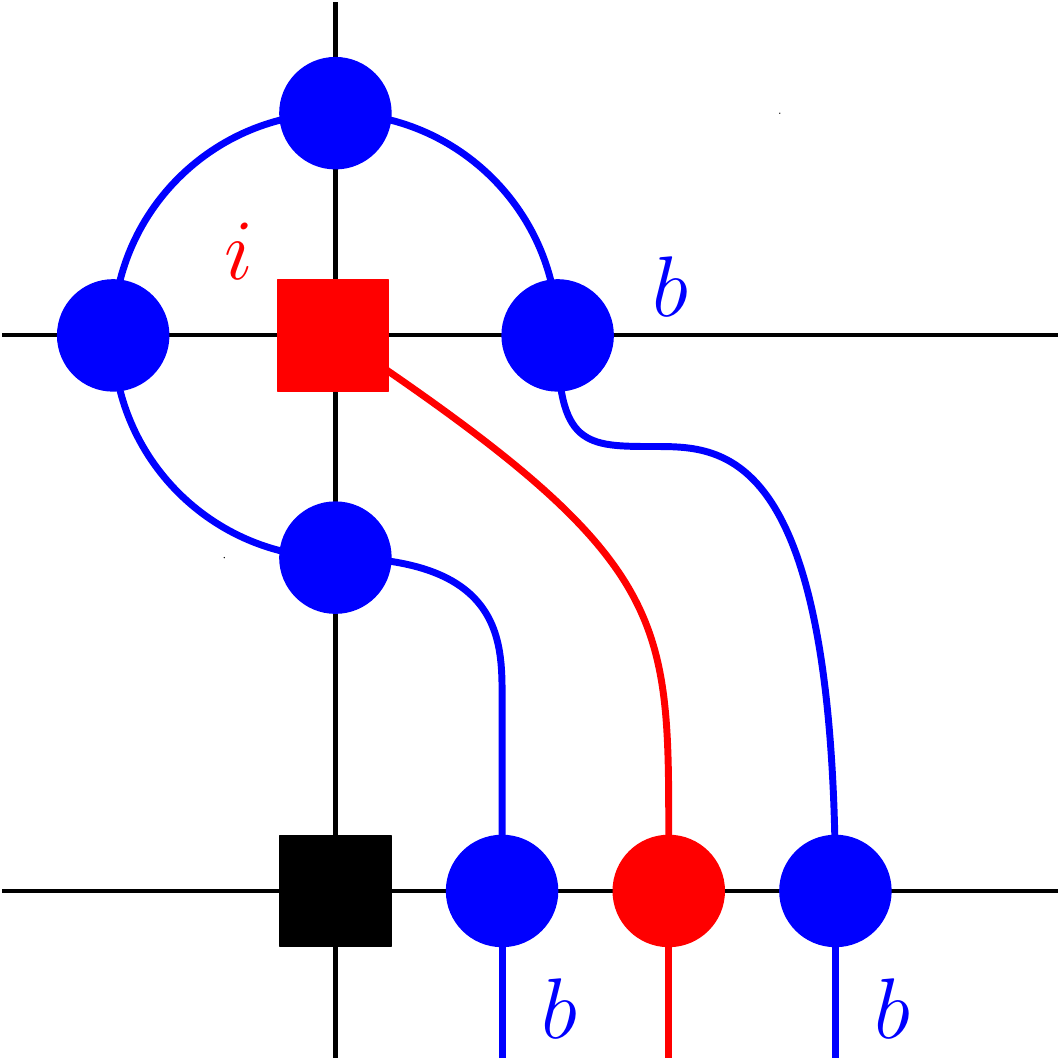}}} =  \sum_{acd\mu\nu}
\vcenter{\hbox{
 \includegraphics[width=0.28\linewidth]{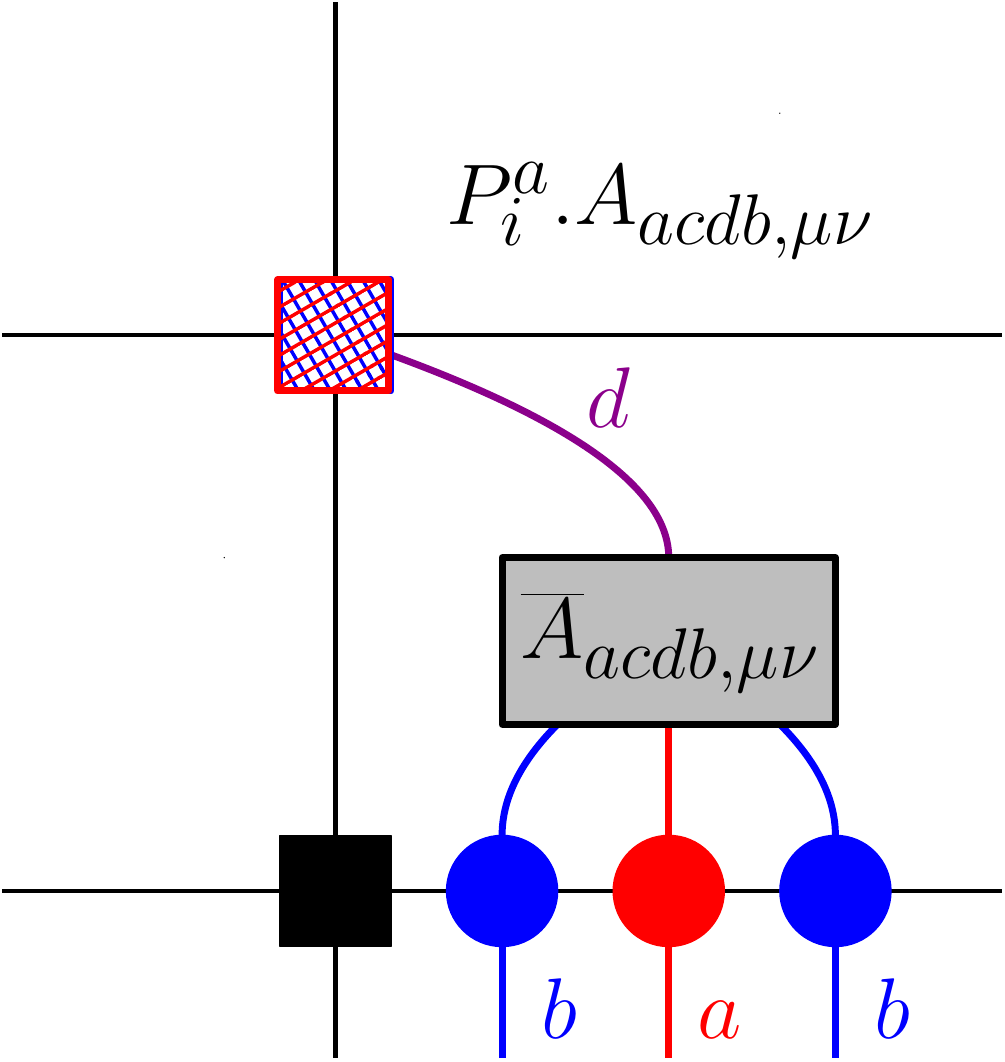}}}
\end{align}
With a slight abuse of notation, the grey rectangle containing $\overline{A}_{acdb,\mu\nu}$ denotes a similar tensor as the algebra object $A_{acdb,\mu\nu}$ in \eqref{algebraobject}, but without the MPO tensors,
$$
\vcenter{\hbox{
 \includegraphics[width=0.35\linewidth]{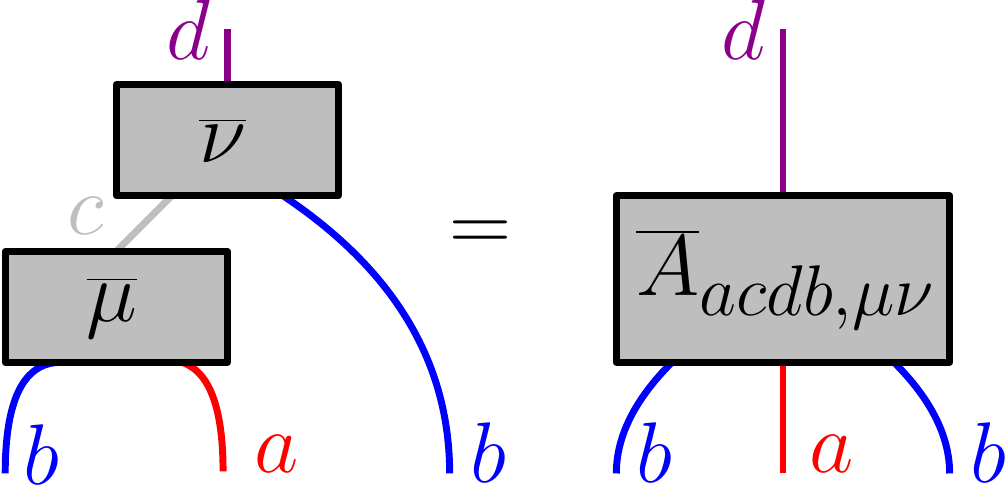}}}.
$$
The tensors $\mathcal{P}_i.A_{acdb,\mu\nu}=P_i^a.A_{acdb,\mu\nu}$ [see Eq.~(\ref{eq:centralidempotentdecomposition})] can easily be determined using the structure constants. Note that all tensors $\mathcal{P}_i.A_{acdb}$  are supported on the subspace determined by $\mathcal{P}_i$, hence they all correspond to the same topological sector. Indeed, braiding an anyon around another one cannot change the topological charges. Remark that after the blue MPO is pulled through the site containing the anyon, the tensor on the site and the braid tensor linking the MPOs are in general entangled, due to the summation over $a,c,d$.

If $\mathcal{P}_i$ is a one dimensional idempotent, the tensor $\mathcal{P}_i.A_{acdb,\mu\nu}$ is only nonzero for a unique choice of $d=a$ and is in that case equal to $\mathcal{P}_i$, up to a constant. Hence, in that case there is no entanglement between the tensor on the site and the tensor that connects the MPOs. 

\begin{figure}[H]
  \centering
    (a)\includegraphics[width=0.3\textwidth]{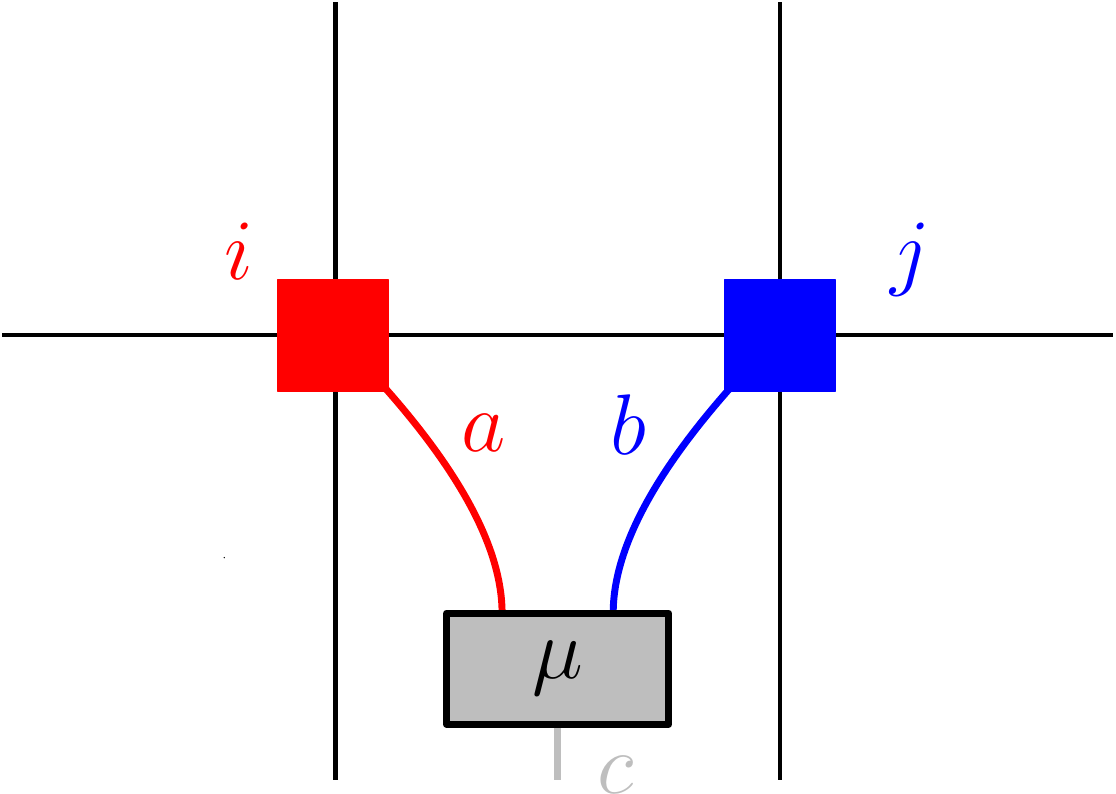}
    (b)\includegraphics[width=0.3\textwidth]{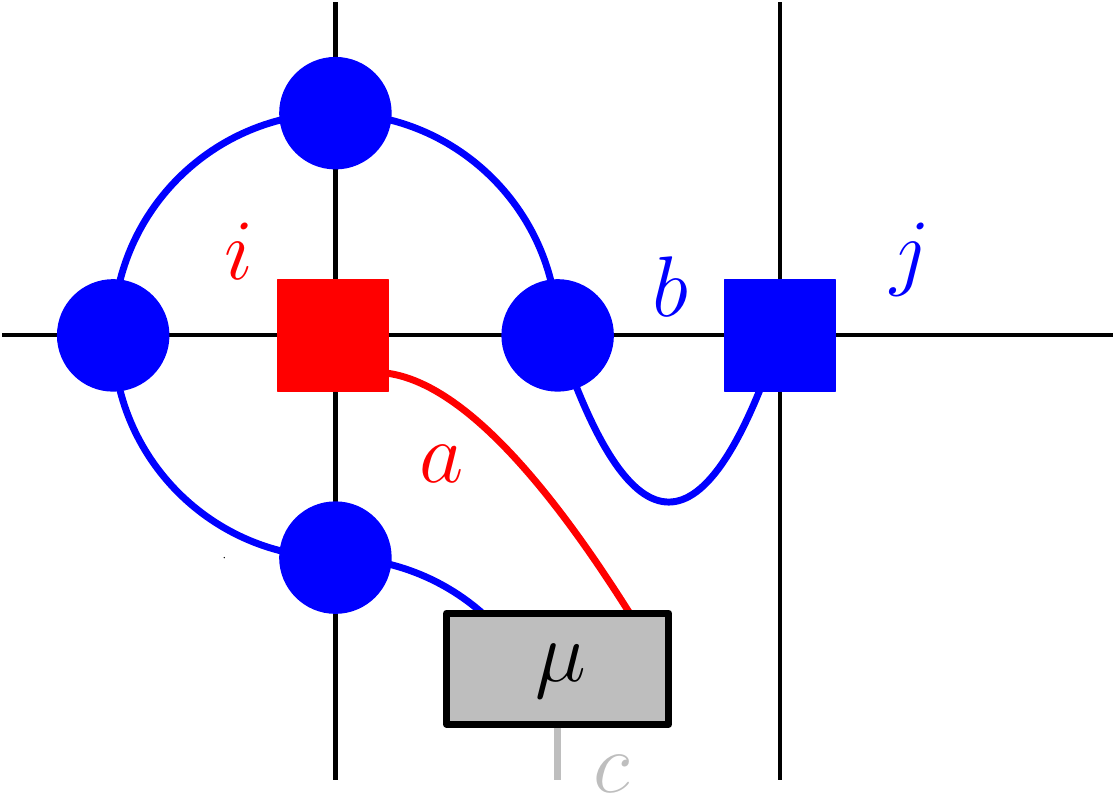}
   \caption{Two anyons, described by idempotents $\mathcal{P}_i,\mathcal{P}_j$, can be fused before exchanging them, as in Figure $(a)$, or after exchanging, as in $(b)$. To compare both diagrams we first use the tensor $\mathcal{R}$ to redraw figure $(b)$. The result is shown in equation \eqref{exchangeR}.}
\label{fig:exchange}
\end{figure}

Once we obtain these tensors $\mathcal{R}_{\mathcal{P}_i,b}$ we know how to resolve the exchange of anyons and we can compute the $R$ matrix (braiding matrix). Suppose we have two anyons, described by idempotents $\mathcal{P}_1, \mathcal{P}_2$ and we want to compare the fusion of these anyons with and without exchanging them. Both situations correspond to figures \ref{fig:exchange}(a) and \ref{fig:exchange}(b) respectively.

All we need to resolve this situation is the tensor $\mathcal{R}_{\mathcal{P}_i,b}$ for all $b$ for which $\mathcal{P}_j$ is non zero. With this tensor we can redraw figure \ref{fig:exchange}(b) in a way similar to the left hand side of \eqref{exchangeR}. It is now clear that the $\mathcal{R}_{\mathcal{P}_i,b}$ tensors encode the $R$ matrices of the topological phase, i.e. the braiding information of the anyonic excitations.

\begin{align}\label{exchangeR}
\vcenter{\hbox{
 \includegraphics[width=0.3\linewidth]{FusionWithBraiding.pdf}}} =
\vcenter{\hbox{
 \includegraphics[width=0.3\linewidth]{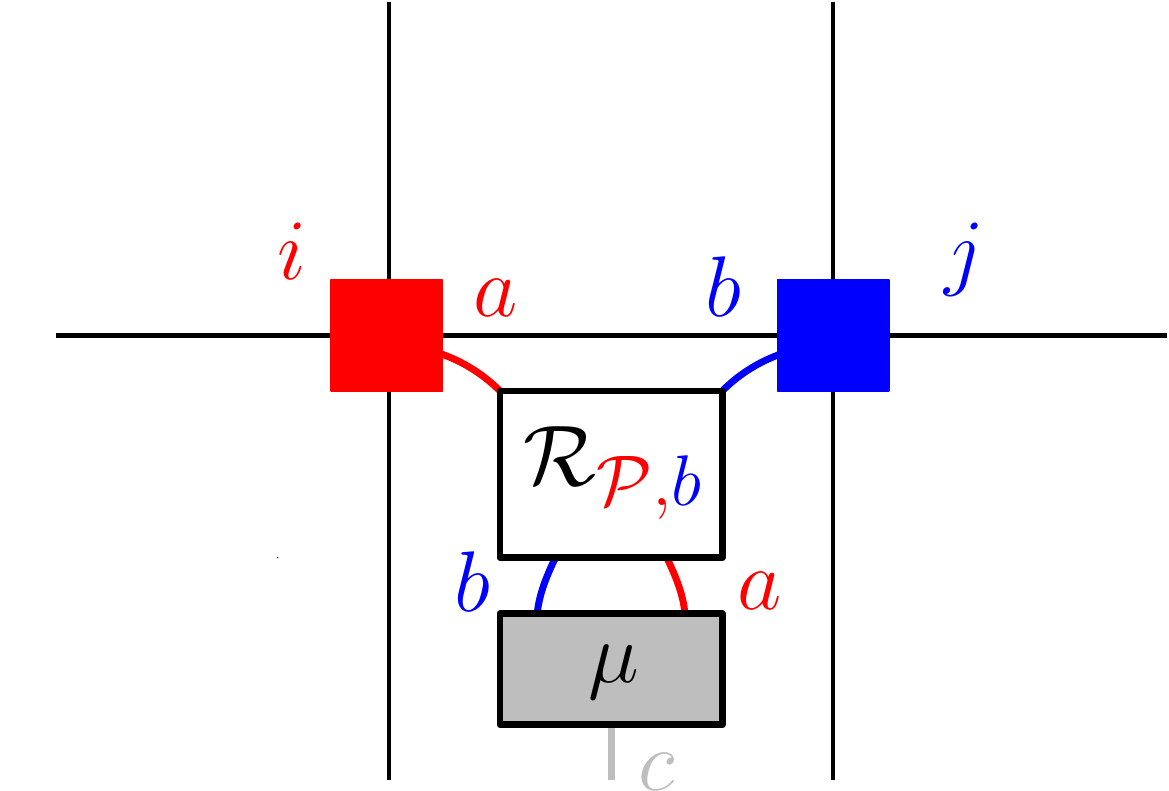}}}
\end{align}

\begin{figure}[H]
  \centering
    (a)\includegraphics[height=0.23\textwidth]{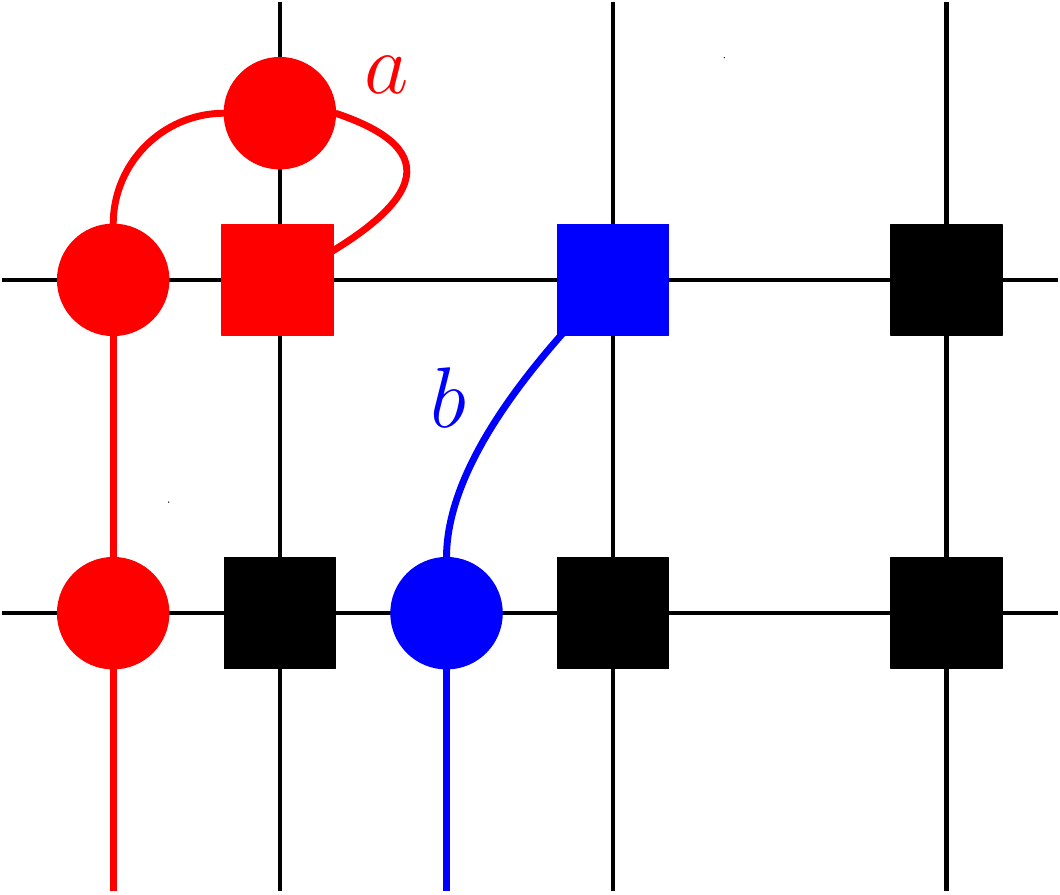} \qquad
    (b)\includegraphics[height=0.23\textwidth]{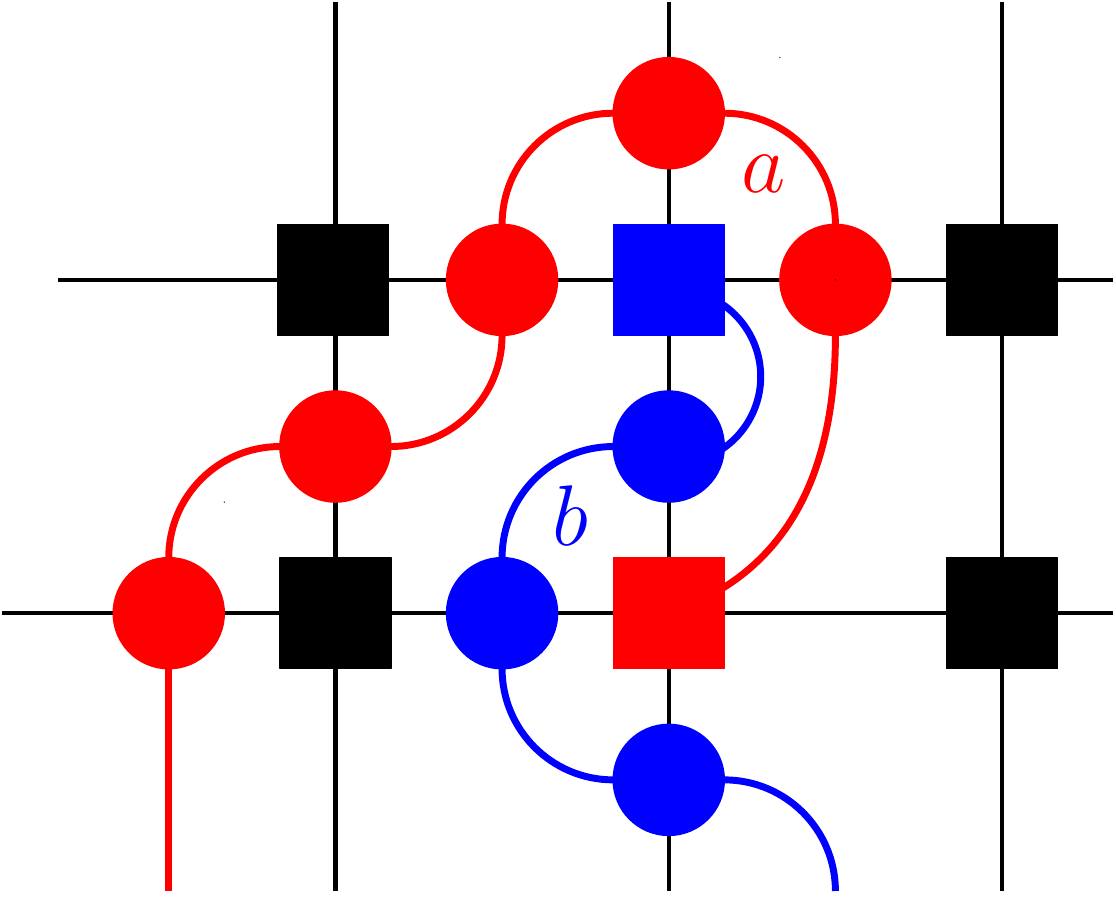} \\
    (c)\includegraphics[height=0.23\textwidth]{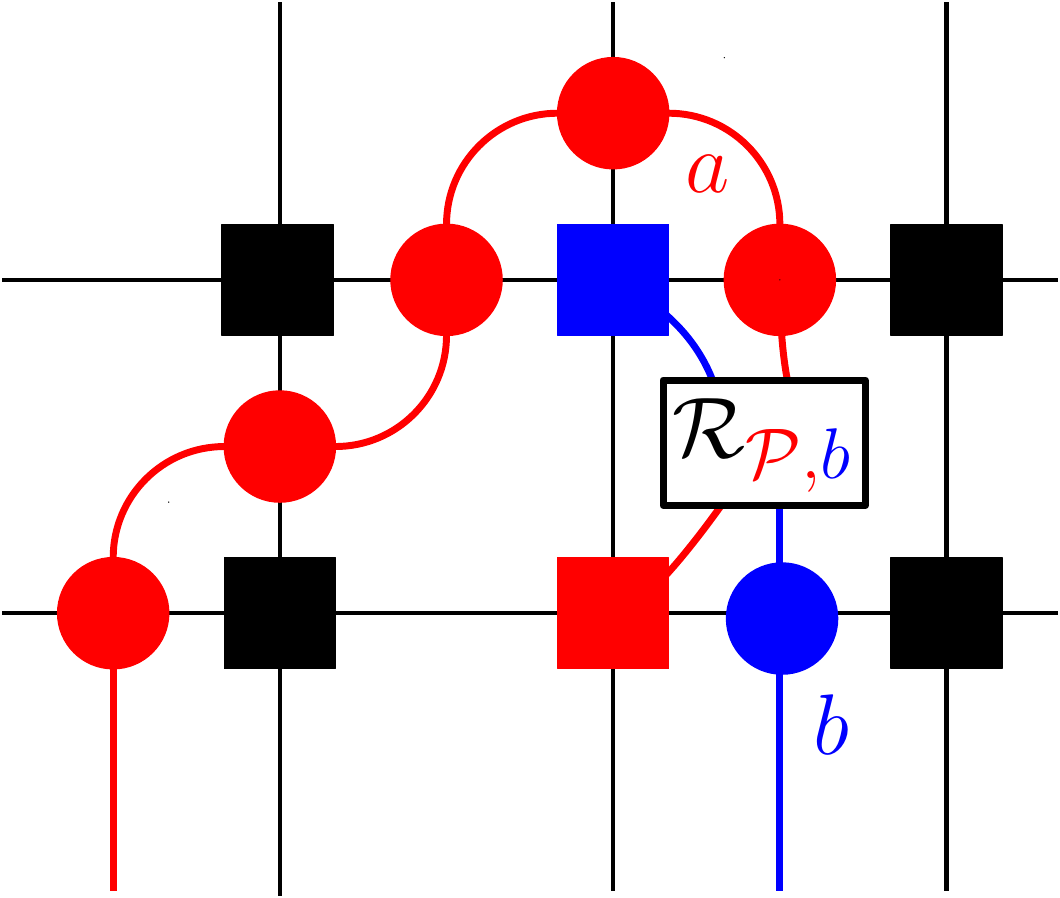} \qquad
    (d)\includegraphics[height=0.23\textwidth]{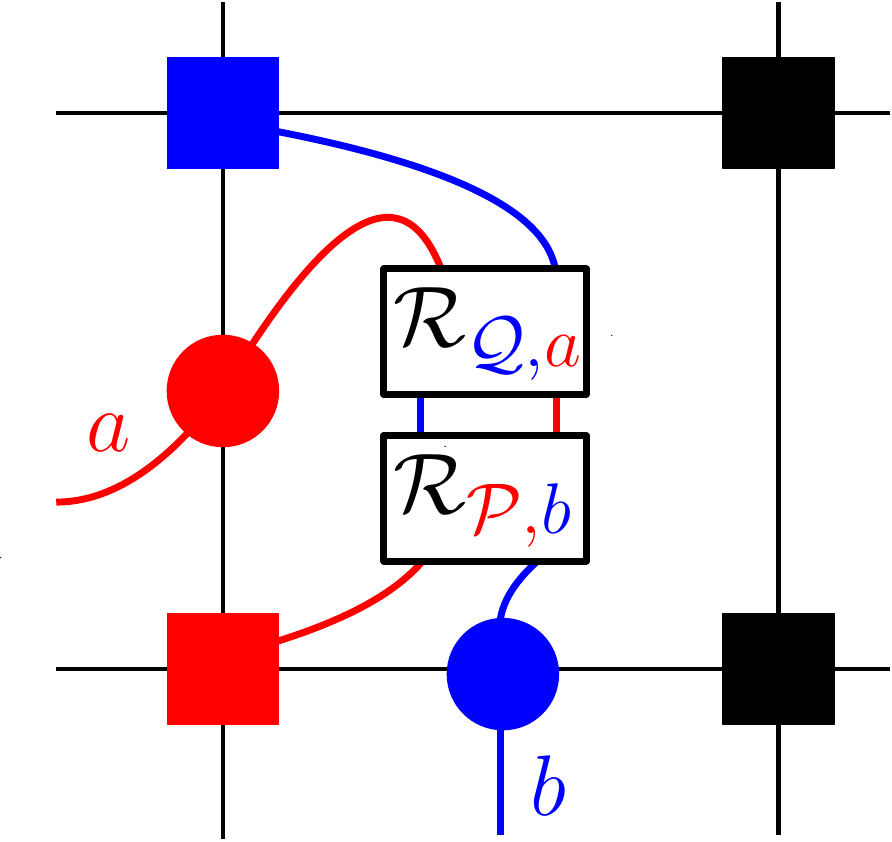}

 \caption{Figure (a): two anyons in a lattice, the lattice sites that contain the central idempotents $\mathcal{P},\mathcal{Q}$ are colored red and blue respectively. Figure (b): we can move the red anyon until the configuration is suited to apply equation \eqref{Rmatrix}. Figure (c): We pull the blue line through the red anyon, using the tensor $\mathcal{R}_{\mathcal{P},b}$ that depends on the red idempotent and the label of the blue line. Figure (d): a similar operation, now with $\mathcal{R}_{\mathcal{Q},a}$.}
\label{fig:fullbraid}
\end{figure}

Analogously, we now show how the full braiding, or double exchange, of one anyon around another can be determined. As before, this information is completely contained within the $\mathcal{R}$ tensors, as shown in figure \ref{fig:fullbraid}. We study the situation where there are two anyon pairs present and we braid one anyon of the first pair completely around an anyon of the second pair. The procedure is shown in figure \ref{fig:fullbraid}. If we compare figures \ref{fig:fullbraid} (a) and (d),  we note that two different changes occurred in the transition between both diagrams. First, the use of relation \eqref{Rmatrix} can induce a non-trivial action on the inner degrees of freedom of the idempotent. While it cannot change the support of the idempotent itself, as this determines the topological superselection sector, the degrees of freedom within a sector can change. This is important if the idempotent corresponding to the anyon is higher dimensional.

\begin{figure}[H]
\begin{align*}
\vcenter{\hbox{
    \includegraphics[width=0.35\textwidth]{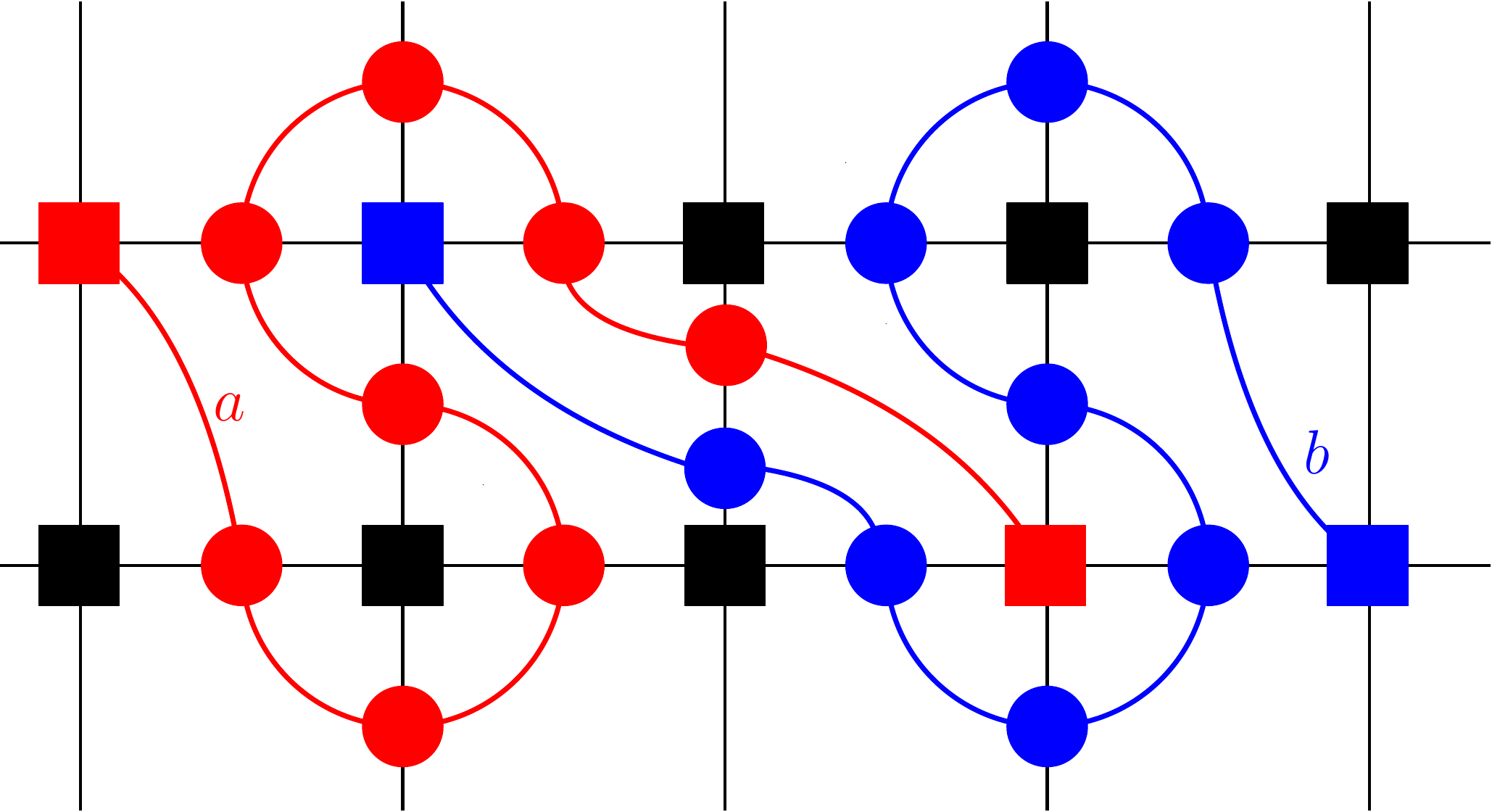}}}
    =
    \vcenter{\hbox{
    \includegraphics[width=0.45\textwidth]{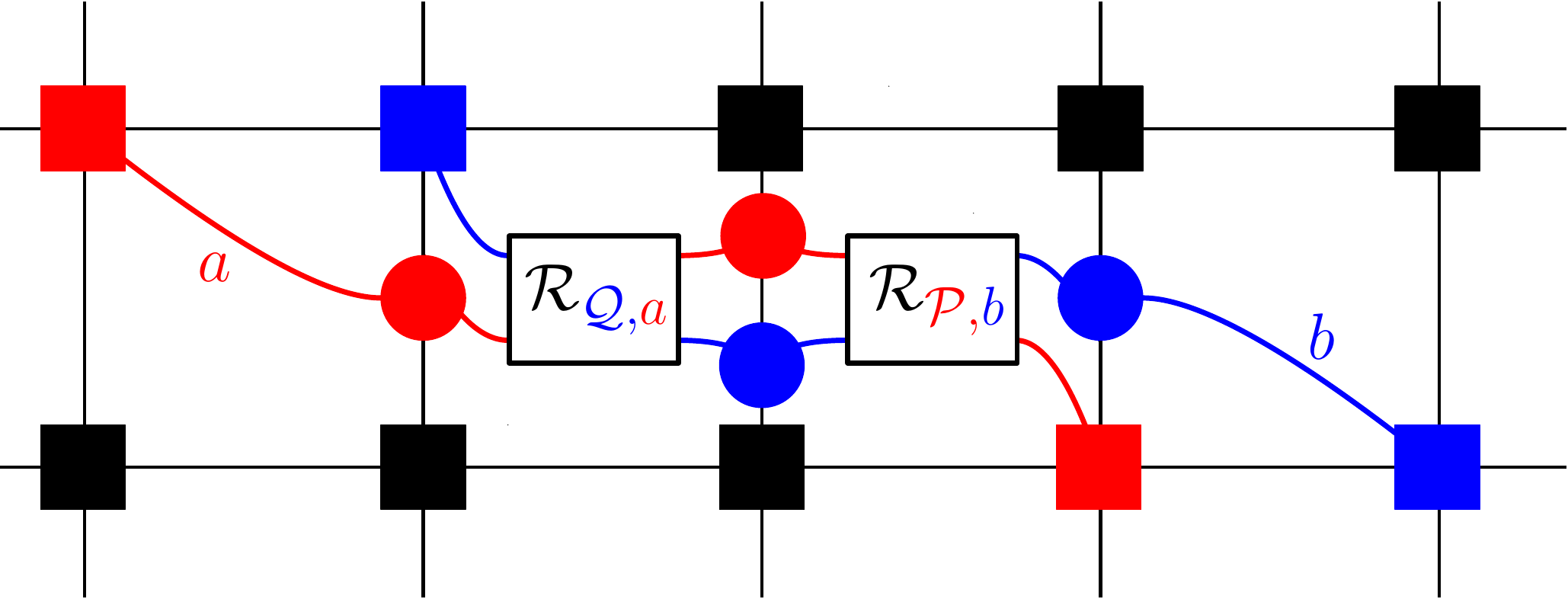} }}
\end{align*}
\caption{A more symmetric version of the braiding process described in figure \ref{fig:fullbraid}.  Completely braiding a red around a blue anyon is described by the contraction of the tensors $\mathcal{R}_{\mathcal{P},b}$ and $\mathcal{R}_{\mathcal{Q},a}$.}
\label{fig:fullbraid2}
\end{figure}

\begin{figure}[H]
  \centering
   (a) \includegraphics[width=0.5\textwidth]{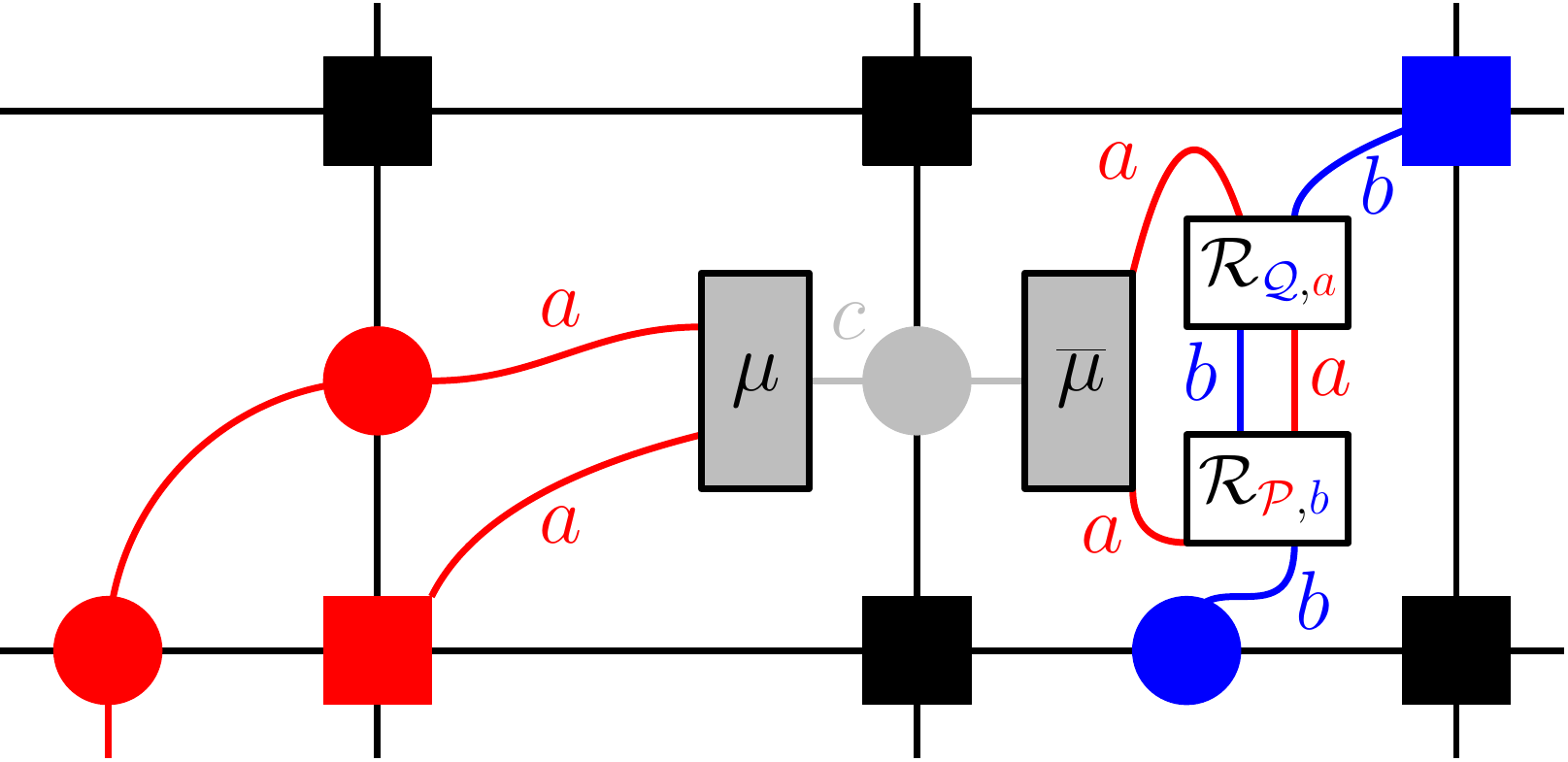}
 \quad  (b) \includegraphics[width=0.20\textwidth]{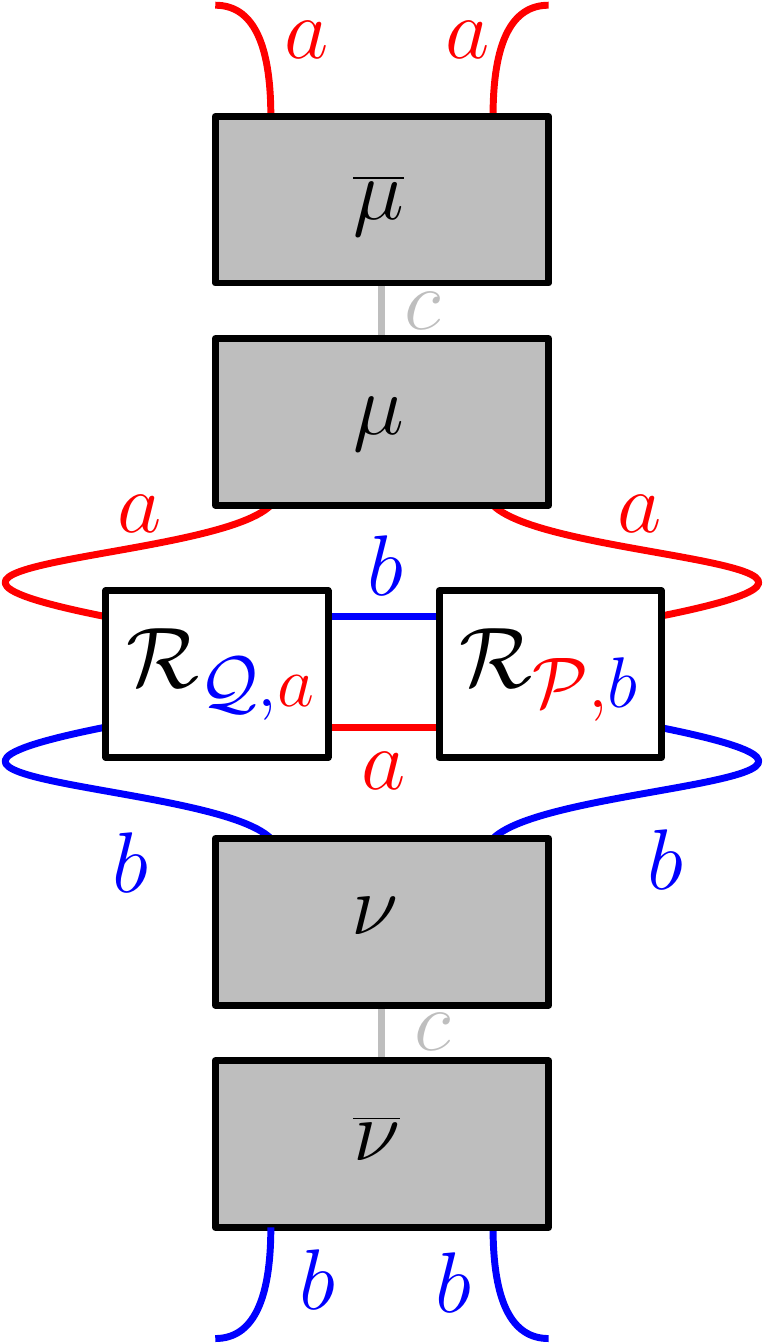}
 \caption{(a) The result of braiding the red anyon around the blue, as in figure \ref{fig:fullbraid}(b). The gray label correspond to the possible fusion channels of the pair of red (or blue) anyons. Before braiding, the pair of red anyons was in the trivial topological sector. After braiding, several fusion results are possible. They can be measured at the gray line. A  sum over the different possible fusion outcome values for these lines is implied. (b) A more symmetric (and rotated) version of (a). Due to the structure of the tensors $\mathcal{R}$, the grey lines  $c$ at the top and bottom are equal.}
\label{fig:fullbraid3}
\end{figure}

Secondly, the fusion channels of the red and blue anyon pair can change. Both pairs were originally in the vacuum sector, but can be in a superposition of sectors after braiding, as is illustrated in Figure \ref{fig:fullbraid3}.

\section{Examples}\label{sec:examples}

We will now illustrate the general formalism of anyons in MPO-injective PEPS with some examples and show that we indeed find all topological sectors. First, we focus on discrete twisted gauge theories \cite{DijkgraafWitten,propitius,Kitaev03,tqd}. After that we turn to string-net models \cite{LevinWen05,TuraevViro}.

\subsection{Discrete gauge theories}

The projector MPO for twisted quantum double models takes the form 
\begin{equation}\label{vacuumMPO}
P = \frac{1}{|\mathsf{G}|} \sum_{g\in \mathsf{G}} V(g),
\end{equation}
where $\mathsf{G}$ is an arbitrary finite group of order $|\mathsf{G}|$ and $V(g)$ are a set of injective MPOs that form a representation of $\mathsf{G}$, i.e. $V(g)V(h) = V(gh)$. $V(g)$ is constructed from the tensors
\begin{align}\label{groupMPO}
\vcenter{\hbox{
 \includegraphics[width=0.13\linewidth]{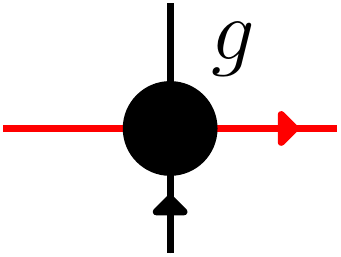}}} \leftrightarrow
\vcenter{\hbox{
 \includegraphics[width=0.35\linewidth]{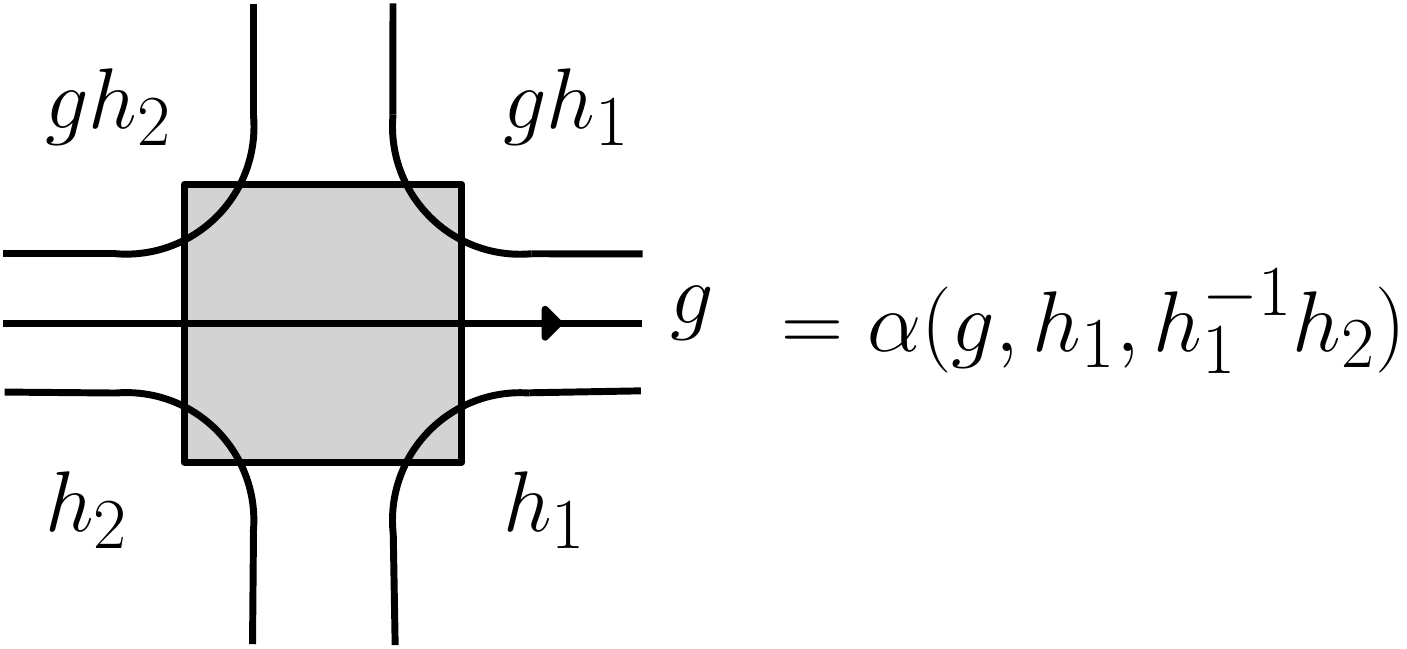}}}
\end{align}
where the internal MPO indices are the horizontal ones. All indices are $|\mathsf{G}|$-dimensional and are labeled by group elements. We use the convention that indices connected in the body of the tensors are enforced to be equal. In \eqref{groupMPO} we only drew the non-zero tensor components, i.e. for lower indices $h_1$ and $h_2$ there is only one non-zero tensor component, namely the one where the upper indices are related by a left group multiplication by $g$. The number $\alpha(g_1,g_2,g_3) \in \mathsf{U}(1)$ is a so-called 3-cocycle satisfying the 3-cocycle condition
\begin{equation}\label{cocyclecondition}
\alpha(g_1,g_2,g_3)\alpha(g_1,g_2g_3,g_4)\alpha(g_2,g_3,g_4) = \alpha(g_1g_2,g_3,g_4)\alpha(g_1,g_2,g_3g_4)
\end{equation}
Without loss of generality one can take the 3-cocycles to satisfy
\begin{equation}\label{gaugechoice}
\alpha(e,g,h) = \alpha(g,e,h) = \alpha(g,h,e) = 1\, ,
\end{equation}
with $e$ the identity group element, for all $g,h \in \mathsf{G}$. For this twisted quantum double MPO we have $g^*=g^{-1}$ and $Z_g = \sum_{h_1}\alpha(g,g^{-1},h_1)\ket{g^{-1}h_1,h_1}\bra{h_1,g^{-1}h_1}$. The specific form of this MPO also allows one to see immediately that the topological entanglement entropy of a contractible region in the corresponding MPO-injective PEPS is given by $\ln|\mathsf{G}|$.

The fusion tensors $X_{g_1g_2}$ for the twisted quantum double MPO take the form
\begin{align}\label{groupfusion}
\vcenter{\hbox{
 \includegraphics[width=0.15 \linewidth]{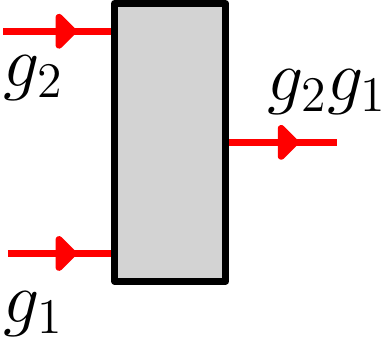}}} \leftrightarrow
\vcenter{\hbox{
 \includegraphics[width=0.37\linewidth]{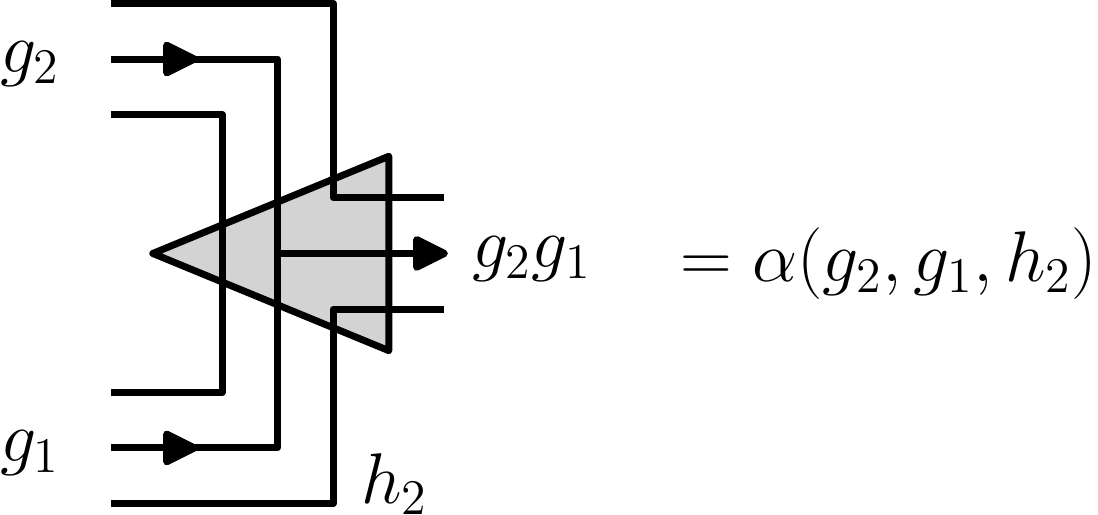}}}\, .
\end{align}
Using \eqref{groupMPO}, \eqref{groupfusion} and \eqref{cocyclecondition} one can check that the injective MPOs $V(g)$ indeed form a representation of $\mathsf{G}$. Using the same data one also sees that the zipper condition \eqref{zippercondition} holds for the tensors of $V(g)$. Again using the cocycle condition the fusion tensors are seen to satisfy the equation
\begin{equation}
(X_{g_3g_2} \otimes \mathds{1}_{g_1})X_{g_3g_2,g_1} = \alpha(g_3,g_2,g_1)(\mathds{1}_{g_3}\otimes X_{g_2,g_1})X_{g_3,g_2g_1}
\end{equation}
So the 3-cocycles $\alpha$ play the role of the $F$ matrices in the general equation \eqref{Fmove}. This connection between MPO group representations and three cocycles was first established in \cite{czxmodel}. For more details about the MPOs under consideration and the corresponding PEPS we refer to \cite{SPTpaper}.

\subsubsection{Topological charge}

Using the MPO and fusion tensors defined above we can now construct the algebra elements $A_{g_1,g_2,g_3,g_4}$ defined by Eq.~\ref{algebraobject}; note that the indices $\mu,\nu$ are always one dimensional in the group case so we can safely discard them. To construct the central idempotents we focus on the following algebra elements
\begin{equation}
A_{g,g^{-1}k^{-1},g,k} = \delta_{[k,g],e} R_g(k^{-1})\, ,
\end{equation}
where $[k,g] = kgk^{-1}g^{-1}$ is the group commutator and $e$ the trivial group element. For convenience, our choice for the basis of the algebra $R_g(k)$ deviates slightly from Eq.~\eqref{algebraobject}. It is constructed by closing a single block MPO \eqref{groupMPO} labeled by group element $k$, satisfying $[k,g] = e$, with a tensor that has as non-zero components
\begin{align}
\vcenter{\hbox{
 \includegraphics[width=0.1\linewidth]{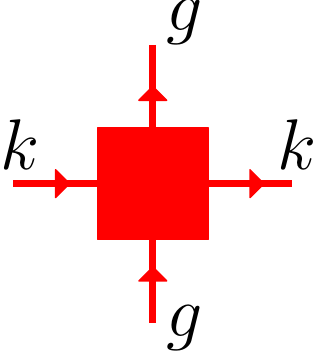}}} \leftrightarrow
\vcenter{\hbox{
 \includegraphics[width=0.25\linewidth]{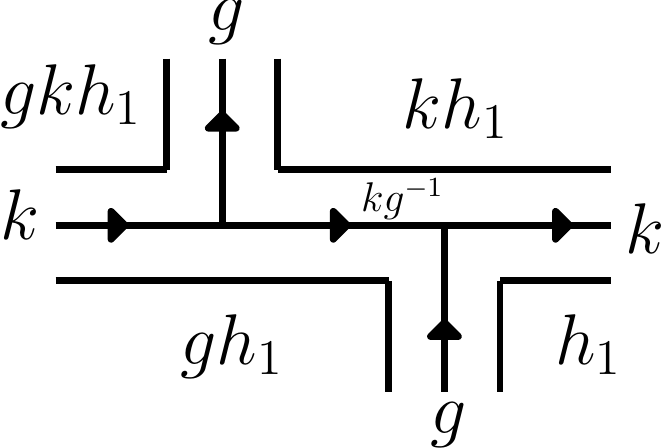}}} = \frac{\alpha(kg^{-1},g,h_1)}{\alpha(g,kg^{-1},gh_1)}
\end{align}
Note that this tensor is chosen slightly different as the one in Eq.~\eqref{algebraobject} and that the direction of $k$ has been reversed. 

By repeated use of the cocycle condition and the fact that $[g,k]=[g,m]=e$ one can now derive the multiplication rule of the algebra elements
\begin{equation} \label{groupmultiplication}
R_g(m)R_g(k) =  \bar{\omega}_g(m,k) \frac{\epsilon_g(mk)}{\epsilon_g(m)\epsilon_g(k)}R_g(mk)
\end{equation}
where
\begin{eqnarray}
\bar{\omega}_g(m,k) &  = &  \frac{\alpha(m,g,k)}{\alpha(m,k,g)\alpha(g,m,k)} \nonumber \\
 \frac{\epsilon_g(mk)}{\epsilon_g(m)\epsilon_g(k)} & = & \frac{\alpha(g,mkg^{-1},g)}{\alpha(g,mg^{-1},g)\alpha(g,kg^{-1},g)}
\end{eqnarray}
One can check that $\omega_g(m,k)$ is a 2-cocycle satisfying the 2-cocycle condition
\begin{equation}
\omega_g(m,k)\omega_g(mk,l) = \omega_g(m,kl)\omega_g(k,l)\, ,
\end{equation}
when $m$, $k$ and $l$ commute with $g$. So the algebra elements $R_g(k)$ form a projective representation of the centralizer $\mathcal{Z}_g$ of $g$. We now define the following projective irreducible representations of $\mathcal{Z}_g$ labeled by $\mu$
\begin{equation}
\Gamma^\mu_g(m) \Gamma^\mu_g(k) = \omega_g(m,k) \Gamma_g^\mu(mk)\, ,
\end{equation}
and the corresponding projective characters $\chi^\mu_g(k) = \text{tr}(\Gamma^\mu_g(k))$. We denote the dimension of projective irrep $\mu$ by $d_\mu$. Using the Schur orthogonality relations for projective irreps one can now check that
\begin{equation}
P_{(g,\mu)} = \frac{d_\mu}{|\mathcal{Z}_g|}\sum_{k\in\mathcal{Z}_g} \epsilon_g(k) \chi^\mu_g(k)R_g(k)
\end{equation}
are Hermitian orthogonal projectors, i.e. $P_{(g,\mu)}^\dagger =  P_{(g,\mu)}$ and $P_{(g,\mu)}P_{(h,\nu)} = \delta_{g,h}\delta_{\mu\nu} P_{(g,\mu)}$. To obtain the central idempotents we have to sum over all elements in the conjugacy class $\mathcal{C}_A$ of $g$, so the final anyon ansatz is
\begin{equation}
\mathcal{P}_{(\mathcal{C}_A,\mu)} = \sum_{g\in \mathcal{C}_A} P_{(g,\mu)}.
\end{equation}
In this way we indeed recover the standard labeling of dyonic excitations in discrete twisted gauge theories: the flux is labeled by a conjugacy class $\mathcal{C}_A$ and the charge is labeled by a projective irrep of the centralizer $\mathcal{Z}_g$ of a representative element $g$ in $\mathcal{C}_A$ \cite{orbifold,DijkgraafPasquierRoche}.

\subsubsection{Anyon ansatz}

\begin{figure}
  \centering
   a)
    \includegraphics[width=0.2\textwidth]{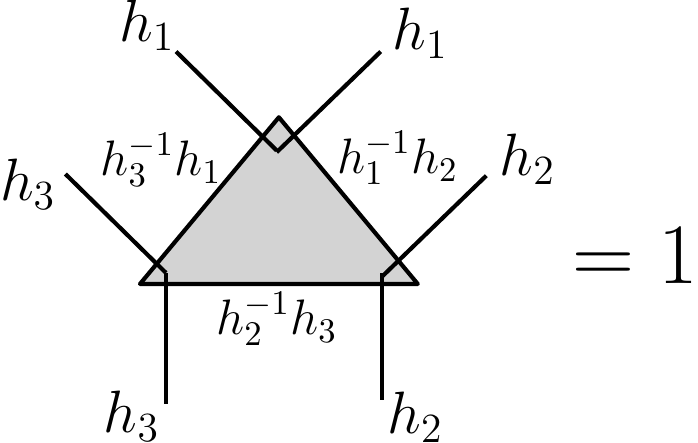}
   b)
    \includegraphics[width=0.2\textwidth]{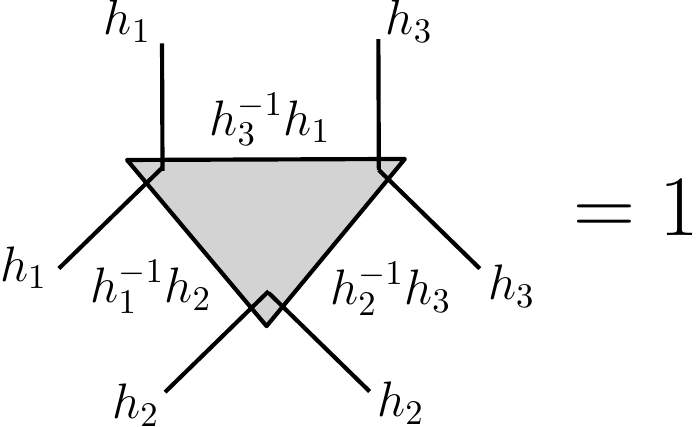}
    c)
    \includegraphics[width=0.2\textwidth]{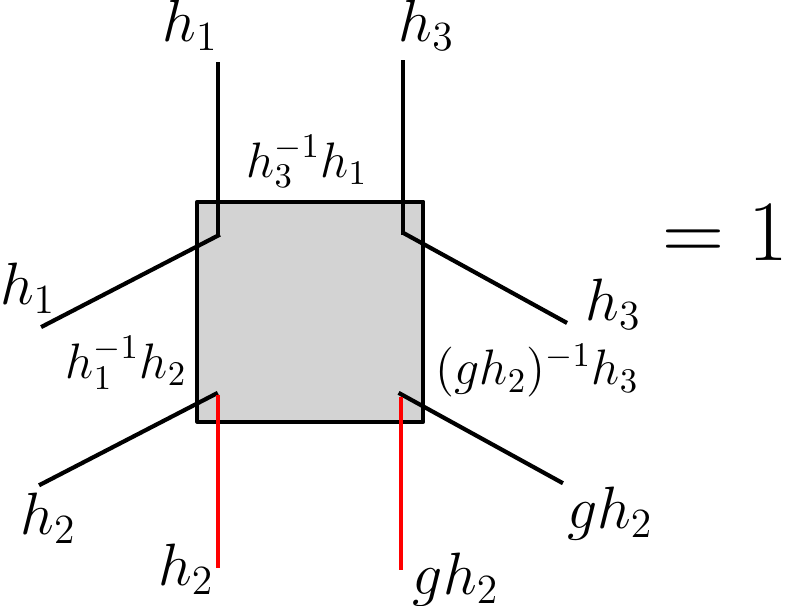}
\caption{Tensors for the non-twisted ($\alpha(g_1,g_2,g_3)\equiv 1$) quantum double PEPS on a hexagonal lattice. a) Ground state tensors for the $A$-sublattice.   b) Ground state tensors for the $B$-sublattice. All indices are $|\mathsf{G}|$-dimensional and are labeled by group elements. There are two virtual indices for every link in the lattice. Indices that are connected in the tensors are enforced to be equal. There are three physical indices on each tensor, of which we write the index value between the virtual indices. All non-zero tensor components have index configurations as indicated in the figures and have value one. The resulting PEPS is MPO-injective with virtual support PMPO \eqref{vacuumMPO}. c) An anyonic excitation tensor for the $A$-sublattice. The extra lower index, which we colored red for clarity, is to be connected to a MPO on the virtual level. Multiplying all the physical indices of an $A$-sublattice ground state tensor counter clockwise gives the identity group element. For the excitation tensor the multiplied value of the physical indices is $h_2^{-1}g^{-1}h_2$, which indicates the presence of a non-trivial flux. The physical indices can distinguish between virtual MPOs corresponding to group elements in different conjugacy classes. However,  the rank-one injectivity structure of the anyonic tensor is reflected in the fact that elements in the same conjugacy class give equivalent configurations of the physical indices.}
\label{fig:QDtensor}
\end{figure}

In this section we will illustrate some aspects of the anyon ansatz that were discussed in section \ref{sec:ansatz}. Firstly, in figure \ref{fig:QDtensor} we show the explicit PEPS ground state and excitation tensors for the non-twisted ($\alpha(g_1,g_2,g_3)\equiv 1$) quantum double PEPS on the hexagonal lattice. This provides an explicit example of an anyonic excitation tensor that has a rank-one injectivity structure on each virtual subspace corresponding to a topological sector.

Secondly, we look at a pair of pure charges. Using similar reasoning as in the previous section we can construct the simple idempotents and nilpotents with diagonal group label as
\begin{equation}
P_{(\mathcal{C}_A,\mu)}^{(g,i),(g,j)} =  \frac{d_\mu}{|\mathcal{Z}_g|}\sum_{k\in\mathcal{Z}_g} \epsilon_g(k) [\Gamma^\mu_g(k)]_{ij}R_g(k)\, ,
\end{equation}
where $i,j$ are matrix indices of the irrep $\Gamma^\mu_g(k)$. The simple idempotents and nilpotents with off-diagonal group label can be obtained via a straightforward generalization, but we will not need them here.

We consider a charge $\mu$ and its anti-charge $\mu^*$. The relevant idempotents and nilpotents are $P_{\mu}^{i,j} \equiv P_{(e,\mu)}^{(e,i),(e,j)}$. To construct a topological PEPS containing the charge pair we start with two tensors $C^i_\mu$ and $C^j_{\mu^*}$, which have as many virtual indices as the coordination number of the lattice and one physical index. Their virtual indices are supported on the subspaces determined by respectively $P_\mu^{i,i}$ and $P_{\mu^*}^{j,j}$. If we interpret the charge tensors as matrices with the physical index as row index and the virtual indices together as the column index then this implies $C^i_\mu P^{k,k}_\nu = \delta_{\mu,\nu}\delta_{i,k}C^i_\mu$ and $C^j_{\mu^*} P^{k,k}_{\nu^*} = \delta_{\mu^*,\nu^*}\delta_{j,k} C^j_{\mu^*}$. We now want to find the complete anyonic excitation tensors $C_\mu$ and $C_{\mu^*}$ such that $C_\mu P_\mu^{i,i} = C_\mu^i$ and $C_{\mu^*}P_{\mu^*}^{j,j} = C^j_{\mu^*}$. For this we proceed as before: we take both tensors and project them in the vacuum sector. We will ignore the PEPS environment and simply work with the tensor product of both charge tensors $C_\mu^i\otimes C_{\mu^*}^j$. The vacuum projector \eqref{vacuumMPO} on this tensor product can be written as
\begin{equation}
\tilde{P} = \frac{1}{|\mathsf{G}|}\sum_{g\in \mathsf{G}} V(g)\otimes V(g) = \frac{1}{|\mathsf{G}|}\sum_{g\in \mathsf{G}} R_e(g)\otimes R_e(g)\, .
\end{equation}
Using the orthogonality relations for irreps we rewrite the vacuum projector as
\begin{equation}
\tilde{P} = \sum_{\nu}\frac{1}{d_\nu}\sum_{p,q=1}^{d_\nu}P^{p,q}_\nu\otimes P^{p,q}_{\nu^*}\, ,
\end{equation}
where $[\Gamma^{\nu^*}(g)]_{pq} = [\bar{\Gamma}^\nu(g)]_{pq}$. We therefore get for the vacuum projected charge pair
\begin{equation}
\left(C_\mu^i\otimes C_{\mu^*}^j\right)\left( \sum_{\nu}\frac{1}{d_\nu}\sum_{p,q=1}^{d_\nu}P^{p,q}_\nu\otimes P^{p,q}_{\nu^*}\right) = \delta_{i,j}\frac{1}{d_\mu}\sum_q C_\mu^i P_\mu^{i,q}\otimes C_{\mu^*}^i P_{\mu^*}^{i,q}\, .
\end{equation}
By taking $C_\mu^q \equiv C_\mu^i P_\mu^{i,q}$ we obtain
\begin{equation}
\left(\sum_{i = 1}^{d_\mu}C_\mu^i\otimes C_{\mu^*}^i\right)\tilde{P} = \left(\sum_{i = 1}^{d_\mu}C_\mu^i\otimes C_{\mu^*}^i\right)\, .
\end{equation}
So we see that for the pair to be in the vacuum state both charges should form a maximally entangled state in the irrep matrix indices. However, as explained in the general discusssion of section \ref{sec:ansatz}, this is purely virtual entanglement that cannot be destroyed by physical operators acting on only one charge in the pair.

\subsubsection{Topological spin}

To calculate the topological spin we first note following relation
\begin{equation}
\Gamma^\mu_g(k)\Gamma^\mu_g(g) = \Gamma^\mu_g(g)\Gamma^\mu_g(k)\frac{\omega_g(k,g)}{\omega_g(g,k)} =  \Gamma^\mu_g(g)\Gamma^\mu_g(k)\, ,
\end{equation}
which holds for all $k \in \mathcal{Z}_g$. Using Schur's lemma this implies that $\Gamma_g^\mu(g) = e^{i2\pi h^\mu_g}\mathds{1}_{d_\mu}$. One can also easily check that
\begin{equation}
\Gamma_g^\mu(g^{-1}) = \omega_g(g,g^{-1})\Gamma_g^\mu(g)^\dagger\, .
\end{equation}
With these observations we now obtain
\begin{align}
P_{(g,\mu)}R_g(g) &=  \frac{d_\mu}{|\mathcal{Z}_g|}\sum_{k\in\mathcal{Z}_g} \chi^\mu_g(k)R_g(kg)\nonumber \\
&= \frac{d_\mu}{|\mathcal{Z}_g|}\sum_{x\in\mathcal{Z}_g} \text{tr}(\Gamma^\mu_g(x)\Gamma^\mu_g(g^{-1})\omega^*_g(x,g^{-1}))R_g(x)\nonumber \\
&= e^{-i2\pi h^\mu_g}\frac{d_\mu}{|\mathcal{Z}_g|}\sum_{x\in\mathcal{Z}_g} \epsilon_g(x) \chi^\mu_g(x)R_g(x) \nonumber\\
&=  e^{-i2\pi h^\mu_g} P_{(g,\mu)} \nonumber
\end{align}
Since $e^{-i2\pi h^\mu_g}$ is the same for all elements in the conjugacy class $\mathcal{C}_A$ of $g$ we obtain the desired result
\begin{equation}
\mathcal{P}_{(\mathcal{C}_A,\mu)}\mathcal{R}_{2\pi} = \mathcal{P}_{(\mathcal{C}_A,\mu)}\sum_{g \in  \mathsf{G}}R_g(g) = \theta_{(\mathcal{C}_A,\mu)}\mathcal{P}_{(\mathcal{C}_A,\mu)}\, ,
\end{equation}
where $\mathcal{R}_{2\pi}$ was introduced in section \ref{sec:topspin}. The phase $\theta_{(\mathcal{C}_A,\mu)} = e^{-i2\pi h^\mu_g}$ gives the topological spin of the corresponding anyon.

\subsubsection{Fusion}

In the group case fusion is easy to calculate analytically because of the following identity for the basis elements of our algebra:

\begin{align}\label{fusionloop}
\vcenter{\hbox{
 \includegraphics[width=0.27\linewidth]{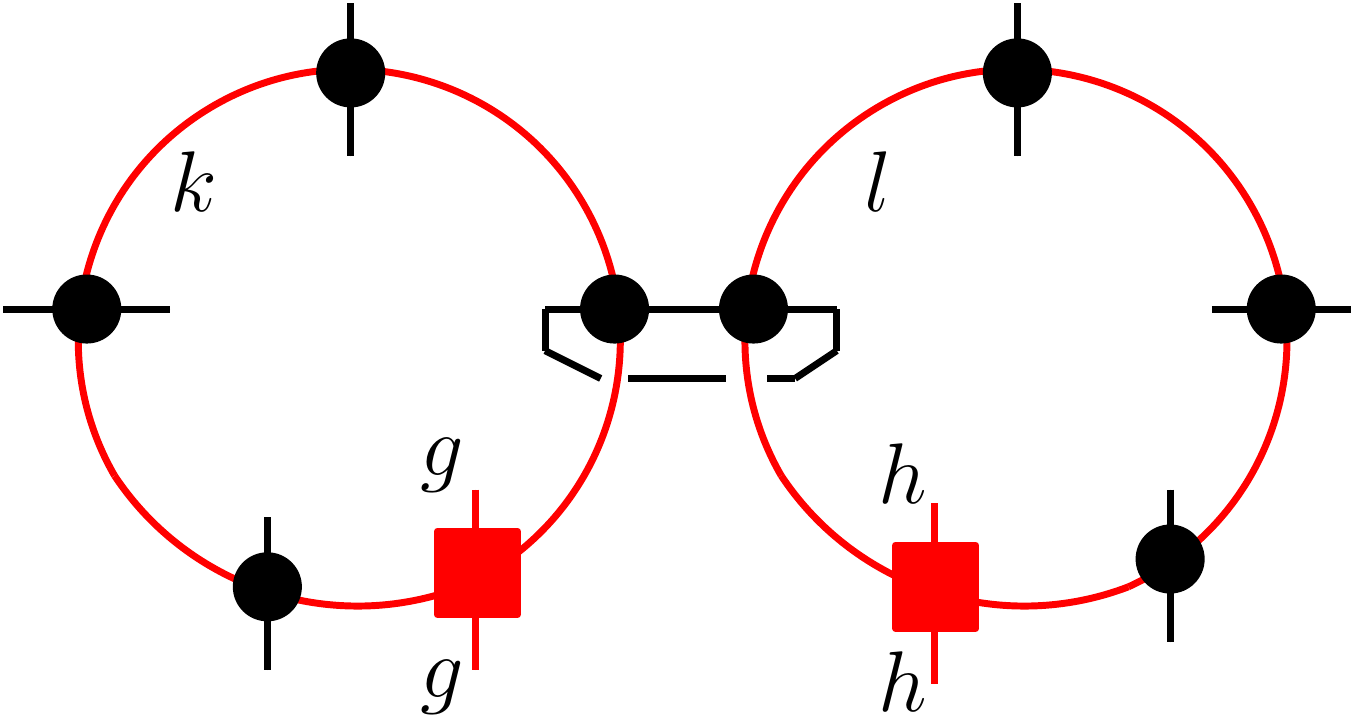}}} = \delta_{k,l}
\vcenter{\hbox{
 \includegraphics[width=0.25\linewidth]{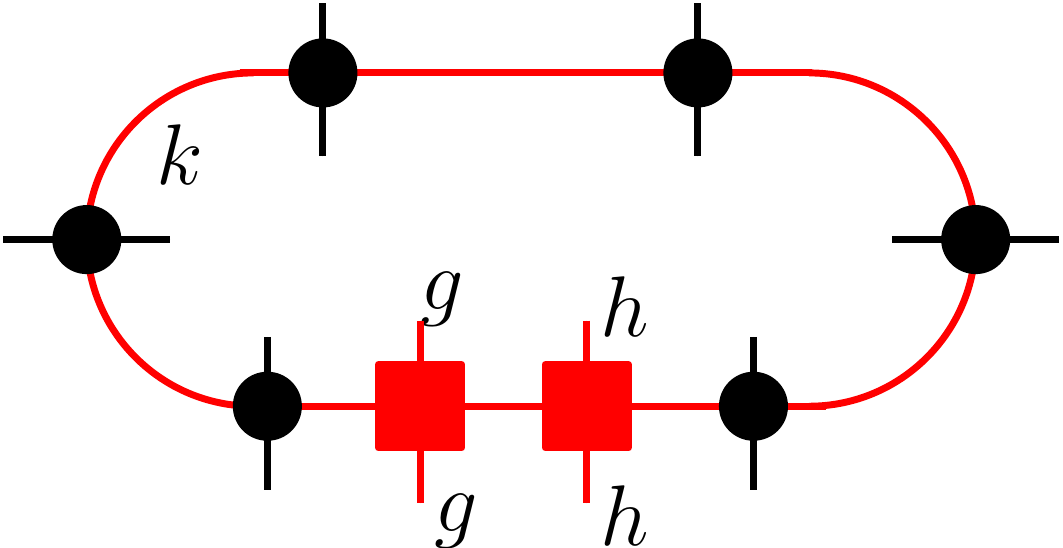}}}
\end{align}
This implies that to calculate fusion relations we can simply trace over the inner indices at the shared boundary of two central idempotents to create a bigger loop. 

We subsequently act with the fusion tensor $X_{gh}$ on the two red inner indices on the right hand side of \eqref{fusionloop}, which acts as a unitary on the support of these indices. We also attach $X^\dagger_{gh}$ to the outer indices in \eqref{fusionloop}, which can obviously be obtained by decomposing the product of the two MPOs $V(g)$ and $V(h)$ that are connected to the central idempotents once we embed them in a MPO-injective PEPS. Using the 3-cocycle condition one can now check that we have

\begin{align}
\vcenter{\hbox{
 \includegraphics[width=0.26\linewidth]{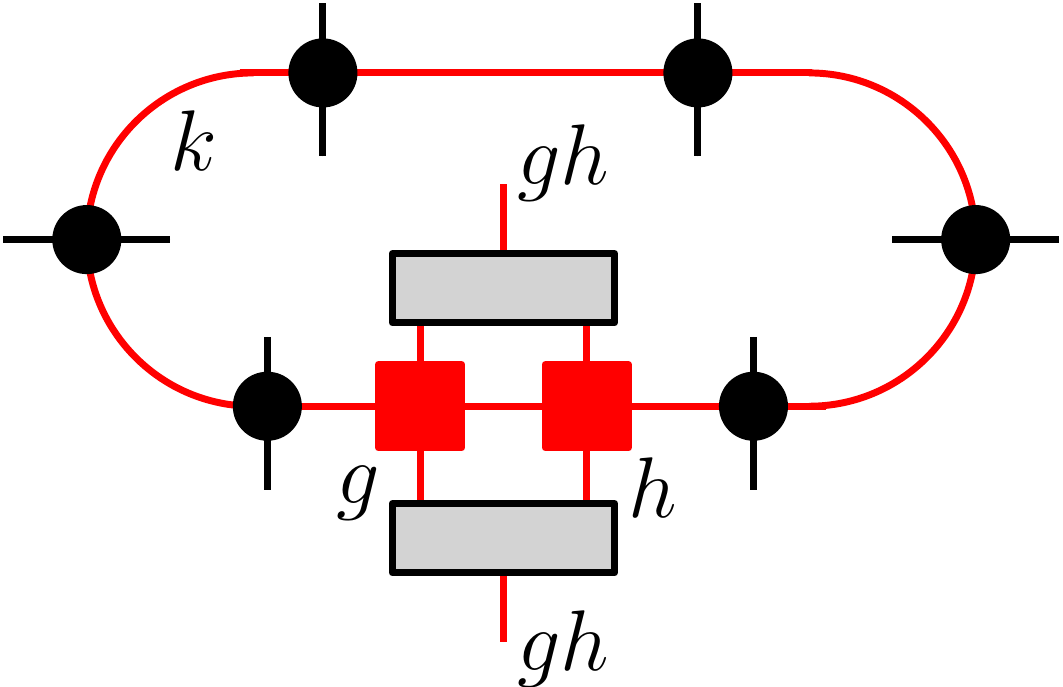}}} = \beta(k,g,h)
\vcenter{\hbox{
 \includegraphics[width=0.26\linewidth]{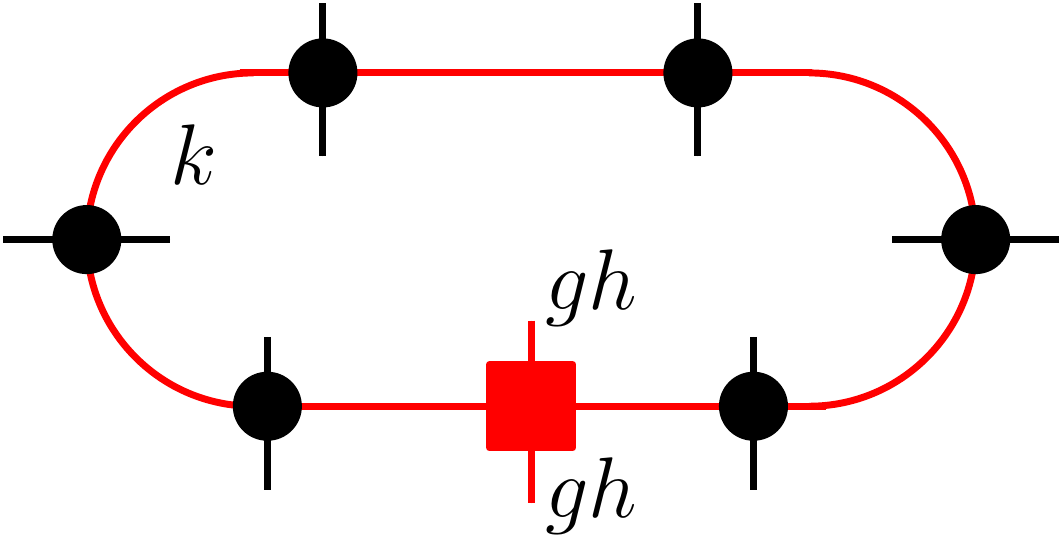}}}\, ,
\end{align}
where $\beta(k,g,h)$ is given by
\begin{equation}
\beta(k,g,h) = \omega_k(g,h)\frac{\epsilon_{gh}(k)}{\epsilon_{g}(k)\epsilon_{h}(k)}\, .
\end{equation}
So we obtain
\begin{equation}
P_{(g,\mu)}\times P_{(h,\nu)} =   \frac{d_\mu d_\nu}{|\mathcal{Z}_{g}||\mathcal{Z}_{h}|}\sum_{k\in\mathcal{Z}_{gh}}  \epsilon_{gh}(k)  \chi^\mu_g(k) \chi^\nu_h(k)\omega_k(g,h)R_{gh}(k) \, .
\end{equation}
We now define $\Gamma^{\mu\nu}_{gh}(k) = \Gamma^\mu_g(k)\otimes \Gamma^\nu_h(k)\omega_k(g,h)$ for all $k$ such that $[k,g]=[k,h] = e$. Then repeated use of the 3-cocycle condition \eqref{cocyclecondition} shows that
\begin{equation}
\Gamma^{\mu\nu}_{gh}(k_1)\Gamma^{\mu\nu}_{gh}(k_2) = \omega_{gh}(k_1,k_2)\Gamma^{\mu\nu}_{gh}(k_1k_2)\, ,
\end{equation}
i.e. $\Gamma^{\mu\nu}_{gh}(k)$ is a projective representation of $\mathcal{Z}_{gh}$. This representation will in general be reducible
\begin{equation}
\Gamma^{\mu\nu}_{gh}(k) \simeq \bigoplus_\lambda \mathds{1}_{W_{\mu\nu}^\lambda}\otimes \Gamma^\lambda_{gh}(k)\, ,
\end{equation}
where the integer $W_{\mu\nu}^\lambda$ denotes the number of times a projective irrep $\lambda$ appears in the decomposition of $\Gamma^{\mu\nu}_{gh}$. From this we get following relation between the projective characters
\begin{equation}
\chi_g^\mu(k)\chi_h^\nu(k)\omega_k (g,h) = \sum_{\lambda} W_{\mu\nu}^\lambda\, \chi_{gh}^\lambda (k)\, .
\end{equation}
So we find
\begin{equation}
P_{(g,\mu)}\times P_{(h,\nu)} = \sum_\lambda W^\lambda_{\mu\nu} P_{(gh,\lambda)}\, ,
\end{equation}
up to some normalization factors. In this way we obtain the final fusion rules
\begin{equation}
\mathcal{P}_{(\mathcal{C}_A,\mu)}\times \mathcal{P}_{(\mathcal{C}_B,\nu)} = \sum_{(\mathcal{C}_C,\lambda)} \mathscr{N}_{(\mathcal{C}_A,\mu),(\mathcal{C}_B,\nu)}^{(\mathcal{C}_C,\lambda)} \mathcal{P}_{(\mathcal{C}_C,\lambda)}\, ,
\end{equation}
where the fusion coefficients can be written down explicitly using the orthogonality relations for projective characters \cite{orbifold,DijkgraafWitten}:
\begin{equation}
\mathscr{N}_{(\mathcal{C}_A,\mu),(\mathcal{C}_B,\nu)}^{(\mathcal{C}_C,\lambda)} = \frac{1}{|\mathsf{G}|}\sum_{g_1\in \mathcal{C}_A} \sum_{g_2\in\mathcal{C}_B}\sum_{g_3 \in \mathcal{C}_C}\sum_{h\in\mathcal{Z}_{g_1}\cap \mathcal{Z}_{g_2}\cap \mathcal{Z}_{g_3}} \delta_{g_1g_2, g_3}\chi_{g_1}^\mu(h)\chi^\nu_{g_2}(h)\bar{\chi}^{\lambda}_{g_3}\omega_h(g_1,g_2)
\end{equation}

\subsection{String-nets}
The next example we consider are the string-net models. For simplicity, we restrict ourselves to models without higher dimensional fusion spaces, i.e. all $N_{ab}^c$ in equation \eqref{fusioncategory1} are either 0 or 1. Also, we only deal with models where each single block MPO is self-dual: $a=a^*$ and $N_{aa}^1=1$. Both restrictions can easily be lifted.

The description of string-nets in the framework presented here was introduced in \cite{MPOpaper}. The string-net models are a prime example of the MPO-injectivity formalism. The PMPO is constructed from the $G$-symbols and the quantum dimensions of a unitary fusion category. The single block MPOs correspond one-to-one with the simple objects of the input fusion category. The fusion matrices $X_{ab}^c$ are also easily constructed from the $G$-symbols and the quantum dimensions. These tensors give rise to an MPO-injective PEPS and they satisfy the properties listed in section \ref{subsec:zipper}. The validity of the general requirements in our formalism follows mainly from the pentagon relation of the $G$-symbols. The properties of section \ref{subsec:zipper} are rooted in the spherical property of unitary fusion categories.

To describe the string-nets as a tensor network, there is one extra technical subtlety we need to take into account. Every closed loop in the PEPS representation of a string-net wave function gives rise to a factor equal to the quantum dimension of the label of this loop. In \cite{MPOpaper}, this was taken care of by incorporating such factors both in the tensors and by adding extra factors for every bend in an MPO. Because of this convention, the MPOs give rise to projectors $P_L$ for every length that are not Hermitian on a closed loop. Luckily, as all these operators are still similar to Hermitian operators via a local, positive similarity transformation, this has no implications for the general theory. For example, we still find that every single block MPO labeled by $a$ has a unique corresponding single block MPO $a^*$ that is obtained by Hermitian conjugation, where $a^*$ is just the categorical dual of $a$. The tensors we describe next are used on a square lattice, similar tensors can be used on different lattices.

First we describe the PMPO. We have
\begin{align}\label{StringnetMPO}
\vcenter{\hbox{
 \includegraphics[height=0.12\linewidth]{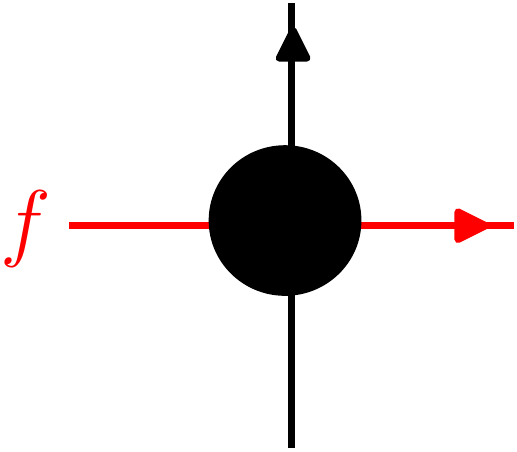}}} \leftrightarrow
\vcenter{\hbox{
 \includegraphics[height=0.15\linewidth]{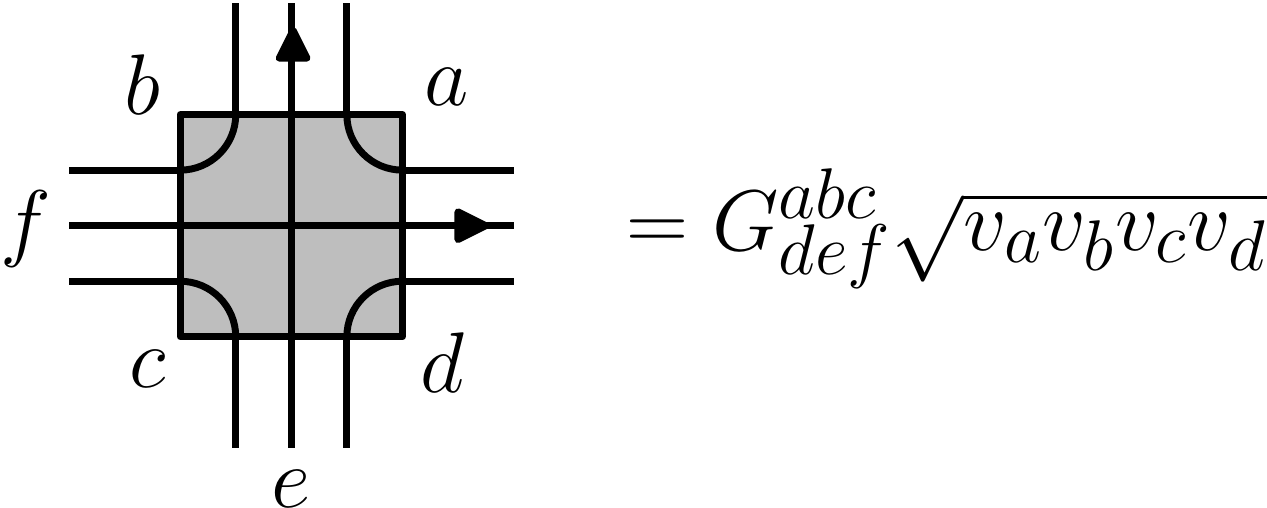}}}
\end{align}
where the internal MPO indices are the horizontal ones and all indices are $N$ dimensional. The single block MPOs are determined by fixing the label $f$. The corresponding weights $w_f$ used to construct a PMPO are given by the quantum dimensions $d_f$ divided by $\mathcal{D}^2 = \sum_a d_a^2$, the total quantum dimension of the fusion category squared. The factors $v_a$ in the definition of the MPO are included to take care of the closed loop factors. They are given by the square roots of the quantum dimensions: $v_a = d_a^{1/2}$. The single block MPOs obtained by fixing $f$ satisfy the algebraic structure of the fusion algebra of the category we used to construct the MPOs.

For the string-net MPOs we consider here the gauge transformations $Z_a$ are all trivial; they amount to simply swapping the double line structure which is present in the virtual indices of the MPO tensor. The fusion tensors $X_{ab}^c$ are given by
\begin{align}
\vcenter{\hbox{
 \includegraphics[height=0.13 \linewidth]{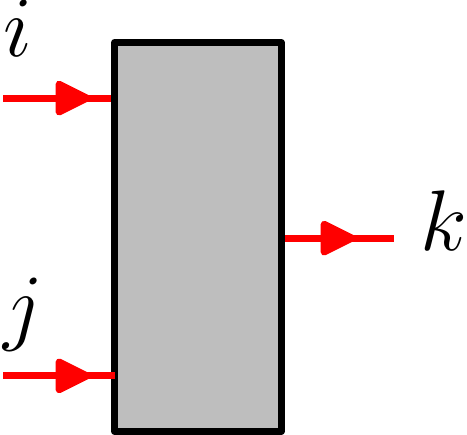}}} \leftrightarrow
\vcenter{\hbox{
 \includegraphics[height=0.16\linewidth]{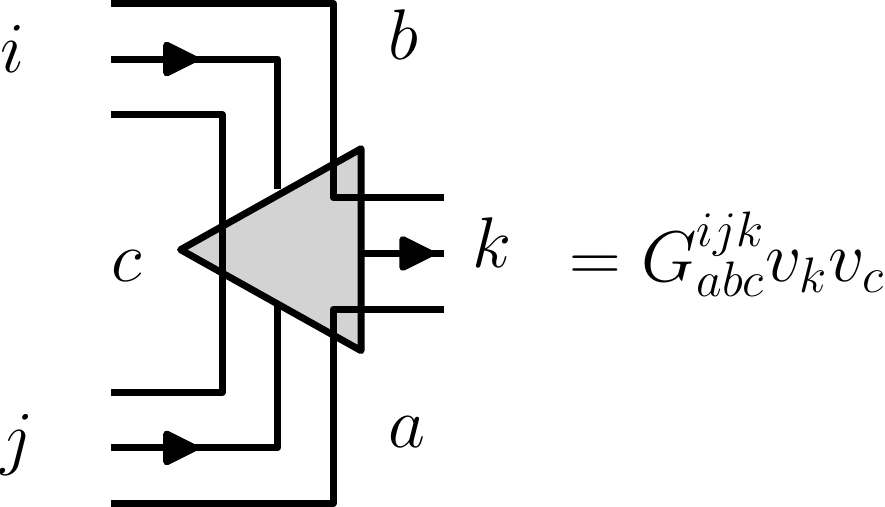}}}.
\end{align}
The factor $v_c$ is only  present for the closed loop condition (and could be taken care of differently).  The pivotal property for these fusion tensors is
\begin{align}
\vcenter{\hbox{
 \includegraphics[height=0.22 \linewidth]{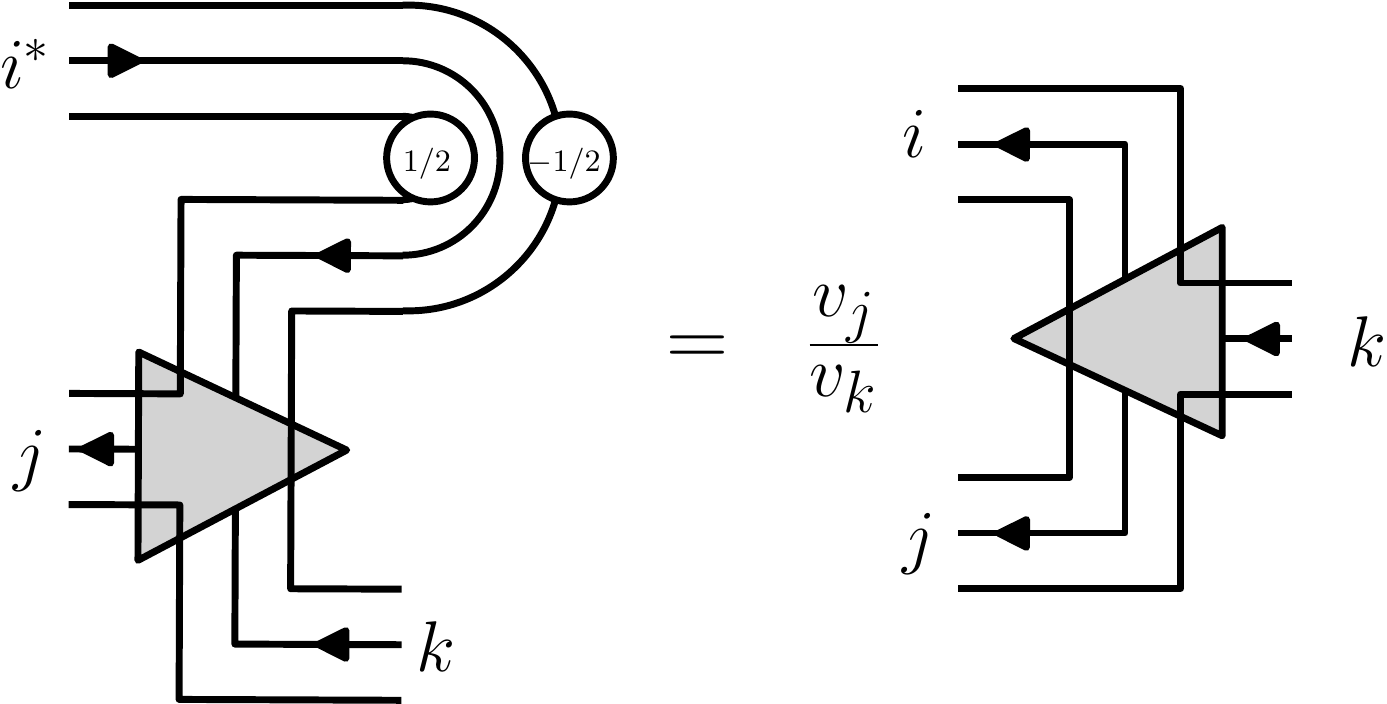}}} \, ,
\end{align}
which is equivalent to \eqref{pivotalnew} up to the diagonal matrices labeled by $1/2$ and $-1/2$, which denote the power of the quantum dimensions that are added to satisfy the closed loop condition. More specifically, these matrices are $\sum_a d_a^{\pm 1/2}\ket{a}\bra{a}$.

With this information, the MPO and fusion tensors can now be used in our framework in order to obtain an ansatz for anyons in string-nets. Unlike in the case of discrete gauge theories, we now need the ansatz \eqref{algebraobject} in full generality. We recall the form of the algebra elements
$$
A_{abcd,\mu\nu} =  \vcenter{\hbox{\includegraphics[width=0.24\linewidth]{algebraobject}}}.
$$
The structure constants that define the multiplication of these objects can be computed analytically with formula \eqref{eq:structureconstants} or numerically. The algebra that describes the anyons is similar to a construction proposed in \cite{Qalgebra}, although obtained from a very different motivation. To obtain the central idempotents of this algebra we use a simple algorithm described in Appendix \ref{app:idempotents}, see also \cite{friedl1985polynomial}. As expected, we obtain both one and higher dimensional central idempotents.

In Appendix \ref{app:snresults} we list the central idempotents and their properties for the Fibonacci, Ising and $\text{Rep}(S_3)$ string-nets. For each of those, we also compute the topological spin using the standard procedure described in subsection \ref{sec:topspin}. For string-nets, these spins can in principle be computed analytically from the central idempotents. Furthermore, we compute the fusion table describing the fusion of two anyons. Thereto, we have numerically performed the procedure explained in subsection \ref{sec:fusion}. We indeed recover the known fusion rules for the anyonic excitations of these theories. Note that there are no fusion multiplicities larger than one in the models we consider. Finally, we explicitly work out the braid tensor $\mathcal{R}$ using the procedure of \ref{sec:braiding} for two anyons in the Fibonacci string-net model in Appendix~\ref{app:fibonacci}.

\section{Discussion and outlook}\label{sec:conclusions}

For all the examples considered here the PEPS anyon construction is equivalent to calculating the Drinfeld center \cite{drinfeld} of the input theory, i.e. the algebraic structure determined by the single block MPOs $O_a$, which was either a finite group (which can be generalized to a Hopf algebra) or a unitary fusion category. This center construction leads to a modular tensor category, which describes a consistent anyon theory \cite{DrinfeldCenter}. When the input theory is already a modular tensor category by itself, the center construction gives a new modular tensor category, which is isomorphic to two copies of the original anyon theory, one of which is time-reversed  \cite{DrinfeldCenter}. It is then clear that the new anyon theory cannot correspond to a chiral phase. This is actually true in general, i.e. the set of modular tensor categories obtained via the center construction cannot describe chiral phases.  In \cite{LagrangianAlgebra} the set of center modular tensor categories was identified with the set of modular tensor categories containing a so-called Lagrangian subalgebra. A physical connection between the existence of a Lagrangian subgroup and the non-chirality of the quantum phase was given in \cite{LevinEdges} for the case of Abelian statistics.

We have found that PMPOs of the form (\ref{mpo}) that can be used to built MPO-injective PEPS give rise to many concepts familiar from the theory of unitary fusion categories: a finite number of simple objects and associated fusion relations, the pentagon equation, a generalized notion of duality and the Frobenius-Schur indicator $\varkappa_a$, pivotal structure and unitarity. However, there is one important property of unitary fusion categories that does not seem to immediately come out of MPO-injectivity, namely the existence of a unique, simple unit element. In other words, we have not found a property of MPO-injectivity that requires the projector MPO to contain a single block MPO $O_1$, satisfying $O_1O_i = O_iO_1 = O_i$ for all $i$. However, if such identity block is not present then we can associate a multi-fusion category to the PMPO. It is known that multi-fusion categories can also be used to construct string-net models \cite{EnrichingLevinWen}. 

So at this point it seems that the only possibilities to have MPO-injective PEPS that describe physics beyond discrete gauge theories and string-nets without having to extend the MPO-injectivity formalism of \cite{MPOpaper} are given by:
\begin{itemize}
\item[(1)] using PMPOs that have no canonical form;
\item[(2)] defining different left handed MPO tensors to construct $\tilde{P}_{C_v}$;
\item[(3)] not imposing the zipper condition (\ref{zippercondition}).
\end{itemize}
Trying option (2) will most likely lead to a violation of unitarity, in which case the algebra $A_{abcd,\mu\nu}$ can no longer be proven to be a $C^*$-algebra. This will lead to non-Hermitian central idempotents, which to some extent obscures their interpretation as topological sectors. Options (1) and (3) are at the moment much less clear to us, so we will not try to speculate on their implications. It would be very interesting to better understand the implications of options (1) - (3) and see if there is any relation between MPO-injective PEPS and the recently constructed tensor network states for chiral phases \cite{chiral1,chiral2}.

To conclude, we have not only established a connection between MPO-injective PEPS and unitary fusion categories as mentioned above but also a formalism to obtain the topological sectors of the corresponding quantum phase. Similar to previous results \cite{Qalgebra,haah} we can relate topological sectors to the central idempotents of an algebra, which in our case is a $C^*$-algebra constructed from the MPO that determines the injectivity subspace of the ground space tensors. The formalism is constructive and gives the correct anyon types for all the examples we worked out. It furthermore allows us to write down explicit PEPS wave functions that contain an arbitrary number of anyons. This gives an interpretation of topological sectors in terms of entanglement. From the PEPS wave functions containing anyons we can extract universal properties such as fusion relations and topological spins in a very natural way. For certain string-net models we also studied the effect of braiding on the PEPS. 

Several open questions concerning topological order in tensor networks remain. As mentioned above, it is not clear if chiral topological phases fit into the MPO-injectivity formalism, or what --if any-- is the correct formalism to describe gapped chiral theories with tensor networks. For non-chiral topological phases the construction presented here defines an equivalence relation for PMPOs, i.e. two PMPOs are said to be equivalent if the resulting central idempotents have the same topological properties. At this point the (Morita) equivalence relation between PMPOs is very poorly understood. It is also known that there is a substantial interplay between the topological order and global symmetries of a quantum system. Some first progress in  capturing universal properties of these so-called symmetry-enriched topological phases with tensor networks was made in \cite{JiangRan}. A direction for future research which enforces itself upon us at the end of this paper is of course the extension of the presented formalism to fermionic PEPS \cite{fPEPS}. We expect that the concept of MPO algebras should also be connected to topological sectors in fermionic tensor networks. It is conceivable that the concepts introduced here might also be relevant for other types of tensor networks, e.g. the Multi-scale Entanglement Renormalization Ansatz (MERA) descriptions of topological phases \cite{MERA1,MERA2}. Besides these theoretical questions there are also a lot of new applications of MPO-injectivity that come within reach, especially the study of topological phase transitions in non-Abelian anyon theories. We hope to make progress on these matters in future work.
\\ \\
\emph{Acknowlegdements -} We acknowledge helpful discussions with Ignacio Cirac, Tobias Osborne, David Perez-Garcia and Norbert Schuch. We also especially like to thank David Aasen for many inspiring conversations and Zhenghan Wang for pointing out to us the possibility of using multi-fusion categories in string-nets. This work was supported by EU grant SIQS and ERC grant QUERG, the Odysseus grant from the Research Foundation Flanders (FWO) and the Austrian FWF SFB grants FoQuS and ViCoM. M.M. and J.H. further acknowledge the support from the Research Foundation Flanders (FWO).

\appendix

\section{Hermitian PMPOs with unital structure and fusion categories}\label{app:equivalence}

In this appendix we will further consider the connection between (unitary) fusion categories and PMPOs that satisfy the zipper condition \eqref{zippercondition} and have a unital structure. As explained in section \ref{subsec:hermiticity} a PMPO is said to have a unital structure if there exists a unique single block MPO labeled by $1$ such that
\begin{eqnarray}
\rho(M_1) \geq \rho(M_b)\;\;\;\forall b\\
N^1_{bb^*} \geq 1\;\;\;\forall b\, ,
\end{eqnarray}
where $M_a = \sum_{i=1}^D B_a^{ii}$ and $\rho(M_a)$ denotes the spectral radius of $M_a$. It was shown in the main text that this definition implies that $\rho(M_1) > \rho(M_b), \forall b\neq 1$ and $N^1_{bb^*} = 1, \forall b$, with in addition $1^* = 1$ and $\varkappa_1 = 1$. From the symmetry of the tensor $N$
\begin{equation}
N^c_{ab} = N^{a^*}_{bc^*} = N_{c^*a}^{b^*}
\end{equation}
and properties of injective MPS it then follows that
\begin{equation}
N_{ab^*}^1 = N^b_{a1} = N^b_{1a} = \delta_{ab}\, ,
\end{equation}
such that $1$ is indeed the trivial element of the algebra of single block MPOs. We now also require that fusing with the unit element 1 is trivial as expressed by the triangle equation \cite{Kitaev06}:

\begin{align}\label{triangleq}
\vcenter{\hbox{
\includegraphics[width=0.5\linewidth]{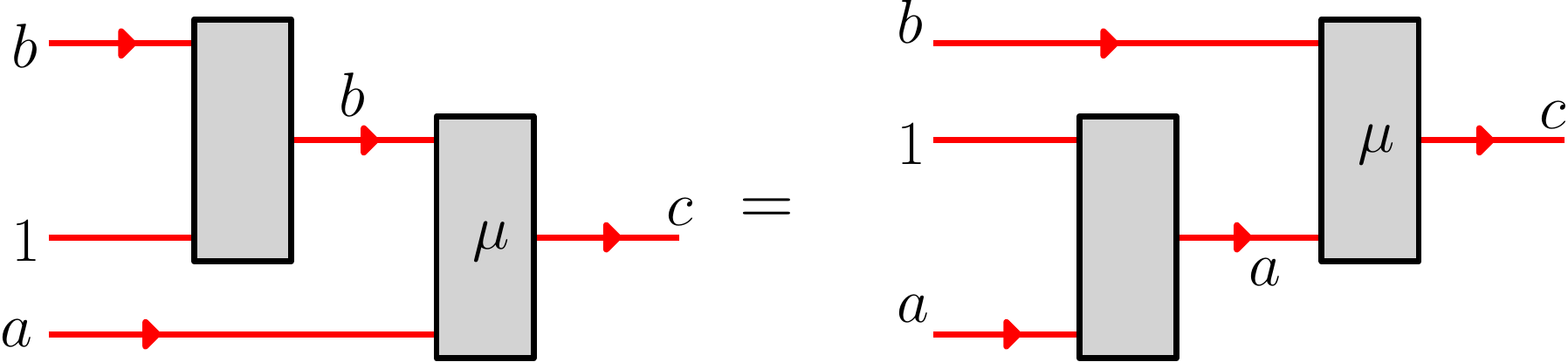}}}
\end{align}
Because the PMPO satisfies the zipper condition we can define the $F$ matrices as in section \ref{subsec:associativity}. The triangle equation can then equivalently be stated as
\begin{equation}
[F^{b1a}_c]_{b1\mu}^{a1\mu} = \delta_{\mu\nu}\, .
\end{equation}
Note that the triangle equation fixes the relative norm and phase of $X_{1a}^a$ and $X_{b1}^b$. Combining the triangle equation \eqref{triangleq} with the pentagon equation \eqref{pentagoneq} gives rise to additional triangle equations:
\begin{equation}
[F^{1ab}_c]_{a1\mu}^{c \nu 1} = \delta_{\mu\nu}
\end{equation}
\begin{equation}
[F^{ab1}_c]_{c\mu 1}^{b 1 \nu} = \delta_{\mu\nu} \, .
\end{equation}
With these assumptions one can use results from the theory of injective MPS to show that there exists a choice of fusion tensors such that:
\begin{equation}\label{property1}
[F^{aa^*a}_a]_1^1 = \frac{\varkappa_a}{d_a}
\end{equation}
\begin{equation}\label{property3}
[F^{a^*aa^*}_a]_1^1 = [(F^{aa^*a}_a)^{-1}]_1^1 = \frac{\varkappa_{a^*}}{d_{a^*}}
\end{equation}
\begin{equation}
\frac{1}{d_a} = \frac{1}{d_{a^*}} > 0\, ,
\end{equation}
where $\varkappa_a = \pm 1$ is defined as in section \ref{subsec:hermiticity} via $Z_a \bar{Z}_{a^*} = \varkappa_a\mathds{1}$. If we now also assume the unitarity condition of section \ref{subsec:zipper}, then we can relate the numbers $d_a$ obtained from the $F$-symbol to the Perron-Frobenius vector of the $N$ coefficients, i.e.:
\begin{equation}\label{property2}
d_a d_b = \sum_c N_{ab}^c d_c
\end{equation}
Identities \eqref{property1} to \eqref{property2} are very tedious to prove. Since they are not the main focus of this paper, the proofs will be given elsewhere.

The positive numbers $d_a$ are called the \emph{quantum dimensions} of the simple objects in the unitary fusion category. Given the quantum dimensions and the $F$ matrices, equations \eqref{property1} and \eqref{property3} are then used to define the Frobenius-Schur indicator $\varkappa_a$ in category theory \cite{Kitaev06}. So we see that our definition of $\varkappa_a$ via the gauge matrices $Z_a$ coincides with that of the Frobenius-Schur indicator in fusion categories for this special class of PMPOs.

\section{$C^*$-Algebra structure of $A_{abcd,\mu\nu}$}\label{app:algebra}

\subsection{Closedness under multiplication}

We consider the objects $A_{abcd,\mu\nu}$ defined in \eqref{algebraobject} and show they form an algebra under matrix multiplication, i.e. we show that
\begin{equation}
A_{hegf,\lambda\sigma}A_{abcd,\mu\nu}=\delta_{ga}\sum_{ij,\rho\tau}\Omega_{(hegf,\lambda\sigma)(abcd,\mu\nu)}^{(hjci,\rho\tau)} A_{hjci,\rho\tau}.
\end{equation}
Using the zipper condition \eqref{zippercondition} we see that we can neglect the MPO tensors and that the object we have to decompose is given by

\begin{align} \sum_{\alpha i}
\vcenter{ \hbox{
\includegraphics[width=0.24\linewidth]{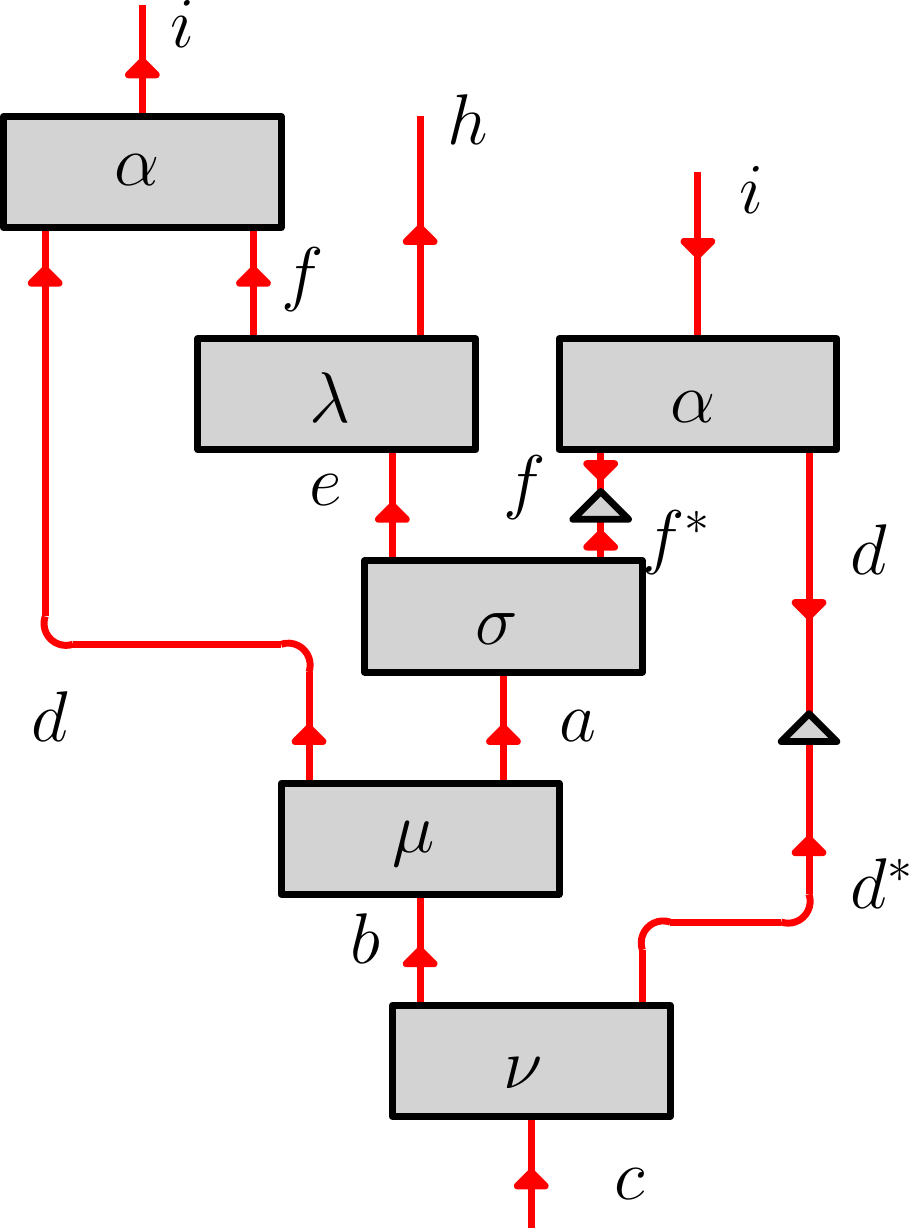}}}
\end{align}
As a first step we use \eqref{pivotalthree} and a `$F$-move' \eqref{Fmove} to obtain

\begin{align} = \sum_{\alpha\beta\gamma\delta ij}\left(C^{i^*\, -1}_{f^*d^*}\right)_{\alpha\beta}\left(F^{def^*}_b \right)^{(a\sigma\mu)}_{(j\gamma\delta)}
\vcenter{\hbox{
\includegraphics[width=0.24\linewidth]{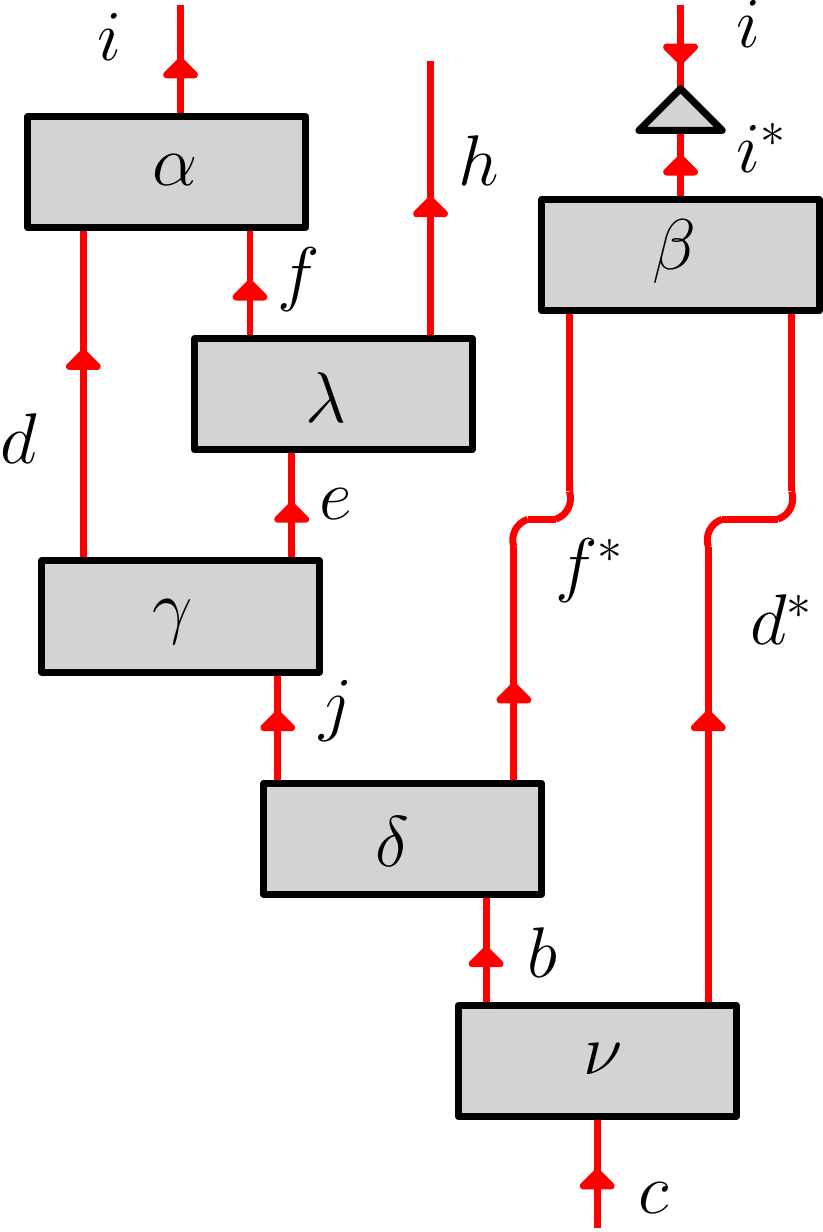}}}
\end{align}
A final two $F$-moves then lead to

\begin{align}\label{eq:structureconstants}
= \sum_{\alpha\gamma\delta j\rho i \kappa\tau} \left(C^{i^*\,-1}_{f^*d^*} \right)_{\rho\kappa}\left(F^{def^*}_b \right)^{(a\sigma\mu)}_{(j\gamma\delta)}\left(F^{dfh}_j \right)^{(e\lambda\gamma)}_{(i\alpha\rho)}\left(F^{jf^*d^*\; -1}_c \right)^{(b\nu\delta)}_{(i^*\beta\tau)} \nonumber\\
\vcenter{\hbox{
\includegraphics[width=0.17\linewidth]{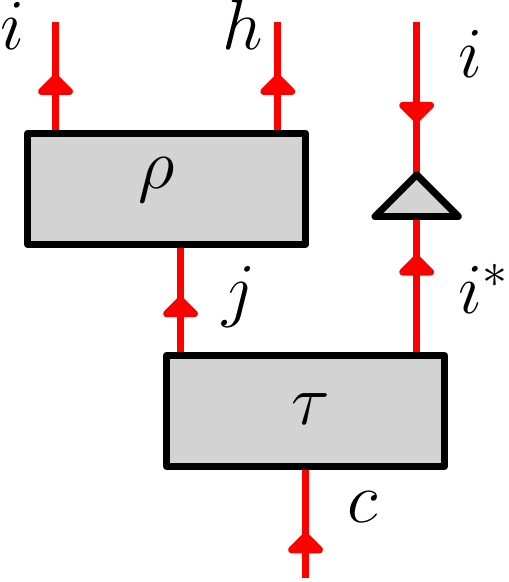}}}
\end{align}
So we indeed obtain a decomposition of the composite object in terms of the original ones, giving the desired algebra structure.

\subsection{Closedness under Hermitian conjugation}\label{subsec:hermitianconjugation}

We now show that the algebra formed by the elements $A_{abcd,\mu\nu}$ is closed under Hermitian conjugation, i.e. $A_{abcd,\mu\nu}^\dagger = \sum_{e,\kappa\lambda}(\Theta_{abcd,\mu\nu})^{e\kappa\lambda}A_{cead^*,\kappa\lambda}$. Starting from the basis element
$$
A_{abcd,\mu\nu} =  \vcenter{\hbox{\includegraphics[width=0.27\linewidth]{algebraobject}}}.
$$
we can implement Hermitian conjugation in the following way:
\begin{itemize}
\item Complex conjugation and exchanging the inner and outer indices of the MPO tensors amounts simply to reversing the arrow on the red line, due to the the relation between left- and right-handed MPO tensors in Eq.~(\ref{eq:lefthandedtensor}).
\item Complex conjugation of the gauge matrix $Z_{d}$, which also simply amounts to reversing the red arrows as discussed in section~\ref{subsec:zipper}.
\item Exchanging the inner and outer indices connected to (the left inverses of) the fusion tensors.
\item Complex conjugation of the fusion tensors, which reverses the red arrows and reconnects them to the other side of the box: the complex conjugate of the fusion tensor is the transpose of the inverse (=hermitian conjugated) fusion tensor. 
\end{itemize}
This gives rise to the following diagram for $A_{abcd,\mu\nu}^\dagger$:
\begin{align}
A_{abcd,\mu\nu}^\dagger = \vcenter{\hbox{
\includegraphics[width=0.3\linewidth]{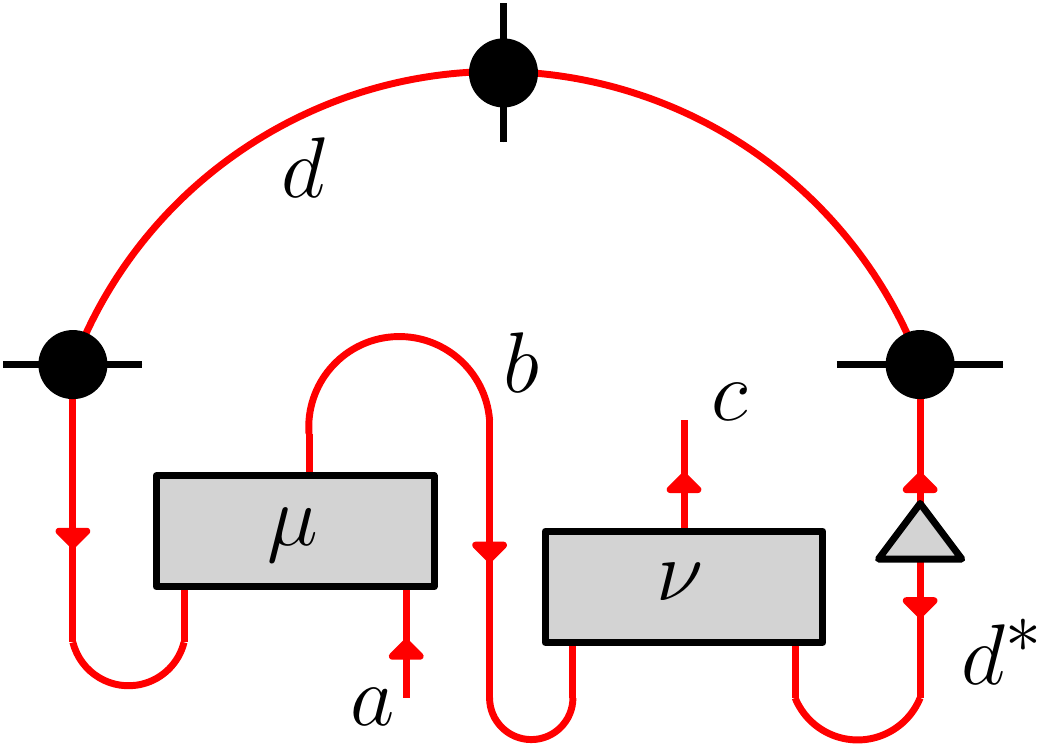}}}
\end{align}
Restoring the orientation of the original loop results in
\begin{align}
A_{abcd,\mu\nu}^\dagger = \vcenter{\hbox{
\includegraphics[width=0.23\linewidth]{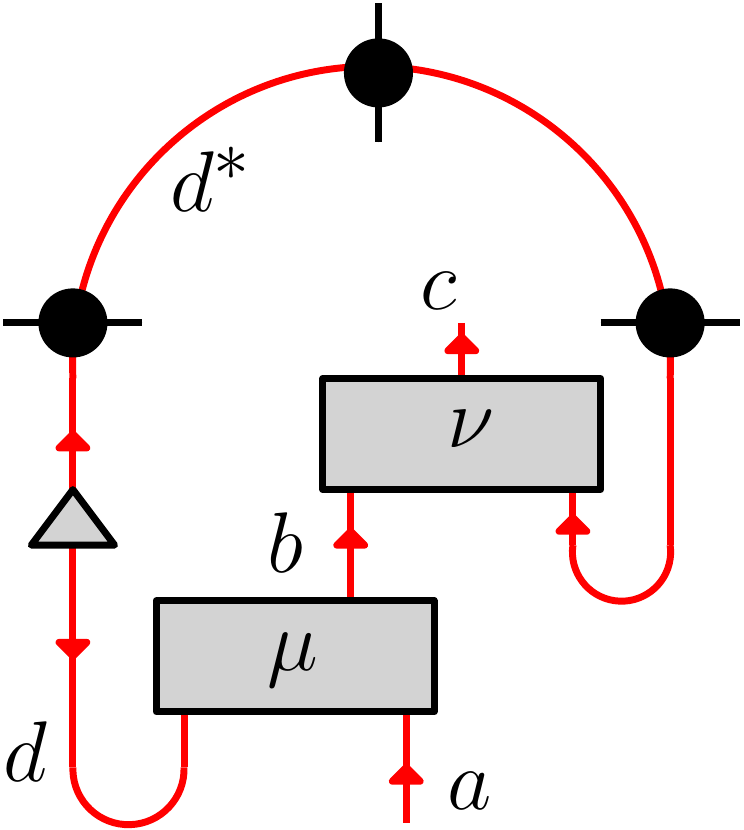}}}
\end{align}
We can now use an $A$ and $A'$ move [see Eq.~\eqref{pivotalnew}] to obtain
\begin{align}
A_{abcd,\mu\nu}^\dagger = \sum_{\rho\sigma} (A^{b}_{d^*a})_{\mu\rho} (A'^{c}_{bd})_{\nu\sigma} \vcenter{\hbox{
\includegraphics[width=0.2\linewidth]{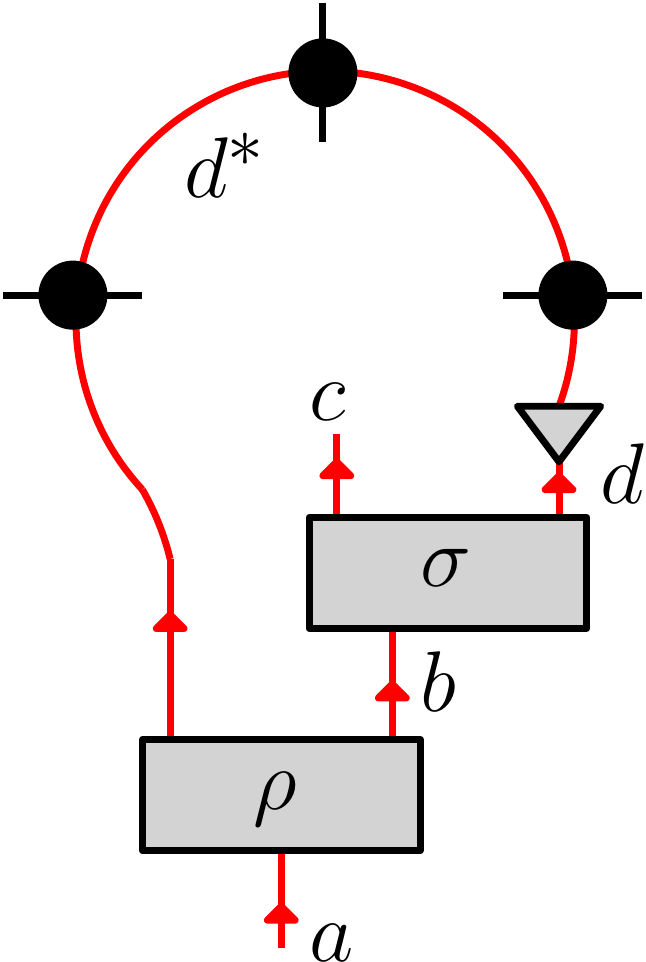}}}
\end{align}
Using a final $F$ move and Eq.~\eqref{FSindicator} we obtain the required relation
\begin{align}
A_{abcd,\mu\nu}^\dagger = \sum_{\rho\sigma e \kappa \lambda} (A^{b}_{d^*a})_{\mu\rho} (A'^{c}_{bd})_{\nu\sigma} \varkappa_{d} (F^{d^*cd}_{a})^{b\sigma \rho}_{e \kappa \lambda} A_{cead^*,\kappa \lambda}.
\end{align}

\section{Finding central idempotents}\label{app:idempotents}
In this appendix we present a simple and constructive algorithm to calculate the decomposition of an algebra $\mathcal{A}$ over $\mathbb{C}$ in primitive central idempotents. The constructive approach to the Artin-Wedderburn theorem is well known in the literature \cite{friedl1985polynomial} and can be generalized to algebras over different base fields. We assume that the Jacobson radical of $\mathcal{A}$ is trivial, one can check this for instance by computing the kernel of a proper matrix, see \cite{friedl1985polynomial} for more details.

The input of the algorithm are the structure constants $d_{ij}^k$ of the algebra $\mathcal{A}$ with respect to a vector space basis $\{b_1,\dots,b_r\}$. We have $b_ib_j=\sum_{k=1}^rd_{ij}^kb_k$. The output of the algorithm are the coefficients in this basis of the minimal central idempotents. These are the elements $p  \in \mathcal{A}$ such that $p\neq 0, p^2=p$, $p$ commutes with every element in $\mathcal{A}$ and $p$ cannot be written as $p=p_1+p_2$ where $p_1,p_2$ also satisfy the previous requirements. Finding the minimal central idempotents is equivalent to determining the block decomposition of a matrix algebra.

We denote the column vector of coefficients of an element $x$ with respect to this basis as $c(x)$. We first calculate the center $Z(\mathcal{A})$ of $\mathcal{A}$. Let $x = \sum_{j=1}^r x_jb_j$. It holds that $x\in Z(\mathcal{A})$ iff $b_ix=xb_i$ for all $i$. It is easy to see that this is equivalent to $\sum_{j=1}^r(d_{ij}^k-d_{ji}^k)x_j=0$ for all $k,i$. We conclude that $x \in Z(\mathcal{A})$ iff $c(x) \in \text{Kern}(Z)$ with $Z_{(i-1)r+k,j}=d_{ij}^k-d_{ji}^k$.

Let $\{z_1,\ldots z_c\}$ be a basis of $\text{Kern}(Z)$. We can easily obtain the structure constants $f_{ij}^k$ with respect to this basis by solving the linear system
$$
\sum_k f_{ij}^k c(z_k) = c(z_iz_j)
$$
for all $i,j$.

We now forget the algebra $\mathcal{A}$ and only work in the commutative algebra $\mathcal{C}:=Z(\mathcal{A})$. From now on, we denote by $c(z)$  the column vector of coefficients of an element $z\in \mathcal{C}$ with respect to the basis $\{z_1,\ldots z_c\}$.

Given an element $z$, recall that the ideal generated by $z$ is defined by $\braket{z}=\text{span}\{xz \: | \: x \in \mathcal{C}\}$. If we take a random element $z \in \mathcal{C}$, we expect that $\braket{z}=\mathcal{C}$. Let us now show how to decompose an ideal as $\braket{z} = \braket{z_1} \oplus \braket{z_2}$.

First, we find a basis of the space $\braket{z}$. This is easily done by computing a basis of the column space of the matrix $[c(zz_1) \dots c(zz_c)]$. Let $\{y_1,\ldots, y_d\}$ be a basis of $\braket{z}$. Second, we compute the identity $I_z$ of the ideal, this is the unique element with $I_{z}y_i=y_i$ for all $i$. After a straightforward calculation we obtain that the coefficients $C_j$ of the identity $I_z$ with respect to the basis $\{y_1,\dots,y_d\}$ are given by the solution of the linear system
$$
\sum_{j=1}^d\left(\sum_{l=1}^c\sum_{m=1}^cc(y_j)_lc(y_k)_m f_{lm}^p\right)C_j = c(y_k)_p
$$
for all $p$.

We can now decompose the ideal $\braket{z}$. The minimal polynomial $P$ with $P(z)=0$ can be calculated as follows. Find the smallest $q$ such that the matrix $[c(z^q) \dots c(z)\: c(I_z)]$ is rank deficient. The zero vector of this matrix gives the coefficients of $P$. Let $n_1,\ldots,n_q$ be the complex roots of $P$, hence $P(x)=\prod_{i=1}^q (x-n_i)$. If $q=1$, the ideal $\braket{z}$ is one dimensional. This implies that $z^2=\lambda z$, hence $z/\lambda$ is an idempotent. If $q>1$ we decompose $P(x)=P_1(x)P_2(x)$ such that $P_1,P_2$ have no common roots.

We claim that $\braket{P_1(z)}\oplus\braket{P_2(z)}:=R(z)$ is the sought after decomposition of $\braket{z}$.
First, we show that the equality holds in $\braket{z}=R(z)$. Clearly the inclusion $\supseteq$ holds. We now show the reverse inclusion. Since $P_1$ and $P_2$ are polynomials over $\mathbb{C}$ and have no common roots, they are coprime. B\'ezout's identity ensures the existence of two polynomials $Q_1,Q_2$ such that $1=Q_1P_1+Q_2P_2$. Evaluating both sides in $z$ gives that $I_z \in R(z)$. Since $R(z)$ is an ideal, $xI_z \in R(z)$ for all $x$, by which we can conclude that $\braket{z} \subseteq R(z)$. It is worth noting that $Q_1(z)P_1(z)$ is the identity of $\braket{P_1(z)}$, hence the calculation of the identity only needs to be performed once at the start of the algorithm.

Second, we show that $\braket{P_1(z)}$ and $\braket{P_1(z)}$ are orthogonal spaces. Take $w_1P_1(z) \in \braket{P_1(z)}$ and $w_2P_2(z) \in \braket{P_2(z)}$, then the equality $w_1P_1(z)w_2P_2(z)=0$ holds since $\mathcal{C}$ is commutative and $P(z)=P_1(z)P_2(z)=0$. This implies that the sum is direct and that $\braket{z}=\braket{P_1(z)}\oplus\braket{P_2(z)}$. Since $\mathcal{C}$ is finite, we can apply this decomposition recursively and after a finite number of steps we find the primitive idempotents of $\mathcal{C}$, from which we can easily obtain those of $\mathcal{A}$.

\section{Results for string-nets}\label{sec:snidempotents}\label{app:snresults}

\subsection{Fibonacci string-net}\label{app:fibresults}
The first example we discuss is the simplest non-Abelian string-net model. As input we use the modular tensor category of Fibonacci anyons. We expect to find central idempotents corresponding to the topological sectors of the doubled Fibonacci theory.

\subsubsection{MPO-tensors}
The categorical data of the Fibonacci theory is well known. The theory has two labels $1$ and $\tau$ that satisfy the non-Abelian fusion rules
$$
N_{11}^{1}= N_{\tau 1}^{\tau}=N_{1\tau }^{\tau}= N_{\tau\tau}^1=N_{\tau \tau}^{\tau}=1\, ,
$$
other multiplicities are zero. The quantum dimensions are given by $d_1=1$ and $d_{\tau}=\frac{1+\sqrt{5}}{2}:= \phi$.
The remaining crucial information are the $F$-symbols of this theory. They are given by
\begin{equation}\label{eq: Fsymboltrivial}
[F^{abc}_{d}]_{e}^{f}=F^{abc}_{def}=\delta_{abe}\delta_{cde}\delta_{adf}\delta_{bcf} F^{abc}_{def},
\end{equation}
where $\delta_{ijk}=1$ if $i,j,k$ can fuse to $1$, i.e. $N_{ij}^k>0$, and $\delta_{ijk}=0$ zero otherwise.
The non-trivial elements of $F$ are given by
$$
F^{\tau\tau\tau}_{\tau 1 1}=\frac{1}{\phi}, \quad
F^{\tau\tau\tau}_{\tau \tau 1}=\frac{1}{\sqrt{\phi}},\quad
F^{\tau\tau\tau}_{\tau 1 \tau}=\frac{1}{\sqrt{\phi}},\quad
F^{\tau\tau\tau}_{\tau \tau \tau}=-\frac{1}{\phi}.
$$
All other non-zero components of $F$ are one. The construction of the tensors is most easily described using the scalars $v_i = \sqrt{d_i}$ and the $G$ symbols,
\begin{equation}\label{eq:Gsymbol}
G^{abc}_{def}=\frac{1}{v_e v_f}F^{abc}_{def}.
\end{equation}

As shown in \cite{MPOpaper}, the Fibonacci string-net state can now be described by a projector MPO constructed from the tensors
\begin{align}\label{StringnetMPO}
\vcenter{\hbox{
 \includegraphics[width=0.35\linewidth]{StringNetMPO_RHS.pdf}}}
\end{align}

After removing the zero rows and columns, this MPO has bond dimension $5$ and consists of two blocks $B_1$ and $B_{\tau}$ of dimension $2$ and $3$ respectively. The blocks $B_1$, $B_{\tau}$ satisfy the Fibonacci fusion rules.
The diagonal matrix $\Delta$ from equation \eqref{mpo} is given by the quantum dimensions of the block labels divided by the square of the total quantum dimension: $w_1=\frac{1}{1+\phi^2}$ and $w_{\tau}=\frac{\phi}{1+\phi^2}$.

For these MPOs we also know the explicit form of the fusion tensors $X$:
\begin{align}\label{StringnetFusiontensor}
\vcenter{\hbox{
 \includegraphics[width=0.3\linewidth]{StringNetFusion_RHS.pdf}}}.
\end{align}

\subsubsection{Central idempotents}

Here we give the central idempotents and their topological spins for the Fibonacci string-net.
Recall that the algebra we decompose is generated by the following basis elements
$$
A_{1111}, A_{\tau\tau\tau 1}, A_{1\tau 1\tau}, A_{1\tau\tau\tau}, A_{\tau 1\tau\tau}, A_{\tau\tau 1\tau}, A_{\tau\tau\tau\tau}.
$$
All other possible elements are zero due to the fusion rules. We find 4 different idempotents, of which $\mathcal{P}_1,\mathcal{P}_2,\mathcal{P}_3$ are one-dimensional and $\mathcal{P}_4$ has dimension two:
\begin{align*}
\mathcal{P}_1&=\frac{1}{\sqrt{5}}\left(\frac{1}{\phi}A_{1111}+\sqrt{\phi}A_{1\tau 1\tau}\right)\\
\mathcal{P}_2&=\frac{1}{\sqrt{5}}\left(\frac{1}{\phi}A_{\tau\tau\tau 1}+\frac{1}{\sqrt{\phi}}e^{-\frac{4\pi i}{5}}A_{\tau 1\tau\tau}+e^{\frac{3\pi i}{5}}A_{\tau\tau\tau\tau}\right)\\
\mathcal{P}_3&=\frac{1}{\sqrt{5}}\left(\frac{1}{\phi}A_{\tau\tau\tau 1}+\frac{1}{\sqrt{\phi}}e^{\frac{4\pi i}{5}}A_{\tau 1\tau\tau}+e^{-\frac{3\pi i}{5}}A_{\tau\tau\tau\tau}\right)\\
\mathcal{P}_4&=\frac{1}{\sqrt{5}}\left(\phi A_{1111}+ A_{\tau\tau\tau 1} -\sqrt{\phi}A_{1\tau 1\tau}+\sqrt{\phi}A_{\tau 1\tau\tau}+\frac{1}{\phi}A_{\tau\tau\tau\tau}\right).
\end{align*}
We recognize $\mathcal{P}_1$ as the vacuum particle. Indeed, when we write out this tensor, we find a diagonal tensor with weights depending on the inner MPO label. These weights correspond exactly to the weights that determine the ground state tensors in the MPO framework \cite{MPOpaper}, denoted by $\Delta$ in equation \eqref{mpo}. More generally, we see in all other examples that we always recover the vacuum particle corresponding to the ground state.

There are some other general remarks we can already see in this example. The vectors $A_{1\tau\tau\tau}$ and $A_{\tau\tau 1\tau}$ are not present in any of the idempotents. These are exactly the vectors $A_{abcd}$ with a different incoming $a$ and outgoing $c$ label. We do not expect them to be present in the decomposition of a central idempotent, as they correspond exactly to off diagonal nilpotent matrices that are not in the center of the algebra. The decomposition of the higher dimensional central idempotent $\mathcal{P}_4$ in irreducible, but not central, one-dimensional idempotents is very simple. The element $\mathcal{P}_4$ contains both terms with $a,c=1$ and $a,c=\tau$. The decomposition of $\mathcal{P}_4$ in two one-dimensional idempotents is obtained by grouping all terms with $a,c=1$ as one idempotent and all terms with $a,c=\tau$ as the second idempotent. This procedure also holds for more general models. All other, one-dimensional, idempotents only contain terms with $a,c=1$ or $a,c= \tau$. Note that a $d$-dimensional idempotent projects onto a $d^2$ dimensional subspace, such that we indeed recover the algebra dimension as $7=1^2+1^2+1^2+2^2$. This is required for our set of central idempotents to be complete.

The topological spins we obtain are given by
$$
h_1=0,\: h_2=-\frac{4}{5},\: h_3=\frac{4}{5},\: h_4=0.
$$
Clearly, we can now make the identification with the well-known anyons from the doubled Fibonacci theory:
$$\mathcal{P}_1=(1,1),\: \mathcal{P}_2=(1,\bar{\tau}),\: \mathcal{P}_3=(\tau,1),\: \mathcal{P}_4=(\tau,\bar{\tau}).$$
We can compare this result with the idempotents obtained in \cite{Qalgebra} and see that both solutions have a similar structure. With a slightly different convention of the basis elements $A_{abcd}$, corresponding to a normalization that depends on $a,b,c,d$, we obtain exactly the same multiplication table and idempotents.

\subsection{Ising string-net}
As a second example we look at the string-net obtained from the Ising fusion category.
\subsubsection{MPO-tensors}
As we saw in the analysis of the Fibonacci model in the previous subsection, we only need the fusion rules, quantum dimensions and $F$-symbols to construct the relevant tensors. The Ising category has three labels $1,\sigma,\psi$ with fusion rules
$$
N_{11}^1=1, N_{1\sigma}^{\sigma}=1, N_{1\psi}^{\psi}=1, N_{\sigma \sigma}^{\psi}=1,
$$
up to the usual allowed permutations of the labels. The only non-trivial fusion rule is $\sigma \times \sigma = 1 + \psi$. The quantum dimension are given by $d_1=1, d_{\sigma}=\sqrt{2}, d_{\psi}=1$.

The $F$-symbols are again given by $F^{abc}_{def}\neq 0$ iff all appearing fusion processes are allowed, see equation \eqref{eq: Fsymboltrivial}. The non-trivial elements are given by
$$
F^{\sigma\sigma\sigma}_{\sigma 1 1}=\frac{1}{\sqrt{2}},\:
F^{\sigma\sigma\sigma}_{\sigma \psi 1}=\frac{1}{\sqrt{2}},\:
F^{\sigma\sigma\sigma}_{\sigma 1 \psi}=\frac{1}{\sqrt{2}},\:
F^{\sigma\sigma\sigma}_{\sigma \psi \psi}=-\frac{1}{\sqrt{2}},\:
F^{\psi \sigma \psi}_{\sigma \sigma \sigma}=-1, \:
F^{\sigma \psi \sigma}_{\psi \sigma \sigma}=-1,
$$
other allowed non-zero components are one. Similarly as for the Fibonacci model, we can now construct the $G$-symbols and from these all necessary tensors, see equations \eqref{eq:Gsymbol},\eqref{StringnetMPO},\eqref{StringnetFusiontensor} in the previous subsection.
\subsubsection{Central idempotents}
We now have all the tensors required to calculate the central idempotents of the Ising string-net.
From the fusion rules we find that there are $12$ non-zero basis elements $A_{abcd}$. The algebra generated by these elements has 9 central idempotents, given by
\begin{align*}
\mathcal{P}_1 &=\frac{1}{4}\left(A_{1111}+2^{3/4}A_{1\sigma 1\sigma}+A_{1\psi 1\psi}\right)\\
\mathcal{P}_2 &=\frac{1}{4}\left(A_{\sigma\sigma \sigma 1}+2^{1/4}e^{\frac{\pi i}{8}}A_{\sigma 1\sigma\sigma}+2^{1/4}e^{-\frac{3\pi i}{8}}A_{\sigma\psi\sigma\sigma}+e^{\frac{\pi i}{2}}A_{\sigma\sigma\sigma\psi} \right)\\
\mathcal{P}_3 &=\frac{1}{4}\left(A_{\sigma\sigma \sigma 1}+2^{1/4}e^{-\frac{\pi i}{8}}A_{\sigma 1\sigma\sigma}+2^{1/4}e^{\frac{3\pi i}{8}}A_{\sigma \psi\sigma\sigma}+e^{-\frac{\pi i}{2}}A_{\sigma\sigma\sigma\psi} \right)\\
\mathcal{P}_4 &=\frac{1}{4}\left(A_{\psi\psi\psi1}+2^{3/4}e^{\frac{\pi i}{2}}A_{\psi\sigma\psi\sigma}-A_{\psi 1\psi\psi}\right)\\
\mathcal{P}_5 &=\frac{1}{4}\left(A_{\psi\psi\psi1}+2^{3/4}e^{\frac{-\pi i}{2}}A_{\psi\sigma\psi\sigma}-A_{\psi 1\psi\psi}\right)\\
\mathcal{P}_6 &=\frac{1}{4}\left(A_{\sigma\sigma\sigma 1}+2^{1/4}e^{-\frac{7 \pi i}{8}}A_{\sigma 1\sigma\sigma}+2^{1/4}e^{\frac{5 \pi i}{8}}A_{\sigma \psi\sigma\sigma}+e^{\frac{\pi i}{2}}A_{\sigma\sigma\sigma\psi}\right)\\
\mathcal{P}_7 &=\frac{1}{4}\left(A_{\sigma\sigma\sigma1}+2^{1/4}e^{\frac{7 \pi i}{8}}A_{\sigma 1\sigma\sigma}+2^{1/4}e^{-\frac{5 \pi i}{8}}A_{\sigma\psi\sigma\sigma}+e^{-\frac{\pi i}{2}}A_{\sigma\sigma\sigma\psi}\right)\\
\mathcal{P}_8 &=\frac{1}{4}\left(A_{1111}-2^{3/4}A_{1\sigma 1\sigma}+A_{1\psi 1\psi}\right)\\
\mathcal{P}_9 &=\frac{1}{2}\left(A_{1111}+A_{\psi\psi\psi 1}-A_{1\psi 1\psi}+A_{\psi 1\psi\psi}\right)
\end{align*}
The corresponding topological spins are found to be
\begin{align*}
h_1&=0,\: h_2=\frac{1}{16},\: h_3=-\frac{1}{16},\: h_4=\frac{1}{2}, \: h_5=-\frac{1}{2}\\
h_6&=-\frac{7}{16},\: h_7=\frac{7}{16},\: h_8=0, \: h_9=0.
\end{align*}
All central idempotents are one dimensional, except for $\mathcal{P}_9$ which is two-dimensional, such that we indeed obtain $12=8 \cdot 1^2 + 2^2$. We can now identify these central idempotents with the anyons in the double Ising model as follows,
\begin{align*}
\mathcal{P}_1&=(1,1),\: \mathcal{P}_2=(\sigma,1),\: \mathcal{P}_3=(1,\bar{\sigma}),\\
\mathcal{P}_4&=(\psi,1),\: \mathcal{P}_5=(1,\psi),\: \mathcal{P}_6=(\sigma,\bar{\psi}),\\
 \mathcal{P}_7&=(\psi,\bar{\sigma}),\: \mathcal{P}_8=(\psi,\bar{\psi}),\: \mathcal{P}_9=(\sigma,\bar{\sigma}).
\end{align*}

\subsection{$\text{Rep}(S_3)$ string-net}
As a final example we consider the string-net with input fusion category the representation theory of $S_3$. As this last category is not modular, the anyons of the string-net are not just doubled versions of the labels of the input data.

\subsubsection{MPO-tensors}
Again, we need to specify the categorical data of the input category and can construct the tensors of the $\text{Rep}(S_3)$ string-net from these. The $\text{Rep}(S_3)$ fusion category has three labels $1,2,3$ with following fusion rules,
$$
N_{11}^1=1, N_{12}^2=1, N_{13}^3=1, N_{33}^2=1, N_{33}^3=1.
$$
up to the allowed permutations of the labels. The non-trivial fusion rule is $3 \times 3 = 1 + 2 + 3$. The quantum dimensions of the labels are $d_1=1, d_2=1, d_3=2$.

As always, the $F$-symbols are given by $F^{abc}_{def} \neq 0 $ if all appearing fusion processes are allowed as in equation \eqref{eq: Fsymboltrivial}. The non-trivial elements are given by
\begin{align*}
F^{323}_{333}&=-1,\quad F^{332}_{333}=-1,\quad F^{233}_{333}=-1,\quad F^{333}_{233}=-1,\\
F^{333}_{311}&=\frac{1}{2},\quad
F^{333}_{312}=\frac{1}{2}, \quad
F^{333}_{313}=\frac{1}{\sqrt{2}},\quad
F^{333}_{321}=\frac{1}{2},\quad
F^{333}_{322}=\frac{1}{2},\\
F^{333}_{323}&=-\frac{1}{\sqrt{2}},\quad
F^{333}_{331}=\frac{1}{\sqrt{2}},\quad
F^{333}_{332}=-\frac{1}{\sqrt{2}},\quad
F^{333}_{333}=0\, ,
\end{align*}
other allowed coefficients are 1. Similarly as for the Fibonacci model, we can now construct the $G$-symbols and from these all necessary tensors, see equations \eqref{eq:Gsymbol},\eqref{StringnetMPO} and \eqref{StringnetFusiontensor}.

\subsubsection{Central idempotents}
The algebra for the given fusion rules is 17-dimensional. We find 8 different central idempotents,
\begin{align*}
\mathcal{P}_1&=\frac{1}{6}A_{3331}-\frac{1}{6}A_{3332}+\frac{1}{3\sqrt{2}}e^{-2\pi i/3}A_{3133}+\frac{1}{3\sqrt{2}}e^{\pi i/3}A_{3233}+\frac{1}{3}e^{2 \pi i/3}A_{3333}\\
\mathcal{P}_2 &= \frac{1}{6}A_{2221}+\frac{1}{6}A_{2122}-\frac{\sqrt{2}}{3}A_{2323}\\
\mathcal{P}_3 &= \frac{1}{2}A_{2221}+\frac{1}{4}A_{3331}-\frac{1}{2}A_{2122}+\frac{1}{4}A_{3332}-\frac{1}{2\sqrt{2}}A_{3133}-\frac{1}{2\sqrt{2}}A_{3233}\\
\mathcal{P}_4 &= \frac{1}{6}A_{3331}-\frac{1}{6}A_{3332}+\frac{1}{3\sqrt{2}}e^{2 \pi i /3}A_{3133}+\frac{1}{3\sqrt{2}}e^{-\pi i/3}A_{3233}+\frac{1}{3}e^{-2 \pi i/3}A_{3333}\\
\mathcal{P}_5&= \frac{1}{6}A_{3331}-\frac{1}{6}A_{3332}+\frac{1}{3\sqrt{2}}A_{3133}-\frac{1}{3\sqrt{2}}A_{3233}+\frac{1}{3}A_{3333}\\
\mathcal{P}_6&=\frac{1}{3}A_{1111}+\frac{1}{3}A_{2221}+\frac{1}{3}A_{1212}+\frac{1}{3}A_{2122}-\frac{\sqrt{2}}{3}A_{1313}+\frac{\sqrt{2}}{3}A_{2323}\\
\mathcal{P}_7&=\frac{1}{2}A_{1111}+\frac{1}{4}A_{3331}-\frac{1}{2}A_{1212}+\frac{1}{4}A_{3332}+\frac{1}{2\sqrt{2}}A_{3133}+\frac{1}{2\sqrt{2}}A_{3233}\\
\mathcal{P}_8&=\frac{1}{6}A_{1111}+\frac{1}{6}A_{1212}+\frac{\sqrt{2}}{3}A_{1313}.
\end{align*}
The idempotents $\mathcal{P}_3,\mathcal{P}_6,\mathcal{P}_7$ are two-dimensional; all other central idempotents have dimension one. We again check the consistency condition $17 = 1+1+1+1+1+2^2+2^2+2^2$, which ensures our set of central idempotents is complete.
 
The only non-zero topological spins are given by
$$
h_1=-\frac{1}{3}, \: h_3=\frac{1}{2}, \: h_4=\frac{1}{3}.
$$

As explained in the main test in subsection \ref{sec:fusion} we can compute the fusion rules of the anyons corresponding to the 8 central idempotents.
We numerically find the following fusion table.
\footnotesize
\begin{center}
  \begin{tabular}{ c || c | c | c | c | c | c | c | c |}
     	   & $\mathcal{P}_1$ & $\mathcal{P}_2$ & $\mathcal{P}_3$ & $\mathcal{P}_4$ & $\mathcal{P}_5$ & $\mathcal{P}_6$ & $\mathcal{P}_7$ & $\mathcal{P}_8$ \\ \hline \hline
     $\mathcal{P}_1$ & $\mathcal{P}_1+\mathcal{P}_2+\mathcal{P}_8$ & $\mathcal{P}_1$ & $\mathcal{P}_3+\mathcal{P}_7$ & $\mathcal{P}_5+\mathcal{P}_6$ & $\mathcal{P}_4+\mathcal{P}_6$ & $\mathcal{P}_4+\mathcal{P}_5$ & $\mathcal{P}_3+\mathcal{P}_7$ & $\mathcal{P}_1$ \\ \hline
     $\mathcal{P}_2$ & $\mathcal{P}_1$ & $\mathcal{P}_8$ & $\mathcal{P}_7$ & $\mathcal{P}_4$ & $\mathcal{P}_5$ & $\mathcal{P}_6$ & $\mathcal{P}_3$ & $\mathcal{P}_2$ \\ \hline
	 $\mathcal{P}_3$ & $\mathcal{P}_3+\mathcal{P}_7$ & $\mathcal{P}_7$ & $\mathcal{P}_1+\mathcal{P}_4+\mathcal{P}_5$ & $\mathcal{P}_3+\mathcal{P}_7$ & $\mathcal{P}_3+\mathcal{P}_7$ & $\mathcal{P}_3+\mathcal{P}_7$ & $\mathcal{P}_2+\mathcal{P}_3+\mathcal{P}_4$ & $\mathcal{P}_3$\\
	       &           &       & $+\mathcal{P}_6+\mathcal{P}_8$     &           &           &           & $+\mathcal{P}_5+\mathcal{P}_6$     &  \\ \hline
	 $\mathcal{P}_4$ & $\mathcal{P}_5+\mathcal{P}_6$ & $\mathcal{P}_4$ & $\mathcal{P}_3+\mathcal{P}_7$ & $\mathcal{P}_2+\mathcal{P}_4+\mathcal{P}_8$ & $\mathcal{P}_1+\mathcal{P}_6$ & $\mathcal{P}_1+\mathcal{P}_5$ & $\mathcal{P}_3+\mathcal{P}_7$ & $\mathcal{P}_4$\\ \hline
	 $\mathcal{P}_5$ & $\mathcal{P}_4+\mathcal{P}_6$ & $\mathcal{P}_5$ & $\mathcal{P}_3+\mathcal{P}_7$ & $\mathcal{P}_1+\mathcal{P}_6$ & $\mathcal{P}_2+\mathcal{P}_5+\mathcal{P}_8$ & $\mathcal{P}_1+\mathcal{P}_4$ & $\mathcal{P}_3+\mathcal{P}_7$ & $\mathcal{P}_5$\\  \hline
	 $\mathcal{P}_6$ & $\mathcal{P}_4+\mathcal{P}_5$ & $\mathcal{P}_6$ & $\mathcal{P}_3+\mathcal{P}_7$ & $\mathcal{P}_1+\mathcal{P}_5$ & $\mathcal{P}_1+\mathcal{P}_4$ & $\mathcal{P}_2+\mathcal{P}_6+\mathcal{P}_8$ & $\mathcal{P}_3+\mathcal{P}_7$ & $\mathcal{P}_6$\\  \hline
	 $\mathcal{P}_7$ & $\mathcal{P}_3+\mathcal{P}_7$ & $\mathcal{P}_3$ & $\mathcal{P}_2+\mathcal{P}_3+\mathcal{P}_4$ & $\mathcal{P}_3+\mathcal{P}_7$ & $\mathcal{P}_3+\mathcal{P}_7$ & $\mathcal{P}_3+\mathcal{P}_7$ & $\mathcal{P}_1+\mathcal{P}_4+\mathcal{P}_5$  & $\mathcal{P}_7$\\
	       &           &       & $+\mathcal{P}_5+\mathcal{P}_6$     &           &           &           & $+\mathcal{P}_6+\mathcal{P}_8$     &  \\ \hline
	 $\mathcal{P}_8$ & $\mathcal{P}_1$ & $\mathcal{P}_2$ & $\mathcal{P}_3$ & $\mathcal{P}_4$ & $\mathcal{P}_5$ & $\mathcal{P}_6$ & $\mathcal{P}_7$  & $\mathcal{P}_8$\\

    \hline
  \end{tabular}
\end{center}
\normalsize
The $S$-matrix can be calculated as explained in subsection \ref{sec:Smatrix}. We find that
$$
S=\frac{1}{6}
\begin{pmatrix}

     1 &    3  &   2   &  1   &  2   &  3   &  2  &   2\\
     3  &   3  &   0  &  -3   &  0  &  -3  &   0  &   0\\
     2  &   0   &  4  &   2  &  -2  &   0  &  -2  &  -2\\
     1  &  -3  &   2   &  1  &   2  &  -3   &  2  &   2\\
     2  &   0   & -2  &   2 &   -2   &  0   & -2  &   4\\
     3  &  -3  &   0  &  -3  &   0   &  3 &    0   &  0\\
     2  &   0  &  -2 &    2  &  -2   &  0  &   4  &  -2\\
     2 &    0  &  -2   &  2  &   4   &  0  &  -2  &  -2
\end{pmatrix}.
$$
These results agree with the theoretical findings in the literature, see for instance \cite{DS3}.

\section{Vacuum $(\tau,\bar{\tau})$-pair} 

Suppose we have a $\mathcal{P}_{\tau\overline{\tau}}$ anyon pair. We can write the tensors that live on the sites containing an anyon as a sum of a tensor with a $1$ MPO attached to it and a tensor with a $\tau$ MPO attached to it,

\begin{align} \label{eq:anyonpair1}
\vcenter{\hbox{
 \includegraphics[width=0.25\linewidth]{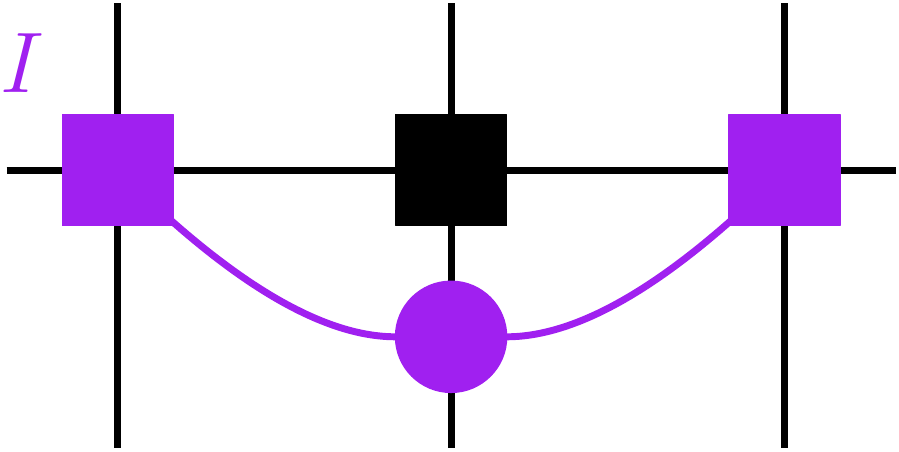}}} =
\vcenter{\hbox{
 \includegraphics[width=0.25\linewidth]{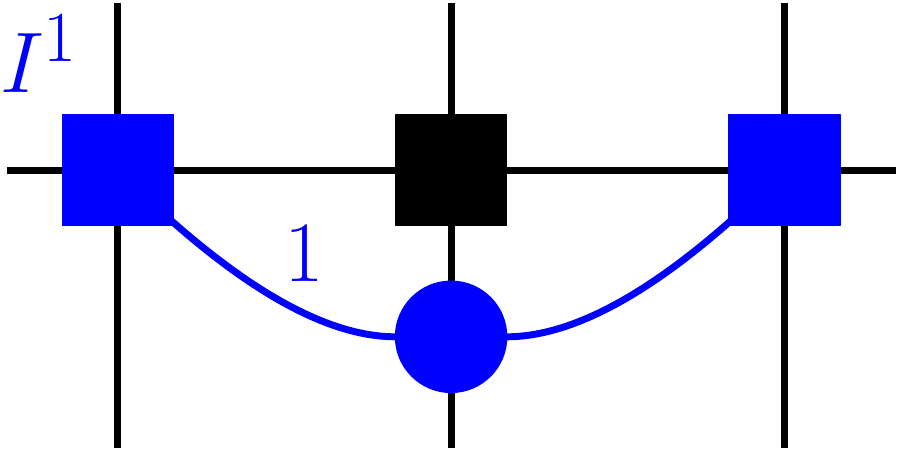}}}+
\vcenter{\hbox{
  \includegraphics[width=0.25\linewidth]{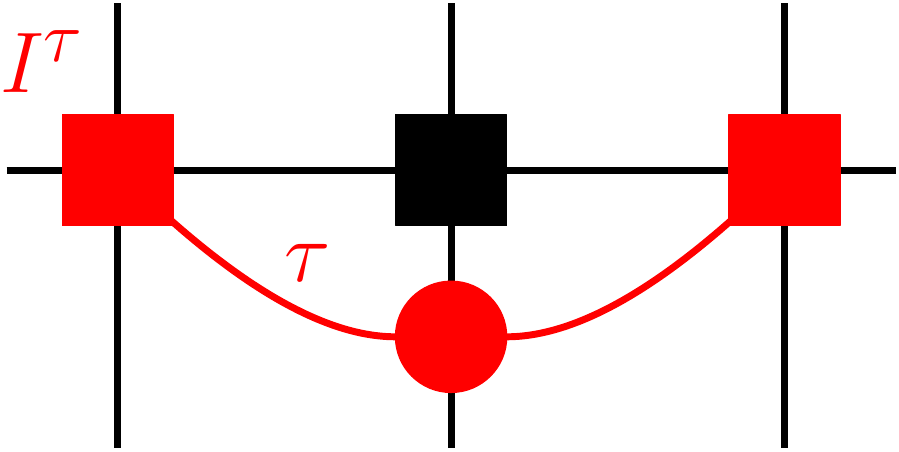}}}.
\end{align}

As we created this pair from the ground state, it is itself in the trivial topological sector, hence invariant under the idempotent of the trivial sector, which is the MPO projector. Hence the tensor network containing this pair is equivalent to the network containing the pair projected on the MPO subspace,
\begin{align*}
\vcenter{\hbox{
 \includegraphics[width=0.25\linewidth]{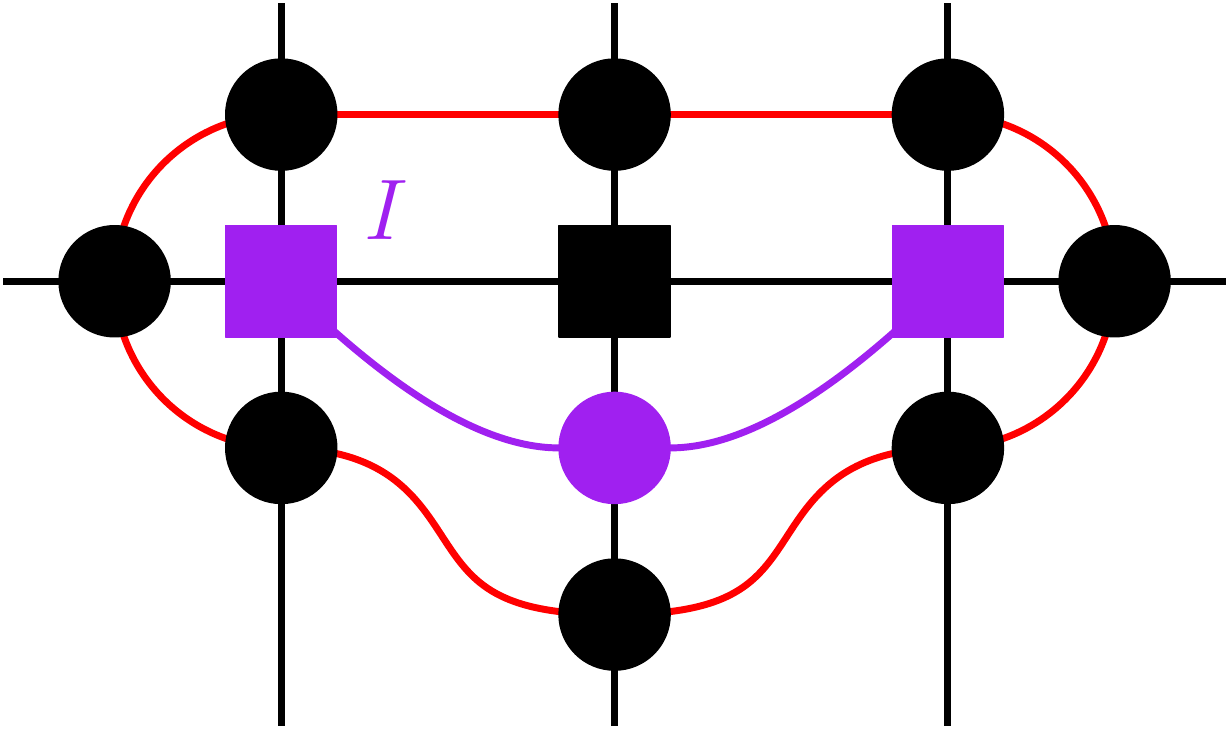}}} &=
\frac{1}{\mathcal{D}^2}\vcenter{\hbox{
 \includegraphics[width=0.25\linewidth]{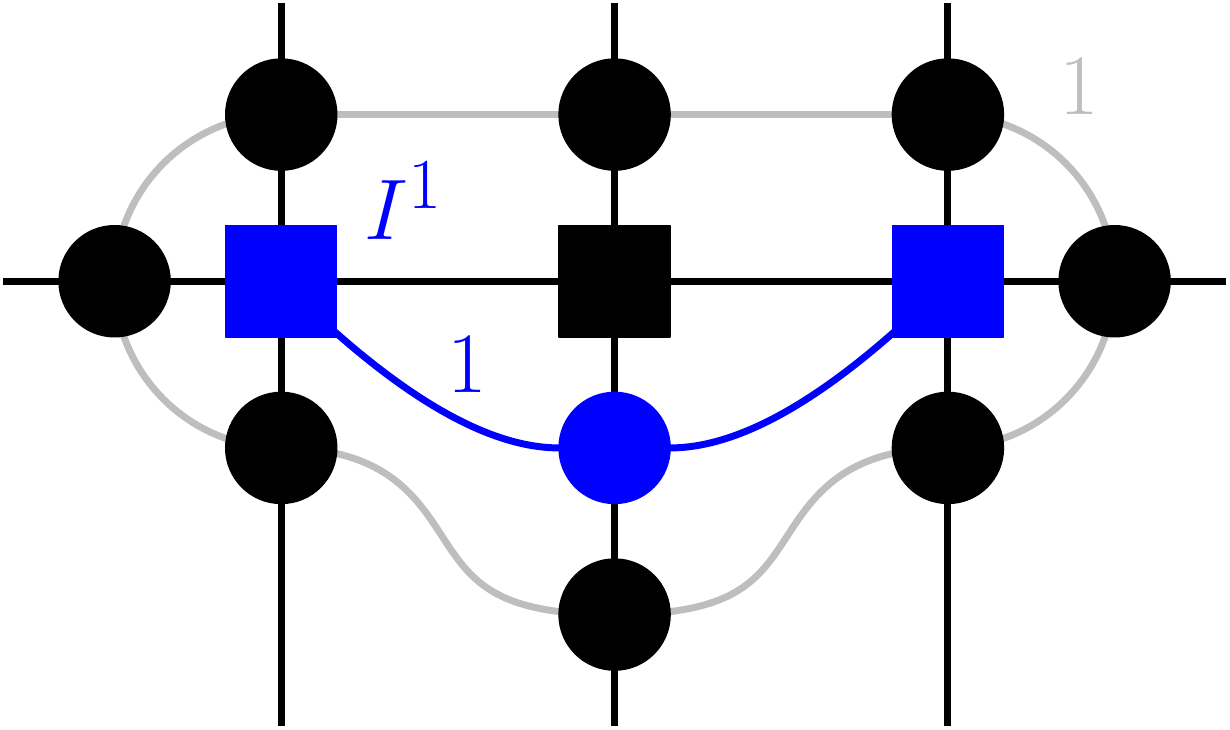}}}+
\frac{\phi}{\mathcal{D}^2}\vcenter{\hbox{
  \includegraphics[width=0.25\linewidth]{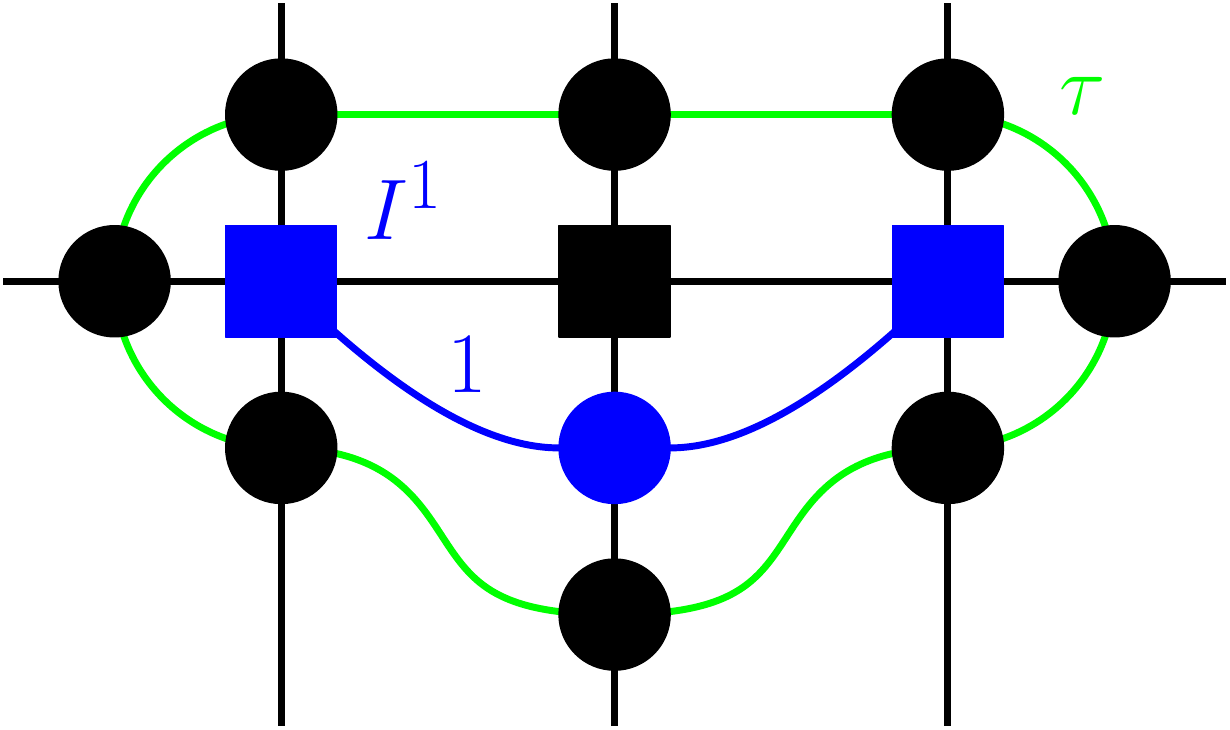}}}\\
  &+
\frac{1}{\mathcal{D}^2}\vcenter{\hbox{
  \includegraphics[width=0.25\linewidth]{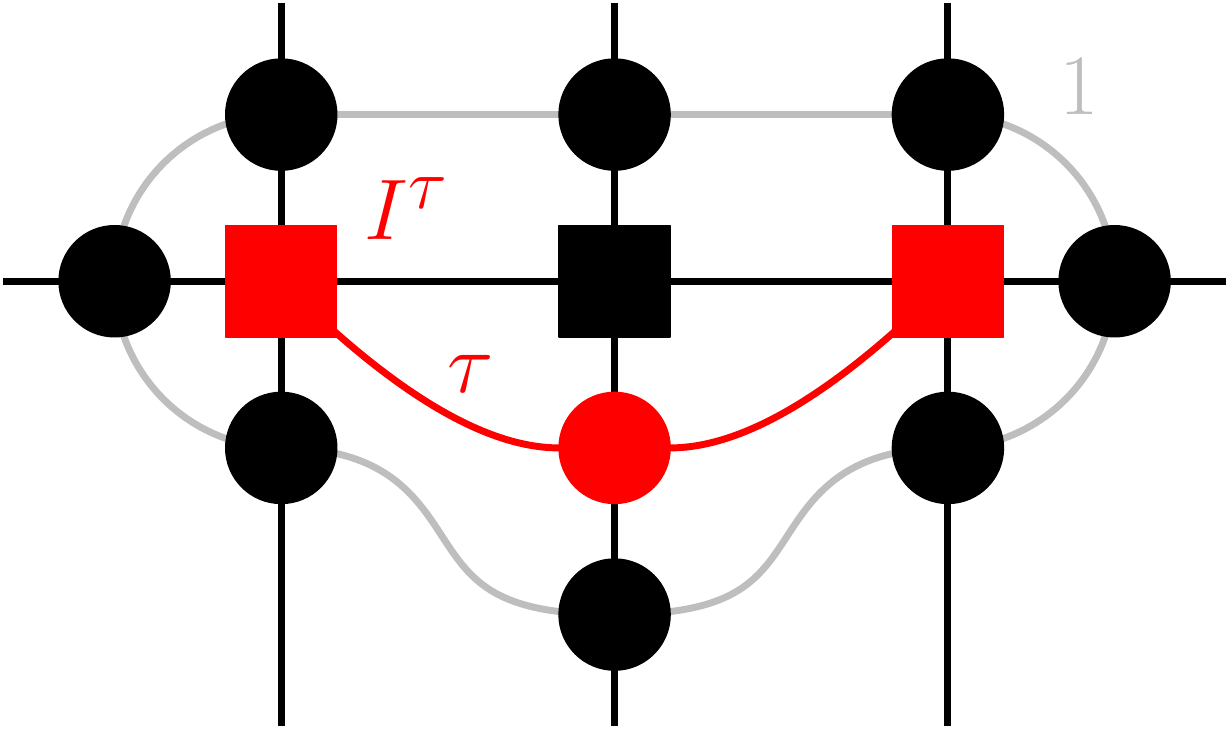}}}+
\frac{\phi}{\mathcal{D}^2}  \vcenter{\hbox{
  \includegraphics[width=0.25\linewidth]{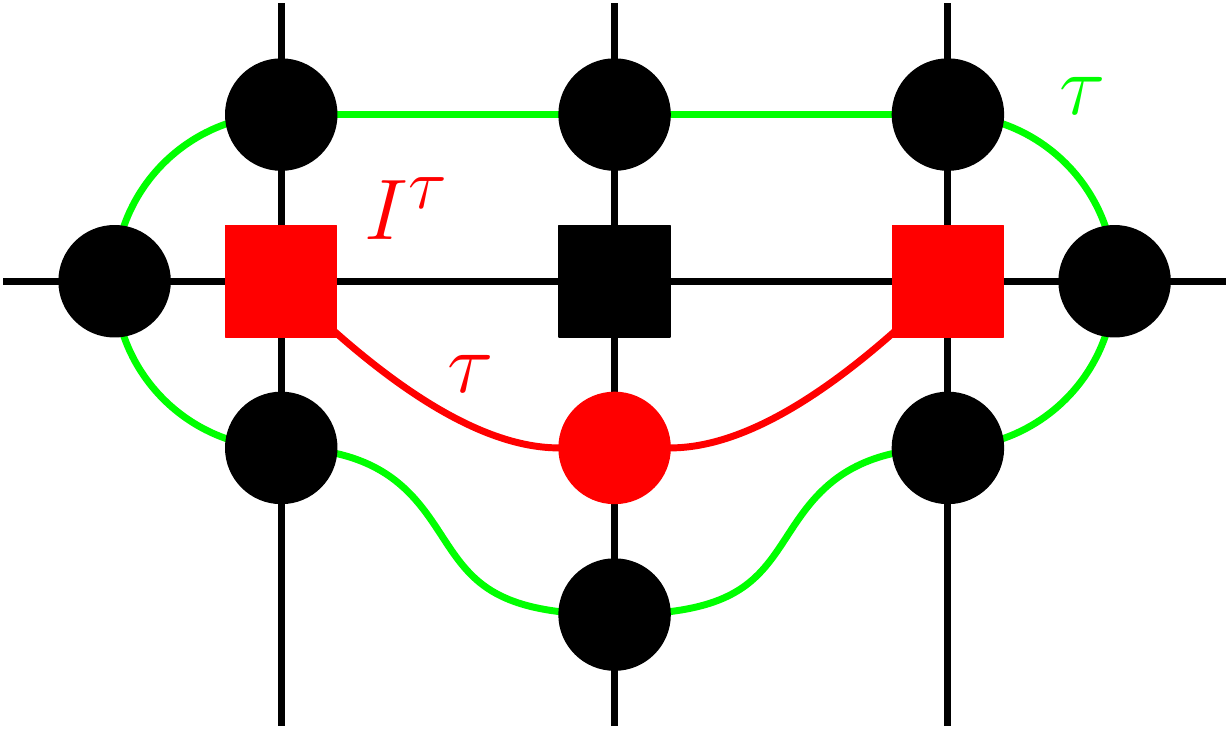}}}.
\end{align*}
Here, $\mathcal{D}^2 = 1+\phi^2$ is the square of the total quantum dimension of the Fibonacci model. We can now apply the same procedure as in the calculation of $\mathcal{R}_{\mathcal{P}_{\tau},\tau}$, which yields,
\begin{equation}
\begin{split}
\vcenter{\hbox{
 \includegraphics[width=0.25\linewidth]{tautauinvariantLHS.pdf}}} &=
\frac{1}{\mathcal{D}^2}\vcenter{\hbox{
 \includegraphics[width=0.25\linewidth]{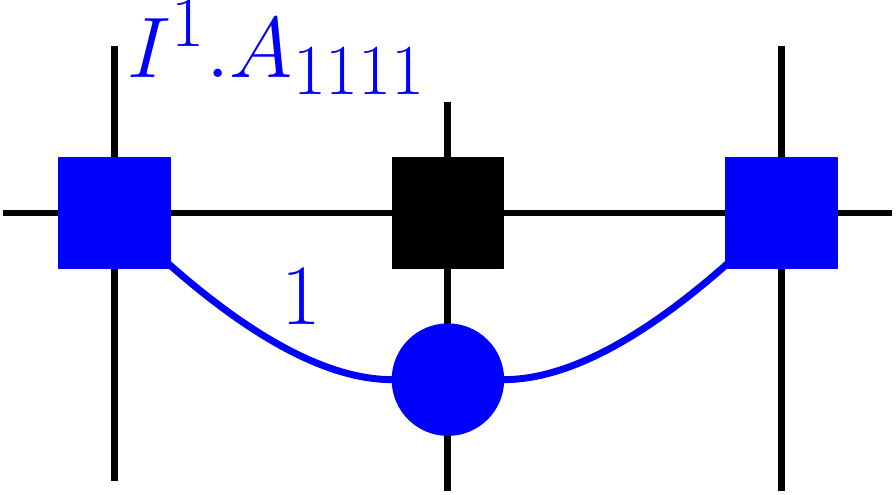}}}+
\frac{\phi}{\mathcal{D}^2}\vcenter{\hbox{
  \includegraphics[width=0.25\linewidth]{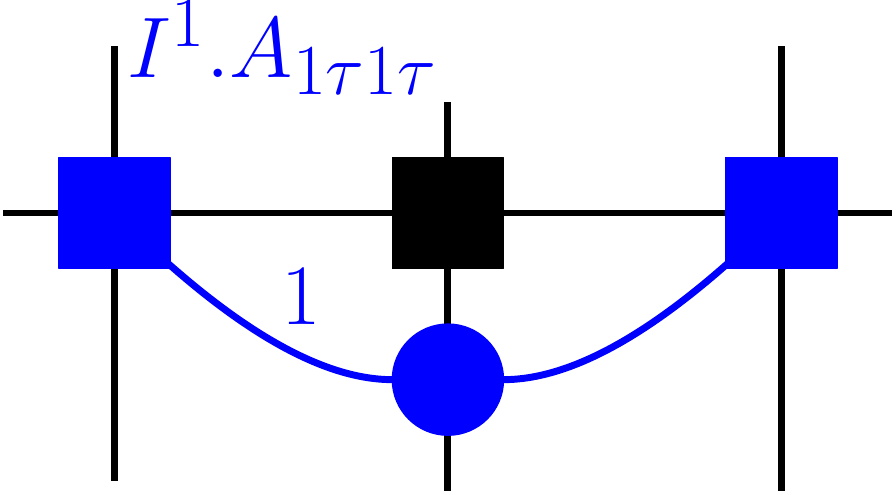}}}\\
  &+
\frac{\phi}{\mathcal{D}^2}\vcenter{\hbox{
  \includegraphics[width=0.25\linewidth]{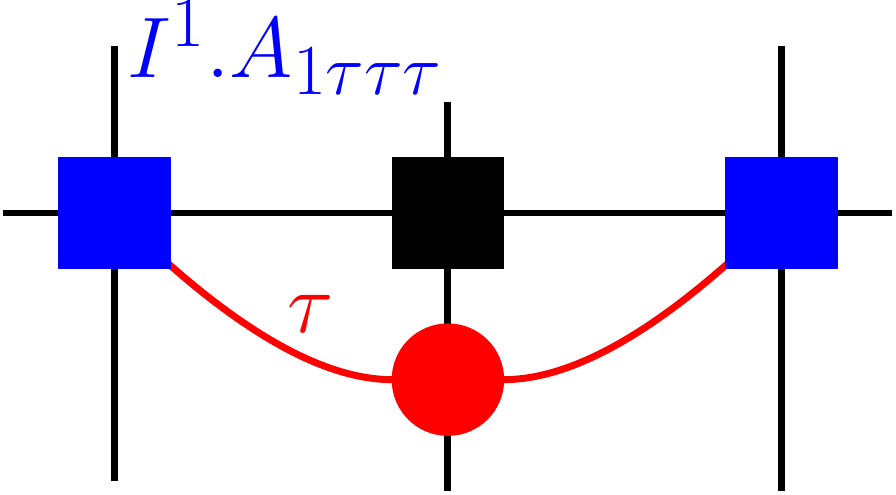}}}+
\frac{1}{\mathcal{D}^2}  \vcenter{\hbox{
  \includegraphics[width=0.25\linewidth]{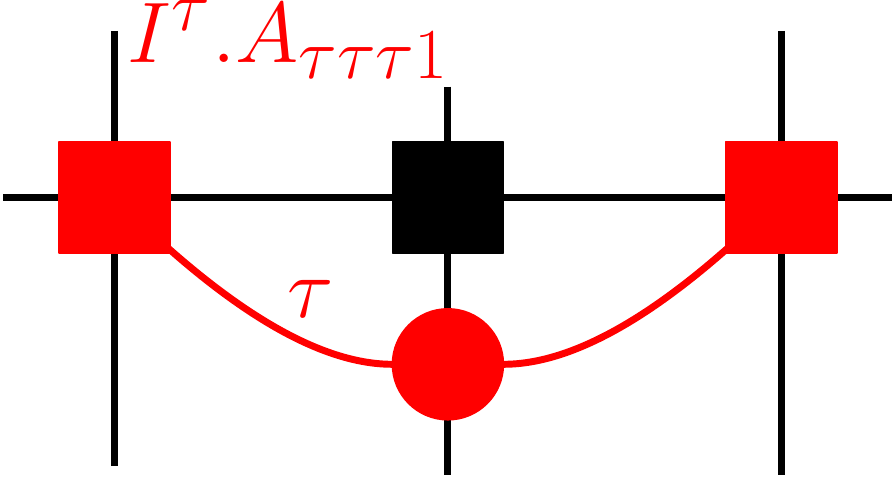}}}\\
  &+
\frac{\phi}{\mathcal{D}^2}\vcenter{\hbox{
  \includegraphics[width=0.25\linewidth]{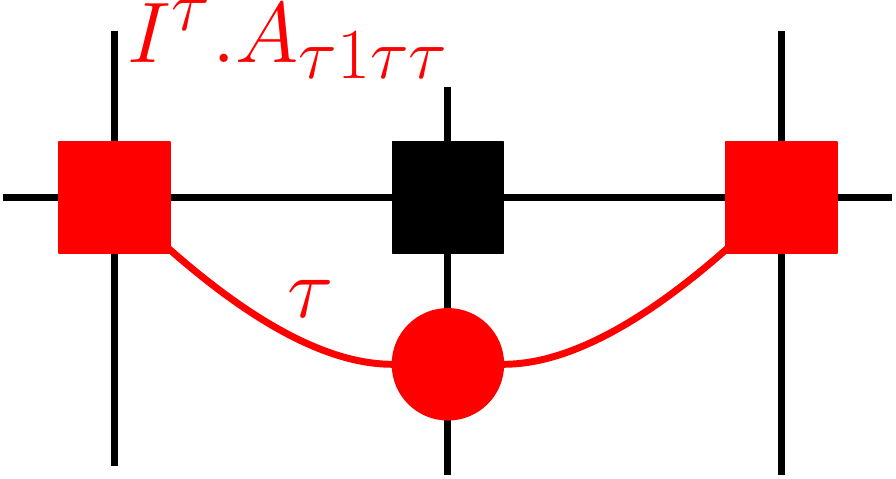}}}+
\frac{\phi}{\mathcal{D}^2}  \vcenter{\hbox{
  \includegraphics[width=0.25\linewidth]{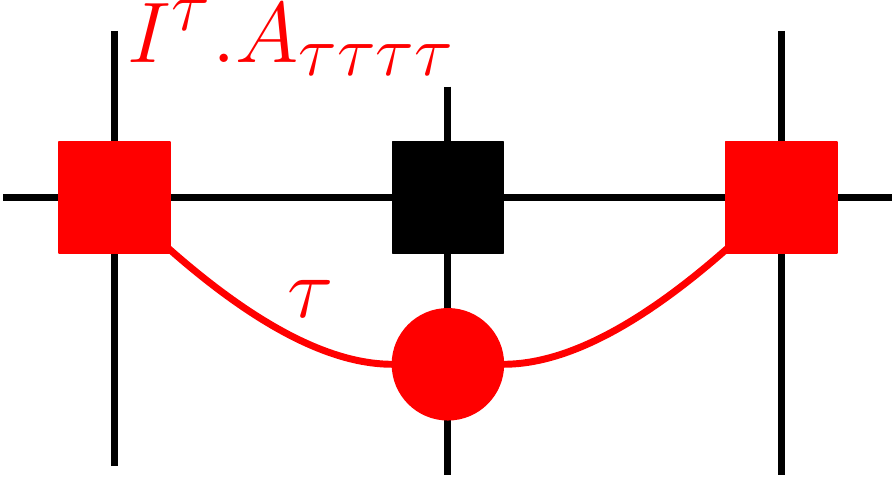}}}\\
  &+\frac{\phi}{\mathcal{D}^2}  \vcenter{\hbox{
  \includegraphics[width=0.25\linewidth]{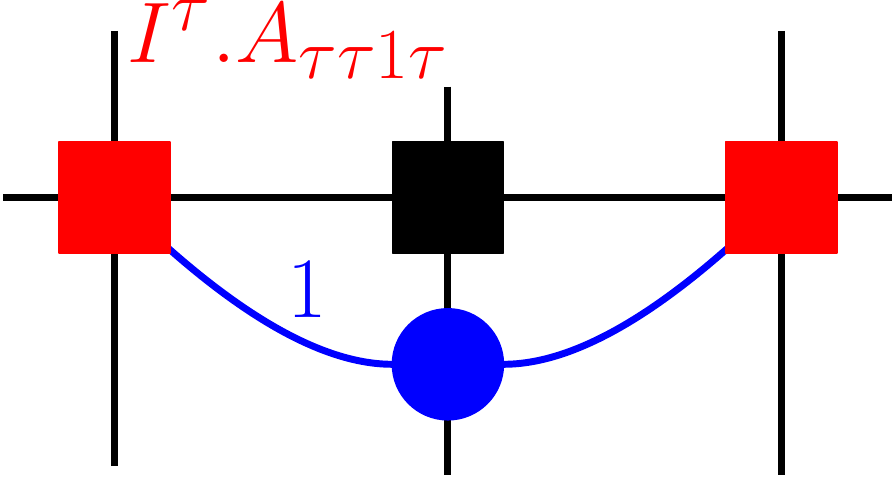}}}.
\end{split}
\end{equation}
The products $\mathcal{P}_{\tau\overline{\tau}}.A_{abcd}$ are easily calculated. We find that
\begin{equation}\label{eq:anyonpair2}
\begin{split}
\vcenter{\hbox{
 \includegraphics[width=0.25\linewidth]{tautauinvariantLHS.pdf}}} &=
\frac{1}{\mathcal{D}^2}\vcenter{\hbox{
 \includegraphics[width=0.25\linewidth]{tautauRHS1.pdf}}}+
\frac{1}{\mathcal{D}^2\phi^2}\vcenter{\hbox{
  \includegraphics[width=0.25\linewidth]{tautauRHS1.pdf}}}\\
  &+
\frac{\phi}{\mathcal{D}^2}\vcenter{\hbox{
  \includegraphics[width=0.25\linewidth]{tautauinvariantRHS33.pdf}}}+
\frac{1}{\mathcal{D}^2}  \vcenter{\hbox{
  \includegraphics[width=0.25\linewidth]{tautauRHS2.pdf}}}\\
  &+
\frac{1}{\mathcal{D}^2}\vcenter{\hbox{
  \includegraphics[width=0.25\linewidth]{tautauRHS2.pdf}}}+
\frac{1}{\mathcal{D}^2\phi^3}  \vcenter{\hbox{
  \includegraphics[width=0.25\linewidth]{tautauRHS2.pdf}}}\\
  &+\frac{\phi}{\mathcal{D}^2}  \vcenter{\hbox{
  \includegraphics[width=0.25\linewidth]{tautauinvariantRHS37.pdf}}}.
\end{split}
\end{equation}

As the pair of anyons is in the vacuum sector we can equate the right hand side of equation \eqref{eq:anyonpair1} to the right hand side of equation \eqref{eq:anyonpair2}. Now, the difference between a $1$ or $\tau$ string can be decided in the presence of other anyons. Hence for this equality to hold it needs to hold for the diagrams with a $1$ and $\tau$ string separately. This gives the following two relations,
\begin{align}\label{eq:anyonpair3}
\vcenter{\hbox{
 \includegraphics[width=0.25\linewidth]{tautauRHS1.pdf}}}
 &=
\frac{1+\frac{1}{\phi^2}}{\mathcal{D}^2}\vcenter{\hbox{
 \includegraphics[width=0.25\linewidth]{tautauRHS1.pdf}}} +
\frac{\phi}{\mathcal{D}^2}\vcenter{\hbox{
  \includegraphics[width=0.25\linewidth]{tautauinvariantRHS37.pdf}}}\nonumber\\
  \vcenter{\hbox{
 \includegraphics[width=0.25\linewidth]{tautauRHS2.pdf}}}
 &=
\frac{2+\frac{1}{\phi^3}}{\mathcal{D}^2}\vcenter{\hbox{
 \includegraphics[width=0.25\linewidth]{tautauRHS2.pdf}}} +
\frac{\phi}{\mathcal{D}^2}\vcenter{\hbox{
  \includegraphics[width=0.25\linewidth]{tautauinvariantRHS33.pdf}}}.
\end{align}
Let us denote by $B_1,B_{\tau}$ the local tensors supported on $P_{\tau,\overline{\tau}}^{1}, P_{\tau,\overline{\tau}}^{\tau}$ respectively. In the figures, $B_1$ is denoted by a blue square and $B_{\tau}$ by a red square.

Clearly the pair of tensors $B_{\tau}.A_{\tau\tau 1\tau}$ give rise to the same tensor network as the pair $B_1$. Similar, the pair $B_{1}.A_{1 \tau\tau\tau}$ and $B_{\tau}$ give rise to the same network. Hence we can choose
\begin{align}\label{eq:B1Btau}
B_1 &=\left(\frac{\sqrt{5}}{\phi}\right)^{-1/2} B_{\tau}.A_{\tau\tau 1 \tau},&B_{\tau} &=\left(\frac{\sqrt{5}}{\phi^2}\right)^{-1/2} B_{1}.A_{1 \tau\tau\tau}.
\end{align}
Both requirements are consistent as $A_{\tau\tau 1 \tau}.A_{1 \tau\tau\tau} = \frac{\sqrt{5}}{\phi\sqrt{\phi}}P^{1}_{\tau\overline{\tau}}$ and
$A_{1 \tau\tau\tau}.A_{\tau\tau 1 \tau}= \frac{\sqrt{5}}{\phi\sqrt{\phi}}P^{\tau}_{\tau\overline{\tau}}$.
We see that, as expected, the requirement that a pair of $\mathcal{P}_{\tau\overline{\tau}}$ anyons is in the topologically trivial sectors, restricts the choice of different $B_1,B_{\tau}$ tensors by enforcing a strict relation between them.

Let us now take such a pair of $\mathcal{P}_{\tau\overline{\tau}}$ anyons and compute the braiding tensor $\mathcal{R}_{\mathcal{P}_{\tau\overline{\tau}},\tau}$. The calculation is the same as we already did for the general pair $B_1,B_{\tau}$, we only need the results for the $\tau$ string in this calculation. Due to the relation \eqref{eq:B1Btau} there is no additional entanglement between the degrees of freedom on the site containing the anyon and the $\mathcal{R}$ tensor. We find that
$$
\mathcal{R}_{\mathcal{P}_{\tau\overline{\tau}},\tau}=
-\frac{1}{\phi^{3/2}}\overline{A}_{1212}+\frac{5^{1/4}}{\sqrt{\phi}}\overline{A}_{1222}+\frac{1}{\sqrt{\phi}}\overline{A}_{2122}+\frac{1}{\phi^2}\overline{A}_{2222}+\frac{5^{1/4}}{\phi}\overline{A}_{2212}.
$$
On the virtual level, this is a $8\times 8$ orthogonal matrix.

\section{Braiding in the Fibonacci string-net} \label{app:fibonacci}
To illustrate the general braiding formalism developed in section \ref{sec:braiding} of the main text we now work out some details for the Fibonacci string-net. As explained above, the crucial information to write down the relevant tensors are the $F$-symbols and the quantum dimensions of this theory. We can use the anyon ansatz to numerically determine the four central idempotents; we listed the outcome of this calculation in Appendix \ref{sec:snidempotents}.

First we focus on the idempotent that describes a $(\tau,1)$ anyon in the PEPS. We denote this idempotent by $\mathcal{P}_{\tau}$. If we recall the definition of the tensors $A_{abcd}$, we can express this idempotent as
\begin{align*}
\mathcal{P}_{\tau}=\frac{1}{\sqrt{5}}\left(\frac{1}{\phi}A_{\tau\tau\tau 1}+\frac{1}{\sqrt{\phi}}e^{\frac{4\pi i}{5}}A_{\tau 1\tau\tau}+e^{-\frac{3\pi i}{5}}A_{\tau\tau\tau\tau}\right).
\end{align*}

Suppose we have two such anyons, then we can determine their possible fusion outcomes. For this we use the fusion procedure explained in Figure \ref{fig:fusionidempotents} in section \ref{sec:fusion}. Clearly, the outgoing $\tau$ strings of the two anyons can be fused to a $1$ or $\tau$ string. The $1$ string can give rise to a fusion product supported in the subspace corresponding to $\mathcal{P}_1$ or $\mathcal{P}_{\tau\overline{\tau}}$, while the $\tau$ string can give rise to a support in all idempotents except $\mathcal{P}_1$. Although it is not easy to determine this analytically, one can readily determine the sectors where the two $\mathcal{P}_{\tau}$ anyons are supported numerically. These sectors are the $\mathcal{P}_1$ and $\mathcal{P}_{\tau}$ sector, as we expect from the fusion rules of Fibonacci anyons.

Let us now concentrate on the exchange of two such $\tau$ anyons and determine the tensor $\mathcal{R}_{\mathcal{P}_{\tau},\tau}$. We first show how one can analytically determine these tensors. This gives insight in the close relation between the idempotents and the $\mathcal{R}$ tensors. The calculation we use to determine the tensor $\mathcal{R}$ resembles the well-known teleportation protocol from quantum information theory. We follow the derivation of section \ref{sec:braiding} in the main text. This gives
\begin{equation} \label{braidfib}
\begin{split}
\vcenter{\hbox{
 \includegraphics[width=0.25\linewidth]{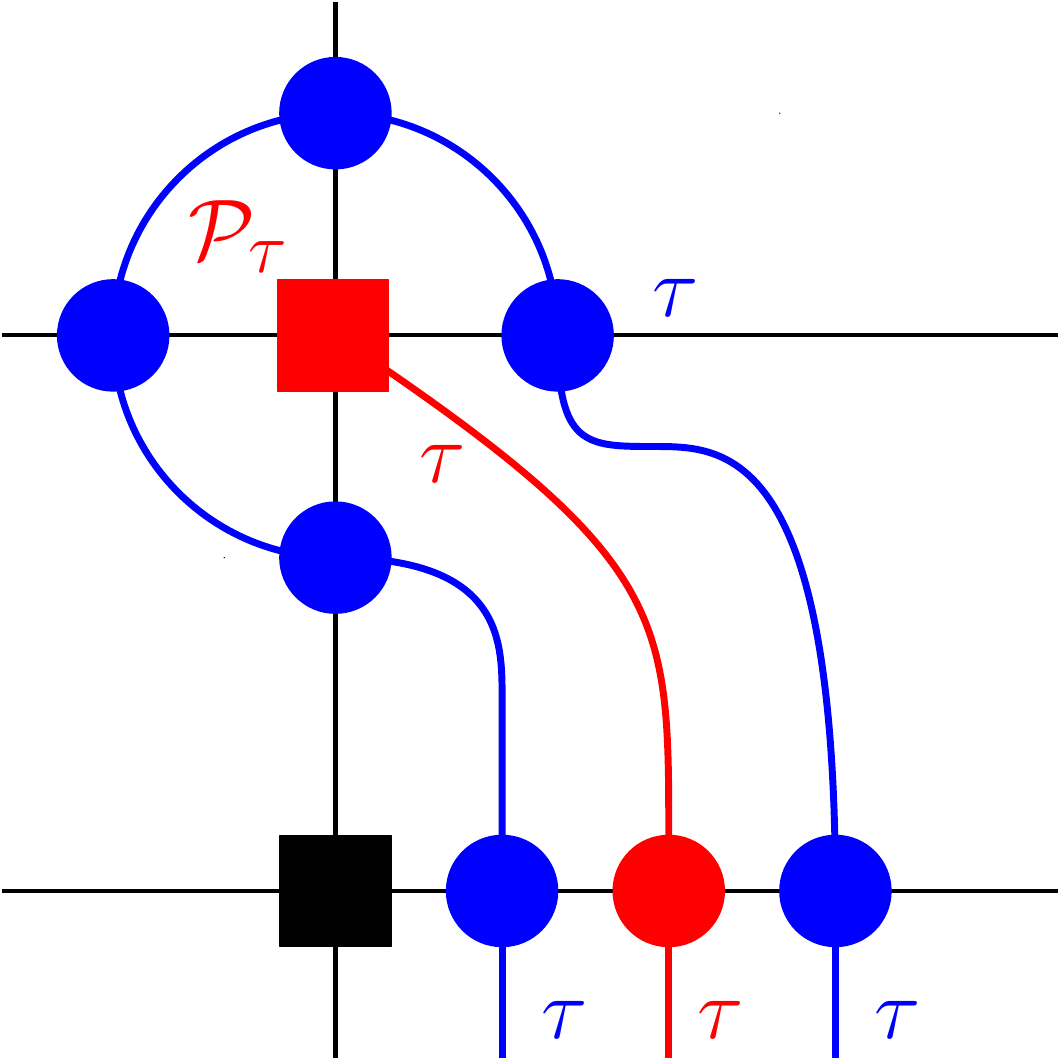}}} =  & \vcenter{\hbox{
 \includegraphics[width=0.25\linewidth]{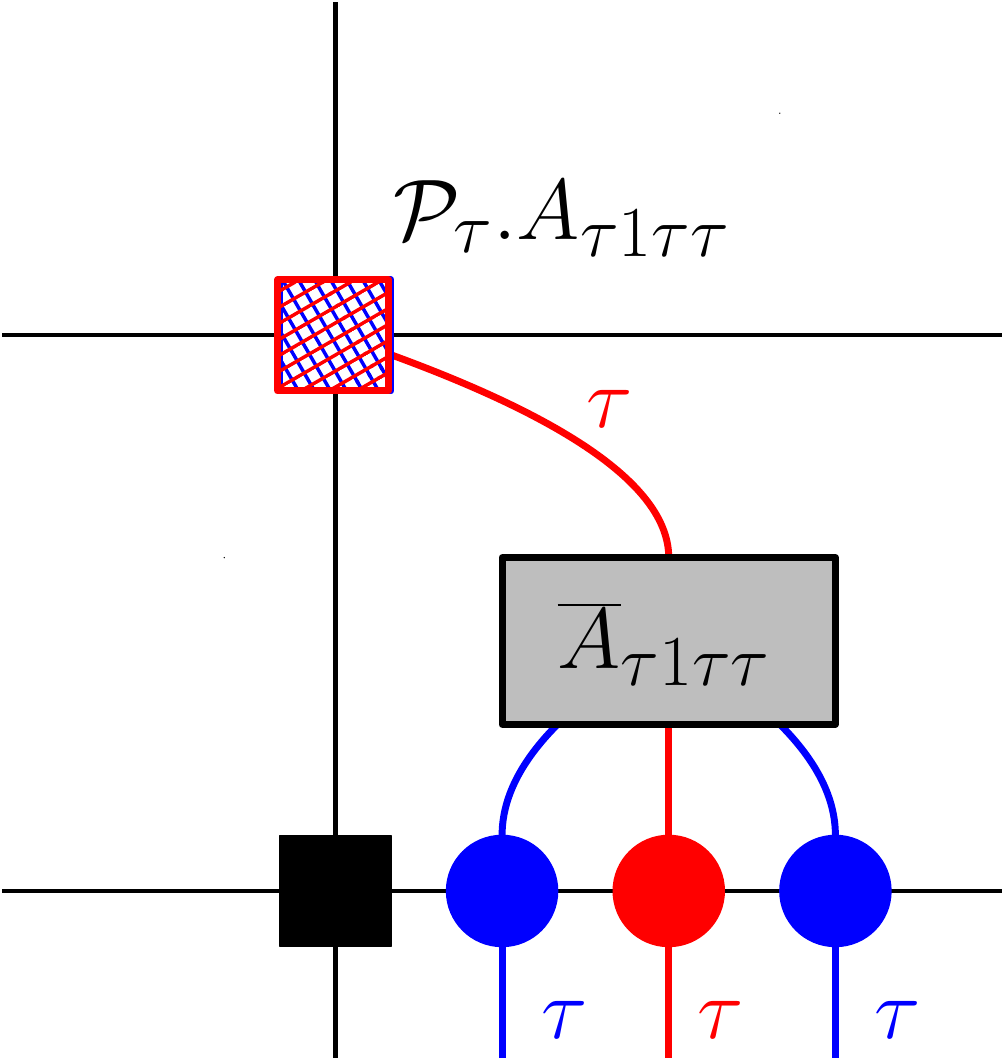}}}+ \vcenter{\hbox{
 \includegraphics[width=0.25\linewidth]{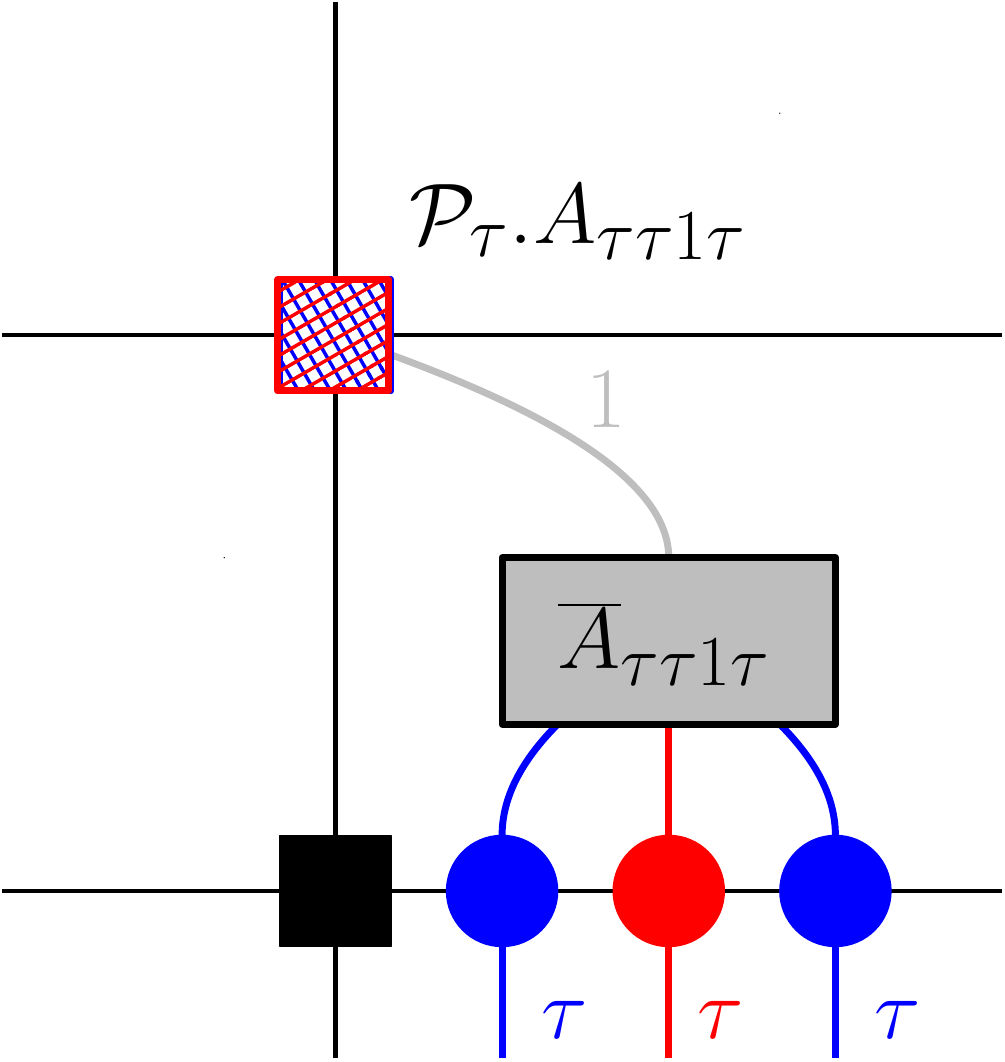}}}\\
 &+  \vcenter{\hbox{\includegraphics[width=0.25\linewidth]{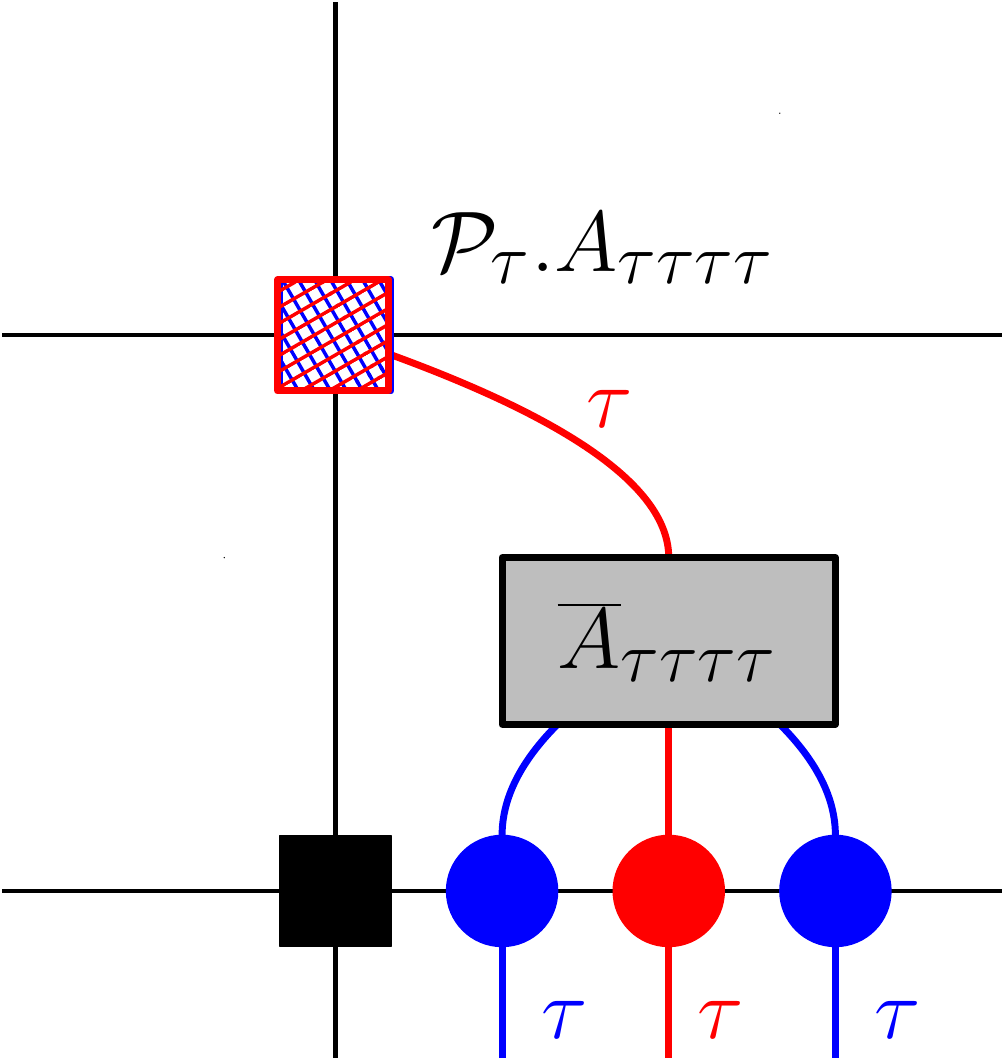}}}
\end{split}
\end{equation}
Since $\mathcal{P}_{\tau}$ is a one dimensional idempotent, $\mathcal{P}_{\tau}  A_{abcd} = \lambda_{abcd} \mathcal{P}_{\tau}$ for complex numbers $\lambda_{abcd}$ that can easily be calculated from the structure constants of the algebra. We find that $\lambda_{\tau 1 \tau \tau}=\frac{1}{\sqrt{\phi}}e^{4\pi i/5}, \lambda_{\tau \tau 1 \tau}=0$ and $\lambda_{\tau \tau \tau \tau}=e^{-3\pi i/5}$, such that Eq.~\eqref{braidfib} is simplified to
\begin{displaymath}
\vcenter{\hbox{
 \includegraphics[width=0.25\linewidth]{RmatrixFib1.pdf}}} =  \frac{e^{4\pi i/5}}{\sqrt{\phi}} \vcenter{\hbox{
 \includegraphics[width=0.25\linewidth]{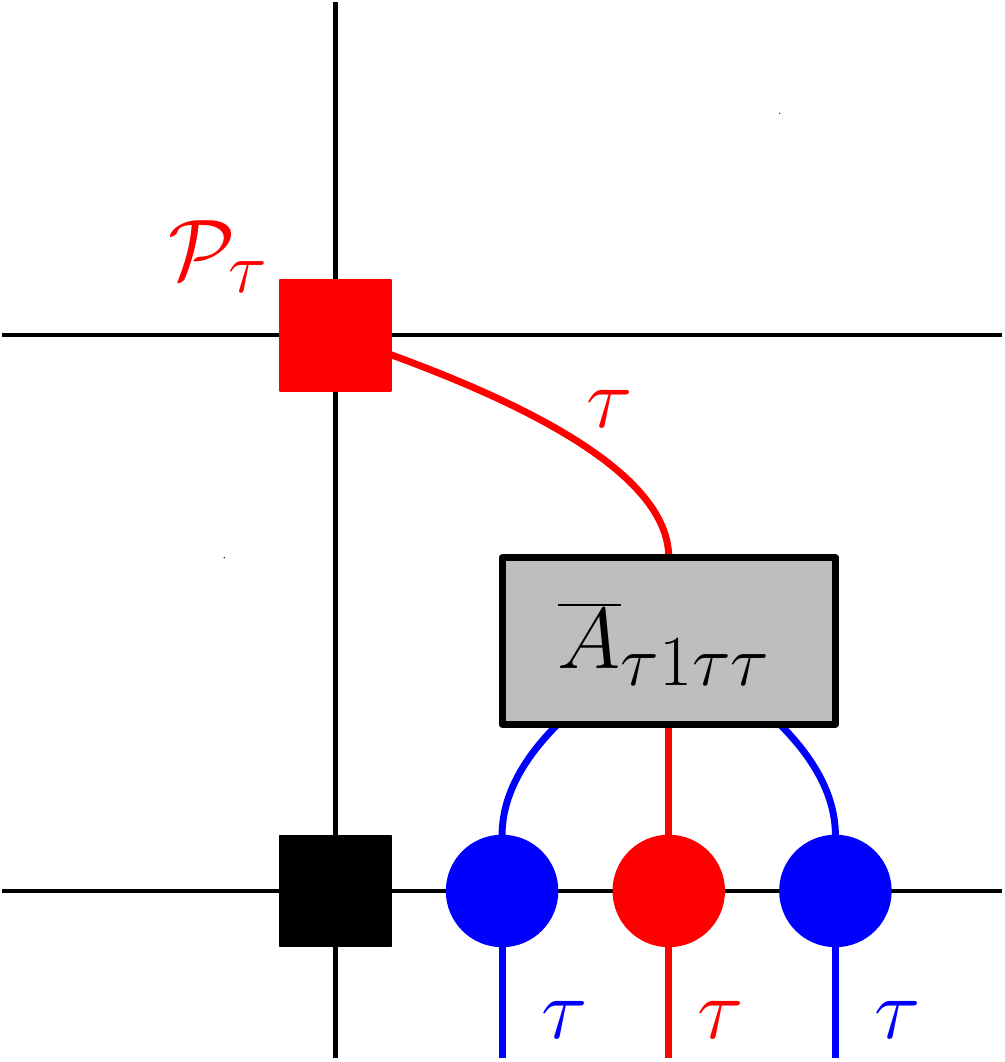}}}+e^{-3\pi i/5}\vcenter{\hbox{
 \includegraphics[width=0.25\linewidth]{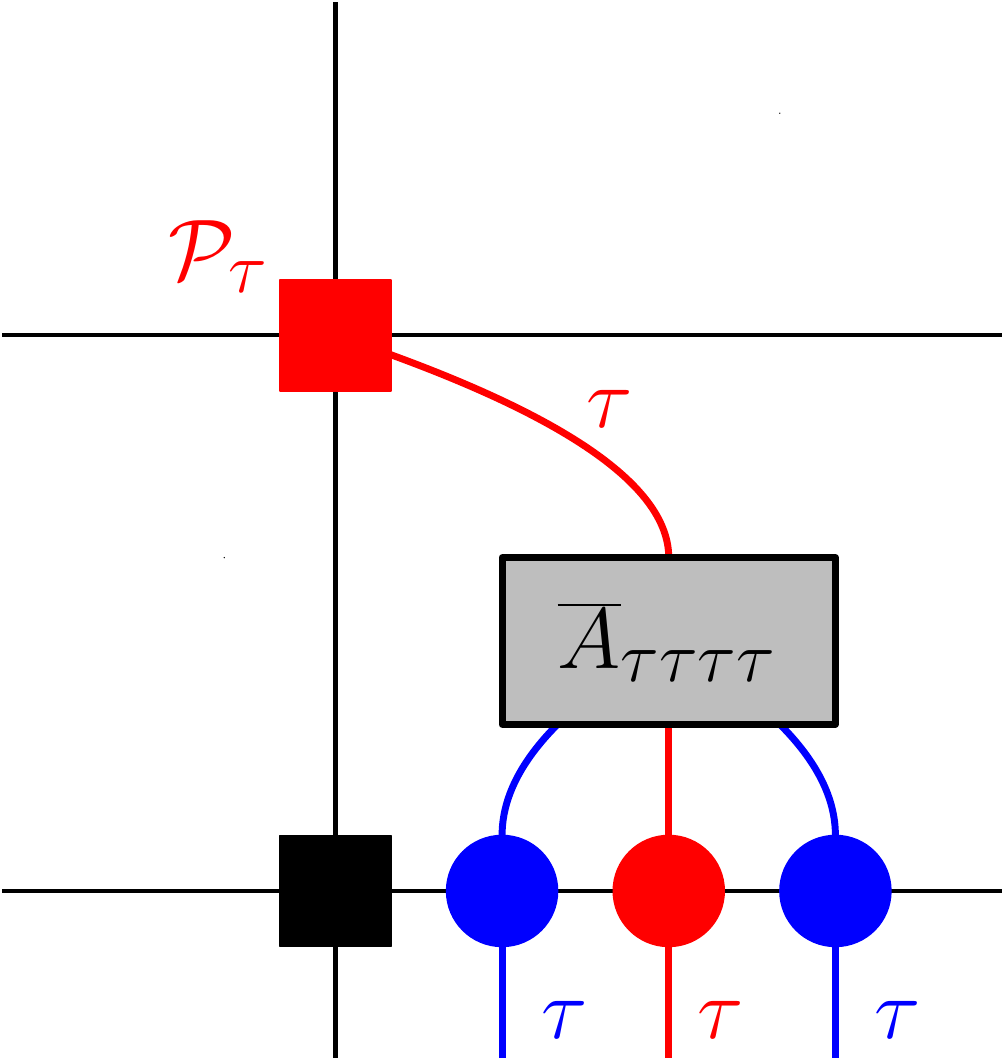}}}.
\end{displaymath}

We conclude from this calculation that the tensor $\mathcal{R}_{\mathcal{P}_{\tau},\tau}$ is  given by
$$
\mathcal{R}_{\mathcal{P}_{\tau},\tau}=\frac{1}{\sqrt{\phi}}e^{4\pi i/5}\bar{A}_{\tau 1 \tau \tau}+e^{-3\pi i/5}\bar{A}_{\tau\tau\tau\tau}.
$$
We can now also look at the contraction of two of these tensors as in figure \ref{fig:fullbraid3}, which describes the full braiding of two anyons. This tensor then describes the monodromy matrix of two Fibonacci anyons. It is known that the elements are given by $e^{2\pi i(h_c-h_{\tau}-h_{\tau})}$ where $c$ is the fusion product of the two anyons, $c=1,\tau$. As the spins of the anyons are $h_1=0$ and $h_{\tau}=2/5$ we expect the tensor  in Figure \ref{fig:fullbraid3} to contain the phases $e^{-4\pi i/5}$ and $e^{\pi i/5}$ in the respective topological sectors. One can readily check that this is indeed the case.

We can also look at the higher dimensional idempotent $\mathcal{P}_{\tau\overline{\tau}}$. In general, the situation gets more complicated due to entanglement between the degrees of freedom on the site where the anyon lives and the  tensors $\mathcal{R}$ in the virtual network. However, if we obtain the anyons by acting on the ground state, such that we have an anyon pair in the trivial sector, we can simplify the expression for $\mathcal{R}_{\mathcal{P}_{\tau\overline{\tau}},\tau}$. The reason this is possible is that by acting locally on the topologically trivial vacuum, we can only create very specific excitation pairs in the $\mathcal{P}_{\tau\overline{\tau}}$ sector. Indeed, the crucial fact that the entire pair is in the trivial sector, determines the relation between the anyon tensors in the different minimal, but non-central, idempotent sectors of a higher dimensional idempotent. For the same reason, one can only create a fluxon pair with zero topological charge in the quantum doubles \cite{Ginjectivity,toriccode,preskill1999lecture}. Such a pair exactly corresponds to the equal superposition of all string types in a given conjugacy class. We now illustrate that this reasoning is still valid in our more general formalism by looking at a $\mathcal{P}_{\tau\overline{\tau}}$ created on top of the trivial sector.

\clearpage

\bibliographystyle{utphys}
\bibliography{AnyonsPEPS}

\end{document}